\documentclass[10pt,twoside]{article}
\usepackage{a4wide}
\usepackage{times}
\usepackage{cite}
\usepackage{latexsym}
\usepackage{amsfonts}
\usepackage{amsmath,amssymb,amsbsy}
\usepackage{bbm}
\usepackage{slashed}
\usepackage{float}
\usepackage{nonfloat}
\usepackage[section]{placeins}
\newif\ifpdf
        \ifx\pdfoutput\undefined
        \pdffalse % we are not running PDFLaTeX
        \else
        \pdfoutput=1 % we are running PDFLaTeX
        \pdftrue
        \fi
\ifpdf
        \usepackage[pdftex]{graphicx}
        \DeclareGraphicsExtensions{.pdf, .jpg}
\else
        \usepackage[dvips]{graphicx}
                \DeclareGraphicsExtensions{.eps, .jpg}
\fi
\renewcommand{\theequation}{\thesection.\arabic{equation}}
\let\vold=\vec
\def\vec#1{\boldsymbol{#1}}
\def\RE{\mathop{\Re{\rm e}}\nolimits}
\def\IM{\mathop{\Im{\rm m}}\nolimits}
\def\Tr{\mathop{\rm Tr}\nolimits}

\def\p{\mathrm{p}}
\def\pb{{\bar{\mathrm{p}}}}
\def\Y{\mathrm{Y}}

\def\L{\Lambda}
\def\Lb{\bar{\Lambda}}
\def\e{\mathrm{e^-}} 
\def\ep{\mathrm{e^+}}
\def\eepp{{\ensuremath{\mathrm{e^-e^+\to \bar{p} p}}}}
\def\ppb{{\ensuremath{\bar{\mathrm{p}}\mathrm{p}}}}
\let\pbp=\ppb
\def\ppnn{\ensuremath{\bar{\mathrm{p}}\mathrm{p}\to \bar{\mathrm{n}}\mathrm{n}}}
\def\LLb{{\ensuremath{\overline{\Lambda}\Lambda}}}
\def\ppLL{\ensuremath{\bar{\mathrm{p}}\mathrm{p}\to\overline{\Lambda}\Lambda}}
\def\N{\mathrm{N}}
\def\Nb{\overline{\mathrm{N}}}

\def\K{\mathrm{K}}
\def\Kb{\overline{\mathrm{K}}}

\makeatletter
\newlength{\earraycolsep}
\def\eqnarray{\stepcounter{equation}\let\@currentlabel%
\theequation
\global\@eqnswtrue\m@th
\global\@eqcnt\z@\tabskip\@centering\let\\\@eqncr
$$\halign to\displaywidth\bgroup\@eqnsel\hskip\@centering
$\displaystyle\tabskip\z@{##}$&\global\@eqcnt\@ne
\hskip 2\earraycolsep \hfil$\displaystyle{##}$\hfil
&\global\@eqcnt\tw@ \hskip 2\earraycolsep
$\displaystyle\tabskip\z@{##}$\hfil
\tabskip\@centering&\llap{##}\tabskip\z@\cr}
\makeatother
\def\CSM{Cross Section Matrix}
\let\CSdM=\CSM

\def\vsigma{\sigma}
\def\vpi{\vec\pi}

\def\SvS{{\cal S}}
\def\PvS{{\bf{\cal P}}}
\def\acc{\check{\rho}}
\def\emit{\hat{\rho}}
\def\d{\mathrm{d}} % straight d for integration
\def\sv{{\vec{s}}}
\def\Av{{\vec{A}}}
\def\Pv{{\vec{P}}}
\def\Sv{{\vec{S}}}
\def\Jv{{\vec{J}}}
\def\ev{{\vec{e}}}
\def\lv{{\vec{l}}}
\def\mv{{\vec{m}}}
\def\pv{{\vec{p}}}
\def\qv{{\vec{q}}}
\def\kv{{\vec{k}}}
\def\nv{{\vec{n}}}

\def\Pv{{\vec P}}

\def\Qv{{\vec Q}}
\def\ku{\hat{\vec{k}}} 
\def\pu{\hat{\vec{p}}}
\def\xu{\hat{\vec{x}}}
\def\yu{\hat{\vec{y}}}
\def\zu{\hat{\vec{z}}}
\def\Mcal{{\mathcal{M}}}         \let\M=\Mcal
\def\H{{\cal H}}
\let\Hcal=\H
\def\Ocal{{\mathcal{O}}}
\def\R{{\mathcal{R}}}
\def\Rt{{\mathcal{\widetilde{R}}}}
\def\Rh{{\mathcal{\widehat{R}}}}
\def\be{\begin{equation}}

\def\ee{\end{equation}}

\def\barr{\begin{eqnarray}}
\def\earr{\end{eqnarray}}
\let\ba=\barr
\let\ea=\earr
\def\la{\label}
\def\ni{\noindent}
\def\ie{{\sl i.e.}}
\def\eg{{\sl e.g.}}
\def\etal{{\sl et al.}}
\def\vs{{\sl vs.}}
\def\phm{\phantom{-}}
\def\ii{\phantom{i}} % good alignment for some tables
\def\sq{\raise -2pt\hbox{\Large $\Box$}}
\def\dis{\scalebox{.9}{$\bigcirc$}}
\def\disp{\raise -1pt\hbox{\large$\otimes$}}
\def\trv{ \raise -2pt \hbox{\large$\nabla$}}
\def\trh{\raise -2pt \hbox{\rotatebox{90}{\hbox{\large$\Delta$}}}}
\begin{document}
%%%%%%%%%%%%%%%
\begin{titlepage}
\title{
{\small \hfill LPSC-08-08}\\[-10pt]
{\smallÊ\hfill LYCEN-2008-1}\\[-10pt]
{\small \hfill To appear in ``Physics Reports''}\\[40pt]
{\Large\bf SPIN OBSERVABLES AND SPIN STRUCTURE FUNCTIONS:}
\\[3pt]
 {\Large\bf INEQUALITIES AND DYNAMICS  
\\[20pt]}}
\author{
Xavier Artru\\
{\small  Universit\'e de Lyon, IPNL}\\[-2pt]
{\small Universit\'e Lyon 1 and CNRS/IN2P3}\\[-2pt]
{\small  69622 Villeurbanne, France}\\[-2pt]
{\small email: \texttt{xavier.artru@ipnl.in2p3.fr}}\\[10pt]
{Mokhtar Elchikh}\\
{\small Universit\'e des Sciences et de Technologie d'Oran}\\[-2pt]
{\small D\'epartement de Physique, BP 1505}\\[-2pt]
{\small El Menauoer, Oran, Algeria}\\[-2pt]
{\small email: \texttt{elchikh@yahoo.com}}\\[10pt]
{Jean-Marc Richard}\\
{\small Laboratoire de Physique Subatomique et Cosmologie}\\[-2pt]
{\small Universit\'e  Joseph Fourier, CNRS/IN2P3, INPG}\\[-2pt]
{\small 53, avenue des Martyrs, 38026 Grenoble, France}\\[-2pt]
{\small email:\texttt{jean-marc.richard@lpsc.in2p3.fr}}\\[10pt]
{Jacques Soffer}\\
{\small Physics Department, Temple University}\\[-2pt]
{\small Barton Hall, 1900 N. 13th Street}\\[-2pt]
{\small Philadelphia, PA 19122 -6082, USA}\\[-2pt]
{\small email: \texttt{jsoffer@temple.edu}}\\[10pt]
{Oleg V. Teryaev}\\
{\small Bogoliubov Laboratory of Theoretical Physics, }\\[-2pt]
{\small Joint Institute for Nuclear Research,}\\[-2pt]
{\small 141980 Dubna, Moscou region, Russia}\\[-2pt]
{\small email: \texttt{teryaev@theor.jinr.ru}}\\[10pt]}
%%%%%%%%
\date{\today \\[12pt]}
\maketitle
\newpage
\begin{abstract}
Model-independent identities and inequalities  which relate the various spin observables of collisions in nuclear and particle physics are reviewed in a unified formalism. Their physical interpretation and their implications for dynamical models are also discussed.
These constraints between observables can be obtained in several ways: from the explicit expression of the observables in terms of a set of helicity or transversity amplitudes, a non-trivial algebraic exercise which can be preceded by numerical simulation with randomly chosen amplitudes, from anticommutation relations, or from the requirement that any polarisation vector is less than unity. The most powerful tool is the positivity of the  density matrices describing the spins in the initial or final state of the reaction or its crossed channels. The inequalities resulting from positivity need to be projected to single out correlations between two or three observables. The quantum aspects of the information carried by spins, in particular entanglement, are considered when deriving and discussing the constraints

Several examples are given, with a comparison with experimental data in some cases. For the exclusive reactions, the cases of the strangeness-exchange proton-antiproton scattering and the photoproduction of pseudoscalar mesons are treated in some detail: all triples of observables are constrained, and  new results are presented for the  allowed domains. The positivity constraints for total cross-sections and for the simplest observables of single-particle inclusive reactions are reviewed.  They also apply to spin-dependent structure functions and parton distributions,  both integrated or transverse-momentum dependent. The corresponding inequalities are shown to be preserved by the evolution equations of quantum chromodynamics.
\end{abstract}
\end{titlepage}
%%%%%%%%%%%%%%
\tableofcontents
\clearpage
\section{Introduction}\label{se:intro}
 The ultimate tests of the dynamics of the standard model and the low-energy hadron interaction require the probing of  the spin degrees of freedom. The  developments on beam, target and detector technology make it now possible to perform scattering experiments with polarised beam and target, and to measure the polarisation of the outgoing particles. 
 For instance, the elastic nucleon--nucleon scattering has been studied with beam and target both polarised. At CERN, the strangeness-exchange reaction  $\ppLL$ has been measured with a polarised hydrogen target, and thanks to the weak decay of the hyperon, the polarisation of both $\overline{\Lambda}$ and $\Lambda$ has been measured.
 
These developments are also applied to inclusive reactions $a+b\to c+X$, with $a$, $b$ and $c$ carrying spin, and measurements of spin observables up to rank three can be envisaged, if the beam $a$ and target $b$ are polarised and the polarisation of the produced particle $c$ is analysed. 

Significant progress has been achieved recently in the investigation of spin observables, spin-dependent structure functions and parton distributions, in particular at RHIC, Jlab, GRAAL, CERN and DESY.  New data are expected from presently running experiments   and further measurements are expected at the future facilities such as the hadron factories at KEK (Japan) and GSI (Germany). 

This experimental progress has stimulated an important theoretical activity. In the eighties, the so-called ``spin crisis'' (one should say ``helicity crisis'') has motivated the high-energy theorists to study the spin aspects of parton distributions and a large effort is now devoted to the quark transverse spin (``transversity'') and the T-odd parton distributions. 

Spin physics will probably benefit from significant progress in the future. As an example, polarised electrons and positron beams are planned for the International Linear Collider  (ILC) to reduce the backgrounds and to improve the sensitivity to new physics. Attempts are made to design polarisation devices for antiprotons which, if successful, will enlarge the physics program of the GSI hadron facility.    

\paragraph{The physics case}
The merit of spin observables for a detailed understanding of the underlying dynamics has been often demonstrated and hardly needs to be stressed once more in great detail (see, for instance, Refs. \cite{Bourrely:1980mr,Leader:2001gr}).
 Let us give one example: in the conventional theory of nuclear forces, pion-exchange has a spin--spin and a tensor component, this latter acting with specific strength on each of the spin--triplet P-wave states of scattering, ${}^3\mathrm{P}_0$, ${}^3\mathrm{P}_1$, ${}^3\mathrm{P}_2$. The pattern induced by pion-exchange is present at very low energy, but, as the incoming energy increases, a  departure is clearly observed. This led Breit and others to introduce a spin--orbit component in the interaction, and to speculate about the existence of vector mesons whose exchange is responsible for this important spin-orbit term in the potential. These $\rho$ and $\omega$ mesons were, indeed, discovered. See for instance, Refs.~\cite{Pais:1986nu,Brown:1996kk}.

In scattering experiments, the integrated cross-section and the main trend of the angular distribution are dominated by simple basic mechanisms, for instance attraction or absorption. To reach more subtle contributions that provide signatures for the more elaborated models, one needs to apply filters.
This is the role of spin observables.

For instance, in the strangeness-exchange reaction  $\ppLL$,   a picture based on the exchange of mesons ($\mathrm{K}$, $\mathrm{K}^*$, etc.) and a model based on $\bar{u}u$ annihilation and $\bar{s}s$ creation were both able to describe the first data taken at LEAR on angular distribution, polarisation and spin-correlation coefficients of  final-state hyperons. However, these models gave different predictions for the depolarisation and spin transfer coefficients that became available once a polarised hydrogen target was installed.

\paragraph{Experimental situation}
Some reactions have received special attention on the experimental side. In the low-energy domain, this is the case for nucleon--nucleon scattering. A comprehensive formalism has been written down \cite{Bystricky:1976jr}.
Delicate experiments, in particular at Saclay (Saturne) and Villigen (PSI), have enabled one to tune the potential models, later on used in nuclear-structure calculations and  phase-shift analyses, and eventually a direct reconstruction of the amplitudes. A summary of most of this work can be found in a review by Leluc and Lehar \cite{Lechanoine-LeLuc:1993be}\footnote{We thank Catherine Leluc for a very informative discussion on the subject}.
More recent measurements have been performed, \eg,  by the PINTEX collaboration at the Indiana Cyclotron\cite{Sarsour:2004xx,vonPrzewoski:2003ig,Meyer:2003fx};
by the EDDA collaboration at COSY \cite{Bauer:2002zm,Hinterberger:2005eg,Altmeier:2004qz};  at JINR, Dubna \cite{Lehar:2001wq};  etc.

Thanks to this detailed knowledge of the scattering observables, models have been constructed for the nucleon--nucleon potential and used  to study more complicated systems from tritium to neutron stars.  A milestone was the Paris potential \cite{Lacombe:1980dr}.  An example of  more recent and more sophisticated development is the so-called ``CD Bonn'' model \cite{Machleidt:2000ge}.
Chiral models have also been developed which hopefully will lead to a unified picture of long- and short-range contributions, see, \eg, \cite{Epelbaum:2004fk}.

Basic mechanisms of the low-energy hadronic interaction are tested in scattering experiments involving polarised light nuclei or polarised photon beams. Spin physics is very much active in particular at Jlab
\cite{DeVita:2004yv,deJager:2006ia,Ripani:2007jd}:
 the CLAS collaboration has just measured the polarisation, and beam-to-recoil spin transfer in the photoproduction process $\gamma\mathrm{N}\to \mathrm{K}\Lambda$, a reaction also studied by the GRAAL collaboration at Grenoble. 
Interestingly, the data saturated the inequalities (which will be reviewed in Section 3), indicating that some of the transversity amplitudes are dominant, which sheds some light on the underlying mechanism. Kloet \etal\  \cite{Kloet:1998js,Kloet:1999wc,Kloet:2000si} have analysed some inequalities on the observables relevant for the photoproduction of vector mesons, and their subsequent decay into two pseudoscalar mesons or a pair of leptons.

The physics of low-energy antiprotons has been particularly active in the 80's and 90's at the LEAR facility of CERN. The analysing power $A_y$ of the annihilation reactions into two pseudoscalars, $\bar{\rm p}+ {\rm p}\to\pi^-\pi^+$ and $\bar{\rm p}+ {\rm p}\to \mathrm{K}^-\mathrm{K}^+$ has revealed a wide kinematical region with nearly maximal values $|A_y|\sim 1$. This will be discussed later in this review.

The reactions $\bar{\rm p}+ {\rm p}\to \overline{\rm Y}+{\rm Y'}$, where $\mathrm{Y}$ denotes a hyperon ($\Lambda$, $\Sigma$, $\Xi$) have been much studied at LEAR. Information on the spin of the final-state particles is provided by their decay. Here several spin observables can be accessed, and some rank-three observables were obtained in the last runs for which a polarised target was installed. The formalism for the reaction $\ppLL$ will be discussed at length in this review. In 1964, \ie, much before the LEAR experiments, some subtleties  of the spin observables were  analysed \cite{iii:B540,Cohen:1964}.  In particular, Cohen-Tannoudji and Messiah noted that (for $\mathrm{Y}=\mathrm{Y}'$) accurate data can indicate whether or not the antiproton beam is polarised \cite{Cohen:1964}.

Several models were proposed for these strangeness-exchange reactions, with a renewed activity when the threshold behaviour was analysed at LEAR, first with an unpolarised target, by the PS185 collaboration. These early LEAR data indicated a sizeable amount of polarisation and spin correlation in the final state, in particular, a vanishing spin-singlet fraction for $\ppLL$. Unfortunately these first data did not enable to distinguish between  the quark-inspired models and the kaon-exchange models \`a la Yukawa, which all gave a fairly good agreement. It was then suggested that a polarised target might help distinguishing among the possible mechanisms.
Holinde \etal, and others (see, \eg, \cite{Klempt:2002ap} for references), claimed that  quark models and kaon-exchange models give different predictions for some observables, such as $D_{nn}$, that become accessible with a polarised target.
 Predictions  on these observables were also made by  Alberg \etal\ with in particular a very large transversity-flip of the baryon, \ie, a large negative depolarisation parameter $D_{nn}\simeq-1$, if the strange quark pair is extracted from the sea of the proton or antiproton \cite{Alberg:1995zp}.  It was then realised \cite{Richard:1996bb} that the existing data already constrain the depolarisation to a small interval near  $D_{nn}=0$. The data taken with a transversely polarised target, gave, indeed, values of $D_{nn}$ close to 0, as it shown in the next section. A more systematic study of the correlation among the observables in this reaction \ppLL\ was  made in  \cite{Elchikh:1999ir,Richard:2003bu,Elchikh:2004ex,Artru:2004jx}.

Spin physics at very high energy  is now investigated thanks to the polarised proton--proton collider in operation at RHIC-BNL since 2002\cite{Bunce:2000uv}. It will allow to test the spin sector of perturbative QCD and to measure the parton distributions, in particular, the transversity distribution whose relevance  was recognised in the early 90's. This is opening a new area which will bring a lot of insight in the nucleon spin structure, a fundamental problem which came up only 20 years ago.

Large single spin asymmetries in hadronic reactions at high energies have been
already observed and they are among the most challenging phenomena to understand
from basic principles in QCD. In particular it was shown recently that initial or final
interactions, due to the exchange of a gluon, generate the theoretical mechanism 
to explain transverse spin asymmetries \cite{Brodsky:2002cx} and outstanding spin
anomalies like the observed violation of the Lam-Tung relation in the angular distribution
of the leptons in the unpolarised $\pi$-N Drell-Yan process \cite{Boer:1999mm,Boer:2002ju}.

\paragraph{Identities and inequalities}

The analysis of experimental data and their interpretation is greatly helped by model-independent constraints (identities and inequalities) between the various observables. These constraints also help in planning the future experimental programs by pointing out the redundant measurements. 

When all possible spin observables are considered, \emph{linear} and \emph{quadratic} identities can be written.  \emph{Linear} identities  reflect the symmetries such as parity, time reversal, identical particles, etc. For instance, in elastic scattering, the analysing power (azimuthal asymmetry of the angular distribution with reference to the target polarisation) coincides with the spontaneous polarisation of the recoiling particle.  Such identities can also be written when the spins of some of the particles are not measured.
\emph{Quadratic}  identities relating observables are known for years, such as $P_n^2+A^2+R^2=1$ for the polarisation and spin-rotation parameters of pion--nucleon scattering, with a notation to be defined more precisely later. Deriving these non-linear equalities can be seen as a purely algebraic (but non trivial) exercise, once the various spin observables are expressed in terms of a set of amplitudes. 
%More systematic approaches can be envisaged, based for instance on Fierz transformations.

Spin observables are also constrained by \emph{inequalities}, which express the positivity of some polarised cross sections, sometimes in a \emph{gedanken} experiment involving \emph{entangled} spin states as we shall show in Section 2. 
For one observable we have ``trivial'' bounds, for instance $|A_n|\le1$ for the analysing power, $|P_n|\le1$ for the spontaneous final polarisation and $|D_{nn}|\le1$ for the so-called depolarisation coefficient. For a pair $\{C_1,C_2\}$ of observables, each of which satisfy $|C_i|\le1$, positivity can restrict the allowed domain to a subset of the square $[-1,+1]^2$. Similarly a triple $\{C_1,C_2,C_3\}$ of observables is often constrained to a subset of the cube  $[-1,+1]^3$, etc.
Such inequalities provide model-independent constraints on observables which are measured independently. They also indicate which of the yet-unknown observables offer the widest range of variation for the further checking of the models. If, for instance, an inequality $C_1^2+C_2^2\le 1$ has been derived, and $C_1$ safely measured to be close to $-1$ in a certain domain of kinematics, it is no longer necessary to device a new experiment for measuring $C_2$. The aim of this review is to show how to derive and understand these constraints, expressed in terms of simple inequalities.

The inequalities discussed in this review precisely indicate whether new observables  are necessary or are already much constrained by the existing ones. 
We shall give other examples in Sec.~\ref{se:incl}, dealing with the inclusive reactions. In this category we  put also the positivity bound for transversity parton distributions, which will be derived in Sec.~\ref{incl:sub:PD}.

\paragraph{A brief history of spin identities and inequalities.}
The idea that spin observables are constrained by positivity conditions is far from being new. Some milestones  are  now presented of the contributions of which we are aware. We apologise  for the important contributions which we may have inadvertently omitted.

In 1958, Lee and Yang \cite{Lee:1958qu} studied the decay symmetry of the $\Lambda^0$ and established constraints in the light of which the experimental data suggested that $\Lambda^0$ has spin 
$1/2$.

Positivity conditions have been written down  for  the density matrix describing spin-one meson resonances  and spin 3/2 baryon resonances \cite{Dalitz:1964bw}.

In 1965, Ademollo, Gatto and Preparata proposed more general tests based on the properties of the density matrix for a two-body inelastic collision, to determine the spins of the final state particles~\cite{PhysRev.140.B192}.

In the 60's,  basic contributions were elaborated by Louis Michel and his collaborators. See, in particular \cite{Doncel:1970,Minnaert:1971,minnaert:672,minnaert:1306} about the ``polarisation domain''. Other references of interest are \cite{Cohen:1964,Daboul:1967}.

In 1975, Delanay and Gammel \cite{Delaney:1975fh} and, independently, Bourrely and Soffer \cite{Bourrely:1974wp} wrote down the quadratic relations among the spin observables  of the proton--proton elastic scattering. Reading the book by Leader on spin physics \cite{Leader:2001gr} tells us that similar results were obtained earlier by  Klepikov, Kogan and Sha\-ma\-nin \cite{Klepikov:1967}.  The art of identities among observables has been further developed, in particular by La France and Winternitz \cite{LaFrance:1981bg}.

\paragraph{Various derivations of the inequalities.}
For the exclusive reactions, many positivity conditions are just a consequence of the \emph{identities} mentioned above. For instance $P_n^2+A^2+R^2=1$ implies $A^2+R^2\le1$ and $|P_n|\le1$. 
Some inequalities can be derived algebraically  starting from the 
 the explicit expression of observables $C_1$, $C_2$, \dots\  in terms of amplitudes $a$, $b$, \dots; however this exercise that looks easy at first sight,  turns out rather involved.
The search for the inequalities can be guided by generating randomly artificial amplitudes and plotting one observable against another one. This gives a first image of the allowed domain for the observables. Once an inequality is guessed from the plot, one may try to derive it rigorously.

Spherical inequalities of the type $C_1^2+C_2^2+...C_n^2\le1$ can be quickly obtained from the anticommutation properties of the observables (see Sec.~\ref{se:basic}). 
More systematically, the inequalities result from the positivity of the density matrix of the set or a subset of initial or final particles for the reaction under study or one of its crossed channel.
This density matrix formalism,  with a strong physics content,  generalises easily to the inclusive reactions. 

For the inclusive reactions, for which there is no quadratic identities, the Cauchy--Schwartz inequality is often used.   

 It is also interesting to interpret the inequalities as constraints on the transfer of the quantum information carried by spins from the initial to the final state. The $S$-matrix can indeed be viewed as a quantum device and the strongest positivity constraints will appear when considering \emph{entangled} spin states in the direct or crossed channels.  

\paragraph{Outline}
The growing number and variety of polarised experiments calls for a permanent effort on the theoretical side 
towards the unification and simplification of the spin formalism, using physically intuitive notations. Powerful methods are also required to derive the model-independent constraints (equalities and inequalities). As an update of the previous reviews \cite{Bourrely:1980mr,Leader:2001gr,Barone:2001sp} the main aim of this report is to present these constraints, first in a general unified manner, then for typical particular reactions. Inequalities are in fact the major topic of this review. Some interpretation in terms of quantum information will be proposed. The analogy between the low- or high-energy hadron physics and the physics of structure functions and parton distributions will be stressed.

For each chosen reaction, the positivity constraints will be compared with the recent data and a short review the underlying dynamics will be given: meson exchange, internal quark--antiquark annihilation, QCD evolution, etc. 
This article is by no means a comprehensive summary of the data on spin measurements and their interpretation. Instead, the examples have been chosen in low-energy hadronic reactions and in the physics of parton distribution to illustrate the role of positivity constraints.  The selection is rather arbitrary, and due to lack of space, entire chapters are omitted, such as hadron form factors.

This review is organised as follows.  In the next section, we briefly recall the basic formalism of spin observables, spin amplitudes, density matrix, etc., and the role of symmetries. General methods for deriving the positivity constraints will be presented. 
In Sec.~\ref{se:excl}, we review the case of some exclusive reactions, such as $\pi\mathrm{N}\to\pi\mathrm{N}$, 
$\overline{\mathrm{N}}\mathrm{N}\to \pi\pi$, the photoproduction of pseudoscalar or vector particles and the hyperon-pair production $\ppLL$, for which various spin observables can be measured. Inequalities constraining the spin observables are derived  and compared to some available experimental data.

This study is resumed in Sec.~\ref{se:incl} for the inclusive reactions. It includes spin observables for the hadronic inclusive reactions, structure functions and parton distributions. 

In Sec.~\ref{se:furth}, we briefly present the aspects related to quantum information, and the ways of adapting the cascade simulations to account for the  spin degrees of freedom.

We consider the future of this physics in Sec.~\ref{se:outl}, with a discussion  of the  forthcoming facilities, detectors, polarised beams and targets.
\newpage\setcounter{equation}{0}
\section{Basic formalism}\label{se:basic}
In this section, the notation is introduced for kinematical variables  and experimental observables. This is a minimal reminder. A more comprehensive formalism is available in previous review articles and textbooks \cite{Bourrely:1980mr,Leader:2001gr,Ohlsen:1972xx}. In particular, the delicate issue of translating the measurements from the laboratory frame to the centre-of-mass frame is omitted here. 

Each particular reaction has its own tradition, that we shall try to follow, at the expense of a slight modification of  the notation from one section to another. 
%
%For instance $\sigma(++)$ sometimes means that the cross section is measured with beam and target polarisation being parallel, while in another context, it might indicate that they have the same helicity, \ie, opposite spins.
%
\subsection{Spin observables}\label{basic:sub:obs}
\subsubsection{Kinematics and frames}\label{basic:sub:kin}
\begin{figure}[!ht]
\centerline{\includegraphics[width=.5\textwidth]{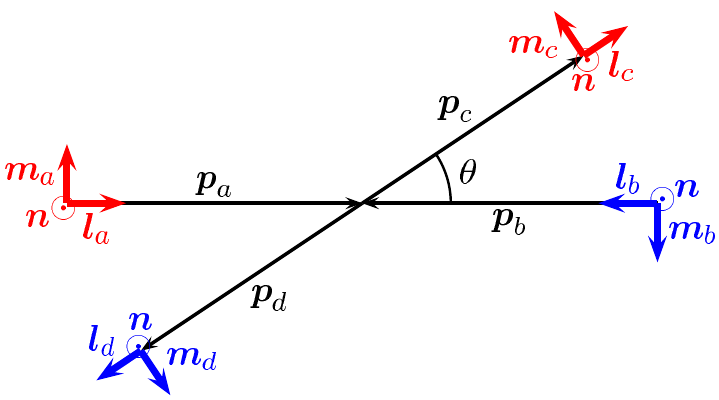}}
\caption{\label{basic:fig:frame} Axes to project out the spin components of each particle in the centre-of-mass frame of the reaction $a+b\to c+d$.}
\end{figure}

Let us consider a reaction $a+b\to c+d$, with four-momenta $\tilde{p}_a,\ldots \tilde{p}_d$, and Mandelstam variables
\be\label{basic:eq:mandel}
s=(\tilde{p}_a+\tilde{p}_b)^2~,\qquad
t=(\tilde{p}_a-\tilde{p}_c)^2~,\quad
u=(\tilde{p}_a-\tilde{p}_d)^2~.
\ee
In the centre-of-mass frame, the three-momenta  are $\vec{p}=\vec{p}_a=-\vec{p}_b$ and $\vec{p}'=\vec{p}_c=-\vec{p}_d$. The particles are ordered as follows: $a$ is the beam, $b$ the target, $c$ the scattered particle, and $d$ the recoiling one, so that for elastic scattering $c$ is identical to $a$, and $d$ to $b$. Except in the case of extreme forward or backward scattering, the scattering plane is well defined, with  a normal vector $\vec{n}=\vec{p}\times \vec{p}'/|\vec{p}\times \vec{p}'|$. A frame is attached to each particle  to project out its spin components, 
\be\label{basic:eq:frame}
\begin{aligned}
&\{\vec{l}_a=\hat{\vec{p}},\,\vec{m}_a,\,\vec{n}\}~,\qquad&
&\{\vec{l}_b=-\hat{\vec{p}},\,\vec{m}_b,\vec{n}\}~,\\
&\{\vec{l}_c=\hat{\vec{p}}',\,\vec{m}_c,\vec{n}\}~,\quad&
&\{\vec{l}_d=-\hat{\vec{p}}',\,\vec{m}_d,\vec{n\}}~,\end{aligned}
\ee
as illustrated in Fig.~\ref{basic:fig:frame}, but the index of the particle is often omitted.  In each case, the sideways axis is defined as $\vec{m}_i=\vec{n}\times \vec{l}_i$. The components $\{V_m, V_n,V_l\}$ of a vector are sometimes denoted $\{V_x,V_y,V_z\}$.
Another set of axes, $\{\vec{S}, \vec{N}, \vec{L}\}$ is used in the literature, following the so-called Argonne convention \cite{Ashkin:1977ek}.  It coincides with $\{\vec{m},\vec{n},\vec{l}\}$ except for particle $b$ where
$\{\vec{S},\vec{N},\vec{L}\}=\{-\vec{m},\vec{n},-\vec{l}\}$. 
\subsubsection{Definition of the spin observables}
Some of the results below hold without limitation. For others, it is implicitly assumed that the spin is either 0 or 1/2, the case of photons or massive vector particles requiring some specific treatment.

For a reaction $a+b\to c+d$ involving spinless particles, the only observable is the angular distribution $I_0$, sometimes called differential cross-section. It is a function of the centre-of-mass energy $\sqrt{s}$ and scattering angle $\theta$.

\begin{figure}
\centerline{\includegraphics[width=.75\textwidth]{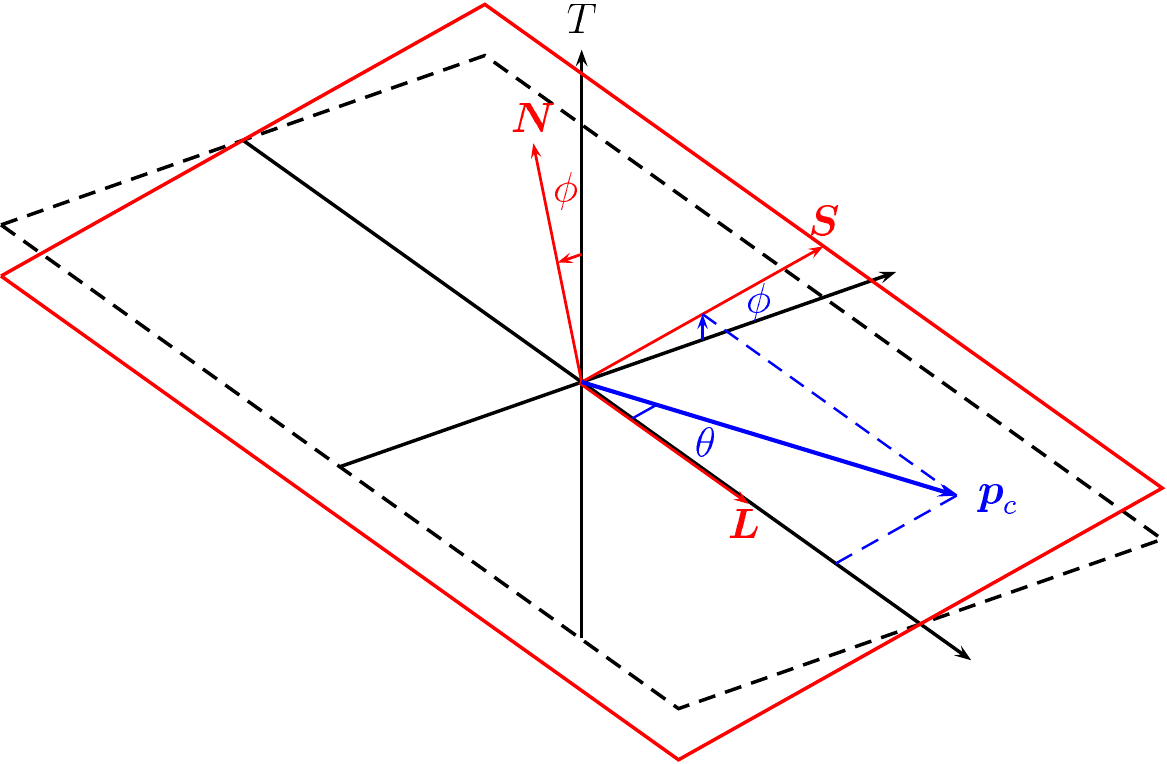}}
\caption{\label{basic:fig:azim} The scattering plane is rotated by $\phi$ with respect to the frame defined by the target polarisation and the beam axis.}
\end{figure}

If the particle $b$ has spin, the target may be polarised. A longitudinal  polarisation $S_z$ does not affect the azimuthal distribution of the scattered particle $c$ or recoil particle $d$.  However, a transverse polarisation $S_T$ gives an azimuthal asymmetry measured by the analysing power $A_n$.  In a coordinate frame attached to $\vec{S}_T$, the differential cross section is
\begin{equation}\label{basic:eq:ana}
I(s,t)={\d\sigma\over\d\Omega}=I_0(s,t)\left(1+A_n(s,t) \vec{S}_T.\vec{n}\right)=I_0\left(1+A_n S_T \cos\phi\right)~,
\end{equation}
where $I_0$ is the differential cross-section in absence of polarisation
and the azimuthal angle $\phi$ of the scattered particle is equal to the
angle between $\vec{S}_T$ and $\vec{n}$ (see Fig.~\ref{basic:fig:azim}).
\noindent
In some of the more systematic notation schemes, the analysing power is denoted $C_{0y00}$, or $(0y|00)$. The positivity of $I$ imposes $|A_n|\le 1$. 

Similarly, if the particle $a$ has spin, the beam can be polarised. A beam asymmetry parameter can be defined by an equation analogous to (\ref{basic:eq:ana}). For the case of a photon beam, see Sec.~\ref{excl:sub:photps} on photoproduction.

If both the beam and the target are polarised, the cross section can be written in the most general form
\begin{equation}\label{basic:eq:Cij00}
I(\vec{S}_a,\vec{S}_b)=I_0\,C_{\lambda\mu00}\,S_a^\lambda\,S_b^\mu~,
\qquad C_{0000}=S^0_a=S^0_b=1\,.
\end{equation}
We have introduced the four-component polarisation vector $S^\mu=(S^0,\vec{S})$ where $\vec{S}=\langle\vec{\sigma}\rangle$ is the usual polarisation vector, with $|\vec{S}|\le S^0=1$. The Greek indices take the values 0, $m$, $l$, $n$, or  0, $x$, $y$, $z$, whereas the Latin  indices $i$, $j$ take only the values $m$, $l$, $n$, or $x$, $y$, $z$. A summation is implied over each repeated index. The components $S^x$, $S^y$ and $S^z$ are measured in the frame (\ref{basic:eq:frame}) associated to each particle. The azimuthal dependence of the differential cross section, like in Eq.~(\ref{basic:eq:ana}), is then due to the variations of $S^x$ and  $S^y$ when the scattering plane is rotated about the beam axis, $\vec{S}$ keeping a fixed direction in the laboratory frame. 

$S^\mu$ is not 
%does \emph{not} transform as a relativistic 4-vector under a change of reference frame. It is linked to, but different from 
the relativistic polarisation 4-vector  $\mathsf{s}^\mu$ which appears in covariant expressions such as (\ref{incl:eq:htensor}). For a particle of 4-momentum $(\gamma m,0,0,\gamma vm)$ moving along the $\vec{z}$ axis, the correspondence is
\be
\left(\mathsf{s}^0, \mathsf{s}^x, \mathsf{s}^y, \mathsf{s}^z\right) = 
\left(\gamma v S^z, S^x, S^y, \gamma S^z\right)~.
\ee
In the rest frame, the $x$, $y$ and $z$ components are the same, but $\mathsf{s}^0=0$ while $S^0\equiv1$. 
%only the space components of these two four-vectors are the same. From $|\vec{S}|\le S0$, one can say that $S^\mu$ is in a sort of  ``future light cone", but not in the usual Minkovski space. Pure polarisation are on the cone surface.

Equation (\ref{basic:eq:Cij00}) applies whether or not discrete symmetries like $P$ and $C$ are conserved. If parity is conserved, the non-vanishing spin parameters
%coefficients are $C_{0000}=1$, 
are the beam asymmetry  $C_{n000}$,  the analysing power $C_{0n00}$ and the initial-state double-spin asymmetry coefficients\,%
\footnote{%
Note that the $C_{ij00}$, often abbreviated as $C_{ij}$ are sometimes named ``correlation coefficients'', though there are not, strictly speaking, defined as correlation factors in statistics.
}
 $C_{ij00}$, with ${ij}={xx}$, ${xz}$, ${zx}$, ${zz}$ and ${yy}$.
 
If a final-state particle $c$ or $d$ has a spin, its average value can be measured. In the case of unpolarised beam and target, this defines the \emph{polarisation}, $P_n^{c}$ or $P_n^{d}$, or, in a more systematic notation, $C_{00n0}$ or $C_{000n}$. Parity conservation requires, indeed,  that this polarisation is normal to the scattering plane.  If both final polarisations are measured, the final-state spin correlation  coefficients $C_{00ij}$ can be determined.  For the most general case where both the beam and the target are polarised, and the final spins analysed, the analogue of (\ref{basic:eq:Cij00}) reads
\be\label{basic:eq:Cijkl}
I(\vec{S}_a,\vec{S}_b,\vec{S}_c,\vec{S}_d)=I_0\,C_{\lambda\mu\nu\tau}
S_a^\lambda S_b^\mu S_c^\nu S_d^\tau~.
\ee
The quantities $C_{\lambda\mu\nu\tau}$ are called the \emph{Cartesian reaction parameters}. A variety of alternative notations can be found in the literature, such as  $(\lambda\mu|\nu\tau)$ also used in this report.

In Eq.~(\ref{basic:eq:Cijkl}), $\vec{S}_c$ and $\vec{S}_d$ are \emph{chosen} (for instance by a selective detector) spin orientations of the final particles. They are \emph{input} variables, not depending on $\vec{S}_a$, $\vec{S}_b$ nor on the reaction mechanism. They have to be distinguished from the (output) \emph{polarisations} of these particles, that we denote $\langle\vec{S}_c\rangle$ and $\langle\vec{S}_c\rangle$ (we have have kept the same letters for simplicity, but with an acute bracket). The latter depend on $\vec{S}_a$, $\vec{S}_b$ and the reaction. For particle $c$, we have
\be\label{basic:eq:polar-c}
\langle \vec{S}_c\rangle ={
\boldsymbol{\nabla}_{\!\vec{S}_c} I(\vec{S}_a,\vec{S}_b,\vec{S}_c,\vec{S}_d=0)
\over 
I(\vec{S}_a,\vec{S}_b,\vec{S}_c=0,\vec{S}_d=0)}~,
\ee
where $\boldsymbol{\nabla}$ stands for the usual gradient operator, and similarly for $d$.

\subsubsection{Centre-of-mass and other frames} Along this review, we restrict ourselves to the centre-of-mass frame, where most experimental results are given, and where the constraints of symmetries such as identical particles or time reversal are most easily expressed.

It is in general rather delicate to translate the spin observables from one frame to another, for instance from the target frame to the centre-of-mass one. See, \eg, 
\cite{Bourrely:1980mr,Leader:2001gr}. However, it could be checked that many of the identities or inequalities survive the various rotations associated to a change of  frame. This is the case, for instance, for the identity $A^2+R^2+D^2=1$ of $\pi\N$ scattering, as discussed in Sec.~\ref{excl:sub:piN}.
\subsection{General properties of density matrices}\label{basic:sub:dens}
The density matrix will be used extensively to derive constraints on spin observables. Here, the properties associated with its rank and positivity are briefly reminded.
\subsubsection{Definitions}
 The density matrix $\rho$ of a quantum system, \eg, the spin state of one or several particles, describes its statistical properties and encodes the relevant information. It has dimension $N\!\times\! N$, where $N$ is the dimension of the quantum Hilbert space of the system, in our case the number of basic spin states: $2s+1$
for one massive particle of spin~$s$, $(2s_a+1)(2s_b+1)$ for two massive particles $a$ and $b$, etc. The expectation value of the observable associated to the operator $\mathcal{O}$  is 
\be\label{basic:eq:observe}
\langle\Ocal\rangle=\Tr\{\rho\,\Ocal\}~.
\ee
The prototype of a matrix density is that of one spin 1/2 particle, which is written as
\begin{equation}\label{basic:eq:prototype}
\rho=(\mathbbm{1}+\vec{S}. \vec{\sigma})/2~, \qquad |\vec{S}|\leq1 ~,
\end{equation} %{\mathbbm{1}+\vec{S}. \vec{\sigma} \over 2}
where $\vec{S}$ is the polarisation vector, $\mathbbm{1}$ the $2\times2$ unit matrix and $\vec{\sigma}=\{\sigma_x,\sigma_y,\sigma_z\}$ the set of Pauli matrices. 

If the system is known to be in a well-defined quantum state $|\psi\rangle$, also called a \emph{pure} state,  $\rho$ is the rank-one projector $\rho=|\psi\rangle\langle\psi|$. If not, the system is in a \emph{mixed} state and $\rho$ can be decomposed as follows 
\be\label{basic:eq:mixed}
\rho=\sum_n w_n|\psi_n\rangle\langle\psi_n|~,\qquad 0\le w_n\le1\,,\qquad \Tr\rho=\sum_n w_n=1~.
\ee
The set of  $|\psi_n\rangle$ in Eq.~(\ref{basic:eq:mixed}) is not unique. Even the number of terms can be varied for a given $\rho$.  However $\rho$ is Hermitian, hence diagonalisable, and one obtains a minimal number of terms by taking the $w_n$ and the $|\psi_n\rangle$ to be the eigenvalues and eigenvectors of $\rho$. These $|\psi_n\rangle$ form an orthonormal basis and $w_n$ is the probability that the system is in the state $|\psi_n\rangle$. From now on $w_n$ will be the $n^{\rm th}$ eigenvalue of $\rho$.  

 The number $r\le N$ of non-vanishing $w_n$ is the \emph{rank} of $\rho$. The rank is one type of \emph{entropy}: the smaller it is, the larger  is the information on the system. The rank-one projector considered above contains the maximum information. This is the case of a fully polarised particle. 
 Other types of entropy are presented in Sec.~\ref{se:furth}.
 The minimum information is given by $\rho=\mathbbm{1}_N/N$, where $\mathbbm{1}_N$ is  the $N\!\times\! N$ unit matrix. It is the case of a completely unpolarised particle. 

 Objects similar to density matrices, but of arbitrary positive trace, describe spin-dependent probabilities, in particular
\begin{itemize}
\item the acceptance matrix of a detector, which can be written as $\check\rho=\sum_n w_n|\psi_n\rangle\langle\psi_n|$, where $w_n$ is the probability of registering the particle if it is in the spin state $|\psi_n\rangle$ (the $|\psi_n\rangle$ form an orthogonal basis). Then $0\le w_n\le1$ but there is no imposed constraint on the trace. For a full-efficiency non-analysing detector, $\check\rho$ is the identity matrix.
\item the non-isotropic decay of an unstable polarised particle can be described by a \emph{decay matrix} $\check\rho(\vec p_1, \vec p_2,...)$ which depend  on the momenta $\vec p_1, \vec p_2,... $ of the decay products. It has the same properties as the acceptance matrix of a detector. Such a matrix can also be introduced for the fragmentation of a polarised quark or gluon. 
\item the Cross Section Matrix (CSM) $\R$, which encodes all the spin dependence of a reaction. Its matrix elements are linear functions of the cartesian parameters $C_{\lambda\mu\nu\tau}$ and its trace is proportional to the unpolarised cross section. It acts on quantum states containing the initial and the final particles together. It will be introduced in Sec.~\ref{basic:sib:csm} and used extensively for deriving general rules and  studying particular reactions.
\end{itemize}

\subsubsection{The positivity conditions}
The density matrix $\rho$, as well as the similar objects $\check\rho$ and $\R$ mentioned above, are \emph{semi-positive}: $\langle\psi|\rho|\psi\rangle\ge0$ for any state $|\psi\rangle$. We shall write it $\rho\ge0$. 
If $\langle\psi|\rho|\psi\rangle>0$ for all states, the matrix is said to be  \emph{positive} or \emph{positive definite} ($\rho>0$). 
The positivity of $\rho$ is equivalent to the positivity of all the  $w_n$'s. For semi-positivity, some eigenvalues are allowed to be zero. $\rho\ge0$ implies $(\Tr\rho)\,\mathbbm{1}_N-\rho\ge0$, where $\mathbbm{1}_N$ is the unit $N\!\times\!N$ matrix. For an acceptance matrix we have the independent constraint $\,\mathbbm{1}_N-\rho\ge0$, owing to $w_n\le1$.

%$\mathbbm{1}_N$ 

Let us consider the symmetric functions of the eigenvalues of an Hermitian  matrix $\rho$ 
\be\label{basic:eq:gendens1}
\Delta_1 = \sum_{i} w_i \,,\quad
\Delta_2 = \sum_{i<j} w_i w_j \,,\quad
\Delta_3 = \sum_{i<j<k} w_i w_j w_k \,, \quad
\cdots \quad
\Delta_N = w_1 w_2 \cdots w_n~.
\ee
A necessary and sufficient condition for $\rho$ to be  semi-positive and of rank $r$ is
\be\label{basic:eq:gendens2}
\Delta_p > 0 \quad {\rm for} \quad  p \le r~,\qquad\Delta_p = 0 \quad {\rm for} \quad  r < p \le N~.
\ee
These conditions can be expressed in terms of determinants. A \emph{minor} is any subdeterminant obtained by removing some rows and the same number of columns. In a \emph{principal minor} the removed rows and columns have the same indices, so that the surviving diagonal elements remain diagonal. $\Delta_p$ is the sum of the \emph{principal minors} of order $p$. In particular, $\Delta_1=\Tr(\rho)$, $\Delta_N=\det(\rho)$.  

Calculating all $\Delta_p$'s is tedious since there are  $N!/[p!(N-p)!]$ principal minors of size $p\!\times\! p$ for each $p$, in total $2^N-1$ determinants. However, deciding  whether or not a matrix is (semi-) positive  requires  much fewer determinants:

\begin{enumerate}
	\item If $\rho$ is positive definite, each principal minor is strictly positive. If $\rho$ is  semi-positive, each principal minor is zero or positive. This gives inequalities generally simpler than (\ref{basic:eq:gendens2}). If anyone principal minor is negative, one can conclude that $\rho$ is non-positive.
	\item If a \emph{nested sequence} of $N$ principal minors is strictly positive, $\rho$ is positive definite. ``Nested'' means that the $p^{\rm th}$ determinant is a principal minor of the $p+1^{\rm th}$ one. One example of nested sequence is the set of ``corner principal minors", the $p^{\rm th}$ one occupying the $p$ first rows and columns of $\rho$. 
        \item If a \emph{nested sequence} of $N$ principal minors is strictly positive except for the largest one which is zero, $\rho$ is semi-positive of rank $N-1$.
\end{enumerate}
 The most used positivity conditions result from applying item 1) to the diagonal elements and the $2\!\times\!2$ principal minors: 
\be\label{basic:eq:gendens2'}
\rho_{kk}\ge0\,,\qquad|\rho_{ij}|^2\le\rho_{ii}\rho_{jj}\qquad\hbox{(without index summation)}~,
\ee
which, combined with $\Tr\rho=1$, imply $|\rho_{ij}|\le1$ (more generally, $|\rho_{ij}|\le\Tr\rho$).

Item 2) cannot be extended to semi-positivity. As a counter-example, the matrix 
\be
\rho=\begin{pmatrix}1&1&1\\ 1&1&1\\ 1&1&a \end{pmatrix}~,
\ee
has its three ``corner principal minors''  positive or zero, but $\rho$ is non-positive for $a<1$. 

\subsubsection{The full domain of positivity} 
Let us release provisionally the trace condition. An Hermitian  matrix $\rho $ depends linearly on $N^{2}$ real parameters, for instance $\RE(\rho_{ii'})$ for $i\le i'$ and $\IM(\rho_{ii'})$ for $i<i'$. In the $N^{2}$-dimensional parameter space, the domain where  $\rho$ is semi-positive is a \emph{convex half-cone} $\mathcal{C}$: if $\rho$ belongs to $\mathcal{C}$, then $a\rho$ also belongs to $\mathcal{C}$ for $a\ge0$. If $\rho_1$ and $\rho_2$ belong to $\mathcal{C}$, then $x\rho_1+(1-x)\rho_2$ also belongs to $\mathcal{C}$ for $0\le x\le1$ (convexity). 
Indeed, for any $|\psi\rangle$ we have $\langle\psi|\rho_1|\psi\rangle\ge0$ and $\langle\psi|\rho_2|\psi\rangle\ge0$, hence $\langle\psi|x\rho_1+(1-x)\rho_2|\psi\rangle\ge0$. 
We can call $\mathcal{C}$ the \emph{positivity cone}. 

The space of Hermitian matrices is endowed by a \emph{Euclidian metric}. The norm of a matrix can be defined as $|\rho|=\sqrt{\Tr\rho^2}$, the distance between two matrices $\rho$ and $\rho'$ is $|\rho-\rho'|$ and their scalar product is $(\rho,\rho')=\Tr(\rho\rho')$. One can decompose $\rho$ as follows 
\be\label{I+traceless}
\rho=\rho_\parallel+\rho_\perp~, \qquad \rho_\parallel=(\Tr\rho/N)\,\mathbbm{1}~, \qquad \Tr\rho_\perp=0 \,.
\ee
$\rho_\perp$ carries the information and $|\rho_\perp|^2$ is a kind of negentropy\,\footnote{Negative of entropy}. One has 
\be\label{basic:eq:rho-dec}
|\rho|^2=|\rho_\perp|^2+|\rho_\parallel|^2~,\qquad |\rho_\parallel|=\Tr\rho/\sqrt{N}~,\qquad
2\Delta_2=(N-1)|\rho_\parallel|^2-|\rho_\perp|^2~.
\ee
For $N=2$, one can parametrise $\rho$ as  $(S^0\,\mathbbm{1}+\vec{S}.\vec{\sigma})/2$, where $(S^0,\vec{S})$  generalises the 4-vector $S^\mu$ introduced in (\ref{basic:eq:Cij00}) for arbitrary $S^0=\Tr\rho$. The positivity cone is simply given by $S^0\ge0$, $4\Delta_2=(S^0)^2-\vec{S}^2\ge0$. It reminds the future cone in the usual Minkowski space. 

\begin{figure}[!!htb]
\begin{center}
\includegraphics[width=.6\textwidth]{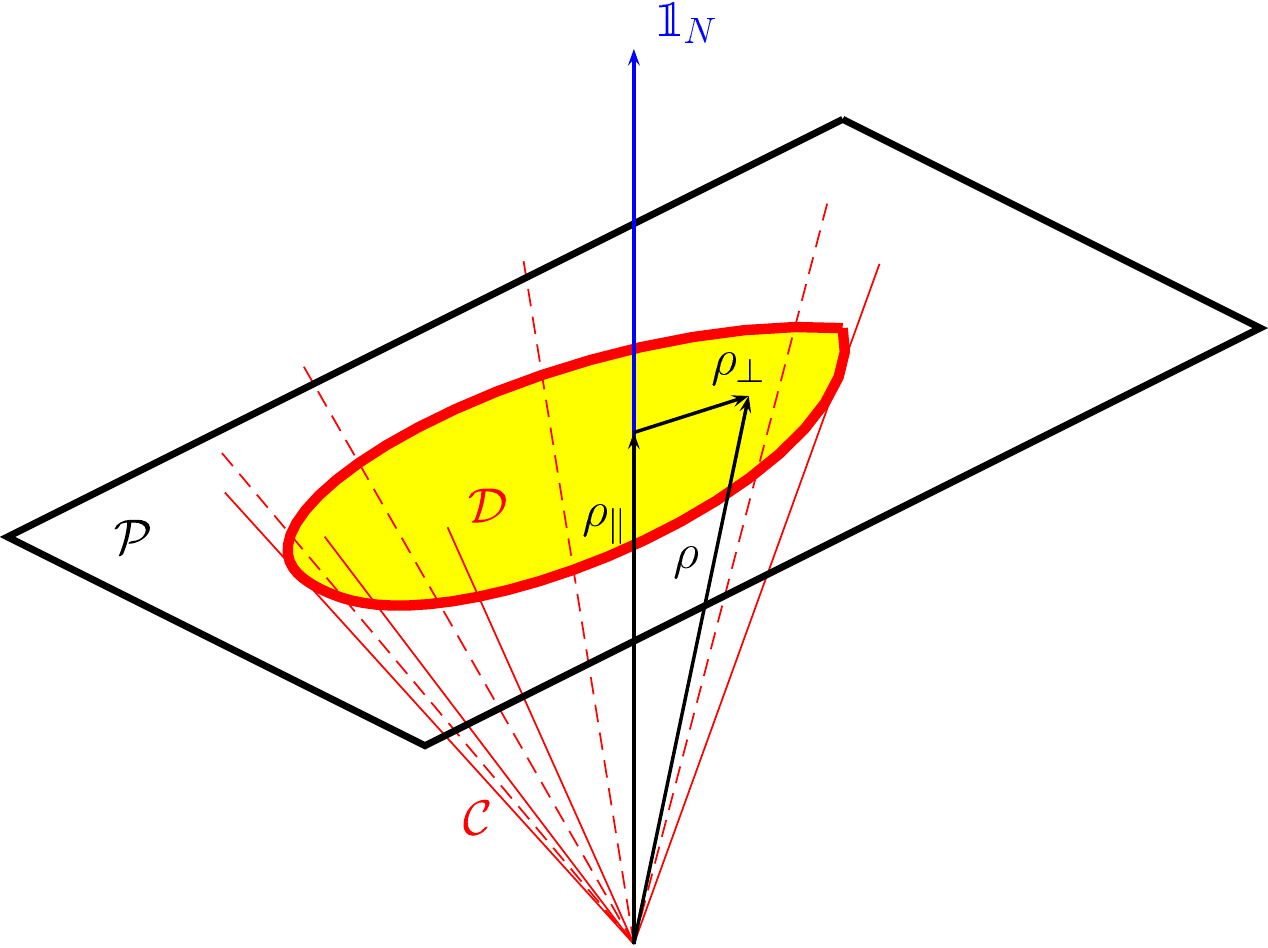}
\end{center}
\vskip -.5cm
\caption{\label{basic:fig:coneXA} Positivity cone $\mathcal{C}$ for matrices of arbitrary trace and positivity domain $\mathcal{D}$ for unit-trace matrices.  $P$ is the hyperplane $\Tr(\rho)=1$. The decomposition $\rho=\rho_\parallel + \rho_\perp$ of Eq.~(\ref{I+traceless}) is also schematically represented.}
\end{figure}

Let us now impose the trace condition $\Tr\rho=1$ which amounts to take the intersection of $\mathcal{C}$ with the hyperplane $P$ of equation $\Tr(\rho)=1$. It gives a  \emph{finite} and \emph{convex} positivity domain $\mathcal{D}$, as represented in Fig.~\ref{basic:fig:coneXA}.
Convexity follows from the same argument as for  $\mathcal{C}$. Finiteness is guaranteed by $|\rho_{ij}|\le1$ for each matrix element. The centre of the domain is the matrix $\rho_\parallel=\boldsymbol{1}_N/N$, for which $\rho_\perp=0$. A similar positivity domain exists for the \emph{normalised} cross section matrix $\hat{\R}=(N/\Tr\R)\,\R$.

The boundary $\partial\mathcal{D}$ of $\mathcal{D}$ is a sheet of the hypersurface $\Delta_N\equiv\det(\rho)=0$. It is a manifold of dimension $(N^{2}-2)$ and degree $N$. On $\partial\mathcal{D}$, $\rho$ is only semi-positive. The other conditions, $\Delta_p\ge0$ for $p=2,\cdots N-1$ define domains which \emph{include} $\mathcal{D}$. These ``auxiliary'' conditions serve to eliminate unphysical sheets of the hypersurface $\Delta_N=0$. The hypersurface where $\Delta_p$ (or any principal minor) vanishes is externally tangent to $\mathcal{D}$. 
The contact zone contains the semi-positive matrices of rank $r<p$. 

%loss of information. 
The distance to $\partial\mathcal{D}$, which grows with $\Delta_N$, is a form of entropy. This is for instance the case of the cross section matrix of an inclusive reaction, where  information is lost through non-detected particles. 

\medskip\ni\emph{Case of rank-one matrices.}
The density matrices we shall encounter are often of rank $r$ smaller than their dimension $N$, therefore sit on $\partial\mathcal{D}$. The density matrix $\rho=|\psi\rangle\langle\psi|$ of a pure state has $r=1$. In this case,  $\Delta_p=0$ for all $p\ge 2$, and all non-linear positivity constraints are saturated. According to (\ref{basic:eq:rho-dec}), such matrices have the largest distance $d_{\rm max}=[(N-1)/N]^{1/2}$ from the centre. One can say that they are at the ``corners'' of $\mathcal{D}$. The same properties apply to the  cross section matrix of  a completely polarised exclusive reaction, which also has $r=1$. For $r>1$, $\Delta_2$ is positive; it is another form of entropy.

\medskip\ni\emph{Symmetries of the positivity domain.}
The (semi-) positivity of a matrix $\rho$ is conserved under a unitary transformation $\rho\to U\rho\;U^\dagger$ and under the transposition $\rho\to\rho^t$. Accordingly, $\mathcal{D}$ is invariant under such transformations. They conserve $\rho_\parallel$, $|\rho|$ and $|\rho_\perp|$, therefore $U$ can be considered as a sort of rotation of $\rho_\perp$ whereas transposition is a reflection about the hyperplane of real matrices. Using unitary transformations, one obtains an infinity of other symmetry hyperplanes.

Let us however point out that for $N>2$, $\mathcal{D}$ is not symmetrical about its centre. Opposite to a pure state matrix $\rho=|\psi\rangle\langle\psi|$ with $|\rho_\perp|=d_{\rm max}$ one has another boundary matrix $\rho'=(\mathbbm{1}-|\psi\rangle\langle\psi|)/(N-1)$ at distance $|\rho'_\perp|=d_{\rm max}/(N-1)$ from the centre.

Symmetries such as parity, charge conjugation, time reversal, etc. studied in Sec. \ref{basic:sub:sym-constraint} impose linear or anti-linear relations to $\rho$, for instance 
$\rho=P\rho P^{-1}$ for parity. These relations together define a hyperplane $H$ which is a symmetry plane of the unrestricted $\mathcal{D}$. The effective domain for $\rho_\perp$ is then the intersection $\mathcal{D}_{{\rm eff}}= \mathcal{D} \cap H$. 

\boldmath\subsubsection{Structure of the boundary of $\mathcal{D}$} \unboldmath
Except for the case  $N=2$ the boundary of $\mathcal{D}$ has a complex structure which is difficult to represent by a drawing and barely compares with familiar three-dimensional bodies. The bulk of $\partial\mathcal{D}$ is made by matrices of rank $N-1$. Decreasing the rank unit by unit, one obtains a hierarchy of ``edges'' of lower and lower dimensions: 
\be 
\mathcal{D}\equiv\mathcal{D}^N_N\supset \mathcal{D}^N_{N-1}\supset \ldots\supset\mathcal{D}^N_{2}\supset\mathcal{D}^N_{1}~.
\ee 
The first domain of this chain is $\mathcal{D}$ itself. $\mathcal{D}^N_{N-1}=\partial\mathcal{D}$ is the whole boundary. $\mathcal{D}^N_{r}$ is the set of semi-positive density matrices of rank less than or equal to $r$, characterized by $\Delta_{r+1}=0$ (which also insures $\Delta_{r'}=0$ and the vanishing of all $r'\!\times\! r'$ subdeterminants for any $r'\ge r$). $\mathcal{D}^N_{1}$, which contains the pure state density matrices, is the ``corner'' of $\mathcal{D}$.

To each semi-positive density matrix $\rho$ of rank $r$ one associates the image Hilbert space $\H_{r}(\rho)=\rho\,\H_N$, of dimension $r$. On this subspace $\rho$ is strictly positive-definite. $\rho$ can thus be defined by the $r2$-1 real parameters of its restriction to $\H_{r}(\rho)$, plus $2r(N-r)$ real angles $\theta_i$ specifying the orientation of $\H_{r}(\rho)$ in $\H_N$. Thus $\mathcal{D}^N_{r}$ has $r^2-1$ flat and $2r(N-r)$ curved dimensions, in total $r(2N-r)-1$ dimensions. The mixture of flat and curved dimensions survives in some of the 3-dimensional projections of Fig.~\ref{excl:fig:lim3d}, where conical or cylindrical pieces of boundaries can be seen. 

Let us apply these results to $N=4$ for illustration. $\mathcal{D}^4_{4}\equiv\mathcal{D}$ has $4^2-1=15$ dimensions. $\mathcal{D}^4_{3}$ has 14 dimensions, 8 flat and 6 curved. $\mathcal{D}^4_{2}$ has 11 dimensions, 3 flat and 8 curved. The ``corner'' $\mathcal{D}^4_{1}$ (which contains the pure states) has only 6 curved dimensions. 

The edges $\mathcal{D}^N_{r}$ are not convex, but \emph{connected}: one can pass continuously from $\rho$ to $\rho'$ in $\mathcal{D}^N_{r}$, first rotating $\H_{r}(\rho)$ to $\H_{r}(\rho')$ (by continuous unitary transformation in $\H_N$), then changing the $r^2-1$ parameters of the restricted density matrix. In particular, the "corner" of $\mathcal{D}$, made of the pure states, is a connected manifold. However, when symmetries such as parity, time reversal, etc. are taken into account, an edge of the effective domain $\mathcal{D}_{\rm eff}$ may not be connex. Let us consider, 
for instance, the density matrix (\ref{basic:eq:prototype}) of a spin one-half. $\mathcal{D}$ is the unit ball and the ``corner'' is the unit sphere, which is connex. If the source of the particle is invariant under reflection about the $xy$ plane, then $S_x=S_y=0$,  $\mathcal{D}_{\rm eff}$ is a diameter and its corner is made of two separate points. 
The corners of squares, triangles, cubes, tetrahedrons, encountered in Sec.~\ref{se:excl} are typical cases of non-connex edges.

\begin{figure}[!!hbtc]
\begin{center}
\includegraphics[width=.7\textwidth]{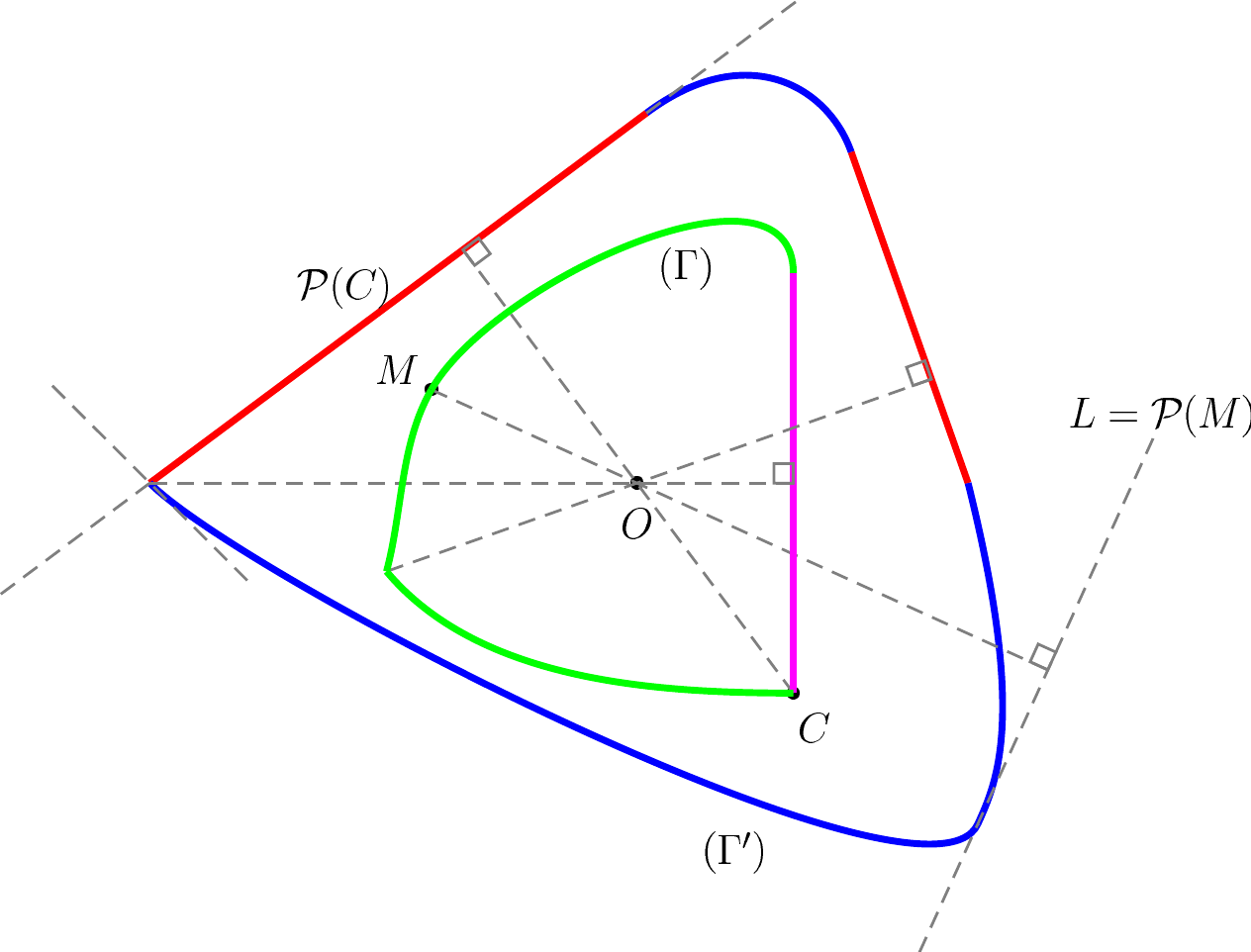}
\end{center}
\caption{\label{basic:fig:polar1} Schematic drawing of a 2-dimensional polar transform. }
\end{figure}

An important property of $\partial\mathcal{D}$ is its
invariance of  under \emph{polar reciprocal transform}, presented below.
\subparagraph{Reminder about the polar reciprocal transform.} 
In the 2-dimensional Euclidian space, the \emph{polar} $L={\cal P}(M)$ of a point $M$ with respect to a circle $C$ is a straight line defined as follows: 
\begin{itemize}
\item 
if $M$ is external to the circle, one draws the lines $MA$ and $MB$ tangent to $C$ in $A$ and $B$. Then $L$ is the whole straight line $AB$, the points $X$ of which satisfy 
\be\label{polar}
\overrightarrow{OX}\cdot\overrightarrow{OM}=p\,,
\ee
where $O$ is the centre and $p=R^2$ the radius squared of the circle. 
\item if $M$ is internal to the circle, $A$ and $B$ become imaginary points, but $L$ is still defined by (\ref{polar}). 
\end{itemize}
\ni Conversely, $M$ is the pole of $L$ and we equally write  $M={\cal P}(L)$. The \emph{reciprocal polar transform} $\Gamma'={\cal P}(\Gamma)$ of a curve $\Gamma$ is the envelope of $L={\cal P}(M)$ when $M$ runs along $\Gamma$. This transformation is reciprocal: $\Gamma={\cal P}(\Gamma')$. 
The correspondence between a \emph{whole} straight line $L$ and its pole $M$ is a particular case of reciprocal polar transform, if $M$ is considered as a \emph{closed} curve of zero length. 
% in that case, $P={\cal P}$. 
If $\Gamma$ is a convex polygon $C_1, C_2, ...C_n$ surrounding $O$,  $\Gamma'$ is a polygon of the same type. Considering a corner $\hat C_i$ as an arc of circle of infinitely small radius, one can say that the corners of $\Gamma$ are reciprocal to the sides of $\Gamma'$ and vice-versa:
\be\label{}
\hat C_i={\cal P}(C'_iC'_{i+1}) \,,\quad C_iC_{i+1}={\cal P}(\hat C'_{i+1}) 
\qquad (C_{n+1}\equiv C_1)\,.
\ee
%
% the border of a convex domain $D$ 
In the continuous limit $n\to\infty$, $\Gamma$ and $\Gamma'$ contain smooth curved parts and possibly corners and rectilinear parts, as in the example shown in Fig.~\ref{basic:fig:polar1}.

The polar transformation generalises easily to the case $p<0$ (imaginary radius) and to higher dimensions: in 3 dimensions, a plane is reciprocal to a point, a straight line is reciprocal to a straight line. In $d$ dimensions, a hyperplane of dimension $d'$ is reciprocal to an hyperplane of dimension $d"=d-d'-1$. The reciprocal polar transform of an ordinary curved $d-1$ dimensional surface is a surface of the same type.
 A corner of one curve is reciprocal to a rectilinear part of the other. Analogous correspondences for convex surfaces in $\mathbbm{R}^3$ are: 
\be\label{edge-edge}
%\left(
\begin{array}{cccc}
(a) & \hbox{cone tip}&  \leftrightarrow & \hbox{planar face} \\
(b) & \hbox{trihedron tip}&  \leftrightarrow & \hbox{planar triangular face} \\
(c) & \hbox{dihedron}  & \leftrightarrow & \hbox{dihedron} \\
(d) & \hbox{planar curved edge}  & \leftrightarrow  & \hbox{cylindrical or conical face} \\
(e) & \hbox{skew edge}  & \leftrightarrow  & \hbox{``locally conical" face}
\end{array}
%\right)
\ee
%In item (d) 
A curved edge (here noted $\gamma$) of $\Gamma$ is the boundary curve between two faces, one of which at least is nonplanar. The ``locally conical" surface of case (e) is the envelope of the planes reciprocal to the points of $\gamma$. The contact lines are tangent to a curve $\gamma'$ and their reciprocal lines are tangent to $\gamma$. In case (d) $\gamma$ is in a plane and $\gamma'$ is reduced to the pole of this plane. 
As an example of case (d), a cylindrical barrel is polar reciprocal to a double-cone buoy.  The equator of the buoy is reciprocal to the cylindrical part of the barrel. A circular edge of the barrel is reciprocal to a cone of the buoy. 

The generalisations in $\mathbbm{R}^d$ of tips, edges, dihedrons, etc., will be called "edge" in a generic way. An edge ${\cal E}_{f,c}$ of $\Gamma$ having $f$ flat dimensions and $c$ curved dimensions is reciprocal to an edge ${\cal E'}_{f',c}$ of $\Gamma'$ with the same number of curved dimensions and
$f'=d-1-c-f$. 

\boldmath\subparagraph{Application to $\mathcal{D}$.}\unboldmath
The traceless part of a density matrix $\rho$ can be represented by an Euclidean vector $\vec\rho_\perp$ of dimension $d=N^2-1$. For two density matrices $\rho$ and $\eta$, one has $\Tr(\rho\,\eta)\ge0$, or %$(\rho,\eta)\equiv
\be\label{basic:eq:polar-recip} %(\ref{basic:eq:polar-recip}) 
\vec\rho_\perp.\vec\eta_\perp\ge-1/N\,.
\ee
The necessary and sufficient condition that a unit trace Hermitian matrix $\eta$ belongs to $\mathcal{D}$ is that it satisfy (\ref{basic:eq:polar-recip}) for any $\rho\in\mathcal{D}$. One can equally say ``for any $\rho\in\partial\mathcal{D}$". Thus, in $\eta$ space, $\mathcal{D}$ is the intersection of the half-spaces defined by the inequality (\ref{basic:eq:polar-recip}) when $\rho$ runs over $\partial\mathcal{D}$. Taking into account the fact that $\mathcal{D}$ is convex, one infers that its boundary $\partial\mathcal{D}$ is the envelope of the planes $\vec\rho_\perp.\vec\eta_\perp=-1/N$ when $\rho$ runs over $\partial\mathcal{D}$. In other words, $\partial\mathcal{D}$ is the reciprocal transform of itself. $\partial\mathcal{D}$ \emph{is invariant under the reciprocal polar transform} of power $p=-1/N$. 

$\mathcal{D}^N_r$ and $\mathcal{D}^N_{N-r}$ are reciprocal edges, like the objects of (\ref{edge-edge}). Furthermore, an hyperplane tangent to $\mathcal{D}^N_r$ is reciprocal to a flat submanifold(prolongated as a whole hyperplane) of $\mathcal{D}^N_{N-r}$. In the example  $N=4$ given earlier, $\mathcal{D}^4_{3}$ is reciprocal to $\mathcal{D}^4_{1}$ and $\mathcal{D}^4_{2}$ is self-reciprocal. A 6-plane tangent to $\mathcal{D}^4_1$ is reciprocal to a flat submanifold (prolongated as a whole 8-plane) of $\mathcal{D}^4_3$. 

When parity, time reversal, etc. have been taken into account, the self-reciprocity property applies to $\mathcal{D}_{\rm eff}$ as well. 

\subsubsection{The projected positivity domains} 
Practically one does not measure the complete set of $N^2-1$ parameters of the density matrix, but a restricted set of mutually orthogonal observables $\{\langle\Ocal_1\rangle, ...\langle\Ocal_n\rangle\}$ with $\Tr\{\Ocal_i\,\Ocal_j\}=g_i\delta_{ij}$ and $n<N^2-1$. The joint domain allowed by positivity,  $\mathcal{D}\{\Ocal_1,...\Ocal_n\}$ is obtained by successive projections of $\mathcal{D}$ on the planes  $\langle\Ocal_k\rangle=0$ with $k=n+1,n+2,\cdots N^2-1$. It is a convex domain, since the $\langle\Ocal_i\rangle$ are linear functions of $\rho$. 

Let us consider, for instance, the projection $\mathcal{D}\{\Ocal_1,\Ocal_2\}$ of $\mathcal{D}\{\Ocal_1,\Ocal_2,\Ocal_3\}$ on the plane $P$ of equation $\langle\Ocal_3\rangle=0$. It contains  the intersection: 
\be
\mathcal{D}\{\Ocal_1,\Ocal_2\}\supset P\cap\mathcal{D}\{\Ocal_1,\Ocal_2,\Ocal_3\}
\ee
If $P$ is a symmetry plane for $\mathcal{D}\{\Ocal_1,\Ocal_2,\Ocal_3\}$, the relation $\supset$ becomes an equality. 
Strict inclusion is obtained for instance in Figs.~\ref{excl:fig:filter-octo} or \ref{excl:fig:7-14-4-3-8-4} when the projection is made on the plane $C_{nn}=0$.

Like the distance to $\partial\mathcal{D}$, the distance to the boundary $\partial\mathcal{D}\{\Ocal_1,...\Ocal_n\}$  of a projected domain is a kind of entropy. It can only increase under successive projections. This increase can be interpreted as the loss of information when part of the observables are integrated over. 

\paragraph{Possible appearance of nonconvex domains.}
As mentioned earlier, one often deals with density matrices of rank $r$ smaller than $N$, therefore sitting on the edge $\mathcal{D}^N_r$, which is not convex. Most often the non-convexity disappears after projection on a subset of observables. For instance a hollow sphere is projected into a full disk. However this requires that the dimension of the effective domain is larger than the number of plotted quantities. A counter-example is given by the set $\{P_n,A,R\}$ in pion-nucleon scattering, whose domain is the hollow unit sphere (\ref{excl:eq:pin4}). Here the cross section matrix $\R$ has dimension $4\times4$, therefore $\mathcal{D}$ has dimension 15, but the effective domain for $\R$ has only 2 dimensions. It comes from $r=1$ together with parity conservation (or more simply from the fact that there are only 2 independent amplitudes).

\paragraph{Polar reciprocity between projection and intersection.}
Let $P$ be a hyperplane which contains the centre $\vec\rho_\perp=0$ of $\mathcal{D}$ and let us compare the intersection $P\cap\mathcal{D}$ and the projection $P(\mathcal{D})$ of $\mathcal{D}$ on $P$. Both domains are convex, and from the inequality (\ref{basic:eq:polar-recip}) one can derive
\be\label{basic:eq:recipro-proj-inter} %(\ref{basic:eq:recipro-proj-inter})
\vec\rho_\perp.\vec\eta_\perp\ge-1/N  \quad\hbox{for all}\quad
\rho\in P(\mathcal{D})\,,\ \eta\in P\cap\mathcal{D}\,.
\ee
This inequality in fact defines $P(\mathcal{D})$ in terms of $P\cap\mathcal{D}$ or vice-versa. It follows that their boundaries $\partial(P\cap\mathcal{D})$ and $\partial P(\mathcal{D})$ are polar reciprocal \cite{Minnaert:1971}. This property applies to $\mathcal{D}_{{\rm eff}}$ as well, when parity, time reversal, etc. have been taken into account. We will explain in Section \ref{basic:sub:reciprocal-method} how it can be used to find $\mathcal{D}\{\Ocal_1,...\Ocal_n\}$.

\subsection{Spin states of a single particle}\label{basic:sub:single}
\boldmath\subsubsection{Spin 1/2}\unboldmath \label{basic:sub:Spin1/2}
The polarisation domain for a single spin 1/2 particle can be reviewed as a warm-up. A pure quantum state for a spin 1/2 oriented along an axis $\vec{u}$  corresponds to an eigenstate of $\vec{\sigma}.\vec{u}$, where $\vec{\sigma}=\{\sigma_x,\sigma_y,\sigma_z\}$. For this state, denoted $|\vec{u}\rangle$, we have $\langle\sigma_i\rangle\equiv\langle\vec{u}|\sigma_i|\vec{u}\rangle=u_i$. The $\zu$ axis is usually chosen along the particle momentum (helicity basis).  
A state polarised toward the transverse vector $\vec\tau$ is the following superposition of the helicity states $|\pm1/2\rangle$: 
\be\label{basic:eq:transv.state}
|\vec\tau\rangle = {|+{1/2}\rangle + \exp[i\phi(\vec\tau)] \,|-{1/2}\rangle\over\sqrt2}~,
\ee
where $\phi(\vec\tau)$ is the azimuth of $\vec\tau$ in the particular frame 
$\{\xu_{\pv},\yu_{\pv},\zu_{\pv}\}$ defined in Sec.~\ref{basic:sub:phase-conv}. 
\emph{Transversity amplitudes} are taken between transversity states with $\phi(\vec\tau)=\pm\pi/2$, \ie, $\vec\tau=\pm\nv$ in Fig.~\ref{basic:fig:frame}.

A partially polarised state is described by the $2\!\times\!2$ density matrix 
\begin{equation}\label{basic:eq:spin12b}
\rho={\mathbbm{1}+\vec{S}. \vec{\sigma} \over 2}~, %\qquad \det(\rho)=1-P2~,
\end{equation}
with expectation values 
\begin{equation}\label{basic:eq:spin12c}
\langle\sigma_i\rangle\equiv\Tr[\sigma_i \rho]=S_i~.%\quadP_x=P_y=0~,\quad P_z=P~.
\end{equation}

The positivity of $\rho$ is ensured by $|\vec{S}|\leq1$. Thus the \emph{polarisation domain} for a spin 1/2 particle is the unit ball. A point on the ball surface corresponds to a pure quantum state. The density matrix is then the rank-one projector $\rho=|\vec{u}\rangle\langle\vec{u}|$. 

The polarisation of outgoing nucleons is measured by rescattering, while for hyperons, it is deduced from the angular distribution of the decay products.
\subsubsection{Generalities on photon polarisation}\label{basic:sub:photpol}
 We now turn to the description of photon polarisation, on which there is a long tradition in optics.
Note that  ``polarisation'' has here a different meaning.
For spin 1/2, the polarisation vector is a real axial vector which is bilinear in the spinor amplitude. A \emph{longitudinally} polarised electron has definite helicity.  For the photon, the polarisation vector is a real or complex polar vector linear in the amplitude. There is no longitudinal polarisation, except for virtual photons. In that case, \emph{longitudinal} polarisation means zero helicity, whereas helicity $\pm1$ states are said to be \emph{transversely} polarised. 

A real photon has two possible helicity states, $|+1\rangle$ and $|-1\rangle$, corresponding to right- and left- circular polarisation, respectively (note that optics textbooks often have the opposite definition of ``right'' and ``left''). Therefore a partial photon polarisation is described by a $2\!\times\!2$ density matrix analogous to that of a spin-1/2 fermion, and can be written as
\be\label{basic:eq:ph1}
\rho_\gamma = \frac{\mathbbm{1}+\vec{S}.\vec{\sigma}}{2}~.
\ee
 $S_1$, $S_2$ and $S_3$ are the \emph{Stokes parameters} of the photon, to be described shortly.
The polarisation domain is again the unit ball $S_1^2+S_2^2+S_3^2\le1$. However the physical significance of the vector $\Sv$ is very different from the vector polarisation of a spin-1/2 fermion. 

Another possible basis consists of two orthogonal \emph{linear} polarisation states. For an exclusive or single-inclusive photon-induced reaction, they can be chosen as $|\vpi \rangle $ and $|\nv \rangle $
where $\nv$ is a unit vector normal to the scattering plane, $\kv$ is the photon momentum and $\vpi = \nv\times\vec{k}/|\vec{k}|$.  These states are related to the helicity ones by
% XA/JUIN:
\be\label{basic:eq:ph2}
|\pm 1\rangle = \frac{\mp|\vpi \rangle-i|\nv \rangle}{\sqrt{2}}~,
\ee
(the state $|+ 1\rangle$ differs in sign from the standard convention (\ref{basic:eq:hel_xyz}) for massive spin one). 

A third basis is made of the two \emph{oblique} states
\be\label{basic:eq:ph3}
|\pm \pi/4\rangle = \frac{|\vpi \rangle \pm |\nv \rangle}{\sqrt{2}}~.
\ee
Knowing the probabilities $p(i)$ of the photon to be in the state $|i\rangle $, and taking as basis vectors 
\be\label{basic:eq:ph4}
|\vpi\rangle = \begin{pmatrix}1\\ 0\end{pmatrix}~,
\qquad
|\nv \rangle = \begin{pmatrix} 0 \\ 1\end{pmatrix}~, 
\ee
which we call the ``straight linear polarisation basis'', the Stokes parameters are defined as in
\cite{Berestetskii:897953}
\begin{itemize}
\item 
$S_3$ or $S_\ominus$ = $ [p(\vpi)-p(\nv)] / [p(\vpi)+p(\nv)]$
(straight linear polarisation),
\item 
$S_1$ or $S_\oslash$ = $ [p(+\pi/4)-p(-\pi/4)] / [p(+\pi/4)+p(-\pi/4)]$
(oblique linear polarisation),
\item
$S_2$ or $S_\odot$ = $ [p(+1)-p(-1)] / [p(+1)+p(-1)]$
(circular polarisation).
\end{itemize}

Other conventions may be used. If the helicity basis was chosen, $S_3$ would be the amount of circular polarisation. The straight linear polarisation basis (\ref{basic:eq:ph4}) is particularly suitable to parity-conserving reactions, in association with the transversity basis for the fermions.

\subsubsection{Massive vector particle} \label{basic:sub:massvec}
\paragraph{Basic states}
A pure spin-state of a spin-$1$ particle (vector meson in our case) is represented by a complex 3-component column vector\,%
\footnote{%
In what follows, it is important to distinguish the spin \emph{state} $|\phi\rangle$ from its column-vector \emph{representation} in a definite basis, $\phi_n=\langle n|\phi\rangle$, and similarly  the density \emph{operator} $\rho$ from the density \emph{matrix}, $\rho_{mn}=\langle m|\rho|n\rangle$.}.
As basic states, one may choose the \emph{magnetic} states $|\lambda\rangle$, eigenstates of $S_z$ ($\lambda=0,\pm1$).
% \emph{magnetic basis} $\{|+1\rangle,|0\rangle,|-1\rangle\}$ 
For $\zu$-axis along the particle momentum, $\lambda$ is the helicity. In this section, however, we will mainly use the \emph{linearly polarised} (LP) basic states $\{|\xu\rangle,|\yu\rangle,|\zu\rangle\}$, also denoted $\{|1\rangle,|2\rangle,|3\rangle\}$. In what follows, generic LP states will be labelled by Latin letters $i,j,k$ or $l\in\{1,2,3\}$, and  magnetic states by Greek letters $\lambda$ or $\mu$. The magnetic and LP bases are related by
\be\label{basic:eq:hel_xyz}
|\lambda=0\rangle=|\hat{\vec{z}}\rangle~, \qquad
 |\lambda=\pm1\rangle=\mp 2^{-1/2}\left(\,|\hat{\vec{x}}\rangle\pm i|\hat{\vec{y}}\rangle\right)~.
\ee
For a partially polarised state, the $3\!\times\!3$ density matrices in the magnetic and LP bases are related by
%$\rho^{M}_{\lambda\lambda'}=\langle \lambda|\rho|\lambda'\rangle$, $\rho^{L}_{ll'}=\langle l|\rho|l'\rangle$
\be
\rho^{M}_{\lambda\lambda'} =\sum_{ll'} \langle\lambda|l\rangle\,\rho^{(LP)}_{ll'} \,\langle l'|\lambda'\rangle~.
\ee

\paragraph{Expansion in basic operators.}
The density operator $\rho$ can be expanded on 9 basic operators $\Sigma_\alpha$, which are mutually orthogonal in the sense of
\be\label{basic:eq:orth-spin1}
\Tr \left(\Sigma_\beta^\dagger \Sigma_{\alpha}\right) = g_\alpha\,\delta_{\beta\alpha}~.
\ee
The squared norms~ $g_\alpha=\Tr\left(\Sigma_\alpha^\dagger\, \Sigma_\alpha\right)$ may be taken different from unity. Writing
\be\label{basic:eq:compact}
\rho= \sum^{}_{\alpha}\,g_\alpha^{-1}\,P_\alpha \,\Sigma_\alpha^\dagger~,
\ee
the \emph{polarisation parameters} $P_\alpha$ are simply the expectation values of the corresponding $\Sigma_\alpha$:
\be\label{basic:eq:expectation}
P_\alpha= \left\langle\Sigma_\alpha\right\rangle=\Tr\left(\rho\,\Sigma_\alpha\right)~.
\ee
The $\Sigma_\alpha$'s can be chosen either self-Hermitian or Hermitian conjugate by pairs (up to a sign). Then the $P_\alpha$'s are either real or complex conjugate by pairs (up to a sign), and hence depend on 9 real parameters (8 if the condition $\Tr\rho=1$ is imposed). Equations~(\ref{basic:eq:compact}-\ref{basic:eq:expectation}) are analogous to (\ref{basic:eq:spin12b}-\ref{basic:eq:spin12c}), with $\sigma_0\equiv \mathbbm{1}$, $P_0\equiv 1$. 

A first explicit form of (\ref{basic:eq:compact}) is simply $\rho=\sum^{}_{i,j}\,\rho_{ij}\, |i\rangle\langle j|$. Then
\be\label{basic:eq:SimpleBasis}
\Sigma_\alpha=\Sigma_{\{ii'\}}=|i'\rangle\langle i| \,,\qquad g_\alpha=1 \,,
\ee
where $\alpha=\{ii'\}$ is a double index and $i,\,i'\in\{x,y,z\}$. For instance, $\Sigma_{\{xy\}}=|y\rangle\langle x|$,  $\ \left\langle\Sigma_{\{xy\}}\right\rangle =\rho_{xy}$. This operator basis will be used in Sec.~\ref{excl:sub:photvec}.

\paragraph{Cartesian parameters.}
Taking the linearly polarised basis, we may decompose $\rho_{ij}$ in a pure trace part ${1\over3}\delta_{jk}$, a real symmetric traceless part $T_{ij}=T_{ji}$ and an imaginary anti-symmetric part,
\be\label{basic:eq:sym+antisym}
\rho=\rho_{ij}|i\rangle \langle j|=\left[{1\over3}\,\delta_{ij}+T_{ij}
-{i\over2}\varepsilon_{ijk} \,P_k\right] |i\rangle \langle j|~,
\ee
with the constraint $T_{xx}+T_{yy}+T_{zz}=0$. 
$\ \varepsilon_{ijk}$ is the fully antisymmetric tensor. Accordingly, for Eq.~(\ref{basic:eq:compact}) one may take the following operators: 
\be\label{basic:eq:decompose}\begin{aligned}
&\Sigma_1= {|\yu\rangle\langle \zu|-|\zu\rangle\langle \yu|\over i}\,,
&&\Sigma_2={|\zu\rangle\langle \xu|-|\xu\rangle\langle \zu|\over i}\,,
&&\Sigma_3={|\xu\rangle\langle \yu|-|\yu\rangle\langle \xu|\over i}\,,\\
&\Sigma_4= |\yu\rangle\langle \zu|+|\zu\rangle\langle \yu|\,,
&&\Sigma_5=|\zu\rangle\langle \xu|+|\xu\rangle\langle \zu|\,,
&&\Sigma_6=|\xu\rangle\langle \yu|+|\yu\rangle\langle \xu|\,,\\
&\Sigma_7=|\xu\rangle\langle \xu|-|\yu\rangle\langle \yu| \,,
&&\Sigma_8=|\zu\rangle\langle \zu| - {|\xu\rangle\langle \xu|+|\yu\rangle\langle \yu|\over 2}\,,
&&\Sigma_0=\mathbbm{1}\,.\end{aligned}
\ee 
$(\Sigma_1,\Sigma_2,\Sigma_3)\equiv\Sv$ is the spin operator and $(P_1,P_2,P_3)=\Pv$ the associated axial polarisation. $P_4, ...P_8$ are the \emph{Cartesian} components of the tensor polarisation, up to normalisation factors. $(P_4,\linebreak[2]{P_5},\linebreak[2]{P_6})=2\RE(\rho_{yz},\rho_{zx},\rho_{xy})$ form the non-diagonal part of it.
%$P_7=T_{xx}-T_{yy}\in[-1,+1]$ generalises the Stokes parameter $S_3$ to massive vector mesons moving along $\zu$. 
$P_3$, $P_6$ and $P_7$, which are in $[-1,+1]$, generalise the Stokes
parameters $S_\odot$,
$S_\oslash$ and $S_\ominus$ to massive vector mesons moving along $\zu$.
$P_8\equiv{3\over2}T_{zz}\in[-{1/2},+1]$ is the \emph{alignment parameter} in the $\zu$ direction.

The operator squared norms are $g_0=3$, $\ g_1,\cdots g_7=2$, $\ g_8={3\over2}$. Apart from ordering and normalisation, $\Sigma_1, ...\Sigma_8$ are the Gell-Mann matrices of SU(3). In literature, the $\Sigma_\alpha$ for $\alpha\ge4$ are often normalised to unit trace: $g_\alpha=1$. We will not do it here, to avoid square root numerical factors. 

The density matrix in the linear polarisation basis reads 
\be
\left(\rho_{ij}\right)={1\over2}
\begin{pmatrix}
{2\over3}(1-P_8)+P_7&P_6-iP_3&P_5+iP_2\\
P_6+iP_3&{2\over3}(1-P_8)-P_7&P_4-iP_1\\
P_5-iP_2&P_4+iP_1&{2\over3}(1+2P_8)\end{pmatrix}~,
\label{basic:eq:rhoSpin1'}
\ee 
taking the constraint $\Tr\rho=1$ into account.

The positivity of $\rho$, written in the form (\ref{basic:eq:rhoSpin1'}), imposes the linear, quadratic and cubic constraints 
\be\label{basic:eq:linquacub}\begin{split}
P_1,...P_7\in[-1,+1]~, \quad&\quad\quad  P_8\in[-{1/2},+1]~,\\
(4/9)\,(1-P_8)^2 &\ge P_3^2 + P_6^2 + P_7^2~,\\
(2/3)\,(1+2P_8)\,\left[(2/3)\,(1-P_8)+P_7\right]&\ge P_2^2 + P_5^2 ~,\\
(2/3)\,(1+2P_8)\,\left[(2/3)\,(1-P_8)-P_7\right]&\ge P_1^2 + P_4^2 ~,\\
\det(\rho_{ij})& \ge 0~.\end{split}
\ee
The symmetric part of $\rho_{ij}$, 
\be\label{basic:eq:symmetric-part}
\Theta_{ij}=\RE\rho_{ij}={1\over3}\,\delta_{ij}+T_{ij}
\ee
is by itself positive semi-definite. 
One can parametrise $T_{ij}$ by two eigenvalues $t'$ and $t''$ (the third one is $t^{'''}=-t'-t''$) and the three Euler angles giving the orientation of the triad of eigenvectors. The remaining parameters of $\rho$ can be the projections $(P',P'',P^{'''})$ of $\Pv$ on this triad. Positivity constrains only the five rotation-invariant quantities $\{t',t'',P',P'',P^{'''}\}$.  
\paragraph{Spherical tensor decomposition}
The \emph{spherical tensor operators}, which transform under the rotation group like eigenstates of $J$ and $J_z$, are convenient when using the magnetic basis for states. Equation~(\ref{basic:eq:compact}) is then written in the fourth form
\be \label{basic:eq:Spherical:rho} 
\rho={1\over3}\mathbbm{1}+{1\over2}\sum^{+1}_{-1}\PvS_\lambda\,\SvS_\lambda^\dagger
+\sum^{+2}_{-2}T_{2\mu}\, \tau_{2\mu}^\dagger~. 
\ee
$\PvS_\lambda$ and $\SvS_\lambda$ are the magnetic components of the vector polarisation and the spin operator, which is a first-rank ($J=1$) tensor. Using the sign convention of (\ref{basic:eq:hel_xyz}), $\PvS_0=P_z$, $~\PvS_{\pm1}=\mp(P_x\pm iP_y)/\sqrt{2}$ and 
\be
\SvS_{0}=S_z=\begin{pmatrix}%
1&0&0\\      0&0&0\\              0&0&-1    \end{pmatrix},\quad
\SvS_{-1}=\frac{S_x-iS_y}{\sqrt{2}}=
\begin{pmatrix}  0&0&0\\   1&0&0\\  0&1&0\end{pmatrix}
=-\SvS_{+1}^\dagger~.
\ee 
The normalised second-rank ($J=2$) spherical tensor operator $\tau_2$ are combinations of  products of the $J=1$ operators: 
\be\label{basic:eq:defin-spheric-tau}
\tau_{2\mu}=[\SvS\otimes
\SvS]^{2}_{\mu}=\sum_{\lambda,\lambda'=0,\mp 1}%\sum_{\lambda'=0,\mp 1}
\langle1,1;\lambda,\lambda'|2,\mu\rangle\,\SvS_{\lambda}\SvS_{\lambda'}~,
\ee
where $\langle1,1;\lambda,\lambda'|2,\mu\rangle$ are the usual Clebsch--Gordan coefficients. In the magnetic basis,
\be
 \tau_{20}={1\over\sqrt{6}} \begin{pmatrix}%
  1 & 0 & 0 \\
  0 & -2 & 0 \\
  0 & 0 & 1 \end{pmatrix}~,\ 
\tau_{21}={1\over\sqrt{2}}\begin{pmatrix}%
  0& -1 & 0 \\
  0 & 0 & 1 \\
  0 & 0 & 0 \end{pmatrix} = - \tau^\dagger_{2-1}~, \ 
\tau_{22}=\begin{pmatrix}
  0& 0 & 1 \\
 0 & 0 & 0 \\
  0 & 0 & 0\end{pmatrix}           =\tau^\dagger_{2-2}~.
\ee
%They are normalised to unity: $\tr (\tau_{2\mu}\ \tau_{2\mu}^\dagger)=1$.
The $\tau_{2\mu}$ operators are related to the $\Sigma_{\alpha}$ of (\ref{basic:eq:decompose}) by %(\ref{basic:hel_xyz}) {basic:eq:traceless}
\be
\tau_{20}=-\sqrt{2/3}\,\Sigma_8~,\quad\tau_{21}=(\Sigma_5+i\Sigma_4)/2~,\quad\tau_{22}=-(\Sigma_7+i\Sigma_6)/2~.
\ee
The same relations hold for the expectation values, $P_\alpha$ and $T_{2\mu}$.

Using the form (\ref{basic:eq:Spherical:rho}) one can write the positivity of $\rho$ in terms of the spherical parameters $\SvS_\lambda$'s and the $T_{2\mu}$'s \cite{Daboul:1967,Kloet:1999wc,Kloet:2000si}. The constraints (not written here) are globally equivalent to (\ref{basic:eq:linquacub}). 
%Writing the matrix $\langle \lambda|\rho|\lambda'\rangle$ in function of the $\SvS_\lambda$'s and the $T_{2\mu}$'s  one can derive the positivity constraints between these parameters \cite{Daboul:1967,Kloet:1999wc,Kloet:2000si}. They are globally, but not one-to-one equivalent to (\ref{basic:eq:linquacub}), except for the cubic one.

\paragraph{Decay of a polarised vector meson in two spinless mesons.}
Each set of basic operators has its advantages and disadvantages. 
%The constraint of parity conservation takes a more simple form when using (\ref{basic:eq:decompose}).  
The angular distribution in the decay $V\to$ 2 spin-0 mesons is most simply expressed in terms of the tensor $\Theta_{ij}$ of (\ref{basic:eq:symmetric-part}). In the $V$ rest frame it reads
\be
f(\ku)= {\d N \over \d^2\ku}={3\over4\pi}\, \Theta_{ij}\,\ku_i \ku_j~, 
\label{basic:eq:angul-distrib}
\ee
where $\kv$ is the relative momentum and $\ku=\kv/|\kv|$. The corresponding expression %in the magnetic basis 
with spherical tensor parameters can be found in \cite{Schilling:1969um}. From experimental data, one can build the moment matrix
\be
f_{mn}= \int \d^2\ku\,f(\ku)\,\ku_m \ku_n = {2\Theta_{mn} + \Theta_{ii} \,\delta_{mn}\over5}~,
\label{basic:eq:Tij-mom}
\ee
from which one gets
\be
\Theta_{ij}= {5f_{ij} - f_{mm} \,\delta_{ij}\over2}~. 
\label{basic:eq:mom-Tij}
\ee
The trace of $f_{ij}$ is equal to that of $\Theta_{ij}$, but its traceless part (related to the asphericity of $f(\ku)$) is $2/5$ that of $\Theta_{ij}$. A too aspherical $f(\ku)$ would violate the positivity of $\Theta_{ij}$ and could reveal an interference with other resonances or a non-resonant background. %

To measure the axial polarisation $\vec P$ of a vector meson, at least a three-body decay is needed, in order to build an axial vector from the momenta. It works for the $\mathrm{a}_1$ meson \cite{Efremov:1992qr}, but not for the $\omega$, $\phi$ or $\mathrm{J}/\psi$ mesons.
\subsubsection{Phase conventions for the helicity and transversity states}
\label{basic:sub:phase-conv}
It is important to define the one- and two-particle states completely, not up to a phase. Let us work in the centre-of-mass frame (the transformation to other frame will not be treated in this report). For the space coordinates one first chooses a \emph{global frame} $\{\xu,\yu,\zu\}$ and introduces the primary helicity states $|p\zu,\lambda\rangle=|p\zu\rangle\otimes|\lambda\rangle$ for a particle moving along $+\zu$. For a particle of arbitrary momentum, $\pv=(p,\theta,\phi)$ in polar coordinates, the helicity state $|\pv,\lambda\rangle$ we take the convention of \cite{Jacob:1959at}:
\begin{align} 
|\pv,\lambda\rangle &={R}(\zu\to\pu) \, |p\zu,\lambda\rangle \,,
\label{basic:eq:helistate-general} \\
{R}(\zu\to\pu) &=
\exp\left(-i\vec\theta.\vec J\right) =
\exp(-i\phi J_z)\,\exp(-i\theta J_y)\,\exp(+i\phi J_z)\,.
\label{basic:eq:helistate}
\end{align}
${R}(\zu\to\pu)$ of (\ref{basic:eq:helistate}) is the rotation  of angle $\theta$ about $\vec\theta=\theta\,(\zu\times\pv)/|\zu\times\pv|$. It is the ``minimal'' rotation which transports $\zu$ to $\pu$. For a spin one-half,
\be
\exp\left(-i\vec\theta.\vec J\right)=(\cos\theta/2)^{-1/2}\,(1+\vec\sigma.\pu\,\vec\sigma.\zu)\,.
\ee
$|\pv,\lambda\rangle$ defined by (\ref{basic:eq:helistate-general}--\ref{basic:eq:helistate}) is a continuous function of $\pu=\pv/|\pv|$ on the whole unit sphere except at the point $-\zu$, where its phase is undefined, unless one fixes $\phi$ with some convention. 

The components of the particle polarisation will be measured in the \emph{particular frame} $\{\xu_{\pv},\linebreak[1]{\yu_{\pv},}\linebreak[2]{\zu_{\pv}\}}$ obtained from $\{\xu,\yu,\zu\}$ by the same rotation ${R}(\zu\to\pu)$. This is important for the validity of formula like (\ref{basic:eq:cs4pol}) or (\ref{basic:eq:corr-param}). 
The particular frames are also necessary to define properly the transversity and linearly polarised states which fulfil (\ref{basic:eq:transv.state}),  (\ref{basic:eq:ph2}--\ref{basic:eq:ph3}) and (\ref{basic:eq:hel_xyz}).
In $2\to2$ reactions, if we are not making the partial wave decomposition, we take $\{\xu,\yu,\zu\}=\{\vec{m}_a,\vec{n},\vec{l}_a\}$ of Fig.~\ref{basic:fig:frame} as global frame, and $\{\vec{m}_i,\vec{n},\vec{l}_i\}$ ($i=a,...d)$ as particular frames.

Another convention, without the $\exp(+i\phi J_z)$ factor in (\ref{basic:eq:helistate}), is also used \cite{Bourrely:1980mr,Leader:2001gr}. The resulting $|\pv,\lambda\rangle$ differs from ours by the phase factor $\exp(-i\phi\lambda)$. For half-integer spin, it is double-valued or discontinuous along some line joining $+\zu$ and $-\zu$ on the unit sphere. The particular axes, called ``helicity frame'', are simply $\{\xu_{\pv},\yu_{\pv},\zu_{\pv}\}=\{\vec{e}_\theta,\vec{e}_\phi,\vec{e}_r\}$, \ie, tangent to the meridian, the parallel and radial vector of the polar coordinate system at point $\pu$.

A  \emph{two-particle helicity state} in its centre-of-mass frame may be written as
 \be\label{2-particle}
% |\kv;A,\alpha;B,\beta\rangle 
 |\kv;a_1,\lambda_1;a_2,\lambda_2\rangle =
% |\kv,A,\alpha\rangle \otimes  |-\kv,B,\beta\rangle 
 |a_1,\kv,\lambda_1\rangle\otimes|a_2,-\kv,\lambda_2\rangle \,
 \ee
where $\kv$ is the relative momentum and $a_1$ and $a_2$ are the particle types. Jacob and Wick \cite{Jacob:1959at} give a definition which differs from ours by the phase factor $(-1)^{s_2+\lambda_2}\,e^{-2i\lambda_2\phi(\kv)}$. Their convention is convenient for the partial wave decomposition but will not be used in this report.

\paragraph{Application to \boldmath$a+b\to c+d$\unboldmath.} 
With $\{\xu,\yu,\zu\}=\{\vec{m}_a,\vec{n},\vec{l}_a\}$ in Fig.\ref{basic:fig:frame}, we have
$(\theta_c,\phi_c)=(\theta_{\rm cm},0)$ and $(\theta_d,\phi_d)=(\pi-\theta_{\rm cm},\pi)$. For particle $b$, we have $\theta_b=\pi$ and we chose $\phi_b=\pi$, \ie, ${\vec{p}}_{b,x}=-0$. For  a boson, it is equivalent to $\phi_b=0$, but not for a fermion. The standard centre-of-mass helicity amplitude is 
\be\label{basic:eq:heli-ampli}
\langle\gamma,\delta|\mathcal{M}^{ab\to cd}(W,\theta_{\rm cm})|\alpha,\beta\rangle 
\equiv
\langle\pv';c,\gamma;d,\delta|\mathcal{M}|\pv;a,\alpha;b,\beta\rangle 
\quad {\rm at}\quad \phi(\pv')=0 \,.
\ee
The helicity amplitudes $f_{\gamma\delta\alpha\beta}$ of Jacob-Wick and $H_{\gamma\delta\alpha\beta}$ of \cite{Bourrely:1980mr,Leader:2001gr} differ from our convention by the factors $(-1)^{s_b-\beta+s_d-\delta}$ and $e^{-i\pi(\beta-\delta)}$ respectively. In order to apply (\ref{basic:eq:cs4pol}) or (\ref{basic:eq:corr-param}) they both require particular frames for $b$ and $d$ which are symmetric about $\vec l$ from those of Fig.\ref{basic:fig:frame}, \ie, 
$\,\xu_{b,d}=-\vec{m}_{b,d}$, $\yu_{b,d}=-\vec{n}$. 

\subsection{Joint density matrix of several particles}\label{basic:sub:several}
A two-particle system $A+B$ is described by a \emph{joint} density matrix $\rho_{AB}$. 
The density matrix of $A$ alone, $B$ being not analysed, is given by the \emph{partial trace} over $B$, denoted 
\be\label{basic:eq:emitA(A+Bu)}
\rho_{A}=\Tr_{B}\,\rho_{AB}~,
\ee
and defined by
\be \label{basic:eq:part-tr}
\langle a|\rho_{A}|a'\rangle=\sum_b\langle a,b|\rho_{AB}|a',b\rangle ~,
\ee
where $|a\rangle$ denotes a spin state of $A$. This can be generalised to more than two particles.
% and the convention of summation over repeated indices is used. 

The spin correlations contained in $\rho_{AB}$ may be of \emph{classical} nature, in which case $\rho_{AB}$ can be decomposed as 
\be\label{basic:eq:sep}
\rho_{AB}=\sum_n w_n\,\rho_n(A)\otimes\rho_n(B) \quad {\rm with}\quad w_n\ge0~,
\ee
and is called \emph{separable}. They may also be \emph{non-classical}, in which case $\rho_{AB}$ cannot be written as (\ref{basic:eq:sep}) and is called \emph{entangled}. A standard example of entangled density matrix is the spin singlet projector
\be\label{basic:eq:singlet} 
\rho_{AB}={\mathbbm{1}-\vec\sigma_A.\vec\sigma_B\over4}~.
\ee
This state is known to violate the Bell inequalities \cite{Bell:1964kc}. There are entangled states which do not violate the Bell inequalities, so that these inequalities are not a sufficient condition of separability. A more severe condition of separability is the positivity of the \emph{partially transposed} matrix $\langle a,b|\rho^{\rm pt}_{AB}|a',b'\rangle=\langle a,b'|\rho_{AB}|a',b\rangle$ \cite{Peres:1996dw,Horodecki:1996nc,Horodecki:2001xx}. This condition is in fact sufficient for two spin-1/2 particles, or a system of a spin 1/2 and a spin 1.

\subsection{Description of the reactions}\label{basic:eq:csm}

Consider the reaction with polarised beam and target and final spin analysed,
\be\label{basic:eq:reac}
\overrightarrow{A}+\overrightarrow{B} \to \overrightarrow{C}+ \overrightarrow{D}~,
\ee
at fixed momenta $\pv_A$, $\pv_B$, $\pv_C$ and $\pv_D$. The scattering amplitude is written as
\be\label{basic:eq:ampl}
\langle c|\otimes\langle d |\,\mathcal{M}%(\pv_A,\pv_B,\pv_C,\pv_D)
\,|a\rangle\otimes|b\rangle \equiv\langle c,d|\,\mathcal{M}\,|a,b\rangle~.
\ee
% $\langle c,d|\ \mathcal{M}\ |a,b\rangle$
Depending on the adopted basis, the  letter $a$ denotes the helicity or the transversity of $A$
%XA/JUIN:
or, in nonrelativistic physics, $s_z(A)$ in a global $\{\xu,\yu,\zu\}$ frame. For a vector particle, it can also denote the direction of its linear polarisation. 
Squaring $\mathcal{M}$, averaging over the initial spin states and summing over the final ones give the unpolarised differential cross section
\be\label{basic:eq:dcs}
{d\sigma\over d\Omega}={1\over n_An_B}\,\sum_{a,b,c,d}|\langle c,d|\,\mathcal{M}\,|a,b\rangle|^2={1\over n_An_B}\,\Tr\{\mathcal{M}\,\mathcal{M}^\dagger\}~.
\ee
where $n_p$ is the number of possible spin states of particle $p$ ($n_p=2s_p+1$ for a massive particle). The \emph{fully polarised} differential cross section (\ref{basic:eq:Cijkl}) reads, for spin-1/2 particles,
\be
{d\sigma\over d\Omega}\left(\Sv_A,\Sv_B,\Sv_C,\Sv_D \right) =
\Tr\{\mathcal{M}\,{\mathbbm{1}+\Sv_A.\vec\sigma_A\over2}\otimes{\mathbbm{1}+\Sv_B.\vec\sigma_B\over2}
\,\mathcal{M}^\dagger {\mathbbm{1}+\Sv_C.\vec\sigma_C\over2}\otimes{\mathbbm{1}+\Sv_D.\vec\sigma_D\over2}\} 
\label{basic:eq:cs4pol}
\ee
%
%XA/JUIN:
(let us recall that the $\Sv_{\rm p}$ and $\vec{\sigma}_{\rm p}$ components for particle p are measured in the particular frame $\{\xu_{\rm p},\yu_{\rm p},\zu_{\rm p}\}$, which generally differs from one particle to another).
More generally, the fully or partially polarised differential cross-section, for any spins, takes the form
\be\label{basic:eq:dpcs1}
{d\sigma\over d\Omega}\left(\rho_A,\rho_B,\acc_C,\acc_D \right) =
\Tr\{\mathcal{M}\,[\rho_A \otimes\rho_B]\,\mathcal{M}^\dagger[\acc_C\otimes\acc_D]\}~.
\ee
The density matrices $\rho_A$ and $\rho_B$ of the initial particles depend on the beam and target preparation and have unit trace. If $A$ is unpolarised, $\rho_A=\mathbbm{1}_A/n_A$.~ $\acc_C$ is not the \emph{outgoing} density matrices of $C$ given by (\ref{basic:eq:reac-rhoC}), but specifies the states accepted by the detector \cite{Landau:101815}. We call it an \emph{acceptance} matrix and put an inverted hat on it. Two extreme cases are $\acc_C=|c\rangle\langle c|$ for an ideal detector selecting only state $|c\rangle$, and $\acc_C=\mathbbm{1}_C$ for a non-analysing detector (or no detector at all for particle $C$). $\acc_C$ has not necessarily unit trace, but its eigenvalues cannot exceed one. The dual roles of density and acceptance matrices will be further discussed in Sec.~\ref{se:furth}.

The \emph{joint outgoing} density matrix of particle $C$ and $D$ is 
\be  \label{basic:eq:reac-rhoCD}   
\rho_{CD}=\mathcal{M}\,[\rho_{A}\otimes \rho_{B}]\,\mathcal{M}^\dagger/
\Tr\{\mathcal{M}\,[\rho_{A}\otimes \rho_{B}]\,\mathcal{M}^\dagger\}~.
\ee
It depends on the initial particle polarisations. The outgoing density matrix of particle $C$ alone is, according to (\ref{basic:eq:emitA(A+Bu)}),
\be  \label{basic:eq:reac-rhoC}   
\rho_{C}\equiv{1\over2}\,\left(\mathbbm{1}+\langle\vec{S}_C\rangle.\vec{\sigma}_C\right)=
\Tr_D\{\mathcal{M}\,[\rho_{A}\otimes \rho_{B}]\,\mathcal{M}^\dagger\,\}/
\Tr\{\mathcal{M}\,[\rho_{A}\otimes \rho_{B}]\,\mathcal{M}^\dagger\}~,
\ee
where $\langle\vec{S}_c\rangle$ is the output polarisation of $C$ given by (\ref{basic:eq:polar-c}).

The correlation parameters entering (\ref{basic:eq:Cijkl}) for a $1/2+1/2\to1/2+1/2$ reaction are given by
\be\label{basic:eq:corr-param}
C_{\lambda\mu\nu\tau} = \Tr\left\{\mathcal{M}\left[\sigma_\lambda(A)\otimes\sigma_\mu(B)\right]\,\mathcal{M}^\dagger
\left[\sigma_\nu(C)\otimes\sigma_\tau(D)\right]\}/\Tr\{\mathcal{M}\,\mathcal{M}^\dagger\right\}~,
\ee
which will be symbolically abbreviated as a sort of expectation value:
\be
(\lambda\mu|\nu\tau)\equiv C_{\lambda\mu\nu\tau}=\langle\sigma_\lambda(A)\sigma_\mu(B)\sigma_\nu(C)\sigma_\tau(D)\rangle~,\label{basic:eq:corr-param'}
\ee
up to transpositions for  $\sigma_\nu(C)$ and $\sigma_\tau(D)$, as discussed in the next section.
\subsubsection{The Cross Section Matrix}\label{basic:sib:csm}
All spin observables of reaction (\ref{basic:eq:reac}) can be encoded in the \emph{cross section matrix} (CSM) $\R$, or its partial transpose $\Rt$, defined by
\be\label{basic:eq:CSdM}\begin{aligned} 
\langle c,d | \mathcal{M}| a,b \rangle\,\langle a',b' | \mathcal{M}^\dagger| c',d' \rangle 
&=\langle a',b';c',d'| \R |a,b\,;c,d\rangle\\
&=\langle a',b';c,d |\Rt|a,b\,;c',d'\rangle
\,.\end{aligned}\ee
The transposition linking $\Rt$ to $\R$ concerns the final particles\,\footnote{%
Alternatively, keeping the same $\Rt$, one may define $\R$ as the full transpose of that given by (\ref{basic:eq:CSdM}). Then the partial transposition between $\Rt$ and $\R$ would apply to the initial particles. This choice was done in Ref.\cite{Artru:2004jx}, where $\R$ is called ``grand density matrix''.
}. 
The diagonal elements of $\R$ or $\Rt$ are the fully polarised cross sections when the particles are in the basic spin states. The knowledge of the non-diagonal elements allows a change of spin basis, thus the CSM describes the spin correlations in any directions.  
By construction, the $\R$ is semi-positive definite and of rank one. This is not generally the case for $\Rt$. On the other hand, $\Rt$ respects the \emph{bra} and \emph{ket} assignments of variables $c,d,c',d'$ in (\ref{basic:eq:CSdM}) and will sometimes lead to simpler formulas. In the following, we shall use either $\R$ or $\Rt$.

Equations (\ref{basic:eq:dcs}), (\ref{basic:eq:dpcs1}), (\ref{basic:eq:reac-rhoCD}) and (\ref{basic:eq:corr-param}) can be rewritten as:
\be\label{basic:eq:dpcsR}\begin{aligned}
{\d\sigma\over \d\Omega} &=\Tr\R\,/(n_An_B)=\Tr\Rt\,/(n_An_B)~,\\[-2pt]
{\d\sigma\over \d\Omega}\left(\rho_A,\rho_B,\acc_C,\acc_D \right) &=
\Tr\{\Rt\,[\rho_A\otimes\rho_B\otimes\acc_C\otimes\acc_D]\,\}~,\\[-1pt]
\rho_{CD}&=\Tr_{A,B}\{\Rt\,[\rho_A\otimes\rho_B]\,\}
/\Tr\{\Rt[\rho_A\otimes\rho_B]\,\}~,\\[3pt]
C_{\lambda\mu\nu\tau}&=\Tr\{{\R}[\sigma_\lambda(A)\otimes 
\sigma_\mu(B)\otimes \sigma_\nu^{t}(C)\otimes
\sigma_\tau^{t}(D)]\}/\Tr{\R}\\[-1pt]
&= \Tr\{\Rt\left[
\sigma_\lambda(A)\otimes\sigma_\mu(B)\otimes\sigma_\nu(C)\otimes\sigma_\tau(D)\right]\}\, /\,\Tr \Rt~.
\end{aligned}
\ee 
The last equation can be generalised to any kind of initial and final observables as
\be\label{basic:eq:R-obs}
\left\langle \Ocal_i \otimes\Ocal_f\right\rangle=
\Tr\{\Rt\left(\Ocal_i \otimes \Ocal_f\right)\}/\Tr \Rt~,\quad{\rm or}\quad
\Tr\{\R\left(\Ocal_i \otimes \Ocal^t_f\right)\}/\Tr \R~.
\ee 
Only the second expression is a \emph{bona fide} expectation value, since $\Rt$ is not necessarily semi-positive.

The partial transposition from $\R$ to $\Rt$ amounts to a $|\emph{ket}\rangle\leftrightarrow\langle \emph{bra}|$ exchange for the final particles, putting these in the same state as the initial particles in $\R$. This evokes the ``crossing'' operation 
\be
\label{basic:eq:crossing'} 
\langle c,d|\mathcal{M}|a,b\rangle\leftrightarrow
        \langle\hbox{\O}|\mathcal{M}|a,b,c,d\rangle ~.   
\ee
Although this operation is not the usual crossing (it does not reverse the charges, spins and four-momenta of particles $C$ and $D$), the right-hand side of (\ref{basic:eq:crossing'}) has the same spin structure as the amplitudes of the true crossed reactions
\be\label{basic:eq:vacuum}
{\rm (a)}\quad A+B+\overline C+\overline D \to \hbox{\O}~, \quad\hbox{or} \quad {\rm (b)}\quad
\mbox{\O} \to \overline A+\overline B+C+D~.\ee
(notwithstanding the fact that (a) and (b) have no physical kinematical region). 
Thus the CSM of (\ref{basic:eq:reac}) has the same structure as the density matrix of the final state in (b). The positivity of $\R$ can be interpreted as the positivity of $\rho_{\bar A\bar B,CD}$ in (\ref{basic:eq:vacuum}). 

Using $\Tr\{\sigma_\lambda\sigma_\mu\}=2\delta_{\lambda\mu}$, the fourth equation of (\ref{basic:eq:dpcsR}) can be inverted as 
\be\label{basic:eq:csdm}
\Rh\equiv(N/\Tr\R)\, \R=C_{\lambda\mu\nu\tau}\,\sigma_\lambda(A)\otimes\sigma_\mu(B)\otimes\sigma^t_\nu(C)\otimes\sigma^t_\tau(D)~.
\ee
$\Rh$ is the \emph{normalised} CSM  which has the same trace as the unit matrix, \ie, $\Tr\Rh=N \equiv \linebreak[4]{n_An_Bn_Cn_D}$, so that $\Rh=\mathbbm{1}_N$ when all Cartesian reaction parameters but $C_{0000}$ are vanishing. As an example, the normalised CSM of the reaction $1/2+0\to1/2+0$ is given in Table \ref{basic:tab:R1/2;1/2} in terms of $(\lambda|\mu)\equiv C_{\lambda\mu}$. The spin variables $b$ and $d$ as well as $\sigma_\mu(B)$ and $\sigma_\tau(D)$ have been removed from preceding equations. This table also describes a reduced CSM of the type (\ref{basic:eq:reducedAC}) and will be repeatedly used later on.
\begin{table}[here]
\caption{\label{basic:tab:R1/2;1/2}%
Normalised cross section matrix $\Rh$ of ${1/2}+0\to{1/2}+0$~ ($\Tr\Rh=4$). Axis labels are $\{1,2,3\}=\{m,n,l\}$ in the helicity basis, $\{l,m,n\}$ in the transversity basis.  $(0-3\,|\,1+i2)$, for instance, is a condensed notation for 
$(0|1)+i\,(0|2)-(3|1)-i\,(3|2)$, which may be seen as 
$\left\langle\left(\mathbbm{1}-\sigma_3\right)\otimes\left(\sigma_1+i\sigma_2\right)\right\rangle$, according to Eq.~(\protect\ref{basic:eq:corr-param'}).}
% La definition de \ii est dans le programme principal
\centerline{$
\begin{array}{c|c c c c}
 & ++ & +- & -+ & -- \\     \hline 
++ & (0+\ii3\,|\,0+\ii3) & (0+\ii3\,|\,1+i2) & (1-i2\,|\,0+\ii3) & (1-i2\,|\,1+i2)
\\ %\hline 
+- & (0+\ii3\,|\,1-i2) & (0+\ii3\,|\,0-\ii3) & (1-i2\,|\,1-i2) & (1-i2\,|\,0-\ii3)
\\ %\hline 
-+ & (1+i2\,|\,0+\ii3) & (1+i2\,|\,1+i2) & (0-\ii3\,|\,0+\ii3) & (0-\ii3\,|\,1+i2)
\\ %\hline 		
-- & (1+i2\,|\,1-i2) & ( 1+i2\,|\,0-\ii3) & (0-\ii3\,|\,1-i2) & (0-\ii3\,|\,0-\ii3)
\\ %\hline 		
 \end{array}
$}
\end{table}
\subsubsection{Reduced cross section matrices}\label{basic:sub:redCSM}
A \emph{reduced CSM} can describe a reaction where only a subset of particles is polarised or analysed. It is obtained by taking the partial trace of the complete CSM over the initial unpolarised particles and over the final particles whose spins are not analysed. For instance, 
\be\label{basic:eq:reducedAC} 
%R_{\overrightarrow{A}+B\to\overrightarrow{C}+D} = n_B^{-1}\ \tr_{B,D}\ 
%R_{\overrightarrow{A}+\overrightarrow{B}\to\overrightarrow{C}+\overrightarrow{D}}
\R_{A,C} = n_B^{-1}\,\Tr_{B}\Tr_{D}\R_{AB,CD}~.
\ee
Similarly, $\R_{CD}$ is the trace over the initial spins, divided by $n_An_B$. $~\Rt_{CD}=\R^t_{CD}$ is proportional  to $\rho_{CD}$ of (\ref{basic:eq:reac-rhoCD}) for unpolarised $A$ and $B$. 
%$\rho_A=\mathbbm{1}_A/n_A$, $\rho_B=\mathbbm{1}_B/n_B$.

The reduced CSM's are (semi-)\,positive and can be treated like the complete CSM, except for the fact that they cannot be factorised like (\ref{basic:eq:CSdM}). Therefore their ranks are generally larger than one. The reduction by a non-polarised or non-analysed particle $X$ decreases the dimension of $ \R$ by a factor $n_X$ and increases its rank by the same factor $n_X$, up to the restriction $\text{rank}\le \text{dimension}$. Thus successive reductions first increase, then decrease the rank of $ \R$.

The CSM formalism generalises to \emph{inclusive} or \emph{semi-inclusive} reactions in a straightforward way: an inclusive CSM is the sum of the reduced CSM of all contributing exclusive channels, after integration over the undetected momenta. The cross section matrix $\R_{AB,C}$ of $A+B\to C+X$ is related by crossing to the $A+B+\overline{C}\to X$ and, by unitarity, to the discontinuity of the forward scattering amplitude of $A+B+\overline{C}\to A+B+\overline{C}$ in the variable $(\tilde p_A+\tilde p_B-\tilde p_C)^2$.  Examples will be given later, see, \eg, Eq.~(\ref{incl:eq:4}),  (\ref{incl:eq:M-rho}) or (\ref{incl:eq:gpos'}).

 The cross section matrix is a systematic but not compulsory tool to obtain the  positivity conditions. These can also be derived from Cauchy--Schwartz inequalities. This will be done, for instance in Sec.~\ref{incl:sub:pdpd} for the Soffer inequality.  
%%%
\subsubsection{Physical meaning of the positivity conditions} 
%%%
The positivity of the cross section matrix of reaction (\ref{basic:eq:reac}) stems from the very general, but non-trivial, requirement that the probability of this reaction, as well as \emph{any reaction involving (\ref{basic:eq:reac}) as a sub-process}, is positive. A first condition is that 
the cross section (\ref{basic:eq:cs4pol}) is positive for arbitrary \emph{independent} $\Sv_A$, $\Sv_B$, $\Sv_C$ and $\Sv_D$ in the unit ball $\Sv^2\le1$. Some geometrical properties of the corresponding domain are discussed in Sec.~\ref{se:furth} and in Ref.~\cite{Artru:2008ax}.
 An equivalent condition is that the polarisation of, for instance, outgoing particle $C$ for given $\Sv_A$, $\Sv_B$, and imposed $\Sv_D$, 
\be\label{outgoingPol}
\langle\Sv_C\rangle = \boldsymbol{\nabla}_{\!\Sv_C} I(\Sv_A,\Sv_B,\Sv_C,\Sv_D) \,/\,I(\Sv_A,\Sv_B,\Sv_C=0,\Sv_D) 
\ee
(note the difference with Eq.~(\ref{basic:eq:polar-c}))
lies in the unit ball $\Sv_C^2\le1$ . 
This condition is called \emph{classical} (see Sec.\ref{furth:sub:quantum}).
As we shall see below,  it is not sufficient.

Suppose, for instance, that the reduced CSM for $A$ and $B$ takes the form $\R_{AB}\propto\mathbbm{1}+\sigma^i_A\otimes \sigma^i_B$, equivalent to $(ij|00)=\delta_{ij}$. The cross section, given by the second line of (\ref{basic:eq:dpcsR}) with $\acc_C=\mathbbm{1}_C$, $\acc_D=\mathbbm{1}_D$, is proportional to $1+\Sv_A.\Sv_B$, therefore zero or positive. If $A$ and $B$ are prepared with correlated spins, one has to replace (\ref{basic:eq:dpcsR}) by 
\be\label{basic:eq:initialcorrel}
{d\sigma\over d\Omega}\left(\rho_{AB}\right) = \Tr\{ \R_{AB}\,\rho_{AB}\}~.
\ee
This is still positive for a \emph{separable} $\rho_{AB}$ (see Eq.~(\ref{basic:eq:sep})). However for the \emph{singlet} state (\ref{basic:eq:singlet}), one obtains $d\sigma/d\Omega\propto\Tr\{(\mathbbm{1}+\sigma^i_A\otimes\sigma^i_B)\,(\mathbbm{1}-\sigma^i_A\otimes \sigma^i_B)/4\}=-2$, which is not acceptable. The occurrence of a negative cross section comes from the non-positivity of $\mathbbm{1}+\sigma^i_A\otimes \sigma^i_B$. This example shows that positivity has to be tested not only with factorised or separable states, but also with \emph{entangled} ones.  The latter impose \emph{non-classical} conditions, stronger than the classical ones. More details are given in Sec.~\ref{furth:sub:quantum}.

Similarly, a reduced CSM of the final particles $\R_{CD}\propto(\mathbbm{1}+\sigma^i_C\otimes\sigma^i_D)/4$ is classically, but not quantum-mechanically acceptable. In fact, quantum mechanics does not allow particles $C$ and $D$ to have their three spin components equal.  

An analogous phenomenon occurs for correlations between initial and final spins. Suppose that the reduced CSM $\R_{A,C}$ defined by (\ref{basic:eq:reducedAC}) takes the form $\R_{A,C}\propto\mathbbm{1}-\sigma_A^i\otimes(\sigma^i_C)^t$, equivalent to $(i0|j0)=-\delta_{ij}$. For independent $\Sv_A$ and $\Sv_C$, $d\sigma/d\Omega\propto1-\Sv_A.\Sv_C\ge0$ or, using (\ref{outgoingPol}), $\langle\Sv_C\rangle=-\Sv_A$. This complete reversal of all spin components is acceptable classically, but not quantum-mechanically. Consider for instance the reaction 
\be
\pi^++{}^4\mathrm{He}\to\pi^0+\mathrm{p}+{}^3\mathrm{He}~, 
\ee
which, in the impulse approximation, can be decomposed into 
\be
(i) \quad {}^4\mathrm{He}\to \mathrm{n}+{}^3\mathrm{He}~, \qquad (ii) \quad \pi^++\mathrm{n}
\to\pi^0+\mathrm{p}~. 
\ee
\begin{figure}[!htb]
\begin{minipage}{.5\textwidth}
\centerline{\includegraphics[width=.6\textwidth]{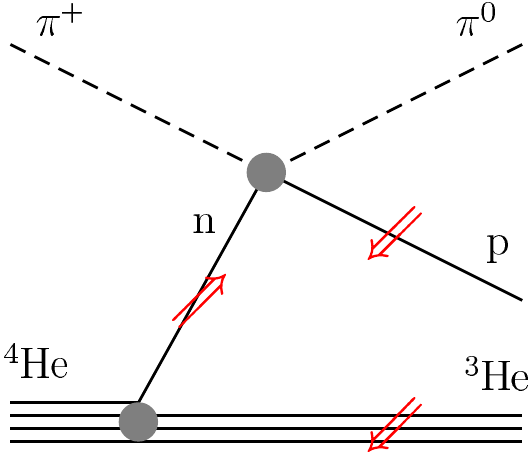}}
\end{minipage}
\begin{minipage}{.49\textwidth}
\caption{\label{basic:fig:Helium}Mechanism for the reaction $\pi^++{}^4\mathrm{He}\to\pi^0+\mathrm{p}+{}^3\mathrm{He}$. The double arrows represent intermediate and final polarisations in the forbidden case $\Sv_p=-\Sv_n$.}
\end{minipage}
\end{figure}
We neglect off mass shell effects, final state interactions, and assume that $ \mathrm{n}+{}^3\mathrm{He}$ is in the singlet $^1\mathrm{S}_0$ wave. Then $\mathrm{n}$ and ${}^3\mathrm{He}$ have all spin components opposite. A complete reversal of the nucleon spin in reaction (ii) would lead to fully equal ${}^3\mathrm{He}$ and $\mathrm{p}$ spins in the final state. This, as shown before, is forbidden. The defect lies in the non-positiveness of $\mathbbm{1}-\sigma_A^i\otimes(\sigma^i_C)^t$. Quantum mechanics does not allow full spin reversal. 

The second counter-example is related to the first one by crossing. A CSM of the form $\mathbbm{1}-\vec\sigma_A.\vec\sigma^t_C$ gives a negative cross section when tested with the entangled ``$t$-channel spin-singlet projector'' $(\mathbbm{1}+\vec\sigma_A.\vec\sigma^t_C)/4$. The lesson of these examples is that positivity  has to be tested with classical and \emph{entangled} states, i.e., which cannot be written as (see Eq.~(\ref{basic:eq:sep})), in the \emph{direct} and \emph{crossed} channels.

\subsection{Search for the positivity domain of a subset of observables}\label{basic:sub:methods}
\subsubsection{Simulation method}
Suppose a reaction described by a set of amplitudes $a$, $b$, \dots, in terms of which the differential cross section is $I=|a|^2+|b|^2+\cdots$ and the spin observables 
$\langle\mathcal{O}_i\rangle$ (times $I$) are given by quadratic expressions such as  $I \langle\mathcal{O}_i\rangle=|a|^2-|b|^2, \cdots$,  $I\langle\mathcal{O}_j\rangle=2\RE(ab^*)+\cdots$, etc.
A simple method to detect inequalities among observables consists of generating random, fictitious amplitudes $a$, $b$, etc. ($a$ can be chosen real and positive), computing the  $\langle\mathcal{O}_i\rangle$, and plotting  the results for  one observable against another one. 
Once the contours revealed, it can be attempted to demonstrate rigorously the corresponding inequalities. 

The method works quite well till about 6 amplitudes, without the need for sophisticated algorithms. This will be illustrated in the case of $\ppLL$. If the number of amplitudes becomes larger, as in the case of the photoproduction of vector mesons, the plots are mostly populated in the centre, and the contour of the domain is less visible. In this case, one should force the random generator to explore the corners.

\subsubsection{Anticommutation and spherical constraints.}\label{basic:sub:antic}
%In the next section, 
Many constraints for reactions with spin one-half particles or photons are of the type
\be
\langle\mathcal{O}_1\rangle^2+\langle\mathcal{O}_2\rangle^2+\cdots\le1~.
\ee
Such disk domains for pairs of observables or spherical domains for triples of observables are, in many cases, straightforward results of anticommutation of the observables. 
Let us recall that the observable $\left\langle\Ocal\right\rangle=C_{\lambda\mu\nu\tau}$
 can be considered as the expectation value of the partially transposed operator $\Ocal^{\rm pt}=\sigma_\lambda(A)\otimes\sigma_\mu(B)\otimes\sigma^t_\nu(C)\otimes\sigma^t_\tau(D)$, according to (\ref{basic:eq:R-obs}). Since $ \sigma_\mu^2=\mathbbm{1}_2$ for each $\mu$, we have $\left(\Ocal^{\rm pt}\right){}^{\!2}=\Ocal^2=\mathbbm{1}$. Furthermore two such operators $\Ocal^{\rm pt}$ and $\Ocal'^{\,\rm pt}$ differing by some indices $\lambda$, $\mu$, $\nu$ or $\tau$ either commute or anticommute. We will forget below the superscript ``${\rm pt}$", since partial transposition does not change the (anti-) commutation properties of $\sigma$-matrices. 

For pairs of anticommuting observables, disk domains result from the following theorem \cite{Artru:2008is}:
\be
\hbox{If } \Ocal^2=\Ocal'^2=\mathbbm{1}\hbox{ and }\Ocal\hbox{ and }\Ocal'\hbox{ are anticommuting, then }
\langle \Ocal\rangle^2+\langle \Ocal'\rangle^2\le1~.
\ee

\ni\textsl{Proof:} set $x=\sqrt{\langle \Ocal\rangle^2+\langle \Ocal'\rangle^2}$, $\langle \Ocal\rangle=a\,x$, $\langle \Ocal'\rangle=b\,x$. Then $a^2+b^2=1$ and $\langle a\,\Ocal+b\,\Ocal'\rangle=x$. From $\Ocal^2=\Ocal'^2=\mathbbm{1}$ and $\Ocal\Ocal'+\Ocal'\Ocal=0$ one gets $(a\Ocal+b\Ocal')^2=\mathbbm{1}$ which means that $a\Ocal+b\Ocal'$ has eigenvalues $\pm1$. Its expectation value $x$ has to be within these eigenvalues, therefore
$x^2\le1$.

\ni\textsl{Generalisation:}  if $\Ocal_1$, $\Ocal_2$, ... $\Ocal_N$ are mutually anticommuting and their squares are the unit matrix,   %$\O\O'+\O'\O=\O'\O"+\O"\O'=\O"\O+\O\O"=0$, 
then $\langle \Ocal_1\rangle^2+\langle \Ocal_2\rangle^2+...+\langle \Ocal_N\rangle^2\le1$. These results apply also to reactions involving photons and gravitons, which are two-components objects. 

 Note that disk or sphere can occur even if the observables commute, for instance if, due to some symmetry, $\Ocal_2$ has the same expectation value as another operator $\Ocal'_2$ which anticommutes with $\Ocal_1$ and $\Ocal_3$. Examples of this situation are indicated by crossed circles in Table~\ref{excl:tab:reca} of Sec.~\ref{excl:sub:empi}.
\subsubsection{Number of amplitudes and existence of constraints}\label{basic:sub:limited}
Suppose that, for instance, 3 observables $\Ocal_1$, $\Ocal_2$, and $\Ocal_3$ of eigenvalues $-1$ and $+1$ are commuting and not related to anticommuting pairs by some symmetry, so that their joint positivity domain $\mathcal{D}_3$ may be larger than the unit ball. Can this domain be the whole cube? A partial negative answer is the following: If the reaction depends on $N$ independent amplitudes, $\mathcal{D}_3$ can reach at most $N$ corners of the cube. The proof is the following: % $\mathcal{D}_3\{\Ocal_1,\Ocal_2,\Ocal_3\}$ 
\begin{enumerate}
\item 
 There is a one-to-one correspondence between matrix amplitudes $\Mcal$ an state vectors $|\Psi\rangle$ of the $\overline A\overline BCD$ channel, defined by
\be\label{basic:eq:crossingM-Psi} 
\langle c,d|\mathcal{M}|a,b\rangle=\langle\Psi|a,b,c,d\rangle ~;   
\ee
the observable $\langle\Ocal\rangle$ reads
\be\label{basic:eq:PsiOPsi} 
\langle\Ocal\rangle=\langle\Psi|\Ocal^{\rm pt}|\Psi\rangle ~.   
\ee
Here again we will forget the superscript ``${\rm pt}$". 
\item 
the number of independent amplitudes is the dimensionality of the Hilbert space of the $|\Psi\rangle$'s.
\item 
Since  $\Ocal_1$, $\Ocal_2$, and $\Ocal_3$ commute, let us take a basis in which they are all diagonal.
% and introduce the 3-component operator $(\Ocal_1,\Ocal_2,\Ocal_3)$. 
Consider a corner $(x_1,x_2,x_3)$ of the cube (each $x_i=\pm1$). If $\mathcal{D}_3$ touches this corner there exists a $\Mcal$- matrix and the corresponding state $|\Psi\rangle$ for which $\langle\Psi|(\Ocal_1,\Ocal_2,\Ocal_3)|\Psi\rangle=(x_1,x_2,x_3)$.
This state is necessarily an eigenstate of $(\Ocal_1,\Ocal_2,\Ocal_3)$, since $x_1$, $x_2$ and $x_3$ are extremal eigenvalues. 
It can be written as $|x_1,x_2,x_3,\nu\rangle$ where $\nu$ stands for possible additional quantum numbers. 
Two states $|x_1,x_2,x_3,\nu\rangle$ and $|x'_1,x'_2,x'_3,\nu'\rangle$ differing by at least one label are orthogonal. 
\item 
if $N_c$ corners are occupied, there are at least $N_c$ mutually orthogonal, hence independent, $|\Psi\rangle$'s. Thus $N_c$ cannot exceed the number of independent amplitudes 
\end{enumerate}
As an example, in the \ppLL\ reaction which depends on 6 amplitudes, only 6 corners of a cube can be reached for a triple of observables. See, \eg, Fig.~\ref{excl:fig:newo} for the case of $\ppLL$.

\subsubsection{Methods of the apparent contour and of the reciprocal polar transform}\label{basic:sub:reciprocal-method}
The method based on anticommutation relations is efficient for finding circular or spherical domains. When it does not give any constraint, the domain may be the square or the cube, or smaller. To get the exact domain, a lengthy but straightforward method consists in determining the \emph{apparent contour} of the positivity domain on the observable plane with the help of differential equations. An example is given in Sec.\ref{excl:sub:underst}. 

A more elegant method is based on the \emph{polar reciprocity between intersection and projection} \cite{Minnaert:1971}, which can be derived from (\ref{basic:eq:recipro-proj-inter}). 
Let us show how it applies to a reaction $1/2+1/2\to 1/2+1/2$. Instead of a unit trace density matrix one works with the cross section matrix normalized to $\Tr\Rh=\Tr(\mathbbm{1}_N)=N$ (here $N=24$).
Following its decomposition (\ref{basic:eq:csdm}) into usual observables, $\Rh$ is represented by a $N2$-dimensional Euclidean vector $\vec C=(C_{0000}, C_{0001},...)\equiv(C_{0000},\vec C_\perp)$ with $C_{0000}=1$ and positivity confines $\vec C_\perp$ in a $N^2-1$ dimensional domain $\mathcal{D}$. The scalar product of two vectors is%
\be\label{}
(1/N)\, \Tr\{\Rh\,\Rh'\}=\vec C\cdot\vec C'=1+ \vec C_\perp\cdot\vec C'_\perp
\ee
and by analogy with (\ref{basic:eq:polar-recip}), we have
\be\label{basic:eq:polar-recip'} %(\ref{basic:eq:polar-recip'}) 
\vec C_\perp\cdot\vec C'_\perp \ge-1\,
%\quad\,\vec C_\perp\in\mathcal{D}\,, \quad\,\vec C'_\perp\in\mathcal{D}\,.
\ee
for $\vec C_\perp$ and $\vec C'_\perp\in\mathcal{D}$. Let us compare the projected domain $\mathcal{D}_3\{\Ocal_1,\Ocal_2,\Ocal_3\}$ of a triplet of observables and the intersection $\mathcal{I}_3$ of $\mathcal{D}$ with the plane 
$(\langle\Ocal_4\rangle,...\langle\Ocal_{N^2-1}\rangle)=0$. In analogy with (\ref{basic:eq:recipro-proj-inter}), (\ref{basic:eq:polar-recip'}) is also satisfied for 
$C_\perp\in\mathcal{D}_3$ and $C'_\perp\in\mathcal{I}_3$, therefore the boundaries $\partial\mathcal{D}_3$ and $\partial\mathcal{I}_3$ are reciprocal under a polar transformation of power $-1$. This is also the case when parity, charge conjugation, etc. have been taken into account. 

This property simplifies the drawing of $\mathcal{D}_3\{\Ocal_1,\Ocal_2,\Ocal_3\}$: we first analyse the different pieces of $\partial\mathcal{I}_3$. This is an easy task since the unobserved variables are just dropped from the equalities and inequalities defining $\mathcal{D}_{{\rm eff}}$.
Then we apply the polar transformation, edge by edge following (\ref{edge-edge}). An example of application of this method will be given in Sec.~\ref{excl:sub:underst}.  

\subsection{Quadratic identities between observables}\label{basic:sub:iqr}
For exclusive reactions described by $n$ independent amplitudes, the matrix $\R$ is of rank one and each $p\!\times\! p$ subdeterminant with $p\ge2$ vanishes. Then all the nonlinear positivity bounds are saturated. The vanishing of the $2\!\times\!2$ determinants (which lead to the vanishing of the higher order ones) gives $\left[n(n-1)/2\right]^2$ real quadratic relations between the $n^2$ real parameters of the density matrix \cite{Delaney:1975fh,Bourrely:1974wp,Klepikov:1967}. These relations are redundant and only $(n-1)^2$ of them can be independent, since the observables depend on $2n-1$ independent real parameters. Consider, for instance, the quadratic relations containing $\rho_{11}$: 
\be\label{basic:eq:quadratic11}
\rho_{11}\,\rho_{ik}-\rho_{i1}\,\rho_{1k} =0~.
\ee
They represent $(n-1)2$ independent real equations. If $\rho_{11}\ne0$, one gets the factorised form of $\rho_{ik}$:
\be\label{basic:eq:quadratic11'}
\rho_{ik} = \rho_{i1}\left(\rho_{1k}/\rho_{11}\right)=f(i)\,g(k)~.
\ee
This factorisation means that $\rho$ is of rank $1$ and all the other $2\!\times\!2$ determinants vanish. Unfortunately, if $\rho_{11}=0$, Eq.~(\ref{basic:eq:quadratic11}) implies $\rho_{i1}=\rho_{1i}=0$ for all $i$ and the only constraint is that the rank is less than $n$. Other subsets of $(n-1)^2$ independent $2\!\times\!2$ relations can be chosen, but again auxiliary inequalities are necessary. For instance the vanishing of the $(n-1)2$ subdeterminants formed by the intersection of two consecutive columns with two consecutive lines induces the vanishing of all the other determinants if $\rho_{22}\,\rho_{33}~...~\rho_{nn}\ne0$. 

\subsection{Symmetry constraints on observables}\label{basic:sub:sym-constraint}
In this section we will present the constraints obtained from various symmetry requirements.

\subsubsection{Rotation invariance (conservation of angular momentum)}\
Rotation invariance is implicitly taken into account in Eqs. (\ref{basic:eq:ana}--\ref{basic:eq:Cij00}), for instance, because the spin components are measured in a coordinate frame linked to the particle momenta and the coefficients only depend on the kinematical invariants. 

Some reactions have cylindrical symmetry about some momentum $\pv$~: forward scattering, total cross section ($A\!+\!B\to X$), two-particle decay, inclusive decay ($A\to B+X$). In these cases the number of independent polarised observables is significantly reduced. As an example, let us treat the decay $\Lambda\to p+ \pi^-$. In the $\Lambda$ rest frame, one may define the asymmetry parameters $(\lambda|\mu)$ with respect to a triad $\{\vec l,\vec m,\vec n\}$ where $\vec l$ is along the direction $\pu$ of the proton, $\vec m$ along $\zu\times\pu$, where $\zu$ is an arbitrarily chosen direction of space, and $\vec n=\vec l\times\vec m$. Due to conservation of $J.\pu$, there is only two non-vanishing helicity amplitudes, $\left\langle+|\mathcal{M}|+\right\rangle$ and $\left\langle-|\mathcal{M}|-\right\rangle$. Then the CSM, which is a $4\!\times\!4$ matrix is of the Table \ref{basic:tab:R1/2;1/2} type, has vanishing elements except in the $2\!\times\!2$ diagonal block acting on the $\{|++\rangle,|--\rangle\}$ subspace. This leads to $(l|l)=1$, $(0|l)=(l|0)\equiv\alpha$ in the notation of the Particle Data Group \cite{Yao:2006px}, $(m|m)=(n|n)\equiv\gamma$ and $-(m|n)=(n|m)\equiv\beta$, all other $(\lambda|\mu)$ being zero. Furthermore, it has rank one, whence the quadratic constraint%
\be
\alpha^2+\beta^2+\gamma^2=1~.  %\Delta A2+D_{mm}2+D_{mn}2=1~.
\ee 
The \emph{analysing power} $\alpha$ governs the angular distribution in the $\Lambda$ frame, $dN/d\Omega\propto1+\alpha \,\pu.\Pv_\Lambda$, where $\pu=\pv/|\pv|$. It is non-zero due to parity non-conservation in weak interactions. The experimental value for $\Lambda\to p+ \pi^-$ is $\alpha=0.642\pm0.013$. The other independent parameter is 
\be\label{basic:eq:twist}
\phi=\arg\left(\gamma+i\beta\right)=
\phi(\Sv_\Lambda)-\phi(\Sv_p)=\arg\left({\Mcal_{++}\over \Mcal_{--}}\right)~.
\ee
The proton transversity is rotated from the $\Lambda$ one by $-\phi~$. Details can be found in \cite{Yao:2006px}.

The same conclusions can be simply obtained using (\ref{basic:eq:corr-param}) and the rotation-invariant form,
\be
\langle\sv_p|\Mcal|\sv_\Lambda\rangle=\langle\sv_p|\,a_S+a_P\,\sigma.\pu\,|\sv_\Lambda\rangle~,
\ee
of the amplitude. $a_S$ and $a_P$ are S- and P-wave amplitudes and, according to Watson's theorem, their phases are the S- and P-wave phase shifts $\delta_S$ and $\delta_P$. A difference between these phase shifts leads to a non-vanishing $\phi$, owing to $\Mcal_{++}=a_S+a_P$,~ $\Mcal_{--}=a_S-a_P$.

Another example is provided by quark distributions, \ie, the reaction $N\to q+X$ (for the definitions, see Sec.~\ref{incl:sub:pdpd}. After integration over the intrinsic transverse momentum, they are invariant by rotation about the nucleon momentum axis ${\vec l}$. The cross section matrix is again of the type of Table \ref{basic:tab:R1/2;1/2}, but $\langle\lambda'_N,\lambda'_q|\R|\lambda_N,\lambda_q\rangle$ is zero unless $\lambda'_N-\lambda'_q=\lambda_N-\lambda_q$ (=$\lambda_X$, due to conservation of $\vec J_l$). This leaves only the diagonal elements and the upper-right and lower-left corners. Furthermore parity conservation imposes $(0|l)=(l|0)=(m|n)=(n|m)=0$. One is left with $(0|0)=1$, $(l|l)=\Delta q(x)/q(x)$ and $(m|m)=(n|n)=\delta q(x)/q(x)$, where $q(x)$ is the unpolarised quark distribution, $\Delta q(x)$ and $\delta q(x)$ are  the helicity and transversity distributions. Part of the parent hadron helicity is taken by other partons and the orbital angular momenta, therefore $(l|l)<1$, unlike in the $\Lambda\to p\pi$ case. Positivity of $\R$ is insured by $\Delta q(x)/q(x)\le1$, $\delta q(x)/q(x)\le1$ and the Soffer inequality, discussed in Sec.~\ref{incl:sub:PD},
\be
2|\delta q(x)|\le q(x)+\Delta q(x)~.
\ee 
%which can be obtained from the positivity of the 2$\times$2 determinant in the $\{|++\rangle, |--\rangle\}$ subspace. 
Similar conclusions hold for the quark fragmentation into baryon $q\to B+X$.
\subsubsection{Parity conservation (in strong and electromagnetic interactions)}
\label{basic:sub:parity}
\paragraph{Case of planar reactions}
The space inversion operator (parity) $P$ transforms a reaction into another one with the same spin vectors but opposite particle momenta. 
In $2\to2$ scattering or in 3-body decay, one can define a \emph{reaction plane} (scattering plane or decay plane). One can also define such a plane in \emph{1-particle inclusive} inelastic scattering or \emph{2-particle inclusive} decay. The set of observed momenta is invariant under the {mirror reflection} with respect to the reaction plane, described by the operator 
\be\label{basic:eq:Pi-parity}
	\Pi = P \,\exp(-i\pi \Jv.\nv)~.
\ee
$\Jv$ is the total angular momentum. $\Pi$ is unaffected by a translation or a boost parallel to the reaction plane. The action of $\Pi$ on one-particle helicity states is given by
%\label{parity-states}
\be
	\Pi\,|\theta,\phi,\lambda\rangle = \eta\, (-1)^{s-\lambda}|\theta,-\phi,-\lambda\rangle~.
	\label{basic:eq:parity-helicity}
\ee
Here $\eta$ is the intrinsic parity of the particle, \eg, $-1$ for the photon and the pion%
\footnote{%
$\eta$ is  a complex number of modulus one. The $\eta$'s of non self-conjugate particles may be multiplied by collective phase factors which, in any reaction, do not change the product of the initial $\eta$'s, divided by the final ones. It is natural to choose the same $\eta$ within a $SU(3)$ multiplet. The intrinsic parities $\eta(a)$ and $\eta(\bar a)$ of a fermion and its antiparticle are such that $\eta(a)\,\eta(\bar a)=-1$
}. 

If parity is conserved, it is advantageous to use the transversity basis or, for spin-1 particles, the states linearly polarised parallel or normal to the scattering plane defined in Secs.~\ref{basic:sub:single}. For a particle in the scattering plane ($\phi=0$ or $\pm\pi$), these states are eigenstates of $\Pi$:
\be\label{basic:eq:P-lin}\begin{aligned}
	\hbox{spin }{1/2}:&&
	\Pi\,|\pm\nv\rangle&=\mp i\eta\,|\pm\nv\rangle~,
	\\
	\hbox{spin } 1:&& \quad\Pi\{|\nv\rangle,|\vpi\rangle,|\vec l\rangle\,\} &= 
	\{\eta\,|\nv\rangle,-\eta\,|\vpi\rangle,-\eta\,|\vec l\rangle\,\}~.
\end{aligned}\ee
By  parity conservation, the amplitude $\Mcal$ of  $A+B\to C+D$ fulfils:
\be
	\Mcal = \left(\Pi_C\otimes\Pi_D\right)^{-1}\Mcal\left(\Pi_A\otimes\Pi_B\right)~.
\label{basic:eq:P-ampli}
\ee
Applying this equation to \emph{both} $\Mcal$ and $\Mcal^\dagger$ in Eq.~(\ref{basic:eq:CSdM}) for the cross section matrix leads to
\be
\label{basic:eq:P-class-R}
\Rt=\left(\Pi_A\otimes\Pi_B\otimes\Pi_C\otimes\Pi_D\right)^{-1}\,\Rt\,\Pi_A\otimes\Pi_B\otimes\Pi_C\otimes\Pi_D~.
\ee
%
%\be\label{basic:eq:P-class-R}\R\, =\, \Pi_A^{-1}\otimes\Pi_B^{-1}\otimes\Pi_C^t\otimes\Pi_D^t \quad \R \quad  \Pi_A\otimes\Pi_B\otimes\left(\Pi_C^{-1}\right)^t\otimes\left(\Pi_D^{-1}\right)^t~.\ee
%
For the observables one has, using (\ref{basic:eq:corr-param}) or (\ref{basic:eq:R-obs}),  
\be
\label{basic:eq:P-class-O}
\left\langle \Ocal_i\,\Ocal_f\right\rangle=
\left\langle \Ocal_i^\Pi\,\Ocal_f^\Pi \right\rangle~.
%\left\langle \left(\Pi_i\,\Ocal_i\,\Pi_i^{-1}\right)\
%\left(\Pi_f\,\Ocal_f\,\Pi_f^{-1}\right)\right\rangle~.
\ee
where $\Ocal^{\Pi} \equiv \Pi\,\Ocal\,\Pi^{-1}$. The transformation $\Ocal\to\Ocal^{\Pi}$
is given below for the operators defined in 
%XA/JUIN:
Sec.~\ref{basic:sub:Spin1/2}--\ref{basic:sub:photpol}: 
\begin{align}
	&\hbox{Spin} \,1/2:&
	(\sigma_l,\,\sigma_m,\,\sigma_n) &\to (-\sigma_l,\,-\sigma_m,\,\sigma_n)~.
	\label{basic:eq:sigma-parity}\\
	&\hbox{Photon, gluon}:&
	(\sigma_\odot,\,\sigma_\oslash,\,\sigma_\ominus) &\to (-\sigma_\odot,\,-\sigma_\oslash,\,\sigma_\ominus)~.
\end{align}
Similar relations for the massive spin-1 particles can be derived from (\ref{basic:eq:P-lin}) and are given in Sec.~\ref{excl:sub:photvec}. Usual one-particle observables $O_\lambda$ are either even or odd under $\Pi$. This is also true for the multi-spin observable $O_{\lambda\mu\nu...}=O_\lambda(A)\otimes O_\mu(B)\otimes O_\nu(C)...$, since parity is multiplicative. From (\ref{basic:eq:P-class-O}) one obtains the following ``classical'' rule:  

\medskip\centerline{ \emph{If parity is conserved, all $\,\Pi$-odd observables vanish.}} %(\ref{classical-parity})

\medskip\noindent
This rule roughly reduces by a factor 2 the number of observables. It does not depend on the intrinsic parity of the particles. It just expresses a classical requirement of reflection symmetry at the level of polarised \emph{cross sections}. It is fulfilled, for instance, by formula (\ref{basic:eq:Cijkl}). A well-known counter-example is $\beta$-decay, \eg, $\mathrm{n\to p+e^- + \bar{\nu}}$,  where the outgoing electron has longitudinal polarisation. 

Less trivial, ``non-classical'', constraints are obtained by applying (\ref{basic:eq:P-ampli}) only to $\Mcal$ or only to $\Mcal^\dagger$ in (\ref{basic:eq:CSdM}) or (\ref{basic:eq:corr-param}-\ref{basic:eq:corr-param'}):
\begin{align}\label{basic:eq:P-quant-R}
\Rt&=\Pi_C^{-1}\otimes\Pi_D^{-1}\,\Rt\;\Pi_A\otimes\Pi_B~,\\
%&&\R= \R \ \  \Pi_A\otimes\Pi_B\otimes\left(\Pi_C^{-1}\right)^t\otimes\left(\Pi_D^{-1}\right)^t~,\\
%
\label{basic:eq:P-quant-O} 
\left\langle \Ocal_i \, \Ocal_f \right\rangle
&=\bigl\langle(\Pi_i \, \Ocal_i) \, (\Ocal_f\, \Pi^{-1}_f)\bigr\rangle~.
\end{align}
For a pseudoscalar meson, $\Pi=-1$. For the photon or gluon, $\Pi=\sigma_\ominus$. For spin-1/2, $\Pi=-i\eta\,\sigma_n$, so
\be\label{basic:eq:P-quant-O-bis}
\Pi\left(\,\sigma_0,\sigma_l,\sigma_m,\sigma_n\right)
=\eta\left(\,-i\sigma_n,\sigma_m,-\sigma_l, -i\sigma_0\right)~.
\ee
The non-classical parity constraint in the case of spin-1/2 particles are known as the \emph{Bohr identities} \cite{Bohr:1959}.

Equation~(\ref{basic:eq:P-quant-R}) or (\ref{basic:eq:P-quant-O}) implies the classical parity rule (\ref{basic:eq:P-class-R}) or (\ref{basic:eq:P-class-O}). In addition, it yields linear identities between the $\Pi$-even observables, which depend on the intrinsic parities and cannot be obtained from purely classical arguments. These identities reduce the number of independent correlation parameters  $(\lambda\mu|\nu...)$ roughly by another factor 2. For instance, in $\pi^0$ decay, the classical rule (\ref{basic:eq:P-class-O}) says that the linear polarisations of the two gamma's are either parallel or orthogonal (not, \eg\  at $\pi/4$). The non-classical rule (\ref{basic:eq:P-quant-O}) selects the orthogonal solution. 

The subdivision in constraints of the (\ref{basic:eq:P-class-O}) and (\ref{basic:eq:P-quant-O}) types, both for parity and time-reversal, was made in Appendix 3.D. of \cite{Bourrely:1980mr}. In this section we  point out the ``classical'' versus ``non-classical'' or ``quantum'' characters of these two types \cite{Artru:2008ax}. As a matter of fact, inclusive reactions have only ``classical constraints'', since the intrinsic parity of the undetected particles takes both signatures. The same is true for partially polarised exclusive reactions. The disappearance of non-classical constraints as information is lost has relationship with decoherence of incompletely controlled quantum systems. This classical vs.\ quantum  distinction will also apply below to charge conjugation and particle indiscernability. 
\paragraph{Case of non-planar reactions}
In $2\to n$ exclusive scattering with $n\ge3$, for instance $\pi+\N\to \pi+\pi+\N$, or in $1\to n$ decay with $n\ge4$ the momenta are not coplanar and cannot be invariant under a mirror reflection like (\ref{basic:eq:Pi-parity}). This is also the case of two-particle inclusive scattering, etc. 
Using the ordinary operator $P$, parity conservation for a $2\to3$ reaction reads
%the amplitude $\Mcal$ of  $A+B\to C+D$
\be\label{basic:eq:P-ampli-nonplan}
	\Mcal(\pv_A,\pv_B,\pv_C,\pv_D,\pv_E) = \left(P_C\otimes P_D\otimes P_E\right)^{-1}\Mcal(-\pv_A,-\pv_B,-\pv_C,-\pv_D-\pv_E)\left(P_A\otimes P_B\right)~.
\ee
Following the same steps as in Eqs.(\ref{basic:eq:P-ampli}-\ref{basic:eq:P-class-O}) one obtains the \emph{classical} constraint
\be
\label{basic:eq:P-class-O-nonplan}
\left\langle \Ocal_i\,\Ocal_f\right\rangle_{\pv_A,\pv_B,\pv_C,\pv_D,\pv_E}=
\left\langle \Ocal_i^ P\,\Ocal_f^ P \right\rangle_{-\pv_A,-\pv_B,-\pv_C,-\pv_D,-\pv_E}~.
\ee
where $\Ocal^{ P} \equiv  P\,\Ocal\, P^{-1}$. In particular $\vec\sigma^P=\vec\sigma$.
Equivalently one can chose an arbitrary plane containing the $z$-axis and use the reflection operator (\ref{basic:eq:Pi-parity}) about this plane. Then,
\be
\label{basic:eq:P-class-O-1}
\left\langle \Ocal_i\,\Ocal_f\right\rangle_{ \phi_A, \phi_B, \phi_C, \phi_D, \phi_E}=
\left\langle \Ocal_i^\Pi\,\Ocal_f^\Pi \right\rangle_{- \phi_A,- \phi_B,- \phi_C,- \phi_D,- \phi_E}~.
\ee
This is the result obtained in \cite{Roberts:2004mn} with the help of helicity amplitudes. 
In practice $\phi_A=\phi_B=0$ and one links the $\phi=0$ plane to a final momentum, for instance to  $(\pv_C+\pv_D)_\perp=-(\pv_E)_\perp$ if $C$ and $D$ can make a resonance.  
Then $\phi_C$ and $\phi_D$ can be replaced by only one azimuthal angle or by $\pv_C.\nv$, which is also $\Pi$-odd. 

Unfortunately, one cannot derive \emph{non-classical} parity constraints like (\ref{basic:eq:P-quant-O}). Indeed, applying (\ref{basic:eq:P-ampli-nonplan}) only to $\Mcal$ or only to $\Mcal^\dagger$ in the analogue of (\ref{basic:eq:corr-param}), amounts to consider an interference between two amplitudes taken at different momenta, a quantity which is not accessible experimentally.  
%
%\subsubsection{Identical particles}
%XA/JUIN:
%
\boldmath\subsubsection{Permutations of $A\leftrightarrow B$ and/or $C\leftrightarrow D$}\unboldmath \label{basic:sub:permut} 
\paragraph{Preliminary: some transformations of helicity states.}
Let us generalise (\ref{2-particle}) by
\be\label{2-particle-creators}
|\kv;A_1,\lambda_1;A_2,\lambda_2\rangle = a^\dagger(\kv;A_1,\lambda_1)\,a^\dagger(-\kv;A_2,\lambda_2)|\O\rangle \,,
\ee
which takes Bose or Fermi statistics into account for identical particles, $A_1=A_2$. The (anti-) commutation relations between creation operators give 
\be
\label{basic:eq:commut}
|\kv;A_1,\lambda_1;A_2,\lambda_2\rangle = \varepsilon_{12}\,|-\kv;A_2,\lambda_2;A_1,\lambda_1\rangle\,,
\ee
with $\varepsilon_{12}=+1$ for a two-boson or boson-fermion pair, $\varepsilon_{12}=-1$ for a two-fermion pair.

If $\kv$ is in the $(x,z)$ half-plane with $\phi(\kv)=0$, a rotation of angle $+\pi$ about $\yu$ yields, according to the convention (\ref{basic:eq:helistate}),
\be
\label{basic:eq:rot/y}
\exp(-i\pi J_y)\,|\kv;A_1,\lambda_1;A_2,\lambda_2\rangle = \varepsilon_1\,|-\kv;A_1,\lambda_1;A_2,\lambda_2\rangle \,,
\ee
where $\varepsilon_i\equiv(-1)^{2s_i}=+1$ for a boson, $-1$ for a fermion.
%Equations (\ref{basic:eq:commut}) and (\ref{basic:eq:rot/y}) also apply in the transversity and linear polarisation bases, provided the basic states are defined as indicated in Sec.\ref{basic:sub:phase-conv}. 

A rotation of a single particle about $\zu$ gives
\be\label{basic:eq:rot/z}
\exp(-iu J_z)\,|\pv(\theta,\phi),\alpha\rangle=\exp(-i\alpha u)\,|\pv'(\theta,\phi+u),\alpha\rangle\,.
\ee
Applying it to the initial particle pair and restoring $\phi_B=\pi$, one obtains as expected
\begin{equation}
\label{basic:eq:rot/z(AB)}
\exp(-i\pi J_z)\,|k\zu;A,\lambda_A;B,\lambda_B\rangle = e^{i\pi(\lambda_B-\lambda_A)}\,
|k\zu;A,\lambda_A;B,\lambda_B\rangle\,.
\end{equation}
%
%with the convention $\phi_B=\pi$ in the primary $\{\xu,\yu,\zu\}$ frame. 
These relations will be used together with rotational invariance,
\be\label{basic:eq:rot-inv}
{\cal M}=\exp(i\vec J.\vec r)\,{\cal M}\,\exp(-i\vec J.\vec r) \,.
\ee

\paragraph{Application to \boldmath$A+B \to C+D$\unboldmath.} From the preceding relations one can derive
%rotational invariance, ${\cal M}=\exp(i\vec J.\vec r)\,{\cal M}\,\exp(-i\vec J.\vec r)$
%
\begin{align}
\langle \gamma,\delta|\mathcal{M}^{AB\to CD}(\theta)|\alpha,\beta\rangle &=
\varepsilon_{A}\,\varepsilon_{AB}\,\varepsilon_{C}\,\varepsilon_{CD}\ 
\langle \delta,\gamma|\mathcal{M}^{BA\to DC}(\theta)|\beta,\alpha\rangle\,, \label{basic:eq:permutABCD}\\
\langle\gamma,\delta|\mathcal{M}^{AB\to CD}(\theta)|\alpha,\beta\rangle &=
\varepsilon_{CD}\,(-1)^{\alpha-\beta-\gamma-\delta}\,
\langle\delta,\gamma|\mathcal{M}^{AB\to DC}(\pi-\theta)|\alpha,\beta\rangle \,,
\label{basic:eq:permutCD}  \\
\langle\gamma,\delta|\mathcal{M}^{AB\to CD}(\theta)|\alpha,\beta\rangle &= \varepsilon_{AB}\,(-1)^{\alpha+\beta+\gamma-\delta}\,
\langle\gamma,\delta|\mathcal{M}^{BA\to CD}(\pi-\theta)|\beta,\alpha\rangle \,.
\label{basic:eq:permutAB}
\end{align}
In Eq.~(\ref{basic:eq:permutABCD}), the product of $\varepsilon$ factors is $+1$ for the 
${\rm bb\to bb}$,
${\rm f{} f\to f{}f}$,
${\rm bb\to f{} f}$,
${\rm f{} f\to bb}$,
${\rm bf\to bf}$,
 and
 ${\rm f{}b\to f{}b}$
  reactions (${\rm b}= \hbox{boson}$, ${\rm f}= \hbox{fermion})$, 
  $-1$ for the ${\rm b f\to f{}b}$ and ${\rm f{}b\to bf}$ reactions. 
  It is obtained from (\ref{basic:eq:rot-inv}) with $\vec r=\pi\yu$ and (\ref{basic:eq:commut})-(\ref{basic:eq:rot/y}) applied to the $A+B$ and $C+D$ pairs.  
Equation (\ref{basic:eq:permutCD}) is obtained with $\vec r=\pi\zu$, (\ref{basic:eq:rot/z(AB)}) and applying (\ref{basic:eq:rot/z}) and (\ref{basic:eq:commut}) to $C$ and $D$. Equation (\ref{basic:eq:permutAB}) results from (\ref{basic:eq:permutABCD}) and (\ref{basic:eq:permutCD}).
\subsubsection{Identical particles}
For a pair of particles of equal spins, one defines the \emph{helicity exchange operator} by
\be\label{basic:eq:rot=perm}
P^{\rm hel.}_{12}\,|\kv,A_1,\lambda_1;A_2,\lambda_2\rangle \equiv
|\kv,A_1,\lambda_2;A_2,\lambda_1\rangle \,.
\ee
Equation (\ref{basic:eq:permutABCD}) applies to $A+A \to B+B$ with a $+$ sign.
It can be rewritten as
\be\label{basic:eq:s-a-p}
P^{\rm hel.}_{34}\,\Mcal\, P^{\rm hel.}_{12} = \Mcal \,.
%\quad \hbox{ \ie,}\quad
%\left\langle\lambda_3,\lambda_4|\Mcal|\lambda_1,\lambda_2\right\rangle %=\left\langle\lambda_4,\lambda_3|\Mcal|\lambda_2,\lambda_1\right\rangle~.
\ee
For the cross section matrix, one obtains the ``classical'' and ``non-classical'' relations 
\begin{align}\label{basic:eq:identic-class}
\R&=(P^{\rm hel.}_{12}\otimes P^{\rm hel.}_{34})\,\R\,(P^{\rm hel.}_{12}\otimes P^{\rm hel.}_{34})~,\\ 
\label{basic:eq:identic-non-class} 
\R&= \R\,(P^{\rm hel.}_{12}\otimes P^{\rm hel.}_{34})~.
\end{align}
For spins $1/2$, these relations give respectively 
\begin{align}\label{basic:eq:P12.P34class}
(\lambda\mu|\nu\tau) &= (\mu\lambda|\tau\nu)~,\\
\label{basic:eq:P12.P34quant}
(\lambda\mu|\nu\tau) &= \Gamma_{\lambda\lambda'\mu\mu'}\, (\lambda'\mu'|\nu'\tau')\,\Gamma_{\nu'\nu\tau'\tau}~,
\end{align}
where repeated indices are summed over and  $\Gamma_{\alpha\beta\gamma\delta}=(1/4)\,\Tr\{\sigma_\alpha\sigma_\beta\sigma_\gamma\sigma_\delta\}$. 
Each of the 256 non-classical relations (\ref{basic:eq:P12.P34quant}) has 16 terms on the right-hand side. These relations are not independent and, by taking appropriate combinations, one obtains the shorter ones, 
\begin{equation}\label{basic:eq:P12.P34quantbis}
\begin{aligned}
(0i|pq)-(0i|qp) &= (jk|r0)-(jk|0r) &\hspace*{-2cm}&(\varepsilon_{ijk}=\varepsilon_{pqr}=+1)~,\\
(ij|pq)-(ij|qp) &= (0k|0r)-(0k|r0)  &\hspace*{-2cm}&(\rm idem)~,\\
(xx|\nu\tau)+(yy|\nu\tau)+(zz|\nu\tau)&=(00|\nu\tau) &\hspace*{-2cm}&(\nu\ne\tau)~,\\
(00|xx)+(00|yy)+(00|zz)&=(xx|00)+(yy|00)+(zz|00)~,\\  
(00-xx\,|\,00-xx) &= (yy+zz\,|\,yy+zz)~,\\
2[(xx|yy)-(yy|xx)] &= (xx-yy\,|\,00-zz)-(00-zz\,|\,xx-yy)~.
\end{aligned}
\end{equation}
In the first two equations, $ijk$ and $pqr$ are even permutations of $xyz$ (here $\{x,y,z\}\equiv\{m,n,l\}$). A symmetric partner of the third equation is obtained for the $(\lambda\mu|\nu\nu)$'s with $\lambda\ne\mu$.  The last two equations plus their permutations of $\{x,y,z\}$ give $3+2$  independent relations. The notation $(xx-yy\,|\,00-zz)$ stands for $(xx|00)-(xx|zz)-(yy|00)+(yy|zz)$. In total one has 36 independent relations.

\paragraph{Scattering at 90$^\circ$.}
Let us consider the case $A+A \to C+D$, where at least the \emph{initial} particles are identical, for instance $\gamma\gamma\to\pi^-\rho^+$. At $\theta=90^\circ$ Eq.~(\ref{basic:eq:permutAB}) gives
\be\label{basic:eq:A+A}
\langle\gamma,\delta|\mathcal{M}^{AA\to CD}(90^\circ)|\alpha,\beta\rangle = (-1)^{\alpha-\beta+\gamma-\delta}\,
\langle\gamma,\delta|\mathcal{M}^{AA\to CD}(90^\circ)|\beta,\alpha\rangle \,.
\ee
For spin one-half fermions, it can be written as
\be\label{basic:eq:A+A'} % (\ref{basic:eq:A+A'})
\mathcal{M}^{AA\to CD}(90^\circ) =
\sigma_l\otimes\sigma_l\,\mathcal{M}^{AA\to CD}(90^\circ)\,
P^{\rm hel}_{12}\,\sigma_l\otimes\sigma_l \,.
\ee
Applying this identity to both $\mathcal{M}$ and $\mathcal{M}^\dagger$ in (\ref{basic:eq:CSdM}) yields the \emph{classical} symmetry relations
\be\label{basic:eq:symA+A-class} % (\ref{basic:eq:symA+A-class})
(\lambda,\mu|\nu,\tau)=(\breve\mu,\breve\lambda|\breve\nu, \breve\tau) \qquad (\theta=\pi/2)
\,.
\ee
where $(\breve\mu,\breve\lambda|\breve\nu, \breve\tau)=
\langle\sigma_\lambda\,\sigma_\nu\,\check\sigma_\mu\,\check\sigma_\tau\rangle$
and $\breve\sigma_\alpha= \sigma_l\sigma_\alpha\sigma_l$, more explicitly 
$\{\breve\sigma_0,\breve\sigma_l,\breve\sigma_m,\breve\sigma_n\}=\linebreak[2]
\{\sigma_0,\sigma_l,\linebreak[2]-\sigma_m,-\sigma_n\}$. 
Equation~(\ref{basic:eq:symA+A-class}) predicts the vanishing of 32 Cartesian parameters like $(00|0m)$, $(nn|ml)$ and yields 96 equivalences such as $(l0|ln)=-(0l|ln)$, $(l0|mn)=+(0l|mn)$. These relations can be understood classically by the fact that the plane normal to the collision axis is a symmetry plane of the reaction.  

Applying (\ref{basic:eq:A+A'}) only to $\mathcal{M}$ in (\ref{basic:eq:CSdM}) yields the \emph{non-classical} relations
\be\label{basic:eq:symA+A-quant} % (\ref{basic:eq:symA+A-quant})
(\lambda\mu|\nu\tau) = 2^{-4}\, (\lambda'\mu'|\nu'\tau') \, 
\Tr(\sigma_\lambda\sigma_{\lambda'}\breve\sigma_\mu\breve\sigma_{\mu'}) \,
\Tr(\sigma_l\sigma_{\nu'}\sigma_\nu) \, \Tr(\sigma_l\sigma_{\tau'}\sigma_\tau)
 \qquad (\theta=\pi/2) \,.
\ee
In particular,
\begin{align}
\label{basic:eq:symA+A-quant1} % (\ref{basic:eq:symA+A-quant})	
2(00|ll) &= +(00|00)+(ll|00)-(mm|00)-(nn|00)  \\
\label{basic:eq:symA+A-quant2}
2(00|mm) &= -(00|nn)-(ll|nn)+(mm|nn)+(nn|nn)  \,.
\end{align}
If parity is conserved and the initial and final products of initial intrinsic parities are equal, 
one may transform (\ref{basic:eq:symA+A-quant1}--\ref{basic:eq:symA+A-quant2}) using 
the Bohr identity (\ref{basic:eq:P-quant-O-bis}) to derive
\be\label{basic:eq:sumCii} % (\ref{basic:eq:sumCii})
C_{ll}+C_{mm}+C_{nn}=1 \,.
\ee
In a $A+A\to A+A$ reaction, $T$-invariance only gives $(00|ll)=(ll|00)$, therefore Eq.~(\ref{basic:eq:symA+A-quant1}) reads	
\be\label{basic:eq:sumAii} % (\ref{basic:eq:sumAii})
A_{ll}+A_{mm}+A_{nn}=1 \,,
\ee
derived in \cite{Brodsky:1979nc} as
$A_{N\!N}-A_{LL}-A_{SS}=1$, in the Argonne notation. 
For instance, in $\mathrm{e}^-\mathrm{e}^-$ scattering, or the scattering of identical colourless quarks  \cite{Chen:1977wg}, one has, at $90^\circ$, $(A_{nn},\,A_{mm},\,A_{ll})=(1/9,\,1/9,\,7/9)$. The quark interchange model \cite{Farrar:1978by} predicts $A_{nn}=A_{mm}=A_{ll}=1/3$ in high-energy $\mathrm{pp}$ scattering at $90^\circ$. When summation is made over the initial and final colours, one gets $(A_{nn},\,A_{mm},\linebreak[2]{\,A_{ll})=(-1/11,\,-1/11,\,5/11)}$ in the scattering of identical quarks at $90^\circ$, and the relation (\ref{basic:eq:sumAii}) is no longer valid.

\boldmath\subsubsection{Charge and $CP$ conjugations in $A+\overline{A}\to B+ \overline{B}$}\unboldmath
Equation (\ref{basic:eq:permutABCD}) also applies to $A+\overline{A}\to B+ \overline{B}$\unboldmath{} with a $+$ sign. If $C$ is conserved, replacing each particle by its antiparticle does not change the amplitude, so that one again obtains Eqs.~(\ref{basic:eq:s-a-p}--\ref{basic:eq:identic-non-class}) and, for one-half spins, 
Eqs.~(\ref{basic:eq:P12.P34class}--\ref{basic:eq:P12.P34quantbis}). 
%$A+\overline{A}\to C+\overline{C}$ the same relations as for $A+A\to C+C$, 

\emph{$CP$ symmetry} can be treated in a similar manner. Applying successively $P$ in the form (\ref{basic:eq:P-ampli}), $C$ and (\ref{basic:eq:permutABCD}) gives
\be
	\Mcal = \left(\Pi_C\otimes\Pi_D\right)^{-1}\,P^{\rm hel.}_{CD}\,\Mcal
	\, P^{\rm hel.}_{AB}\,\left(\Pi_A\otimes\Pi_B\right)~.
\label{basic:eq:CP-inv}
\ee
%
%using the operator $PC=\Pi\,C\,\exp(-i\pi J_y)$. 
For a fermion-antifermion pair, $\Pi_1\,\Pi_2=\sigma(1)_y\otimes\sigma(2)_y$. 
%= {1\over2}\,\left[1-\sigma_x(1)\otimes\sigma_x(2)+\sigma_y(1)\otimes\sigma_y(2) -\sigma_z(1)\otimes\sigma_z(2)\right]~.
The classical $CP$ constraints may be obtained applying successively the transformations  (\ref{basic:eq:sigma-parity}) %(\ref{sigma-parity}) 
and (\ref{basic:eq:P12.P34class}). The non-classical ones are of the form (\ref{basic:eq:P12.P34quant}) with $\Gamma$ replaced by 
$\Gamma^{CP}_{\lambda\lambda'\mu\mu'}=(1/4)\Tr\{\sigma_\lambda\sigma_{\lambda'} \,
\sigma_y\,\sigma_\mu\sigma_{\mu'}\,\sigma_y\}$. 
Alternatively, one may keep the formulas (\ref{basic:eq:P12.P34quant}--\ref{basic:eq:P12.P34quantbis}) but transforming the 
$(\lambda\mu|\nu\tau)$'s according to 
\be\label{hat-mu}
(\lambda\mu|\nu\tau) \to (\lambda\hat\mu|\nu\hat\tau) =
\langle\sigma_\lambda\,\hat\sigma_\mu\,\sigma_\nu\,\hat\sigma_\tau\rangle \,,
\ee
with $\hat\sigma_\alpha\equiv\sigma_y\sigma_\alpha\sigma_y$.
For instance, the fourth Eq.~(\ref{basic:eq:P12.P34quantbis}) becomes
$-(00|xx)+(00|yy)-(00|zz)=-(xx|00)+(yy|00)-(zz|00)$ (here again $\{x,y,z\}\equiv\{m,n,l\}$).
These constraints hold whether or not $P$ and $C$ are separately conserved. 
\boldmath\subsubsection{Time reversal invariance in $A+B\to A+B$}\unboldmath
We consider the two time-reversed {elastic} reactions
\begin{align}
 A(\pv_i,\alpha_i)+B(\kv_i,\beta_i) &\to A(\pv_f,\alpha_f)+B(\kv_f,\beta_f)~,\label{basic:eq:non-revers}\\
 A(-\pv_f,\alpha_f)+B(-\kv_f,\beta_f) &\to A(-\pv_i,\alpha_i)+B(-\kv_i,\beta_i)~.\label{basic:eq:revers}
\end{align}
Here $\alpha$ and $\beta$ are \emph{helicities}. At the level of \emph{amplitudes}, time reversal invariance reads
\begin{equation}\label{basic:eq:T-quant}
	\langle f|\mathcal{M}|i\rangle=\langle i|T^\dagger\,\mathcal{M}\,T|f\rangle~.
\end{equation}
A one-particle helicity state, for instance $|\pv,\alpha\rangle$, transforms as 
\begin{equation}
	T\,|\pv,\alpha\rangle=\xi_A\,\exp(-2i\alpha\phi_\pv)\,|-\pv,\alpha\rangle~,
\label{basic:eq:T-action}
\end{equation}
where $\phi_\pv$ is the azimuth of $\pv$ and $\xi_A$ is a phase factor intrinsic to particle $A$.

The centre-of-mass momenta of reactions (\ref{basic:eq:non-revers}) and (\ref{basic:eq:revers}) are related by a symmetry about the momentum transfer $\Qv=\pv_f-\pv_i$. We take the $z$-axis along $\Qv$ and use rotation invariance to replace (\ref{basic:eq:T-quant}) by 
\begin{equation}\label{basic:eq:T+rot}
	\langle f|\mathcal{M}|i\rangle=\langle i|T^\dagger\,\exp(i\pi J_z)\,\mathcal{M}\,\exp(-i\pi J_z)\,T|f\rangle~.
\end{equation}
This equation relates two amplitudes of different helicities but \emph{identical momenta}. It is valid in the centre-of-mass frame and any frame where $Q^0=0$, \eg, the Breit frames of $A$ and $B$. Using
\be
\exp(-iu J_z)\,|\pv(\theta,\phi),\alpha\rangle=\exp(-i\alpha u)\,|\pv'(\theta,\phi+u),\alpha\rangle~,
\ee
equation (\ref{basic:eq:T+rot}) takes the explicit form
\begin{equation}
	\langle\alpha_f,\beta_f|\mathcal{M}|\alpha_i,\beta_i\rangle
	=(-1)^{\alpha_i-\alpha_f-\beta_i+\beta_f}\,\langle\alpha_i,\beta_i|\mathcal{M}|\alpha_f,\beta_f\rangle~,
\label{basic:eq:T+rot(M)}
\end{equation}
omitting the momentum arguments. Applying this equation to both factors of (\ref{basic:eq:CSdM}), we obtain the ``classical'' time-reversal relation
\begin{equation}
\langle\alpha'_i\beta'_i\alpha'_f\beta'_f|\R|\alpha_i\beta_i\alpha_f\beta_f\rangle
=(-1)^{\alpha_i-\alpha'_i+\alpha_f-\alpha'_f+\beta_i-\beta'_i+\beta_f-\beta'_f}\
\langle\alpha'_f\beta'_f\alpha'_i\beta'_i|\R|\alpha_f\beta_f\alpha_i\beta_i \rangle~.
\label{basic:eq:T-inv/R-class}
\end{equation}
%
%in analogy to (\ref{P-class}) 
For observables $\mathcal{P}$ and $\mathcal{Q}$ dealing with the initial and final states separately, we have
\begin{equation}\label{basic:eq:T-class-O}
\left\langle \mathcal{P}(i)\,\mathcal{Q}(f)\right\rangle=
\left\langle \mathcal{Q}^{RT}(i)\,\mathcal{P}^{RT}(f) \right\rangle~,
\end{equation}
where $\Ocal^{RT}\equiv \exp(-i\pi J_z)\,T\Ocal^t\,T^\dagger\,\exp(i\pi J_z)$. For one-particle observables,
\be\label{basic:eq:T-class-Oa}
\begin{aligned}
&\hbox{any spin:}&\left\langle\alpha|\Ocal^{RT}|\alpha'\right\rangle &= (-1)^{\alpha-\alpha'}
\left\langle\alpha'|\Ocal|\alpha\right\rangle~,\\
&\hbox{spin 1/2:}&\left(\sigma_x,\sigma_y,\sigma_z\right)^{RT} &= \left(-\sigma_x,\sigma_y,\sigma_z\right)~,\\
&\hbox{photon:}&\left(\sigma_\ominus,\,\sigma_\oslash,\,\sigma_\odot\right)^{RT} &= 
\left(\sigma_\ominus,\,-\sigma_\oslash,\,\sigma_\odot\right)~.
\end{aligned}
\ee
%
%For spin one-half particle, $\left(\sigma_x,\sigma_y,\sigma_z\right)^R=\left(-\sigma_x,\sigma_y,\sigma_z\right)$.

Applying (\ref{basic:eq:T+rot(M)}) only to the second factor of (\ref{basic:eq:CSdM}) gives the ``non-classical'' relation 
%(analogous to \ref{P-quant})
\begin{equation}
\langle\alpha_i'\beta'_i\alpha'_f\beta'_f|\R|\alpha_i\beta_i\alpha_f\beta_f\rangle
=(-1)^{\alpha_i-\alpha_f+\beta_i-\beta_f}\
\langle\alpha_i'\beta'_i\alpha'_f\beta'_f|\R|\alpha_f\beta_f\alpha_i\beta_i \rangle~,
\label{basic:eq:Tquant-R}
\end{equation}
or  %$\R=\R\ K_{13}\,K_{24}$ with 
\begin{equation}\label{basic:eq:R-Kif}
\R=\R\,K_{if}(A)\,K_{if}(B)~,\quad{\rm with}\quad K_{if}=P^{\rm hel.}_{if}\,\exp(i\pi[J_l(i)-J_l(f)])~.
\end{equation}
In terms of observables, 
\be\label{basic:eq:Tquant-O}
\left\langle \Ocal\right\rangle =
\left\langle\left[K_{if}(A)\,K_{if}(B)\,\Ocal^{\rm pt}\right]^{\rm pt}\right\rangle~,
\ee
where  $pt$ means partial transposition, applied to final-particle observables only. 
For a spin 1/2, $K_{if}=P^{\rm hel.}_{if}\,\sigma_z(i)\,\sigma_z(f)
= {1\over2}\, [1-\sigma_x(i)\,\sigma_x(f)-\sigma_y(i)\,\sigma_y(f)+\sigma_z(i)\,\sigma_z(f)]$. 

For the $1/2+1/2\to1/2+1/2$ case, the classical relations (\ref{basic:eq:T-class-O}-\ref{basic:eq:T-class-Oa}) lead to simple relations like $(xz|y0)=-(y0|xz)$. The non-classical constraints (\ref{basic:eq:R-Kif}) or (\ref{basic:eq:Tquant-O}) read%
\footnote{This is equivalent to Eq.~(4) of Appendix 3.D of  \cite{Bourrely:1980mr}, except for a discrepancy for the factor (1/4) in front of $\Gamma^T$.
}
\be\label{basic:eq:Tquant-O1/2}
(\lambda\mu|\nu\tau) = \Gamma^T_{\lambda\lambda'\nu\nu'}\, (\lambda'\mu'|\nu'\tau')\,\Gamma^T_{\mu\mu'\tau\tau'}~,
\ee
with $\Gamma^T_{\alpha\beta\gamma\delta}=(1/4)\, 
\Tr\{\sigma_\alpha\sigma_\beta\,\sigma_z\,\sigma_\gamma^t\sigma_\delta^t\,\sigma_z\}$. 
Alternatively, one may keep the formulas (\ref{basic:eq:P12.P34quant}--\ref{basic:eq:P12.P34quantbis}) but transforming the 
$(\lambda\mu|\nu\tau)$'s according to 
\be\label{check-mu}
(\lambda\mu|\nu\tau) \to (\lambda\nu|\check\mu\check\tau) \equiv
\langle\sigma_\lambda\,\sigma_\nu\,\check\sigma_\mu\,\check\sigma_\tau\rangle \,,
\ee
with $\check\sigma_\alpha=\sigma_z\sigma^t_\alpha\sigma_z$, more explicitly 
$\{\check\sigma_0,\check\sigma_x,\check\sigma_y,\check\sigma_z\}=
\{\sigma_0,-\sigma_x,\sigma_y,\sigma_z\}$. 
For instance, the fourth Eq.~(\ref{basic:eq:P12.P34quantbis}) becomes
$-(0x|0x)+(0y|0y)+(0z|0z)=-(x0|x0)+(y0|y0)+(z0|z0)$ (with $\{x,y,z\}\equiv\{m,n,l\}$ again).

For the simpler case $1/2+0\to 1/2+0$, for instance $\p+\pi \to \p+\pi$, one should remove the $\mu$ and $\tau$ variables and the second $\Gamma^T$ in (\ref{basic:eq:Tquant-O1/2}). From (\ref{basic:eq:T-class-O}--\ref{basic:eq:T-class-Oa}) and (\ref{basic:eq:Tquant-O1/2}), one obtains
\begin{gather}\begin{aligned}
(0|m)=-(m|0), \quad (0|n) &= +(n|0), \quad (0|l)=+(l|0)~,\\
(m|n)=-(n|m), \quad (m|l) &= -(l|m), \quad (n|l)=+(n|l)~,\label{basic:eq:Tclass.pi-N}\end{aligned}\\
%\intertext{and}
(0|0) = (l|l)-(m|m)+(n|n)~,\label{basic:eq:Tquant.pi-N}
\end{gather}
the last constraint being non-classical. These constraints can also be obtained applying (\ref{basic:eq:T-inv/R-class}) and (\ref{basic:eq:Tquant-R}) to the cross section matrix given in Table \ref{basic:tab:R1/2;1/2}, with $a=\alpha_i$, $~c=\alpha_f$. 

\subsubsection{Hermitian scattering matrix}

In the Born approximation, $\mathcal{M}=\mathcal{M}^\dagger$. This is also the case when the intermediate states in the unitarity relation $\mathcal{M}-\mathcal{M}^\dagger\propto i\mathcal{M}\mathcal{M}^\dagger$ are kinematically forbidden, for instance in the quark-distribution amplitudes $h\to q+X$ or  in the Weissz{\"a}cker--Williams amplitude for $\e \to\gamma+\e$. 
%For elastic 2$\to$2 reactions, it yields relations analogous to those of $T$-invariance. 
Here we consider the combination of hermiticity of $\mathcal{M}$ and $PT$-invariance, which gives
\be\label{basic:eq:PT+Born}
\left\langle i',f'|\R|i,f\right\rangle=\left\langle\,PT(i),PT(f)\,|\R|\,PT(i'), PT(f')\,\right\rangle~.
%\left(\left\langle\,PT(i'),\,PT(f')\,|\R|\,PT(i),\,PT(f)\,\right\rangle\right)^*~.
\ee
Using the helicity states transformation,
\be
PT|\theta,\phi,\lambda\rangle=\eta\xi\,(-1)^{s+\lambda}|\theta,\phi,-\lambda\rangle~,
\ee
we obtain for the CSM 
\be\label{basic:eq:PTH-R}
\langle \lambda'_1,\lambda'_2;\lambda'_3,...|\R|\lambda_1,\lambda_2;\lambda_3,...\rangle=
\prod_i(-1)^{\lambda_i-\lambda'_i}
\left\langle -\lambda_1,-\lambda_2;-\lambda_3,...|\R|-\lambda'_1,-\lambda'_2;-\lambda'_3,...\right\rangle~,
\ee
and for the observables
\be\label{basic:eq:PTH-O}
\left\langle \Ocal_i\,\Ocal_f\right\rangle=
\left\langle \Ocal_i^{PT}\,\Ocal_f^{PT} \right\rangle~,
\ee
where $\Ocal^{PT}$, for one-particle observables,  is given  by
%$H$ means the hermitian conjugation and $\Ocal^{PTH}=(PT)^{-1}\,\Ocal^\dagger\,PT$  
$\left\langle \lambda|\Ocal^{PT}|\lambda'_1\right\rangle = (-1)^{\lambda-\lambda'}
\left\langle -\lambda'|\Ocal|-\lambda\right\rangle$. For spin 1/2, $\vec\sigma^{PT}=-\vec\sigma$.

This result is valid for non-planar reactions as well and any number of particles. Equation~(\ref{basic:eq:PTH-O}) tells that %single-spin one-half asymmetries, or 
asymmetries involving an odd number of vector polarisations vanish under the $PT$ + hermiticity hypothesis. Therefore, a non-zero asymmetry is due either to $PT$ violation or to intermediate states in unitarity relation, or both.  
Interesting exceptions are the Sivers \cite{Sivers:1989cc} and Boer--Mulders \cite{Boer:1997nt}  effects, which are $(n|0)$ and $(0|n)$ asymmetries in $\overrightarrow{N}\to \overrightarrow{q}+X$. In fact, 
final- or initial-state interaction \cite{Brodsky:2002cx} results in an effective $h\to q+X$ amplitude which has a spin-dependent phase. 
Within QCD, one cannot isolate the quark emission process in a gauge invariant way. To satisfy gauge invariance, one has to include the interaction of the hard-scattered quark with spectator partons. 
\subsubsection{Crossed reactions}
Let us consider the two crossed reactions:
\be\label{basic:eq:cross-reac}
(1)\quad X+A(\tilde p,\alpha)\to Y~,\qquad(2)\quad X\to\overline{A}(\tilde p,\beta)+Y~,
\ee %({\rm a})
where $X$ and $Y$ represent the remaining sets of initial and final particles. $\tilde p$ is the four-momentum of particle $A$ or $\overline{A}$. Crossing symmetry can be formally written as
\be\label{basic:eq:cross-sym}
\langle s_Y|\,\mathcal{M}_{1}(\tilde k_X,\,-\tilde p\,;\,\tilde k_Y)\,(PT)_A|s_X,\alpha\rangle=
\langle \alpha,s_Y|\mathcal{M}_{2}(\tilde k_X\,;\,\tilde p,\,\tilde k_Y)|s_X\rangle~.
\ee
$(PT)_A$ is the product of parity and time reversal operators acting on particle $A$ only. It reverses the spin of $A$ but not the momentum. $\tilde k_X$, $\tilde k_Y$, $s_X$ and $s_Y$ stand for the 4-momenta and spins of the remaining particles. The left-hand side amplitude is evaluated at non-physical kinematical variables, but can defined by analytic continuation. This equation has a simple interpretation in the Dirac hole theory, at least in the fermionic case: the annihilation of $A$ in the negative energy state $PT|-\tilde p,\alpha\rangle$ is equivalent to the creation of $\overline{A}$ in the state $|\tilde p,\alpha\rangle$.  
For the CSM (Eq.~\ref{basic:eq:CSdM}), the crossing relation reads
\be\label{basic:eq:R-cross}
\R_2(\tilde k_X;\tilde p,\tilde k_Y)=(PT)_A^{-1}\,\R_1(\tilde k_X,-\tilde p;\tilde k_Y)(PT)_A~.
\ee
The single-spin observables $\left\langle\Ocal_A\right\rangle_{1}$ of (1) and 
$\left\langle\Ocal_{\bar A}\right\rangle_{2}$ of (2) are related by
\be\label{basic:eq:O-cross}
\left\langle\Ocal_{\bar A}(\tilde k_X;\tilde p,\,\tilde k_Y)\right\rangle_{2}=
\left\langle (PT)_A\,\Ocal^t_A(\tilde k_X,-\tilde p;\tilde k_Y)\,(PT)_A^{-1}\right\rangle_{1}~.
\ee
These relations can be iterated to cross several particles, for instance from $\pi^-\p\to\pi0\mathrm{n}$ to $\bar{\mathrm{n}}\p\to\pi^+\pi0$. Relation (\ref{basic:eq:O-cross}) also generalises to multi-particle observables: in the right-hand side the transposition of $\Ocal$ concerns only the crossed particles. The crossing transformation 
%\be\label{O-cross-gener}\ee
is given below for the operators defined in this section:
\be\label{basic:eq:crossing}
\begin{aligned}
& & \Ocal &\to PT\,\Ocal^t\,(PT)^{-1}~,\\
\hbox{Spin}\,1/2:& &\vec\sigma &\to -\vec\sigma~,\\
\hbox{Photon, gluon}:& &(\sigma_\odot,\,\sigma_\oslash,\,\sigma_\ominus) &\to (-\sigma_\odot,\,\sigma_\oslash,\,\sigma_\ominus)~.\\[2pt]
\hbox{Vector particle -- Cartesian operators:}& &(\Sigma_1,\,\Sigma_2,\,\Sigma_3) &\to -(\Sigma_1,\,\Sigma_2,\,\Sigma_3)~,\\[-1pt]
&&(\Sigma_4,\cdots\Sigma_8) &\to (\Sigma_4,\cdots\Sigma_8)~,\\
\hbox{Vector particle -- magnetic operators:}&  &
\mathcal{S}_\lambda& \to -\mathcal{S}_\lambda\\[-2pt]
 &&\tau_{2\mu} &\to \tau_{2\mu}~.
\end{aligned}
\ee

As an example of application of (\ref{basic:eq:O-cross}), the quark transversity correlation in $\ep \e\to q\,\bar q$ is $(00|nn)=\langle\sigma_n(q)\,\sigma_n(\bar q)\rangle=-2tu/(t2+u2)$. The crossed observable in $\e q\to \e q$ is the depolarisation parameter $(0n|0n)=\langle\sigma_n(q_i)\,\sigma_n(q_f)\rangle=2su/(s2+u2)$, which can be obtained from $(00|nn)$ by crossing $t\to s$ and a change of sign owing to (\ref{basic:eq:crossing}).  

If one knows constraints for the observables  of $A+B\to C+D\,(+X)$, one can obtain analogous constraints for $A+\overline{C}\to\overline{B}+D\,(+X)$, using the crossing relations (\ref{basic:eq:crossing}). For instance, in the inclusive case (see Sec.~\ref{se:incl}), Eq.~(\ref{incl:eq:constr}) can be deduced from Eq.~(\ref{incl:eq:DM}).   

%$(\lambda,\mu|\nu\tau)$

\subsubsection{Chiral invariance}
In hard scattering processes, helicity is conserved along a fermion line, up to correction factors of the order of $|m^2/\tilde q^2|^{1/2}$, where $\tilde q=\tilde p'-\tilde p$ is the four-momentum transferred to this line by the other particles. In an annihilation or creation process, $\tilde q$ is time-like and the helicities of the fermion and the antifermion add up to zero. This conservation rule comes from the chiral invariance of the Lagrangian of the standard model in the $m\to0$ limit. Thus, in this limit, if fermions $f_1$ and $f_2$ or $\bar f_2$ are on the same line, $D_{ll}=+1$ in the $f_1\to f_2$ case and  $A_{ll}$ or $C_{ll}=-1$ in the $f_1+\bar f_2$ case. 

In many cases, however, different Feynman diagrams connect external particles by fermion lines in different ways. For instance in $\ep\,\e\to\mathrm{e'^+\,e'^-}$, the $t$-channel pole diagram connects $\e$ to $\mathrm{e'^-}$ whereas the $s$-channel pole connects $\e$ to $\ep$. Let us first consider a process 
\be
X\to f_1+f_2...+f_N+Y~,
\ee 
where the spin-1/2 fermions $f_1,...f_N$ are not connected to the remaining particles, which are globally denoted  as $X$ and $Y$. 
% $e^-\,e^-\to e^-\,e^-$ or , Then $D_{ll}<1$, or $A_{ll}$ or $C_{ll}>-1$. 
The longitudinally polarised cross section has the form
\be
\sigma(\lambda_1,\lambda_2, ...\lambda_N)=I_0\,(1+A_1\lambda_1+A_2\lambda_2+...+A_N\lambda_N+
A_{12}\lambda_1\lambda_2+...)~,
\ee
where $\lambda=\pm1$ is twice the helicity. Due to helicity conservation, this cross section is non-vanishing only if $\lambda_1+\lambda_2+\cdots+\lambda_N=0$, which can be expressed by 
\be
(\lambda_1+\lambda_2+\cdots+\lambda_N)\,\sigma(\lambda_1,\lambda_2, \ldots,\lambda_N)=0 ~.
\ee
Before using this identity, one must replace every $\lambda_i^2$ by 1. Then the coefficient of each monomial, \eg of $\lambda_2\lambda_5\lambda_7$, has to vanish. On gets the following sets of identities
\begin{align}
A_1+A_2+\cdots+A_N&= 0~, \\
1+A_{12}+A_{13}+\cdots+A_{1N}&=0\quad\hbox{and permutations~,}\\
A_1+A_2+A_{123}+A_{124}+\cdots+A_{12N}&=0\quad\hbox{and permutations~,}\\
A_{12}+A_{23}+A_{13}+A_{1234}+A_{1235}+\cdots+A_{123N}&=0\quad\hbox{and permutations~,}
\end{align}
and so on. The first equation tells that the average total final helicity is zero. The second tells that if particle 1 has $\lambda=+1$, the average total helicity of the remaining particles is $-1$, etc. (these equations take simpler homogeneous forms if one defines $A_{11}=1$, $A_{121}=A_2$, $A_{1231}=A_{23}$, etc.).

Identities for the more general reaction 
\be \label{basic:eq:f1f2...}
f_1+f_2...+f_M+X\to f_{M+1}...+f_N+Y~,
\ee
are obtained from the preceding one by crossing, using (\ref{basic:eq:crossing}). Thus, in $\ep\,\e\to \ep\,\e$ at large momentum transfer, we have $(l0|00)+(0l|00)=(00|l0)+(00|0l)$,~ $1+(ll|00)=(l0|l0)+(l0|0l)$, etc.

Chiral invariance has also major consequences on transverse spin observables. First of all, the CSM matrix $\langle\lambda_1,...\lambda_N|\R|\lambda_1,...\lambda_N\rangle$ of (\ref{basic:eq:f1f2...}) cannot contain an odd number of helicity flips $\lambda'_i=-\lambda_i$. Therefore $\left\langle \Ocal_{\lambda\mu...}\right\rangle=0$ for an odd number of transverse indices. In particular single transverse spin asymmetries vanish, at least in a hard sub-process. The quark \emph{transversity}\,%
\footnote{%
In the parton terminology, transversity means transverse polarisation of any azimuth, \ie, not necessarily normal to the scattering plane. }
distribution $\delta q(x)$ in the nucleon cannot be measured in ordinary deep-inelastic scattering, since it involves only one helicity flip along the quark line \cite{Artru:1989zv}.
But it can be measured in the \emph{semi-inclusive} reaction $\e+\overrightarrow{N}\to \mathrm{e}'^{-}+\overrightarrow{\Lambda}+X$ if the $\Lambda$ is analysed \cite{Artru:1990wq}.

Transversity correlations (T.C.) between two fermions of a hard subprocess exist only if they belong  to the same connected subset. In QED and QCD, there is no T.C. between quarks and leptons or between quarks of different flavour. Photons or gluons do not mediate transverse spin information. Under $Q^2$ evolution, $\delta q(x)$ does not mix with any polarised gluon distribution, unlike $\Delta q(x)$. Note that T.C. can exist between two fermions $f_1$ and $f_2$ not directly connected in any Feynman diagram, for instance the initial $\e$ and the final $\ep$ of Bhabha scattering. It suffices that a third fermion $f_3$ connects to $f_1$ in a diagram $D_1$ and to $f_2$ in a diagram $D_2$. In fact $f_1$ and $f_2$ are connected in the unitarity diagram representing the $D_1\times D_2$ interference. %two initial  $e^-$'s    $f_1$ and $f_2$ 

The azimuthal dependence of allowed transverse spin correlations are strongly constrained as we shall see below. 
In the massless limit, longitudinal polarisation $\vec{S}_{\rm L}$ is a Lorentz pseudoscalar and transverse polarisation $\vec{S}_{\rm T}$ can be promoted to a four-vector $\mathcal{T}^\mu$ defined up to a ``gauge" transformation \cite{Artru:2002pu}
\begin{equation}
\mathcal{T}^\mu \rightarrow \mathcal{T}^\mu + C \, p^\mu
\,,
\label{basic:eq:gauge}
\end{equation}
and constrained by~ $\tilde p.\mathcal{T} = 0$ and  $\vec{S}_{\rm L}^2-\mathcal{T}^2\equiv\vec{S}_{\rm L}^2+\vec{S}_{\rm T}^2 \le 1$. 
The ``gauge'' $\mathcal{T} = (0, \vec{S}_{\rm T})$ is generally used. A pure state with $\vec{S}_{\rm T}\ne0$ can be expressed, up to an over-all phase, as the following superposition of helicity states 
\begin{equation}\label{basic:eq:SL+ST}
|\vec{S}_{\rm L}, \vec{S}_{\rm T} \rangle = %2^{-{1\over2}} \, 
\left|{\vec{S}_{\rm T}\over2}\right|^{1\over2}\,\left[\exp\left({\eta-i\phi\over2}\right)\,\left|+{1\over2}\right\rangle + \exp\left({-\eta+i\phi\over2}\right)\,\left|-{1\over2}\right\rangle\right]
\,,
\end{equation}
where $\phi$ is the azimuth of $\vec{S}_{\rm T}$ in the $\{\vec m,\vec n,\vec l\}$ frame of the particle, $\vec{S}_{\rm L} \equiv \tanh \eta$ and $|\vec{S}_{\rm T}|=1/\cosh\eta$. The chiral transformation $\Psi\to \exp(-i\Delta\phi\gamma_5/2)\,\Psi$ multiplies the right-handed component by $\exp(-i\Delta\phi/2)$ and the left-handed one by $\exp(i\Delta\phi/2)$, therefore rotates $\vec{S}_{\rm T}$ by $\Delta\phi$ about the momentum.  

Applying chiral symmetry to the reaction (\ref{basic:eq:f1f2...}), the polarised cross section is invariant under the simultaneous rotation of the transversity vectors of $f_1, ...f_N$ by a common angle $\Delta\phi$ about the respective particle momenta. %This can be checked using (\ref{SL+ST}) and helicity conserving amplitudes. 
We can call this a "Cardan transformation", by analogy with rotations transmitted by a mechanical shaft drive transmission. 
In  $\e\,q\to \e\,q$, for instance, where the quarks are on the same fermion line, the ``Cardan'' invariance gives $D_{nn}=D_{mm}$, $D_{mn}=-D_{nm}$. 
%$e^-\,q\to e^-\,q$ is the depolarisation parameter $D_{nn}=\langle\sigma_n(q_i)\,\sigma_n(q_f)\rangle=2su/(s2+u2)$, which can be obtained from $C_{nn}$
In the inclusive section $f_a+f_b\to C+X$, one has $A_{nn}=A_{mm}$, $A_{mn}=-A_{nm}$. No transverse spin correlation survives if one integrates over the azimuth of particle $C$ (Hikasa theorem \cite{Hikasa:1985qi}). Indeed, an anticlockwise rotation of the scattering plane about $\pv_a$ is equivalent to a clockwise rotation of both $\vec{S}_{\rm T}(a)$ and $\vec{S}_{\rm T}(b)$ about $\pv_a$, which is at variance with Cardan transformation (clockwise rotations of $\vec{S}_{\rm T}(a)$ and $\vec{S}_{\rm T}(b)$ about $\pv_a$ and $\pv_b$ respectively). 

In a similar manner, we can apply the imaginary chiral rotation $\exp(\Delta\eta\gamma_5)$ to initial fermions and $\exp(-\Delta\eta\gamma_5)$ to the final fermion, that is to say %$\mathcal{M}=\Gamma^{-1}\mathcal{M}\Gamma$, 
\begin{equation}
\eta_i \to \eta_i + \Delta\eta\quad\hbox{for initial fermions ,} 
\qquad  \eta_j \to \eta_j - \Delta\eta\quad\hbox{for final fermions.} 
\end{equation}
This ``see-saw'' transformation leaves invariant the rescaled cross section
\be
\tilde\sigma \equiv \sigma\,\prod_1^N |\vec{S}_{\rm T}(i)|^{-1}~.
\ee
%\emph{(\`a continuer)}

%
\subsubsection{Inequalities relating isospin-partner reactions}
Consider for instance the nucleon--antinucleon elastic and charge-exchange scattering reactions, in the limit of isospin symmetry. The isospin $I=1$ amplitude $\mathcal{M}_1$ corresponds to $\bar{\mathrm{n}}\mathrm{p}\to\bar{\mathrm{n}}\mathrm{p}$, while $\ppb$ elastic scattering is governed by
the combination  $(\mathcal{M}_1+\mathcal{M}_0)/2$ of $I=1$ and $I=0$, and the charge exchange (c.e.) $\ppnn$, to $(\mathcal{M}_1-\mathcal{M}_0)/2$ \cite{Klempt:2002ap}.  Each isospin amplitude $\mathcal{M}_i$ includes five spin amplitudes (this is a limiting case of the $\ppLL$, with $g\to 0$ when the baryon masses become equal, see Sec.~\ref{excl:sub:hyp-pair} for details), and can be written as
\begin{equation} \label{basic:eq:isoa}
(a+b)\mathbbm{1}
+(a-b)\, {\vec \sigma}_1.\vec{n} \, {\vec \sigma}_2 .\vec{n}  
+(c+d)\, {\vec \sigma}_1. \vec{m} \, {\vec \sigma}_2 . \vec{m}+(c-d)\, {\vec \sigma}_1 .\vec{l} \, {\vec \sigma}_2. \vec{l} 
+i e\,( {\vec\sigma}_1+{\vec \sigma}_2 ). \vec{n}~. 
\end{equation}
Hence, 
\begin{equation}\label{basic:eq:iso1}
a(\bar{\mathrm{n}}\mathrm{p})=a(\bar{\mathrm{p}}\mathrm{p})+ a(\mathrm{c.e.})~,
\end{equation}
and similarly for the four other amplitudes $b$, $c$, $d$, $e$ entering $\mathcal{M}$, as per Eq.~(\ref{basic:eq:isoa}).

Since $I_0=|a|^2+|b|^2+\cdots |f|^2$, it is a textbook exercise to derive from Eq.~(\ref{basic:eq:iso1})
\begin{equation}\label{basic:eq:iso2}
\left[\sqrt{\sigma(\bar{\mathrm{p}}\mathrm{p}})-\sqrt{\sigma(\mathrm{c.e.})}\right]^2\le
\sigma(\bar{\mathrm{n}}\mathrm{p})\le
\left[\sqrt{\sigma(\bar{\mathrm{p}}\mathrm{p}})+\sqrt{\sigma(\mathrm{c.e.})}\right]^2~,
\end{equation}
either for the angular distribution $I_0$ or the integrated cross-section $\sigma$.

However, it is less known that inequalities can also be derived for spin observables. At first sight, nothing simple can be written for, \eg, the polarisation $P$ and analysing power $A_n$  given by $I_0 P=I_0 A_n=2\IM(ae^*)$. If it is rewritten in the transversity basis as $I_0 P=|a'|^2-|e'|^2$, it is readily seen that the inequality of type (\ref{basic:eq:iso2}) written for $|a'|^2$ cannot be safely combined with those written for $|e'|^2$. The difficulty was overcome by Michel \etal\ \cite{Bourrely-iso,Bourrely:1980mr} (see, also, \cite{Korkea-aho:1971,Doncel:1972rk,Doncel:1973rh,Tornqvist:1973is}). The quantities $I_0(1\pm P)$ are easily seen to consist of a sum of positive terms in the transversity basis
\begin{equation}\label{basic:eq:iso3}
I_0(1-P)=|b'|^2+|c'|^2+|d'|^2+2|e'|^2~,
\qquad
I_0(1+P)=2|a'|^2+|b'|^2+|c'|^2+|d'|^2~.
\end{equation}
Hence both  quantities $X_{\mp}=I_0(1\mp P)$ verify
\begin{equation}\label{basic:eq:iso4}
\left[\sqrt{X(\bar{\mathrm{p}}\mathrm{p})}-
\sqrt{X(\mathrm{c.e.})}\right]^2\le
X(\bar{\mathrm{n}}\mathrm{p}) \le
\left[\sqrt{X(\bar{\mathrm{p}}\mathrm{p})}+\sqrt{X(\mathrm{c.e.})}\right]^2~.
\end{equation}

This result is rather general. For  any spin observable $\Ocal$, there always exists a basis for the amplitudes such that 
\begin{equation}\label{basic:eq:iso4a}
I_0 \Ocal=\pm |a|^2\pm |b|^2+\cdots
\end{equation}
so that $I_0(1\pm \Ocal)$ is a sum of squared amplitudes.
\newpage\setcounter{equation}{0}
\section{Exclusive reactions}\label{se:excl}
In this section, we review the formalism for some exclusive reactions, with various  number of amplitudes and spin observables. More details are given for the hyperon-pair production reaction \ppLL\ and for the photoproduction of pseudoscalar mesons, for which measurements have been performed recently.
\subsection{Pion--nucleon elastic scattering}\label{excl:sub:piN}
As described in many textbooks, see, \eg, \cite{JacobChew:1964}, there are  two independent amplitudes $\mathcal{A}(s,t)$ and $\mathcal{B}(s,t)$ to describe $\pi\mathrm{N}$ elastic scattering. They are defined by
\begin{equation}\label{excl:eq:pin1}
\Mcal_{\rm fi}= -2m \bar{u}(\tilde{p}')\left[-\mathcal{A}+i\gamma.{\tilde{q}+\tilde{q}'\over 2}\mathcal{B}\right]u(\tilde{p})~,
\end{equation}
acting between the Dirac spinors of initial and final nucleons of four-momenta $\tilde{p}$ and $\tilde{p}'$, respectively, while  $\tilde{q}$ and $\tilde{q}'$ are  the pion four-momenta. The amplitude can be rewritten in the centre-of-mass as 
\begin{equation}\label{excl:eq:pin1a}
\Mcal_{\rm fi}=8\pi\sqrt{s}\,\chi_f^\dagger \left(f+ i g\,\vec{\sigma}.\vec{n}\right)\chi_i~,
\end{equation}
acting on the  Pauli spinors $\chi_i$ and $\chi_f$, with the relations \cite{JacobChew:1964}
\begin{equation}\label{excl:eq:pin1b}
f = {p^0+m\over 8\pi\sqrt{s}}\left[\mathcal{A}+\mathcal{B}(\sqrt{s}-m)\right]~,\qquad 
g=-{p^0-m\over8\pi\sqrt{s}}\left[\mathcal{A}-\mathcal{B}(\sqrt{s}+m)\right]~,
\end{equation}
where the normalisation is such that the angular distribution (differential cross section) reads
\begin{equation}\label{excl:eq:pin2}
{\d \sigma\over \d \Omega}=I_0=|f|^2+ |g|^2~.
\end{equation}
The first spin observable is  the polarisation $P_n$ of the recoiling nucleon. It  is equal to the analysing power $A_n$, accessible if the target is transversally polarised. They are given by
\begin{equation}\label{excl:eq:pin3}
I_0P_n=I_0A_n=2\IM(fg^*)~. 
\end{equation}
Most phase-shift analyses, which led to valuable information on pion dynamics and nucleon resonances were based on the data available on $I_0$ and $P_n$. See, \eg, \cite{Arndt:1990bp} and Refs.\ therein.

Two additional observables are actually measurable if the target is polarised. The scattering process conserves the polarisation modulus, 
%($|\langle \Sv'\rangle|=|\langle \Sv\rangle|$), ($|\Sv'|=|\Sv|$), 
since the pion carry no spin information, and rotates the $(\vec{n},\Sv)$ plane by an angle $\alpha$ about $\vec{n}$ (here $\Sv'$ stands for $\langle \Sv'\rangle$ in the notation of (\ref{basic:eq:polar-c})). If all spins are measured in the frame 
$\{\vec{l}_{\rm b}=-\hat{\vec{p}},\,\vec{m}_{\rm b},\vec{n}\}$ associated with the beam,  the depolarisation parameters $D_{ij}=C_{i0j0}$ which do not vanish are
\begin{equation}\label{excl:eq:pin3a}
D^{\rm (b)}_{nn}=1~,\quad D^{\rm (b)}_{mm}=D^{\rm (b)}_{ll}=(|f]^2-|g|^2)/I_0~, \quad
D^{\rm (b)}_{ml}=-D^{\rm (b)}_{lm}=2\RE(fg^*)/I_0~.
\end{equation}
These parameters fulfil the Bohr identity (\ref{basic:eq:P-quant-O}).
% and the time reversal relations (\ref{basic:eq:Tclass.pi-N}-\ref{basic:eq:Tquant.pi-N}). 
If each particle has its spin  measured is its own frame, this gives the 
so-called spin-rotation parameters $A=D_{ll}=D_{mm}$ and $R=D_{lm}=-D_{ml}
%=A\tan(\alpha-\theta)
$, given by 
\be\label{excl:eq:pin3b}\begin{aligned}
I_0 A&=(|f|^2-|g|^2)\cos\theta-2\RE(fg^*)\sin\theta~,\\
I_0 R&=-(|f|^2-|g|^2)\sin\theta-2\RE(fg^*)\cos\theta~.\end{aligned}
\ee

From Eqs.\ (\ref{excl:eq:pin3}) and (\ref{excl:eq:pin3b}), it is obvious that
\begin{equation}\label{excl:eq:pin4}
P_n^2+A^2+R^2=1~.
\end{equation}
This can be interpreted as follows: given a purely sideways  or purely longitudinal target polarisation, the sum of the square of the three components of the polarisation of the final nucleon is unity. In practice, this means that two of these observables are needed, as well as the sign of the third one, for a complete information.

A value $D^{\rm (b)}_{mm}=0$, for instance, indicates that $|f|=|g|$, but does not inform about 
the relative phase of the two amplitudes.
If, instead, it happens that $D^{\rm (b)}_{mm}=+1$, then $g=0$, and the measurement of $D^{\rm (b)}_{ml}$ or $P_n$  is not necessary, except for cross-checking. 

\begin{figure}
\begin{center}
\includegraphics[width=.4\textwidth]{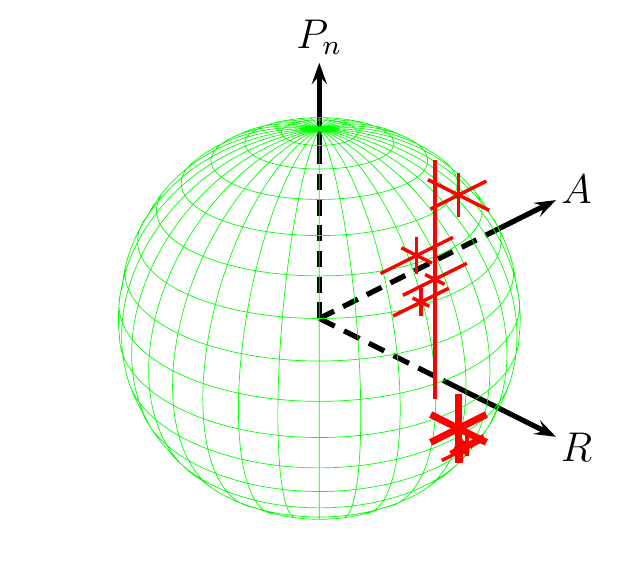}\qquad
\includegraphics[width=.4\textwidth]{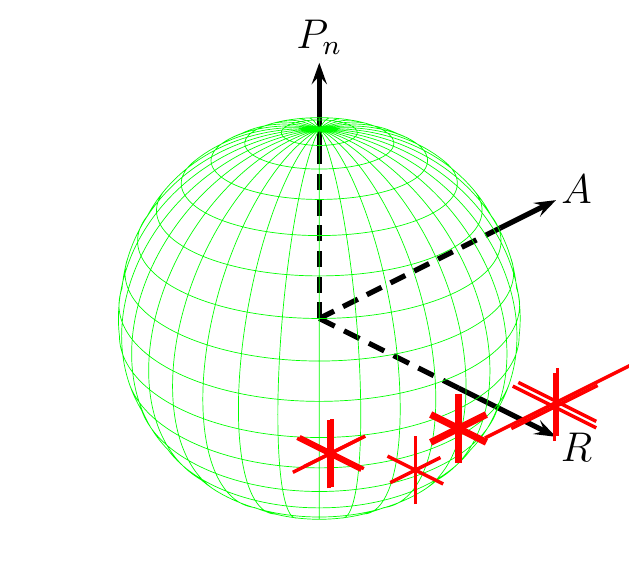}
\end{center}
\caption{Spin observables  for $\pi^-\mathrm{p}$ elastic scattering at 0.573 
and 0.685 GeV$/c$ (left), and for $\pi^+\mathrm{p}$ elastic scattering at 0.657 GeV$/c$ (right), in axes $\{R,A,P_n\}$ compared to the unit
sphere. The data are from Abaev \etal\ \protect\cite{Abaev:1988kn} for $\pi^-\mathrm{p}$ and Supek \etal\ \protect\cite{Supek:1993qa} for $\pi^+\mathrm{p}$, and are shown with the error bars on the three observables.
\label{excl:fig:abaev}}
\end{figure}

Results for the $\pi\mathrm{N}$ observables are listed in the Durham data base of the Particle Data Group.  Note that in some cases, the constraint (\ref{excl:eq:pin4}) is imposed in the analysis. In Tables
\ref{excl:tab:pimp} and \ref{excl:tab:pipp}, examples are given of simultaneous measurements of $A$, $R$ and $P_n$. The corresponding data are plotted in  $A$, $R$, $P_n$ coordinates in Fig.\ \ref{excl:fig:abaev}, against the unit sphere. The constraint (\ref{excl:eq:pin4}) is reasonably satisfied with regard to  the error bars.
\begin{table}
\caption{Spin observables for $\pi^-\mathrm{p}$ elastic scattering, as measured
by Abaev \etal\ \protect\cite{Abaev:1988kn} at 0.573 and 0.685 GeV$/c$.\label{excl:tab:pimp}}
\begin{center}\renewcommand{\arraystretch}{1.5}
\begin{tabular}{ccccc}
$p_{\mathrm{lab}}$ & $\cos(\vartheta_{\mathrm{cm}})$ & $R$  &  $A$  & $P_n$ \\
\hline
& $-0.63$ & $\phm 0.47\pm0.22$ & $\phm 0.53\pm0.20$ & $\phm0.70\pm0.13$ \\
0.573& $-0.72$ & $\phm 0.73\pm0.11$ & $-0.03\pm0.26$ & $\phm0.68\pm0.11$ \\
& $-0.80$ & $\phm 0.82\pm0.07$ & $\phm 0.01\pm0.23$ & $\phm0.56\pm0.10$ \\
& $-0.63$ & $\phm 0.86\pm0.05$ & $\phm 0.16\pm0.14$ & $-0.49\pm0.07$ \\
0.685& $-0.72$ & $\phm 0.87\pm0.04$ & $\phm0.19\pm0.12$ & $-0.46\pm0.07$ \\
& $-0.80$ & $\phm 0.86\pm0.06$ & $-0.13\pm0.20$ & $-0.50\pm0.08$ \\
\hline
\end{tabular}
\end{center}
\end{table}

%\newpage

\begin{table}
\caption{Spin observables for $\pi^+\mathrm{p}$ elastic scattering, as measured
by Supek \etal\ \protect\cite{Supek:1993qa} at 0.657~GeV$/c$.\label{excl:tab:pipp}}
\begin{center}\renewcommand{\arraystretch}{1.5}
\begin{tabular}{cccc}
$\cos(\vartheta_{\mathrm{cm}})$ & $R$  &  $A$  & $P_n$ \\
\hline
 $\phm0.150\pm0.012$ & $\phm 1.04\pm0.30$ & $\phm 0.65\pm0.31$ & $-0.36\pm0.20$ \\
 $\phm0.079\pm0.012$ & $\phm 1.05\pm0.28$ & $\phm 0.66\pm0.52$ & $-0.33\pm0.20$ \\
 $-0.704\pm0.008$    & $\phm 1.03\pm0.20$ & $-0.34\pm0.18$     & $-0.33\pm0.20$ \\
 $-0.755\pm0.008$    & $\phm 0.68\pm0.23$ & $-0.61\pm0.26$      & $-0.27\pm0.20$ \\
 $-0.796\pm0.008$    & $\phm 0.74\pm0.23$ & $-0.65\pm0.20$      & $-0.22\pm0.20$ \\
\hline
\end{tabular}
\end{center}
\end{table}

The reactions $\mathrm{K}^\pm\mathrm{p}\to \mathrm{K}^\pm\mathrm{p}$ have the same spin algebra than $\pi^\pm\mathrm{p}\to \pi^\pm\mathrm{p}$.  No spin transfer data are available in the case of $\mathrm{K}^+$. In the case of $\mathrm{K}^-$, some measurements have been done for $R$ at $p_{\mathrm{lab}}=40\;$GeV \cite{Bruneton:1975ar}, and $A$ was computed from $A=(1-R^2-P_n^2)^{1/2}$, using the polarisation data of the same group \cite{Bruneton:1975aq}.
\subsection{Antiproton--proton annihilation into two pseudoscalar mesons}
The reaction $\ppb\to\pi\pi$ is related by crossing to  $\pi\mathrm{N}$ elastic scattering and thus has the same number of independent amplitudes. The formalism is given, \eg, by Frazer and Fulco 
\cite{PhysRev.117.1603}.
The amplitude can be written as 
\be\label{excl:eq:psa}
\Mcal_{\rm fi}=-2 m \bar{u}(\tilde{p}')\left[-\mathcal{A} + i\gamma.{\tilde{q}-\tilde{q'}\over 2}\mathcal{B}
\right]v(\tilde{p})~,
\ee
in terms of the Dirac spinors and four-momenta of the antinucleon: $\tilde{p}=(p^0,\vec{p})$, nucleon $\tilde{p}'=(p^0,-\tilde{p})$ and pions $\tilde{q},\tilde{q}'=(q^0,\pm\vec{q})$.
It can be rewritten as
\begin{equation}\label{excl:eq:2ps1}
\Mcal_{\rm fi}=8 \pi \sqrt{s}\sqrt{p\over q}\Hcal~,
\quad
\Hcal=\chi^\dagger(\mathrm{N})(h_1 \vec{\sigma}.\vec{p}_1+ h_2 \vec{\sigma}.\vec{p}_2)\chi(\overline{\mathrm{N}})~,
\end{equation}
where $\chi$ is the spinor of the nucleon or antinucleon, and  $m$ the nucleon mass.
If the spinors are chosen as spin states along the initial momentum $\vec{p}_1$,  the helicity amplitudes 
\be\label{excl:eq:2ps2}
\Hcal _{++}=h_1 p_1+ h_2 p_2\cos\theta~,\qquad \Hcal_{+-}=h_2 p_2 \sin\theta~,
\ee
are obtained, to which the transversity amplitudes are related by
\be\label{excl:eq:3ps2a}
\Hcal^+={\Hcal _{++}+i\,\Hcal _{+-}\over\sqrt2}~,\qquad
\Hcal^-={\Hcal _{++}-i\,\Hcal _{+-}\over\sqrt2}~.
\ee

The spin observables are the analysing power, the beam asymmetry and the spin correlations in the initial state, and are submitted to an identity of type (\ref{excl:eq:pin4}).
Only the  differential cross-section $I_0$ and the analysing power $A_n$ have been reached, as polarised antiprotons are not yet available. They are given by 
\be\label{excl:eq:2ps3}
\begin{aligned}
I_0&=|\Hcal_{++}|^2+|\Hcal_{+-}|^2=|\Hcal^+|^2+|\Hcal^-|^2~,\\
I_0 A_n&=2 \IM(\Hcal_{++}\Hcal^*_{+-})=|\Hcal^+|^2-|\Hcal^-|^2~.
\end{aligned}\ee

\begin{figure}
\centerline{\includegraphics[width=.75\textwidth]{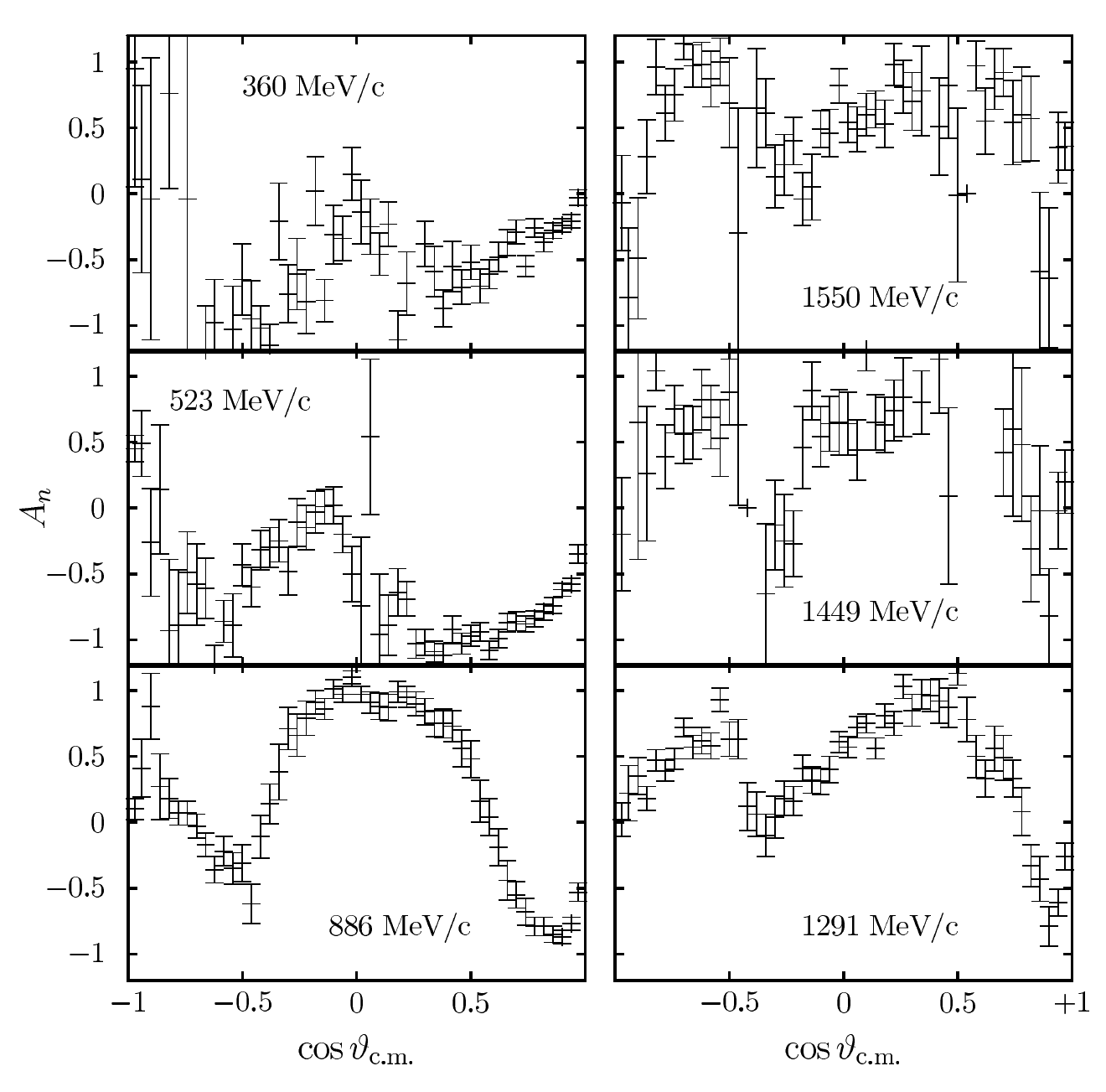}}
\caption{\label{excl:fig:Polpipi} %
Analysing power for $\bar{\rm p}{\rm p}\to \pi^-\pi^+$ at selected values of the antiproton momentum in the target frame. The data are from the PS172 collaboration  \protect\cite{Hasan:1992sm}.}
\end{figure}

The analysing power $A_n$ has been measured by the PS172 collaboration working with the LEAR facility of CERN \cite{Hasan:1992sm}.  They got remarkable results, with $A_n$ saturating or nearly saturating the limit $|A_n|\le 1$ in a wide domain of energies and angles. See Fig.~\ref{excl:fig:Polpipi}.

These results triggered several discussions. An amplitude analysis indicated the possibility of a series of broad resonances \cite{Oakden:1993vn}. In Ref.~\cite{Takeuchi:1992si}, it was stressed that the PS172 results suggest a simple relation $\Mcal_{++}\propto i\Mcal_{+-}$ between the helicity amplitudes.

In \cite{Elchikh:1993sn}, it was pointed out that the  transversity states are eigenstates of the tensor operator, which is expected to be very large in the nucleon--antinucleon initial state, especially for isospin $I=0$ \cite{Dover:1978br}. Both eigenstates experience a strong annihilation, but if the real potential is attractive in one state and repulsive in the other one, the amplitudes acquire different phases, and this produces a large value for $A_n$, as in  Eq.~(\ref{excl:eq:2ps3}). Hence the strong spin effect seen by PS172 is partly due to the strong spin dependence of the nucleon--antinucleon interaction.

\begin{figure}
\centerline{\includegraphics[width=.75\textwidth]{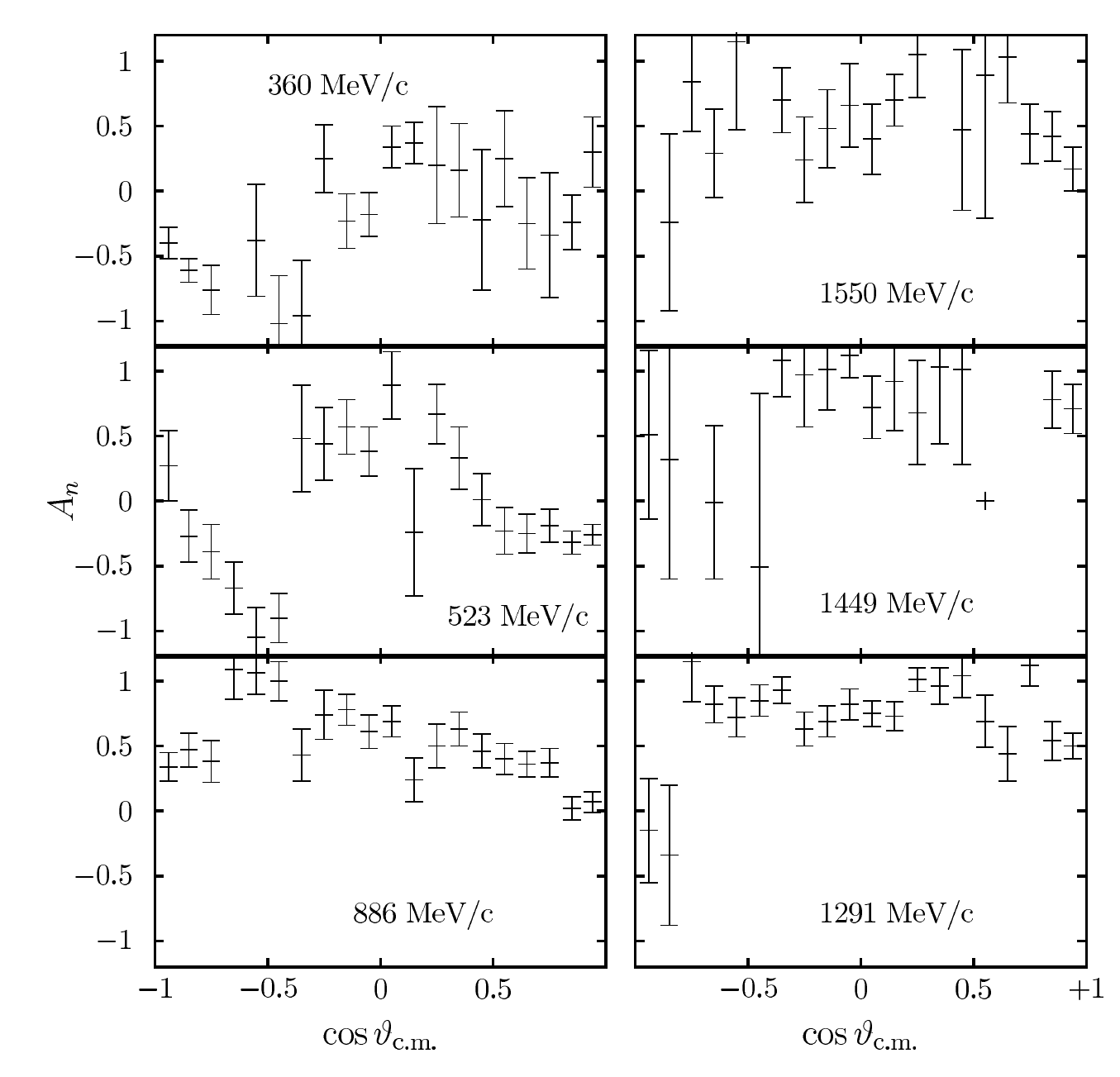}}
\caption{\label{excl:fig:PolKK}%
Analysing power for $\bar{\rm p}{\rm p}\to \mathrm{K}^-\mathrm{K}^+$ at selected values of the
antiproton momentum in the target frame. The data are from the PS172 collaboration  \protect\cite{Hasan:1992sm}.}
\end{figure}

The PS 172 collaboration \cite{Hasan:1992sm} also measured the analysing power for $\ppb\to\mathrm{K}^+\mathrm{K}^-$, and found  that $|A_n|$ is also rather large at some angles, see  Fig.~\ref{excl:fig:PolKK}.

If annihilation into two pseudoscalar mesons is ever measured with a polarised antiproton beam hitting  a polarised target, then it could be checked that the beam asymmetry $C_{n000}=-A_n$, and that the spin correlation coefficients $A_{mm}$ and $A_{ml}$ vanish when $|A_n=1|$. Indeed, from \cite{Leader:2001gr},
\be
I_0 A_{mm} =|\Hcal_{++}|^2-|\Hcal_{+-}|^2~,\quad
I_0 A_{ml}=-2 \RE(\Hcal_{++}\Hcal^*_{+-})~,
\ee
and one easily derives $A_n^2+A_{mm}^2+A_{ml}^2=1$, which is the analogue of (\ref{excl:eq:pin4}) for the crossed reaction.
\subsection{Hyperon-pair production}\label{excl:sub:hyp-pair}
\subsubsection{Experimental situation}
The reaction $\ppLL$ and other strangeness-exchange scattering processes with an antihyperon--hyperon pair in the final state has been measured at various energies.
The most precise and systematic study has been carried out  by the PS185 collaboration \cite{Pas2001,Bassalleck:2002sd,Bassalleck:2004dz,Paschke:2006za} at the LEAR facility of CERN.

The aim was to study how strangeness is produced. 
Schematically, two classes of models existed when the PS185 experiment was proposed. The first one is based on $\mathrm{K}$ and  $\mathrm{K^*}$ exchange, in analogy with the exchange of charged mesons such as  $\pi^+$ or $\rho^+$ mediating the charge-exchange reaction $\mathrm{\bar{p} p\to\bar{n}n}$. Alternatively, hyperon-pair production can be described at the quark level, with a pair of light quarks annihilated and a pair of strange quarks, $\bar{s}s$, created out of the vacuum, for instance in the ${}^3\mathrm{P}_0$ model \cite{LeYaouanc:1988fx}.

\begin{figure}[!!ht]
\begin{center}
\includegraphics[width=.7\textwidth]{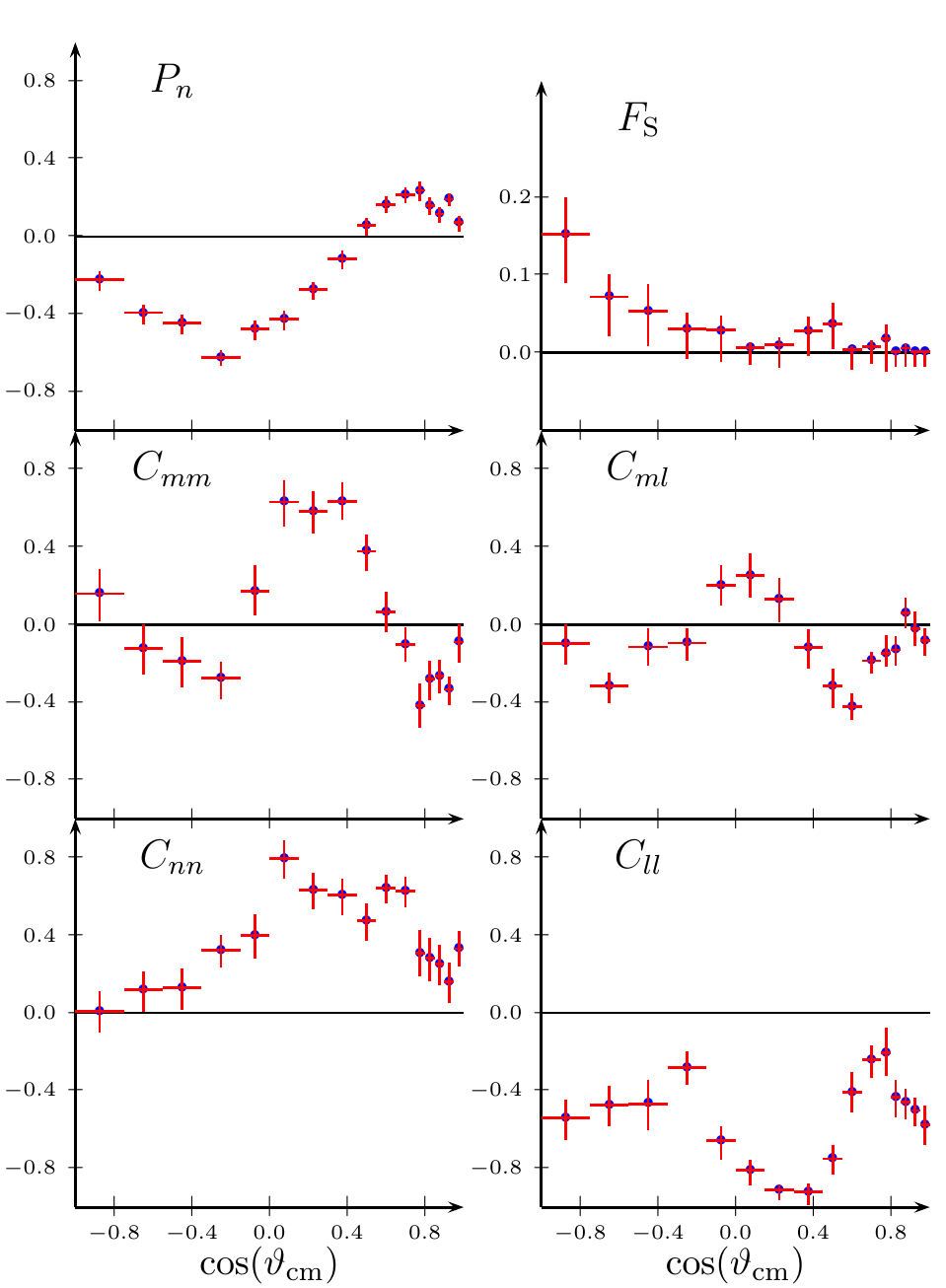}
\end{center}
\caption{\label{excl:fig:PS185a} Results of PS 185 \protect\cite{Paschke:2006za} on $\ppLL$ for the polarisation, and initial-state spin correlations, out of which the spin-singlet fraction $F_S$ can be estimated.  This corresponds to one of the last runs, at $p_{\mathrm{lab}}=1.637\;$GeV$/c$. Earlier measurements, obtained without polarised target, are in good agreement.}
\end{figure}

Both models were rather successful to describe the first data of PS185, taken with an unpolarised proton target at various values of the incoming antiproton momentum. This corresponds to the polarisation $P_n$ and final-state spin-correlation coefficients $C_{ij}$, out of which the singlet fraction can be estimated. Examples of observables accessible without polarised target are given in Fig.~\ref{excl:fig:PS185a}\,\footnote{They actually correspond to the last run, with a polarised target, but are equivalent to the earlier data obtained without polarisation of the target.}.

To better distinguish among the models, the experiment was improved as to include a polarised target.
An estimate by Holinde \etal\ \cite{Haidenbauer:1992wp}, for instance, indicated that the depolarisation $D_{nn}$ is predicted differently by the kaon-exchange model and the quark-pair creation model. Meanwhile a third type of model was proposed by Alberg \etal\ \cite{Alberg:1995zp}, in which the $\bar{s}s$ pair is not created out the vacuum, but extracted out the polarised sea of the incoming proton or antiproton, leading to a pronounced negative 
$D_{nn}$\,\footnote{Remember that $D_{nn}=+1$ in absence of spin-dependent forces.}.
This is in this context that a warning was set that the existing data already constrain the allowed range of $D_{nn}$ \cite{Richard:1996bb,Elchikh:1999ir}, and the problem of the spin observables of the $\ppLL$ reaction was further studied \cite{Richard:2003bu,Artru:2004jx,Elchikh:2004ex,%
Elchikh:2005wg}.

\begin{figure}[!!ht]
\begin{center}
\includegraphics[width=.7\textwidth]{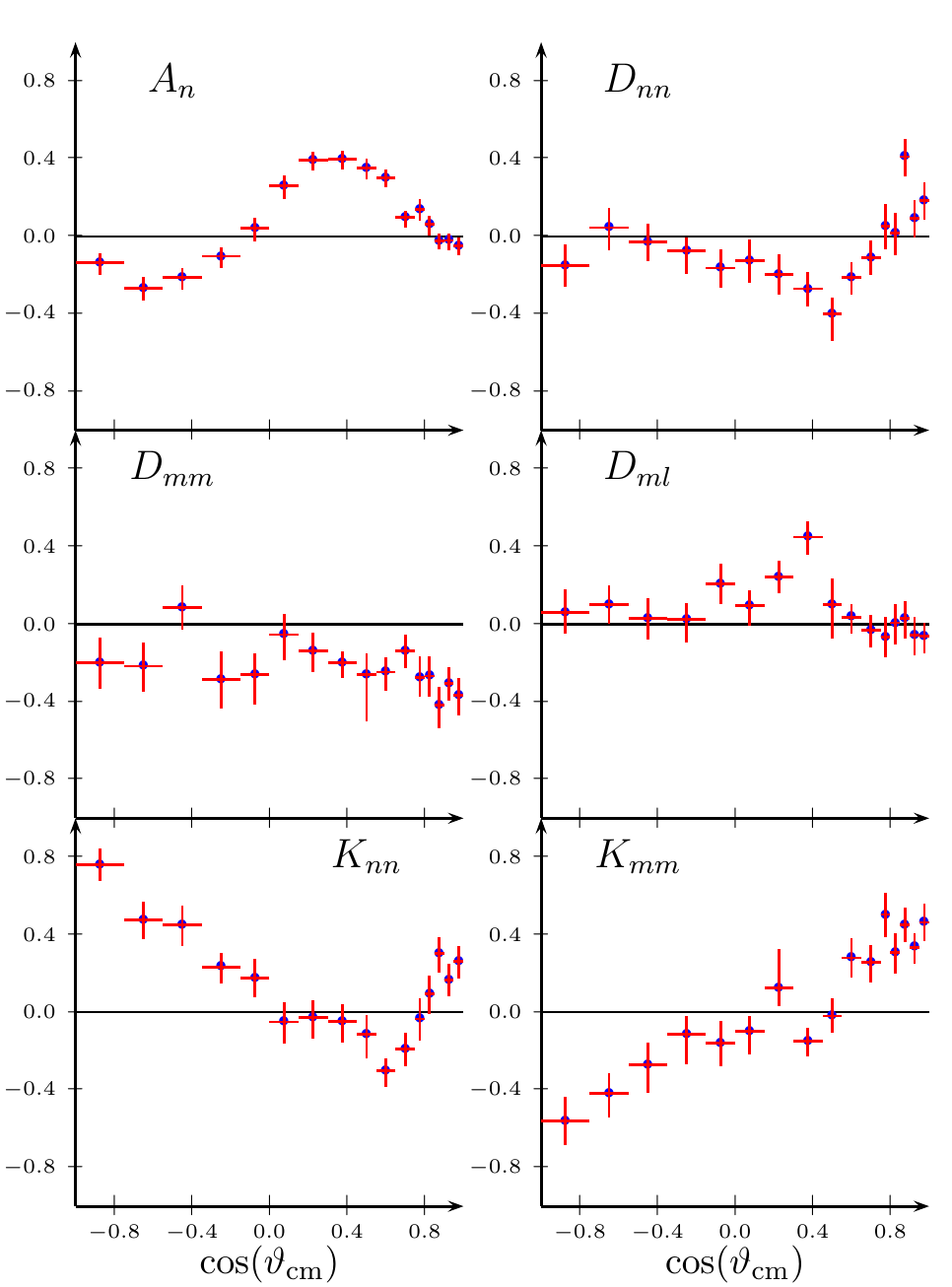}
\end{center}
\caption{\label{excl:fig:PS185b} Some of the results of PS 185 \protect\cite{Paschke:2006za}  for the analysing power, depolarisation and transfer of polarisation of the reaction  $\ppLL$ at $p_{\mathrm{lab}}=1.637\;$GeV$/c$.}
\end{figure}

The PS185  data of the at incident momentum $p_{\mathrm{lab}}=1.637\;$GeV$/c$ have been  analysed and published in great detail \cite{Pas2001,Bassalleck:2002sd,Paschke:2006za}. Some of them are given in Figs.~\ref{excl:fig:PS185a} and \ref{excl:fig:PS185b}.  Preliminary results of the $1.525\;$GeV$/c$ measurements have been presented at the LEAP03 conference \cite{Bassalleck:2004dz}, but never published, to our knowledge.

\subsubsection{Formalism of hyperon-pair production}
There are six independent amplitudes for \ppLL, once symmetries are enforced. 
We follow here the notation adopted in the recent  literature. See, for 
example
%\footnote{Sometimes the $m$ and $l$ axes are defined differently for each outgoing particle, 
%but the formulas remain identical if the operators in front of the amplitudes 
%$b$, $c$ and $g$ are modified accordingly. 
%Otherwise, there are some changes of sign.}
\cite{Holzenkamp:1989tq,Haidenbauer:1992ws,Paschke:2000mx}.
The amplitude is denoted%
\begin{multline} \label{excl:eq:amp-ppLL}
\mathcal{M}=(a+b)\mathbbm{1} 
+(a-b)\, {\vec \sigma}_1.\vec{n} \, {\vec \sigma}_2 .\vec{n}  
+(c+d)\, {\vec \sigma}_1. \vec{m} \, {\vec \sigma}_2 . \vec{m}\\
 {}+(c-d)\, {\vec \sigma}_1 .\vec{l} \, {\vec \sigma}_2. \vec{l} 
+i e\,( {\vec\sigma}_1+{\vec \sigma}_2 ). \vec{n} 
+g\,( {\vec \sigma}_1 . \vec{l}  \, {\vec \sigma}_2 . \vec{m} +  
{\vec \sigma}_1 . \vec{m}  \, {\vec \sigma}_2.  \vec{l} )\ ,
\end{multline}
in terms of the  unit vectors $\vec{l}$, $\vec{m}$ and $\vec {n}$ adapted to each particle, as per Eq.~(\ref{basic:eq:frame}).
If $ I_{0}=\Tr[\mathcal{M} \mathcal{M}^{\dag}]$ is the differential cross section, the spin observables 
are defined as usual by
\be\begin{aligned}
\label{Gen-Obs}
C_{00n0}I_0=P_{n}I_{0}&=\Tr[\vec{\sigma}_1. \vec{n} \mathcal{M} \mathcal{M}^{\dag}]~,\\
C_{0n00}I_0=A_{n}I_{0}&=\Tr[\mathcal{M} {\vec{\sigma}}_{2}. \vec{n} \mathcal{M}^{\dag}]~,\\
C_{00i0j}I_0=D_{ij}I_{0}&=
       \Tr[\vec{\sigma}_2. \vec{\jmath} \mathcal{M}  {\vec{\sigma}}_2 .
           \vec{\imath} \mathcal{M}^{\dag}]~,\\
C_{0ij0}I_0=K_{ij}I_{0}&=
  \Tr[\vec{\sigma}_1. \vec{\jmath} \mathcal{M} {\vec{\sigma}}_2 . \vec{\imath} \mathcal{M}^{\dag}]~,\\
C_{00ij}I_0=C_{ij}I_{0}&= \Tr[\vec{\sigma}_1. \vec{\imath}\,\vec{\sigma}_2 .
           \vec{\jmath} \mathcal{M} \mathcal{M}^{\dag}]~,\\
C_{0\alpha i j}I_0=C_{\alpha ij}I_{0}&=\Tr[\vec{\sigma}_1.\vec{\imath} \, 
\vec{\sigma}_2. \vec{\jmath}\, \mathcal{M}\, \vec{\sigma}_{2} .\vec{\alpha}\, \mathcal{M}^{\dag}]~.             
\end{aligned}\ee
The spin observables (times $I_{0}$) which are accessible with a transversally polarised 
proton target are given by the quadratic relations 
\be\begin{aligned}
\label{excl:eq:ppllobs}
 I_0&=|a|^2+|b|^2+|c|^2+|d|^2+|e|^2+|g|^2~,\quad
& P_{n}I_0&=2\IM (a{e}^*+d{g}^*)~,\\
C_{nn}I_0&=|a|^2-|b|^2-|c|^2+|d|^2+|e|^2+|g|^2~,
 &A_{n}I_0&=2\IM (a{e}^*-d{g}^*)~,\\
D_{nn}I_0&=|a|^2+|b|^2-|c|^2-|d|^2+|e|^2-|g|^2~,
&C_{ml}I_0&=2\RE (a{g}^*-d{e}^*)~,\\
K_{nn}I_0&=|a|^2-|b|^2+|c|^2-|d|^2+|e|^2-|g|^2~, 
&D_{mm}I_0&=2\RE (a{b}^*+c{d}^*)~,\\
C_{mm}I_0&=2\RE (a{d}^*+b{c}^*-g{e}^*)~,
&D_{ml}I_0&=2\RE (c{g}^*+b{e}^*)~,\\
C_{ll}I_0&=2\RE (-a{d}^*+b{c}^*+g{e}^*)~,
&K_{mm}I_0&=2\RE (a{c}^*+b{d}^*)~,\\
C_{nlm}I_0&=2\IM (g{e}^*-a^*{d}+b^*{c})~,
&K_{ml}I_0&=2\RE (b{g}^*+c{e}^*)~,\\
C_{nml}I_0&=2\IM (g{e}^*-a^*{d}-b^*{c})~,&C_{nmm}I_0&= 2\IM (d{e}^*-a{g}^*)~,\\
C_{mnl}I_0&=2\IM(ab^*-cd^*)~,&C_{mln}I_0&=2\IM(ac^*-bd^*)~,\\
C_{mnm}I_0&=2\IM(be^*+cg^*)~,&C_{mmn}I_0&=2\IM(ce^*+bg^*)~,
\end{aligned}\ee
supplemented by  (see, \eg, \cite{Pas2001})
\be\begin{aligned}
C_{000n}&=C_{00n0}~,\qquad &C_{nnn}&=A_n~,\\
C_{lm}&=C_{ml}~, &C_{nll}&=C_{nmm}~,
\end{aligned}\ee
the first identity meaning that $\Lb$ and $\L$ have the same polarisation,

A purely algebraic $b\leftrightarrow c$ symmetry is observed, which, for instance, leaves $C_{nn}$ invariant and exchanges $D_{nn}$ and $K_{nn}$. The pairs $\{D_{nn},C_{nn}\}$ and $\{K_{nn},C_{nn}\}$ are, indeed, submitted to the same constraints, which will be studied in the next subsection.
This symmetry corresponds to   a single particle interchange: the observables for $\bar{\mathrm{p}}\mathrm{p}\to \Lambda\overline{\Lambda}$ are just a relabelling of these for $\ppLL$.
More permutation properties of (\ref{excl:eq:ppllobs}) are discussed in \cite{LaFrance:1981bg}.
\subsubsection{Empirical approach}\label{excl:sub:empi}
Richard and Elchikh \cite{Richard:1996bb,Elchikh:1999ir}
 have studied the inequalities relating pairs of \ppLL\ observables in a 
 empirical but systematic way.  The investigation has been extended to triples of observables
 \cite{Richard:2003bu,Artru:2004jx}. The method simply consists of generating random, fictitious amplitudes $a$, $b$, etc. ($a$ can be chosen real and positive), computing the observables (\ref{excl:eq:ppllobs}), and plotting  the results the one against the other. 
Once the contours revealed, it can be attempted to demonstrate rigorously the corresponding inequalities.
 
 Whilst for pairs of observables, there are only three possibilities, for triples of observables, many more cases have been identified. Some of them have already been studied \cite{Richard:2003bu,Artru:2004jx,Elchikh:2005wg}\,\footnote{The ``twisted cushion'' shown in the above references turns out  to be, by more careful investigation,  a straight tetrahedron whose volume is slightly smaller.}.
 A more comprehensive survey is done below.
 
\begin{figure}[!ht]
\centerline{
\includegraphics[width=.3\textwidth]{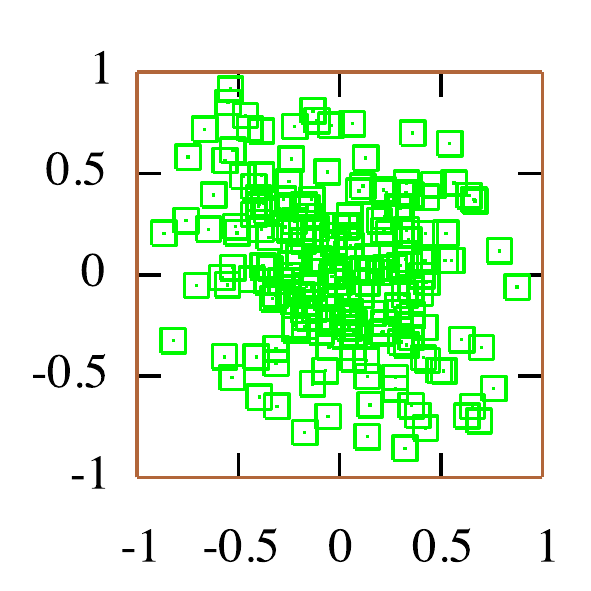}
\includegraphics[width=.3\textwidth]{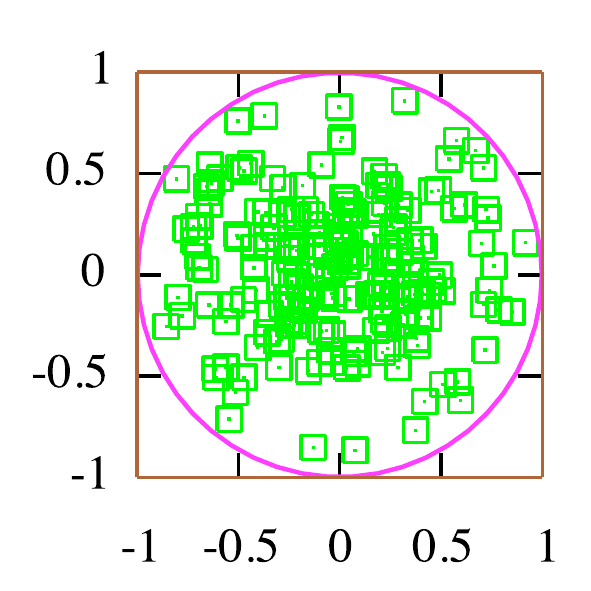}
\includegraphics[width=.3\textwidth]{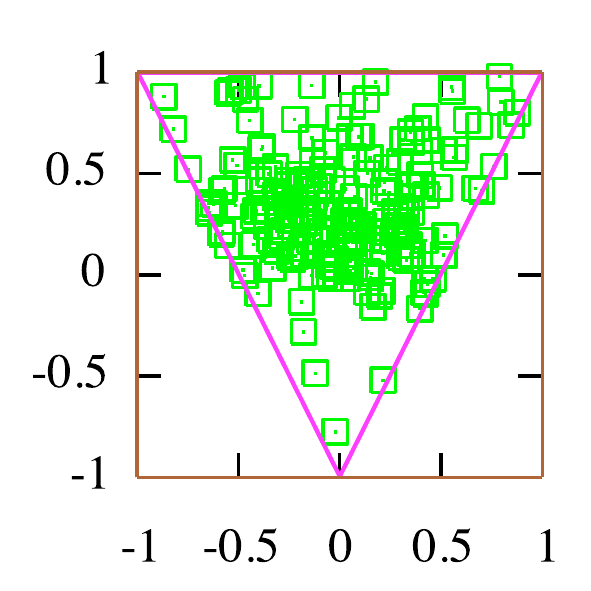}}
\caption{\label{excl:fig:ex2} Random simulation of  $P_n$ vs.\ $A_n$ (square), $A_n$ vs.\ $D_{mm}$ (disk) and $P_n$ vs.\ $C_{nn}$ (triangle)}
\end{figure}

\paragraph{Pairs of observables} For pairs of observables, either the full square $[-1,+1]^2$ is covered  or a subdomain, which is the unit disk or a triangle, as shown in the examples given  in Fig.~\ref{excl:fig:ex2}: $P_n$ vs.\ $A_n$ (square), $A_n$ vs.\ $D_{mm}$ (disk) and $P_n$ vs.\ $C_{nn}$ (triangle). These plots correspond to 2000 random choices of amplitudes. To lighten the figure, some points inside the convex domain of the previous ones are omitted.
The results are summarised in Table~\ref{excl:tab:reca} for the sixteen observables listed in  (\ref{excl:eq:ppllobs}).
\begin{table}[!ht]
\caption{\label{excl:tab:reca} Domain allowed for pairs of observables: the entire square
$\left(\protect\sq\right)$, the unit disk $ \left(\protect\dis\right)$, the triangle $|2\Ocal_1| \le\Ocal_2+1$ $\left(\protect\trv\right)$, or
$|2\Ocal_2| \le \Ocal_1+1$ $\left(\protect\trh\right)$, where $\Ocal_1$ is horizontal and $\Ocal_2$ vertical. The symbol \protect\disp\ indicates that the pair of observables is constrained in the unit disk, but the corresponding operators do not anticommute. }
\begin{center}
\scalebox{.9}{
\begin{tabular}{ c c c c c c c c c c c c c c c c c c| c}
%\cline{1-13}
\rotatebox{90}{$A_{n}$}&
\rotatebox{90}{ $C_{nn}$}&
\rotatebox{90}{$D_{nn}$}&
\rotatebox{90}{$K_{nn}$}&
\rotatebox{90}{$C_{ml}$}&
\rotatebox{90}{$D_{mm}$}&
\rotatebox{90}{$C_{mm}$}&
\rotatebox{90}{$C_{ll}$}&
\rotatebox{90}{$D_{ml}$}&
\rotatebox{90}{$K_{mm}$}&
\rotatebox{90}{$K_{ml}$}&
\rotatebox{90}{$C_{nlm}$}&
\rotatebox{90}{$C_{nml}$}&
\rotatebox{90}{$C_{nmm}$}&
\rotatebox{90}{$C_{mnl}$}&
\rotatebox{90}{$C_{mln}$}&
\rotatebox{90}{$C_{mnm}$}&
\rotatebox{90}{$C_{mmn}$}\\
 \hline
\sq 	 &\trh  &\sq   &\sq  & \dis  & \disp  &\dis &\dis & \disp &\dis  &\dis   &\dis  & \dis & \dis & \dis &\dis &\dis & \dis&$P_{n}$  \\
%\hline    
       & \trh  & \sq & \sq &  \sq & \dis   &\sq &\sq  & \dis  &\dis  &\dis   & \sq  &\sq   &\sq   &  \dis &\dis &\dis & \dis&$A_{n}$ \\
%\cline{2-14}
       &         &  \sq& \sq & \trv &   \dis &\sq   &\sq   &\dis & \dis &\dis  &\sq   & \sq  & \trv  & \dis &\dis &\dis & \dis&$C_{nn}$ \\
%\cline{3-14}
      &          &       &  \sq & \dis &\sq   & \dis  &\dis  & \sq  &\dis &  \dis &\dis  &  \dis  &\dis &\sq&\dis&\sq&\dis &$D_{nn}$ \\
%\cline{4-14}
     &          &        &         &\dis  & \dis &\dis  & \dis &\dis  &\sq  &\sq   & \dis  & \dis   &\dis &\dis&\sq&\dis&\sq&$K_{nn}$  \\
%\cline{5-14}
    &           &        &          &      &\dis   & \dis  & \dis &\disp &\disp&\dis  &\sq   & \sq  &    \dis   &  \dis &\dis &\dis & \dis&$C_{ml}$\\
%\cline{6-14}
    &          &        &           &      &      &  \sq&  \dis &  \dis     &  \sq&\disp   &\dis    & \sq &   \dis  & \dis &\dis &\dis & \dis&$D_{mm}$ \\
%\cline{7-14}  
    &          &        &          &       &      &      &   \sq&\dis      &\sq&  \dis  &\dis   & \dis  & \sq   &\sq&\sq&\dis&\dis& $C_{mm}$ \\
%\cline{8-14}
    &         &        &          &       &       &      &       &\sq     &\dis &\sq &  \dis   & \dis  & \sq  &\dis&\dis&\sq&\sq&$C_{ll}$ \\
%\cline{9-14}
    &        &          &          &      &      &      &       &          &\disp&\sq  &\sq  &  \dis   &  \disp   & \dis &\dis &\dis & \dis&$D_{ml}$ \\
%\cline{10-14}
    &        &         &          &       &     &       &      &          &      &   \dis     & \sq    &  \dis   &  \dis  & \dis &\dis &\dis & \dis&$K_{mm}$ \\
%\cline{11-14}     
    &   &   &   &   &   &   &   &   &   &           & \dis  &\sq   & \disp&  \dis &\dis &\dis & \dis&$K_{ml}$  \\
%\cline{12-14}
    &   &   &   &   &   &   &   &   &   &   &            &\sq   & \dis& \dis&\sq&\sq&\dis&$C_{nlm}$  \\
      &   &   &   &   &   &   &   &   &   &   &            &  & \dis& \sq&\dis&\dis&\sq&$C_{nml}$  \\     
%\cline{13-14}
    &   &   &   &   &   &   &   &   &   &   &            &  & &  \dis &\dis &\disp & \disp& $C_{nmm}$  \\
       &   &   &   &   &   &   &   &   &   &   &            &  & & &\sq &\dis&\disp & $C_{mnl}$\\
  &   &   &   &   &   &   &   &   &   &   &            &  & &     & & \disp &\dis & $C_{mln}$\\
     &   &   &   &   &   &   &   &   &   &   &            &  & & & & & \sq& $C_{mnm}$\\
\end{tabular}}
 \end{center}
\end{table}

\begin{figure}[!ht]
\centerline{
\includegraphics[width=.30\textwidth]{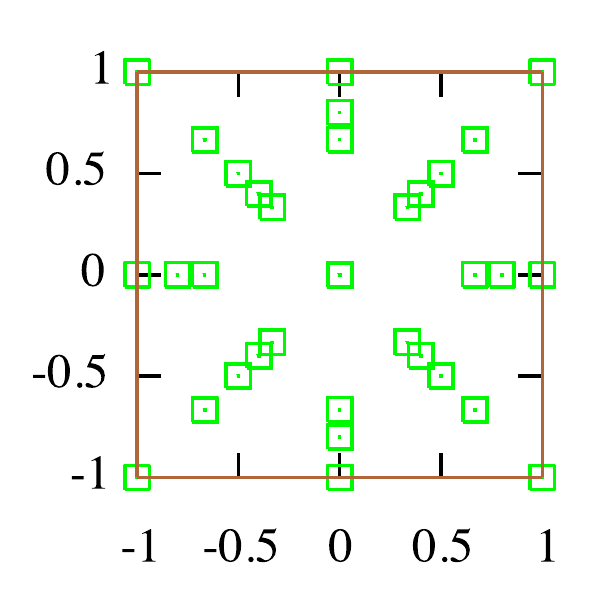}
\includegraphics[width=.30\textwidth]{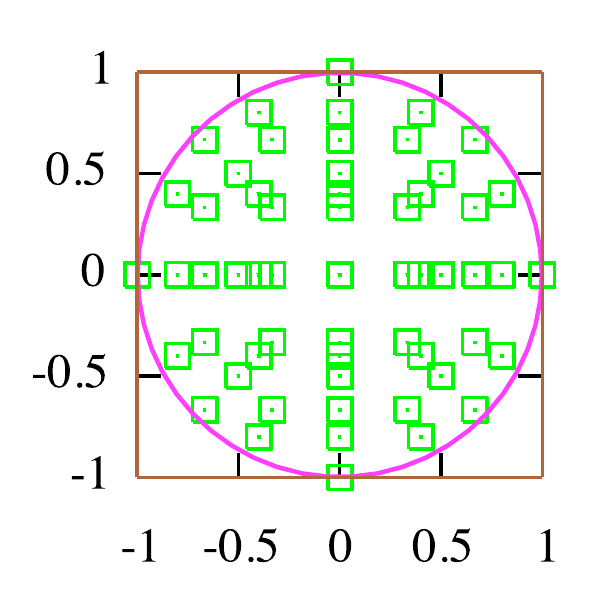}
\includegraphics[width=.30\textwidth]{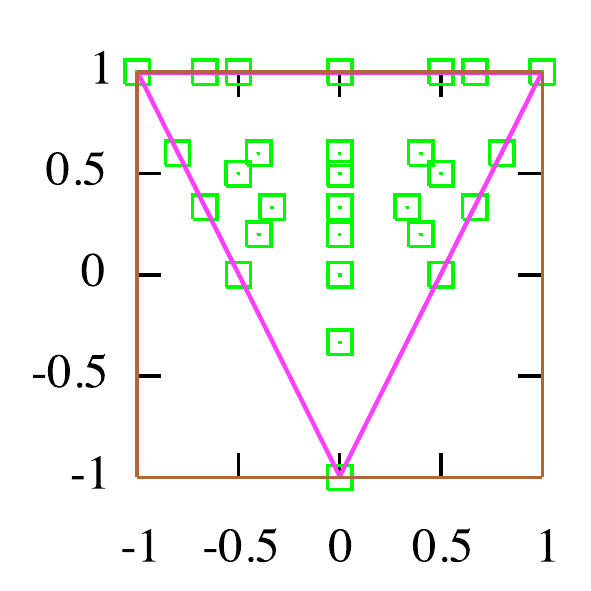}}
\caption{\label{excl:fig:ex2a} Same as Fig.~\protect\ref{excl:fig:ex2}, but each real or imaginary part amplitude is forced to be either 0 or $\pm1$.}
\end{figure}

Note that the randomly generated amplitudes favours the centre of the plot, and doubt can be cast on whether the boundaries are actually reached.  Instead , one can force the real or imaginary part of the amplitudes to be either $0$ or $\pm1$, as in Fig.~\ref{excl:fig:ex2a}, where the border of the square, disk or triangle are clearly seen to be reached for special values of the observables.

\clearpage
\paragraph{Triples of observables}
For three observables, there is a larger variety of situations.
Some of them are direct consequences of the observations made for pairs, while other indicate new patterns that cannot be anticipated from the projections.

\begin{figure}[!ht]
%\vspace{-.3cm}
\centerline{\includegraphics[width=.85\textwidth]{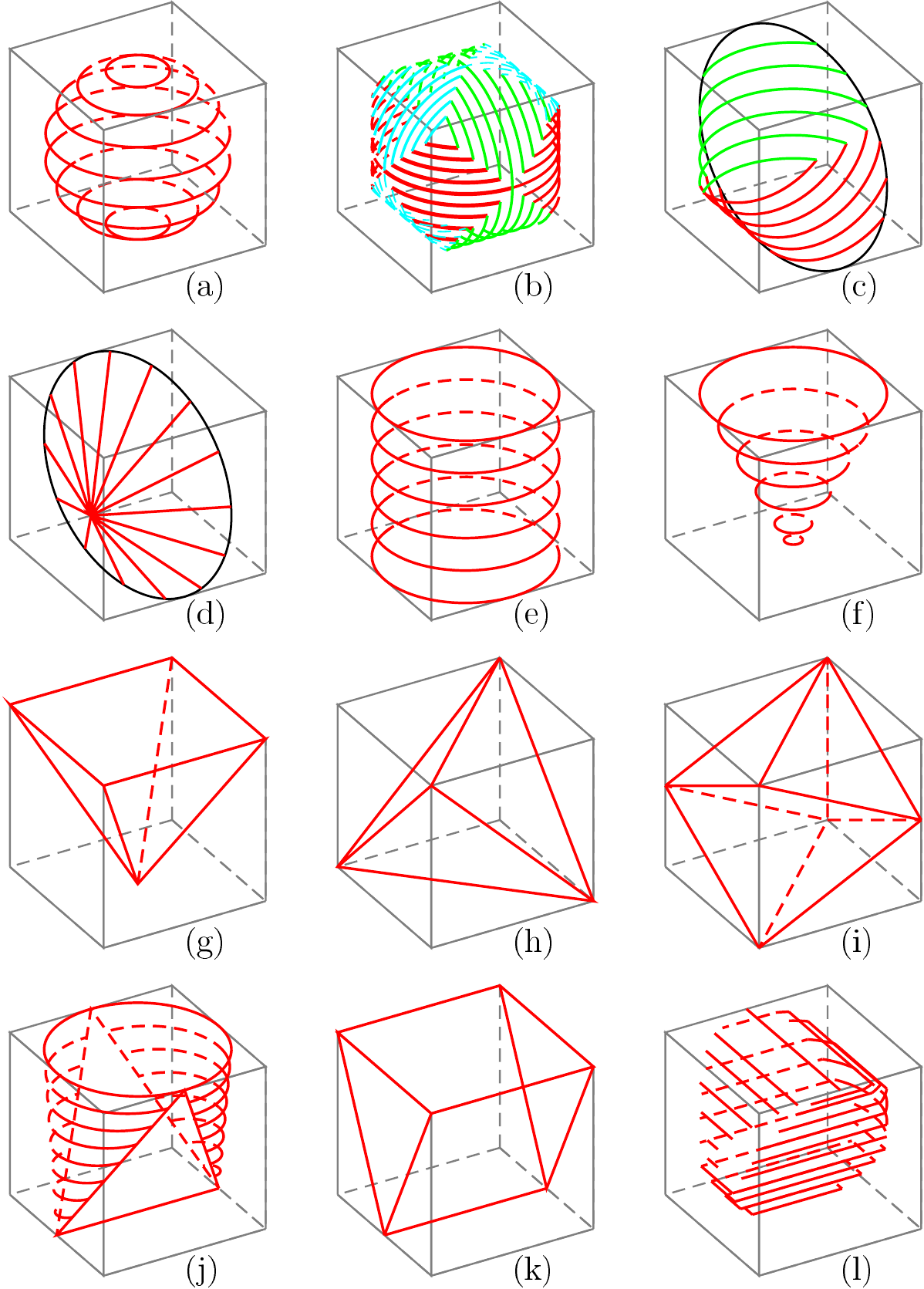}}
\caption{\label{excl:fig:lim3d}  Some allowed domains encountered in  simulating randomly three observables: 
the unit sphere (a), 
the intersection of three orthogonal cylinders of unit radius (b),
the intersection of two cylinders (c),
  or a slightly smaller  double cone (d), 
  a cylinder (e),
  a cone (f), 
a pyramid (g), 
  a tetrahedron (h),
  an octahedron (i),
  a ``coffee filter'' (j),
an inverted tent (k), 
and the intersection of two cylinders and a dihedral (l).  For clarity, part of the limiting surface is sometimes removed. Some figures transformed by parity with respect to the centre of the cube or by interchange of the axes are also obtained.}
\end{figure}

Figure~\ref{excl:fig:lim3d} shows the boundary of the domains we have identified for the observables listed in (\ref{excl:eq:ppllobs}):  the unit sphere, a pyramid, an upside-down tent, a cone, a cylinder, the intersection of two orthogonal cylinders or a double cone which is slightly smaller than this intersection,  a combination of the disk, square and triangle projections delimiting a volume similar to a ``coffee filter'', the intersection of three orthogonal cylinders (larger than the unit sphere!), a tetrahedron,  the intersection of two cylinders and two planes, an octahedron, or figures deduced by mirror symmetry.

\begin{figure}[!ht]
\unitlength=.1\textwidth
\begin{center}\begin{picture}(10,4)(0,0)
\put(0,0){\includegraphics[width=4.5\unitlength]{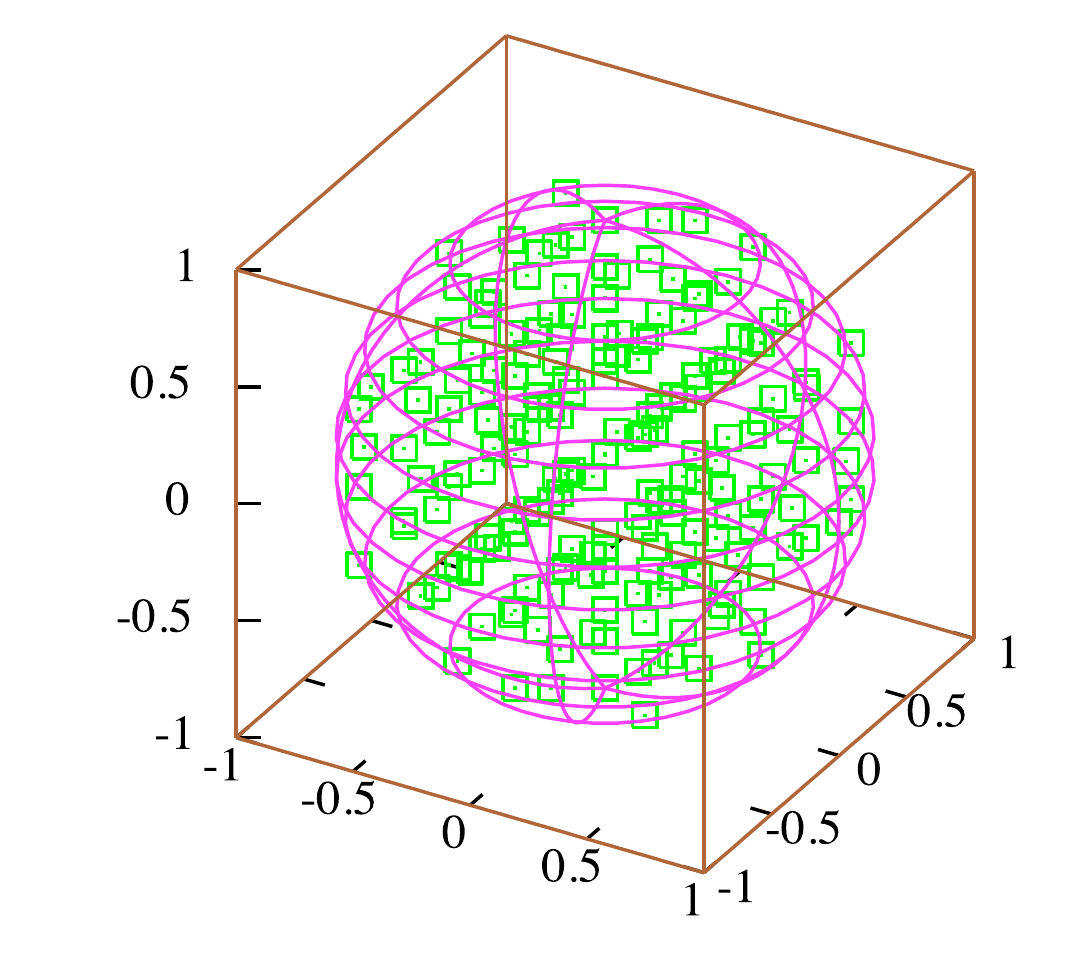}}
\put(5,0){\includegraphics[width=4.5\unitlength]{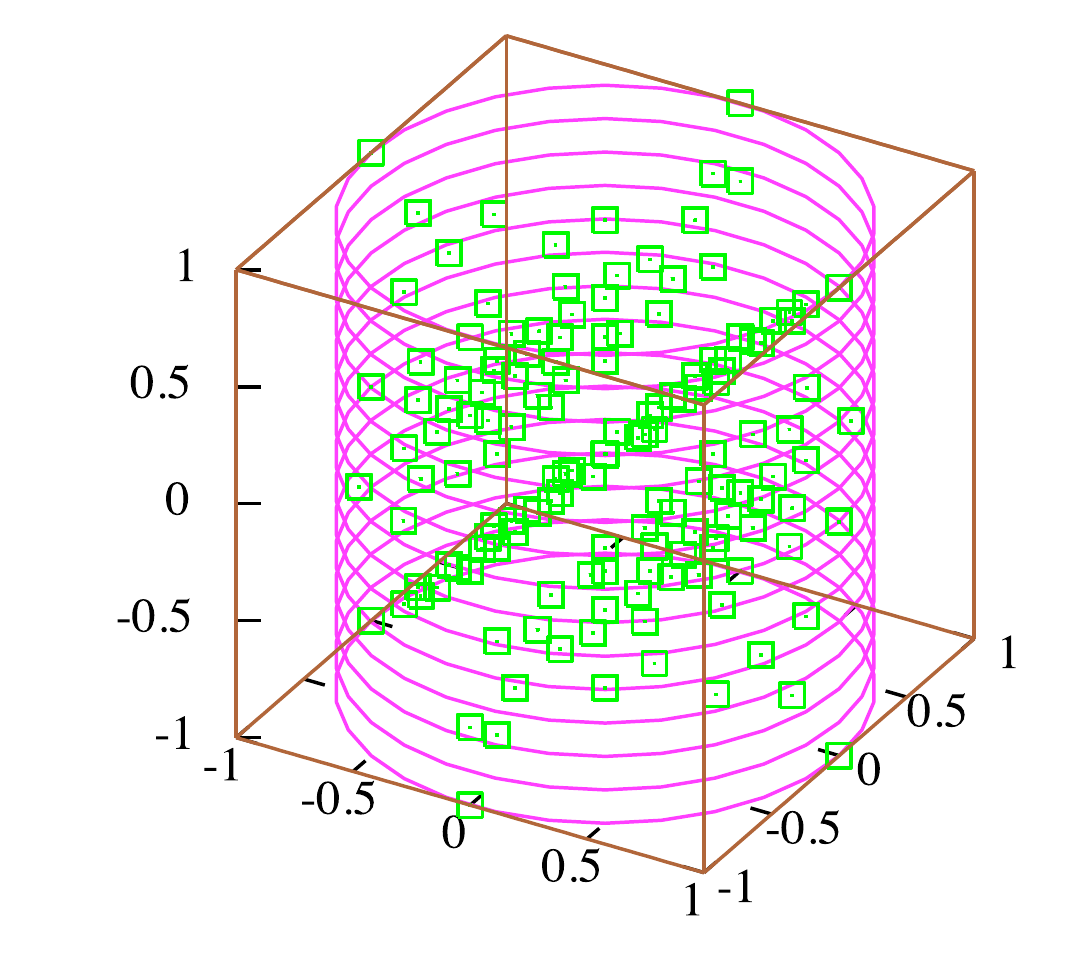}}
\put(4.65,.5){%
\put(0,0){\vector(4,-1){1}}\put(1.05,-.4){$x$}
\put(0,0){\vector(1,1){.6}}\put(.7,0.5){$y$}
\put(0,0){\vector(0,1){1}}\put(-0.05,1.1){$z$}}
\end{picture}\end{center}
\caption{\label{excl:fig:2-8-10-3-7-9} Simulation  of the domain allowed  for the observables $\{P_n,D_{mm},C_{ll}\}$ (left) and $\{C_{ml},C_{mm},A_n\}$ (right), using amplitudes whose real and imaginary parts are chosen to be either $0$ or $\pm 1$.}
\end{figure}

There are many cases  where the three pairs are limited to the unit disk, and the three observables constrained inside the unit sphere, as for  $\{P_n,D_{mm},C_{ll}\}$, shown in Fig.~\ref{excl:fig:2-8-10-3-7-9}.

\begin{figure}[!ht]
\unitlength=.1\textwidth
\begin{center}\begin{picture}(10,4)(0,0)
\put(0,0){\includegraphics[width=4.5\unitlength]{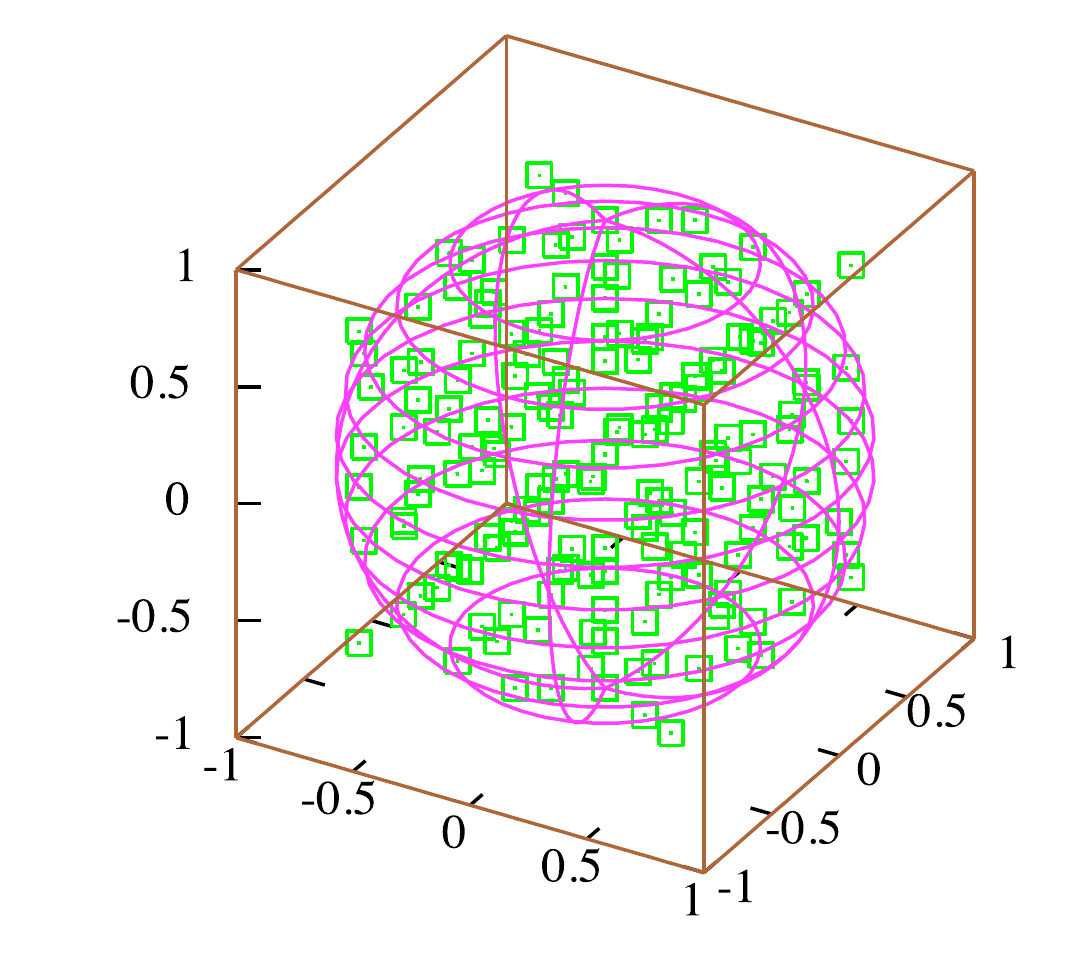}}
\put(5,0){\includegraphics[width=4.5\unitlength]{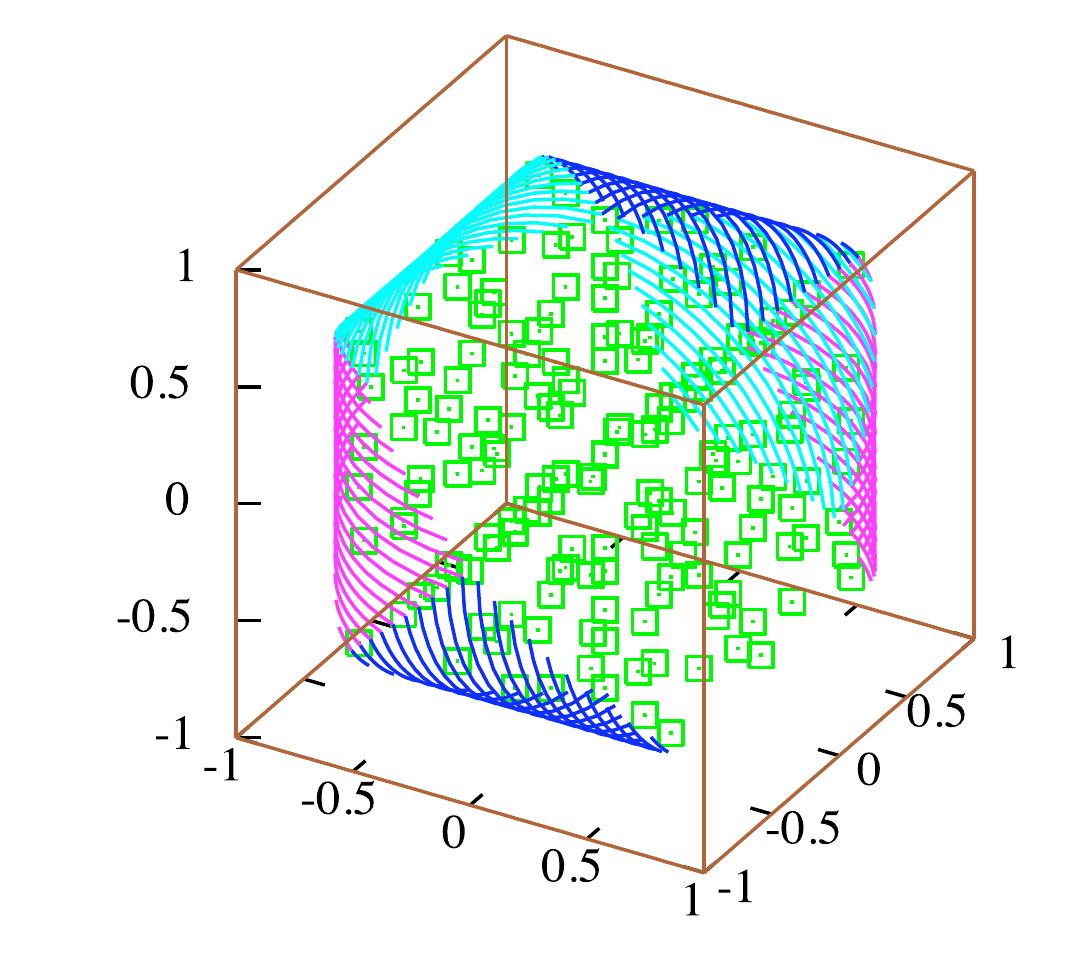}}
\put(4.65,.5){%
\put(0,0){\vector(4,-1){1}}\put(1.05,-.4){$x$}
\put(0,0){\vector(1,1){.6}}\put(.7,0.5){$y$}
\put(0,0){\vector(0,1){1}}\put(-0.05,1.1){$z$}}
\end{picture}\end{center}
%\centerline{\includegraphics[width=.5\textwidth]{FigExcl/P3-3-8-13c}
%\includegraphics[width=.5\textwidth]{FigExcl/P3-3-8-13d}}
\caption{\label{excl:fig:AnDmmKml}
Fictitious observables $\{A_n,D_{mm},K_{ml}\}$ obtained with  amplitudes whose real and imaginary parts are chosen to be either $0$ or $\pm 1$, shown against the unit sphere  (left) or the intersection of the three unit cylinders.}
\end{figure}

There are a few cases, 
$\{A_n,D_{mm},K_{ml}\}$,
$\{A_n, D_{ml}, K_{mm}\}$, 
$\{C_{ml}, D_{mm}, D_{ml}\}$, 
$\{C_{ml}, D_{mm},$ $K_{ml}\}$, 
$\{C_{ml}, D_{ml}, K_{mm}\}$, 
$\{C_{ml}, K_{mm}, K_{ml}\}$, 
$\{D_{mm}, D_{ml}, C_{nmm}\}$,  and
$\{K_{mm}, K_{ml}, C_{nmm}\}$,
where each projection is contained in the unit disk, but nevertheless, the domain is slightly larger than the unit sphere. The example of  $\{A_n,D_{mm},K_{ml}\}$ is shown in Fig.~\ref{excl:fig:AnDmmKml}.

\begin{figure}[!ht]
\unitlength=.1\textwidth
\begin{center}\begin{picture}(10,4)(0,0)
\put(0,0){\includegraphics[width=4.5\unitlength]{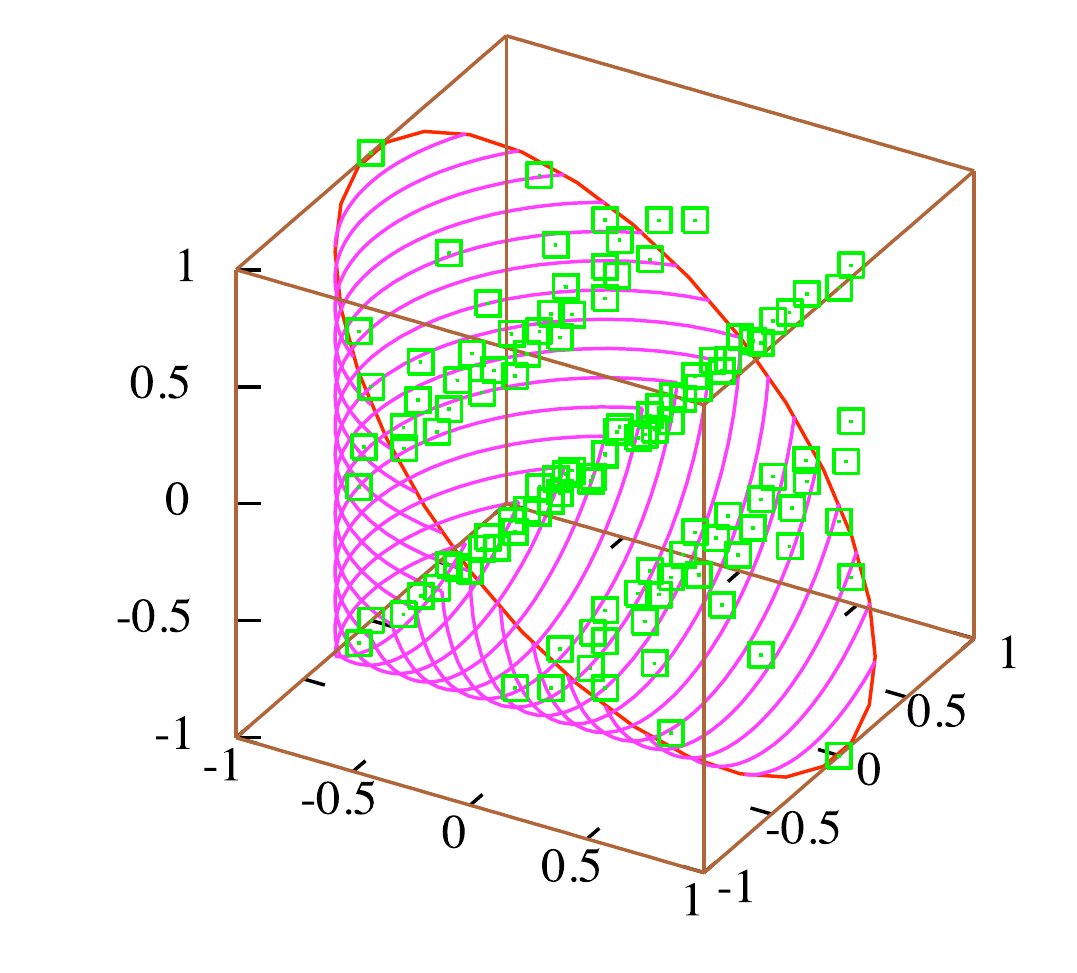}}
\put(5,0){\includegraphics[width=4.5\unitlength]{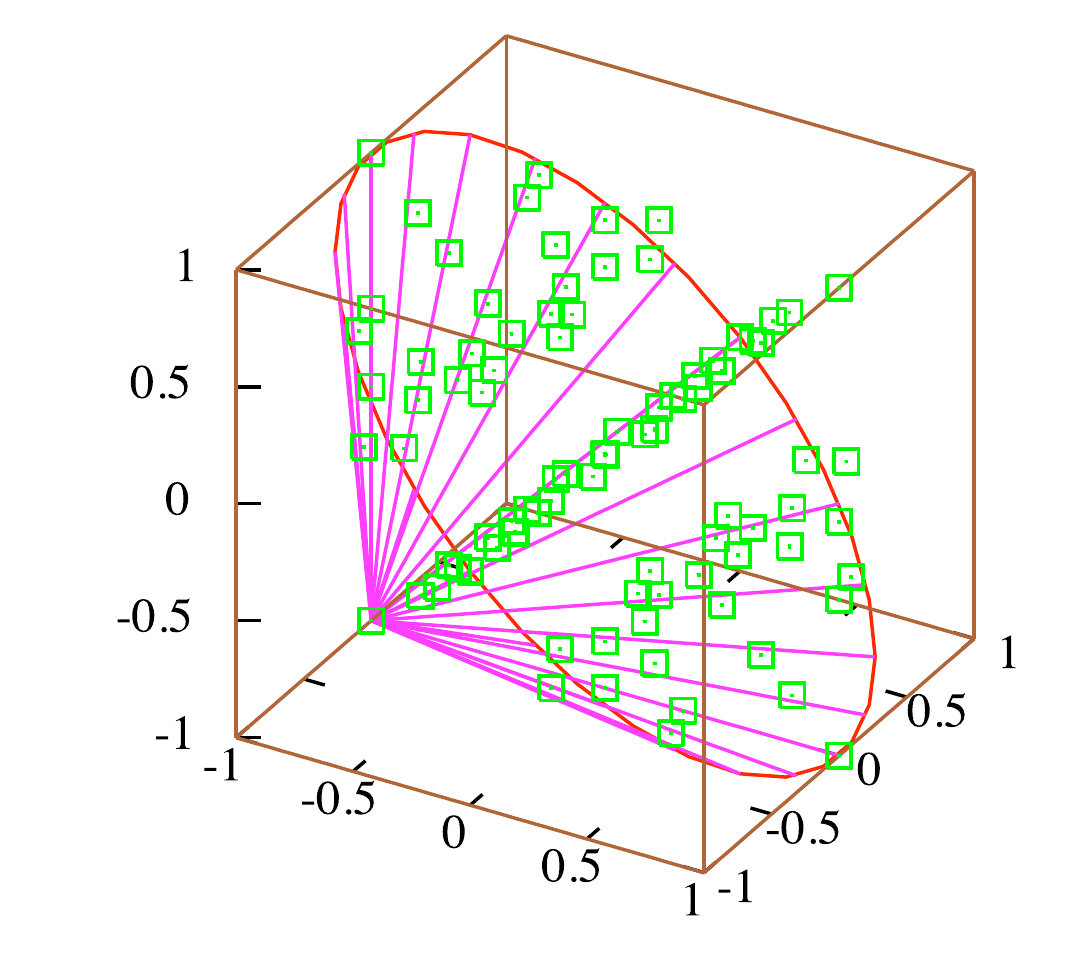}}
\put(4.65,.5){%
\put(0,0){\vector(4,-1){1}}\put(1.05,-.4){$x$}
\put(0,0){\vector(1,1){.6}}\put(.7,0.5){$y$}
\put(0,0){\vector(0,1){1}}\put(-0.05,1.1){$z$}}
\end{picture}\end{center}
%
%\unitlength=.1\textwidth
%\begin{center}\begin{picture}(10,5)(0,0)
%\put(-1,0){\includegraphics[width=5\unitlength]{FigExcl/P3-2-3-8}}
%\put(5.5,0){\includegraphics[width=5\unitlength]{FigExcl/P3-7-9-10d}}
%\put(4.65,.4){%
%\put(0,0){\vector(4,-1){1}}\put(1.05,-.4){$x$}
%\put(0,0){\vector(1,1){.6}}\put(.7,0.5){$y$}
%\put(0,0){\vector(0,1){1}}\put(-0.05,1.1){$z$}}
%\end{picture}\end{center}
\caption{\label{excl:fig:2-3-8-7-9-10} The domain for  $\{x,y,z\}=\{P_n,D_{mm},A_n\}$ (left) is inside the intersection of two cylinders, while the domain for $\{C_{ml},C_{mm},C_{ll}\}$ (right) is inside a slightly smaller double cone. For clarity, only the lower part of the boundary is shown..}
\end{figure}

There are many examples where, in a triple, two pairs are constrained in the unit disk, while the projection of the third pair occupies the entire square. The domain is in general the intersection of the two cylinders.  The case of  $\{D_{mm},P_n,A_n\}$ is displayed in  Fig.~\ref{excl:fig:2-3-8-7-9-10}.
For $\{C_{ml},C_{mm},C_{ll}\}$, also shown in  Fig.~\ref{excl:fig:2-3-8-7-9-10}, a smaller inner  domain has been identified: the cone of vertex $(-1,0,-1)$ and joining the ellipsis which is at the intersection of the two cylinders, and the symmetric cone.

The next category is the simple cylinder. An example is $\{A_n,C_{ml},C_{mm}\}$, see Fig.~\ref{excl:fig:2-8-10-3-7-9}: there is no constraint on $\{A_n,C_{ml}\}$  nor on 
$\{A_n,C_{mm}\}$, while $\{ C_{ml},C_{mm}\}$ is constrained in a disk. 

\begin{figure}[!ht]
\unitlength=.1\textwidth
\begin{center}\begin{picture}(10,4)(0,0)
\put(0,0){\includegraphics[width=4.5\unitlength]{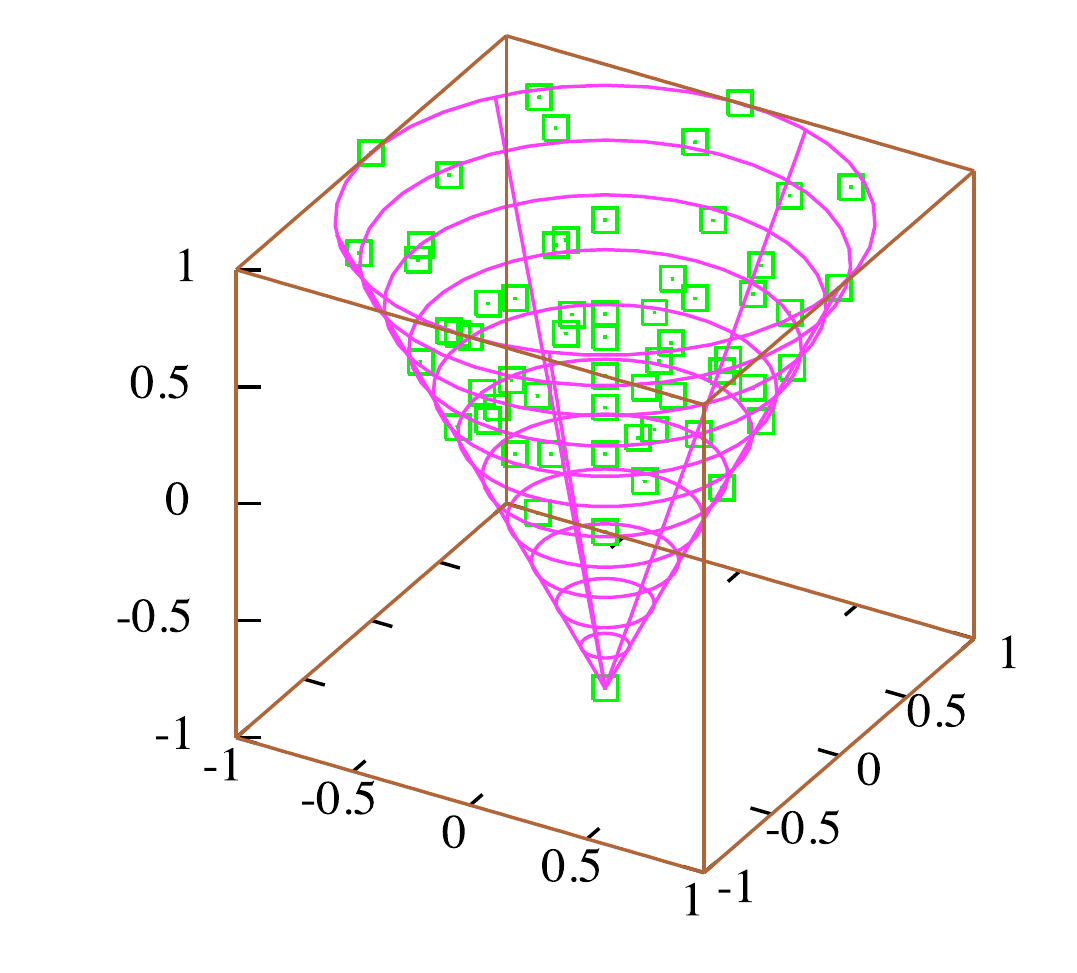}}
\put(5,0){\includegraphics[width=4.5\unitlength]{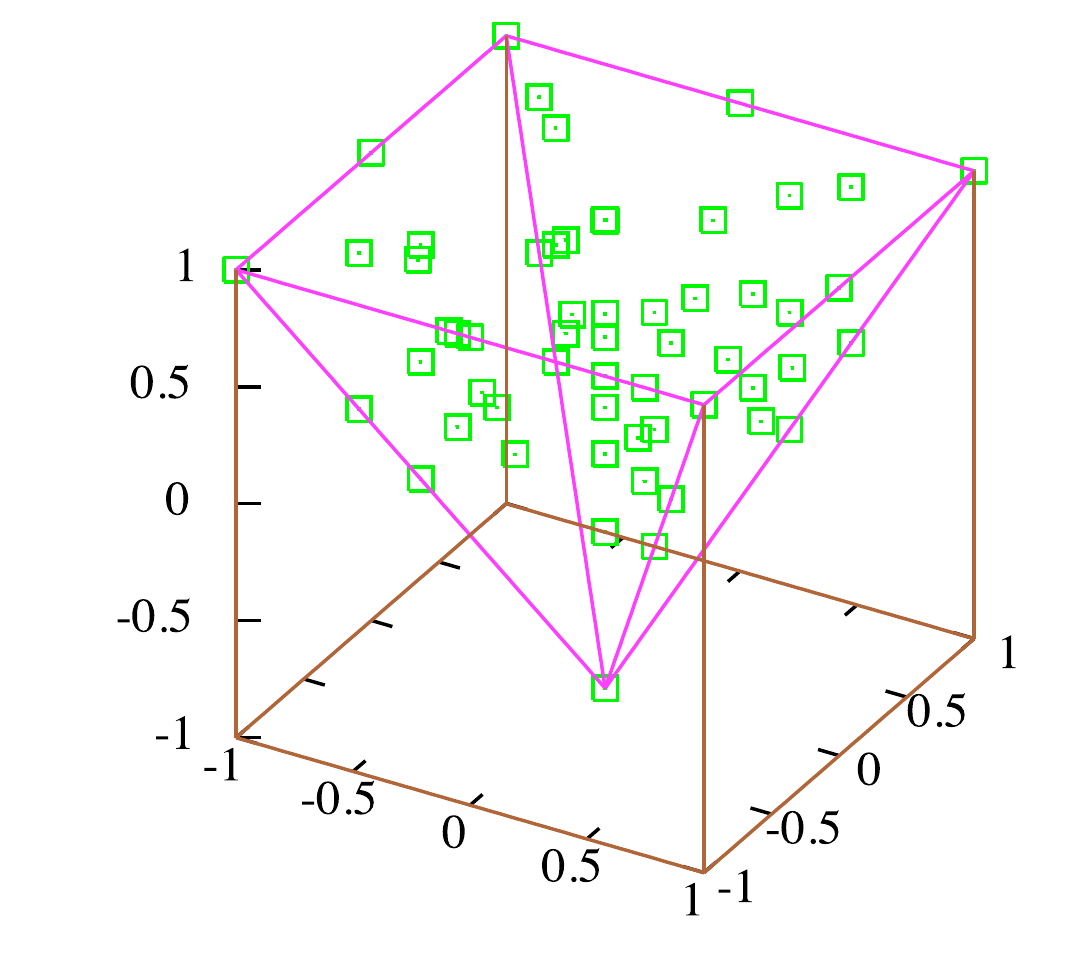}}
\put(4.65,.5){%
\put(0,0){\vector(4,-1){1}}\put(1.05,-.4){$x$}
\put(0,0){\vector(1,1){.6}}\put(.7,0.5){$y$}
\put(0,0){\vector(0,1){1}}\put(-0.05,1.1){$z$}}
\end{picture}\end{center}
%\centerline{%
%\includegraphics[width=.5\textwidth]{FigExcl/P3-2-4-7}
%\includegraphics[width=.5\textwidth]{FigExcl/P3-2-3-4}}
\caption{\label{2-4-7-2-3-4} Domain for $\{P_n,C_{ml},C_{nn}\}$ (left) and 
$\{P_n,A_n,C_{nn}\}$ (right).}
\end{figure}

As seen in Table \ref{excl:tab:reca}, there are just a few cases of triangular constraint on pairs. Hence it is rather rare to find a triple with two triangles combined with a disk, to make a cone, as in
the example of $\{P_n,C_{ml},C_{nn}\}$, shown in Fig.~\ref{2-4-7-2-3-4}. The two other cases are $\{P_n, C_{nmm},C_{nn} \}$, and 
$\{C_{ml},C_{nmm}$, $C_{nn} \}$.

Another type of cone, with square section, \ie, a pyramid, can be found with two pairs in a triangular constraint, and no constraint on the third pair, as for 
$\{P_n,A_n,C_{nn}\}$, shown in  Fig.~\ref{2-4-7-2-3-4}.

\begin{figure}[!ht]
\unitlength=.1\textwidth
\begin{center}\begin{picture}(10,4)(0,0)
\put(0,0){\includegraphics[width=4.5\unitlength]{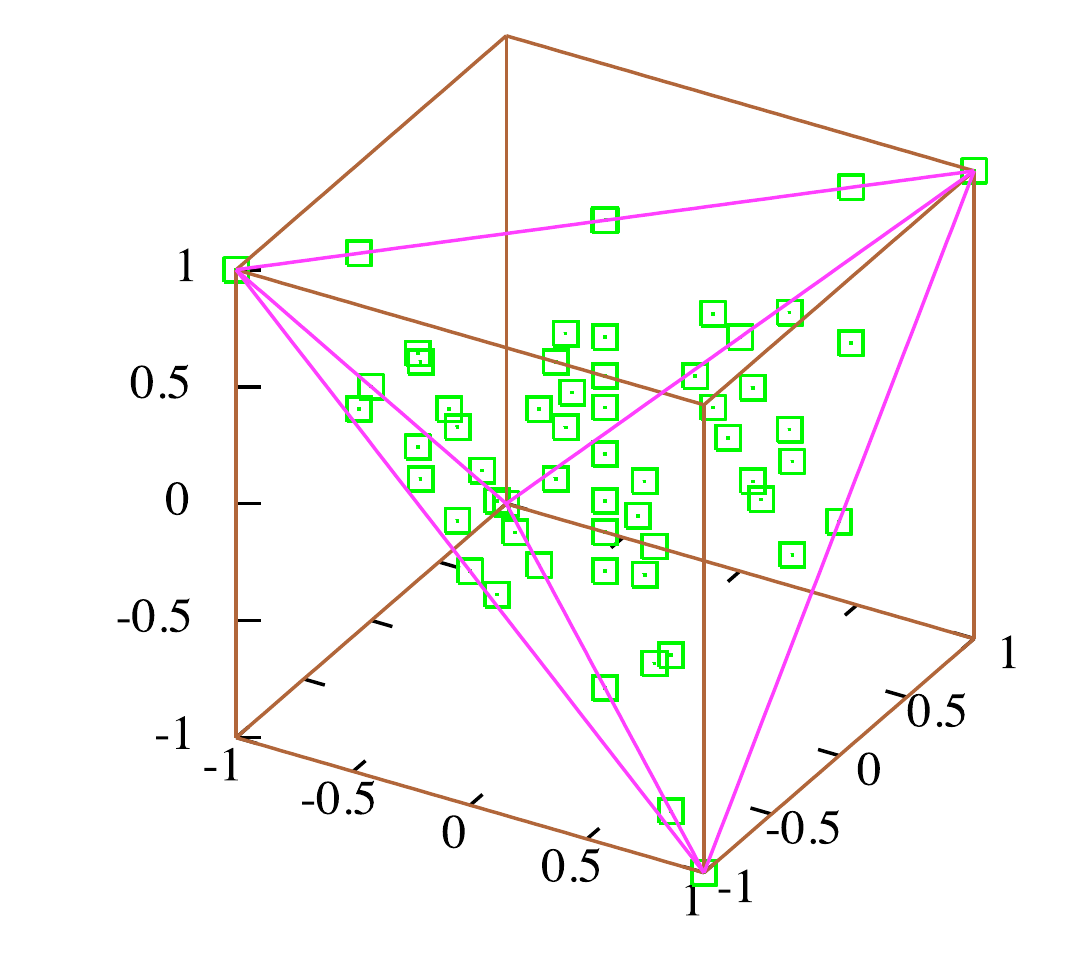}}
\put(5,0){\includegraphics[width=4.5\unitlength]{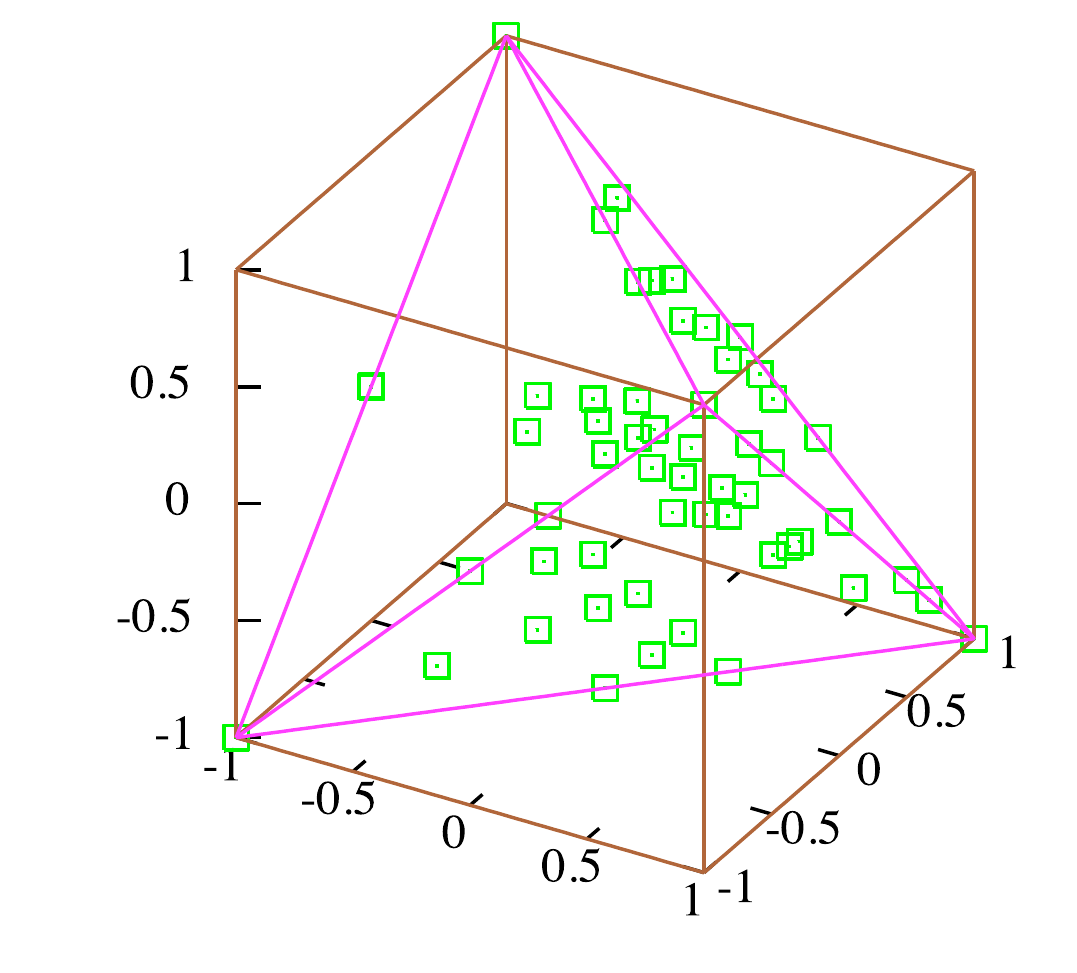}}
\put(4.65,.5){%
\put(0,0){\vector(4,-1){1}}\put(1.05,-.4){$x$}
\put(0,0){\vector(1,1){.6}}\put(.7,0.5){$y$}
\put(0,0){\vector(0,1){1}}\put(-0.05,1.1){$z$}}
\end{picture}\end{center}
%\centerline{
%\includegraphics[width=.5\textwidth]{FigExcl/P3-2-3-5}
%\includegraphics[width=.5\textwidth]{FigExcl/P3-4-9-10}}
\caption{\label{2-3-5-4-9-10}Domain for  $\{P_n,A_n,D_{nn}\}$ (left) and  $\{C_{nn},C_{mm},C_{ll}\}$ (right).}
\end{figure}

In the case of $\{P_n,A_n,D_{nn}\}$, there is no restriction on each pair, but the three observables are restricted inside a tetrahedron, which occupies only 1/3 of the volume of  the cube. A similar tetrahedron, but inverted, is obtained for $\{C_{nn},C_{ll}, C_{mm}\}$, see Fig.~\ref{2-3-5-4-9-10}.

\begin{figure}[!ht]
\unitlength=.1\textwidth
\begin{center}\begin{picture}(10,4)(0,0)
\put(0,0){\includegraphics[width=4.5\unitlength]{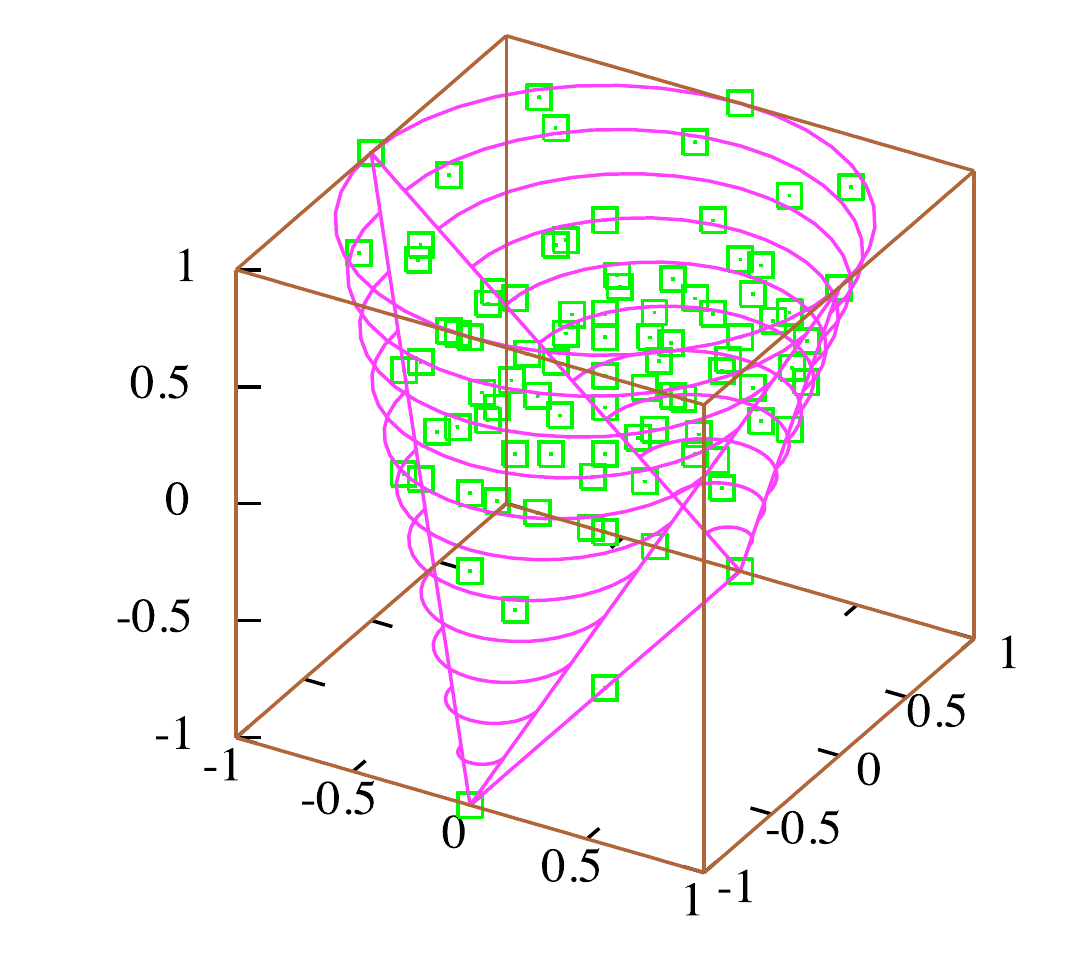}}
\put(5,0){\includegraphics[width=4.5\unitlength]{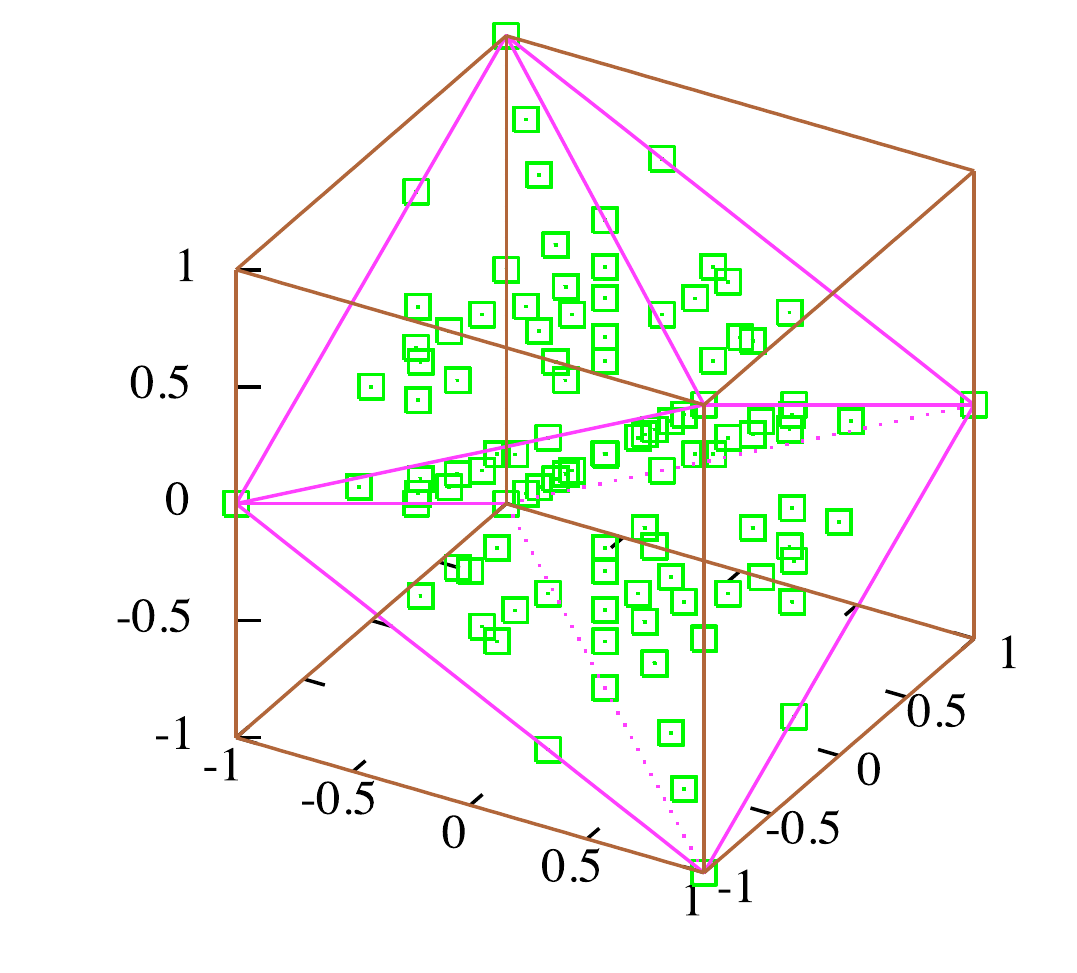}}
\put(4.65,.5){%
\put(0,0){\vector(4,-1){1}}\put(1.05,-.4){$x$}
\put(0,0){\vector(1,1){.6}}\put(.7,0.5){$y$}
\put(0,0){\vector(0,1){1}}\put(-0.05,1.1){$z$}}
\end{picture}\end{center}
%
%\centerline{
%\includegraphics[width=.5\textwidth]{FigExcl/P3-2-9-4}
%\includegraphics[width=.5\textwidth]{FigExcl/CmmCllCnmm}}
%
\caption{\label{excl:fig:filter-octo}
Observables $\{P_n,C_{mm},C_{nn}\}$ (left) and $\{C_{mm},C_{ll},C_{nmm}\}$ (right).}
\end{figure}

Besides these tetrahedrons, the  six triples of observables $\{P_n, D_{nn}, K_{nn}\}$, 
$\{A_{n},D_{nn},K_{nn}\}$,
$\{A_{n},C_{mm},C_{ll}\}$, 
$\{A_{n},C_{nlm}, C_{nml}\}$,
$\{C_{ml}, C_{nlm},C_{nml}\}$, and
$\{C_{mm},C_{ll},C_{nmm}\}$,
 also do not show any constraint on each pair, but do not occupy the entire cube.  Their domain is an octahedron, as shown in Fig.~\ref{excl:fig:filter-octo}, or an octahedron deduced by interchange of axes or parity.

The case of observables $\{P_n, C_{mm}, C_{nn}\}$ is remarkable:  a pair has a disk constraint, another pair a triangular constraint, and the third one, no constraint. The result looks like a paper filter for coffee, see Fig.~\ref{excl:fig:filter-octo}, left.
%
%\begin{figure}[!ht]
%\begin{minipage}{.5\textwidth}
%\centerline{
%\includegraphics[width=1\textwidth]{FigExcl/P3-2-9-4}}
%\end{minipage}\hfill
%\begin{minipage}{.4\textwidth}
%\caption{\label{2-9-4} Simulation by random amplitudes of the observables $\{P_n,C_{mm},C_{nn}\}$.}
%\end{minipage}\end{figure}
%

\begin{figure}[!ht]
\centerline{
\includegraphics[width=.5\textwidth]{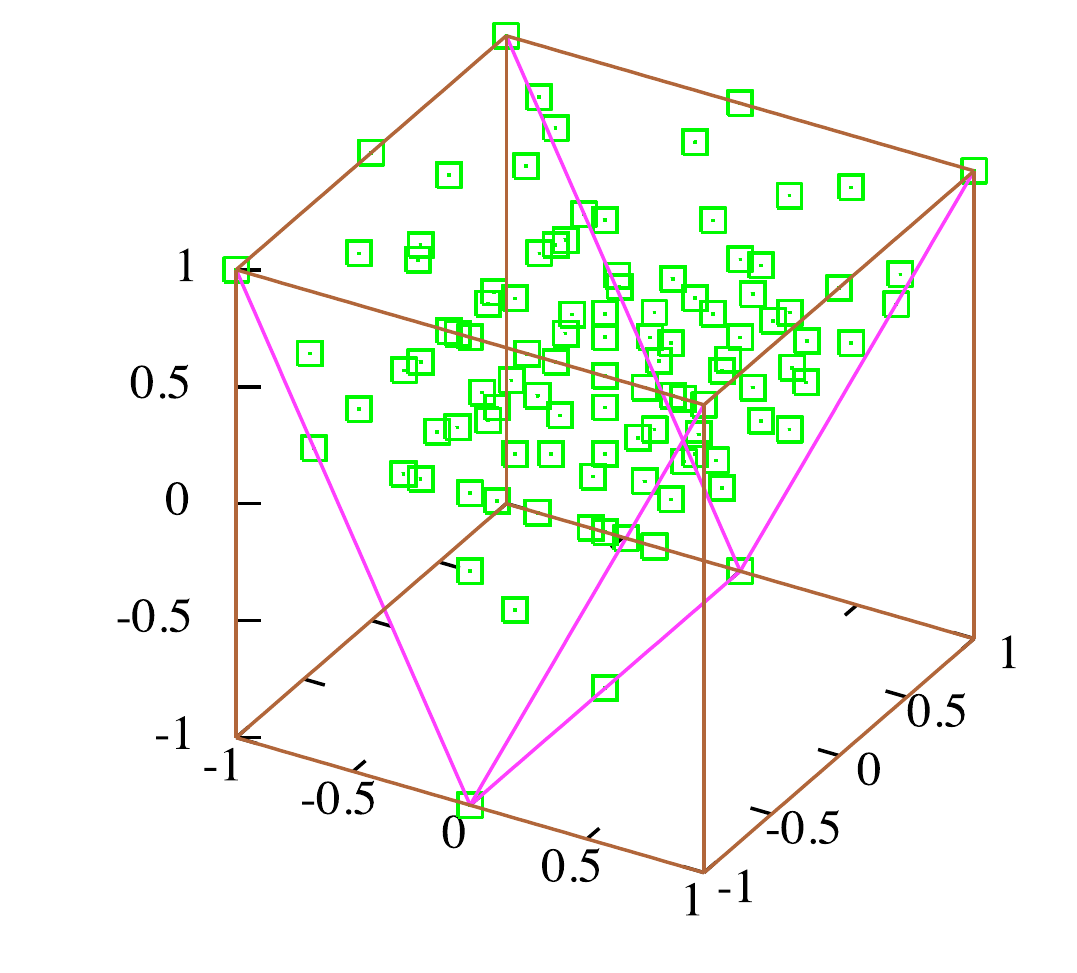}
\includegraphics[width=.5\textwidth]{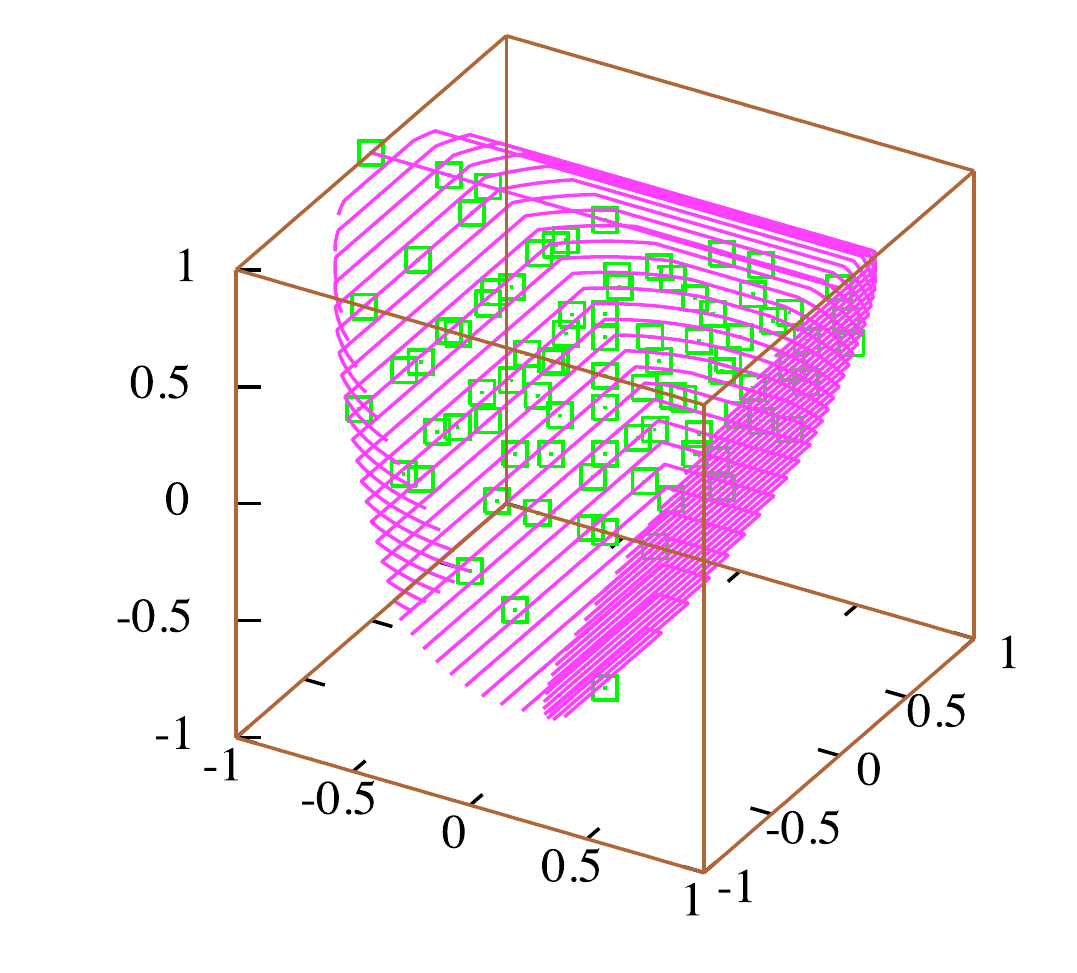}}
\caption{\label{excl:fig:7-14-4-3-8-4} Simulation by random amplitudes of the observables $\{C_{ml},C_{nlm},C_{nn}\}$ (left) and $\{A_n,D_{mm},C_{nn}\}$ (right).}
\end{figure}

The case of $\{C_{ml},C_{nlm},C_{nn}\}$ is simpler, with a triangle and two squares seen in projection. The domain is the inverted tent seen in Fig.~\ref{excl:fig:7-14-4-3-8-4}.

The triple $\{A_n, D_{mm},C_{nn}\}$, shown in Fig.~\ref{excl:fig:7-14-4-3-8-4}, has a boundary made by the intersection of two cylinders and a set of two planes (dihedral).

\begin{figure}[!ht]
\begin{minipage}{.6\textwidth}
\centerline{\includegraphics[width=.7\textwidth]{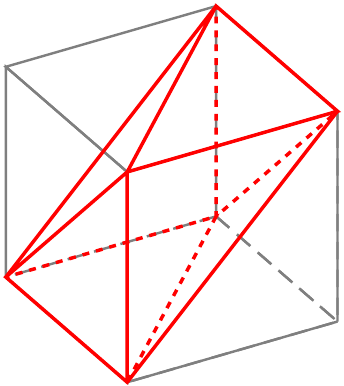}}
\end{minipage}
\begin{minipage}{.39\textwidth}
\caption{\label{excl:fig:newo} Any triple out of  $\{C_{mm},C_{ll}, A_{mm}, A_{ll}\}$ is constrained within an octahedron obtained by slicing off two opposite corners of the unit cube.}
\end{minipage}
\end{figure}

It should be stressed again that the figures shown in Fig.~\ref{excl:fig:lim3d} do not exhaust all possibilities. If longitudinal polarisation is allowed, and if the antiproton beam is polarised, new observables can be reached. In particular, for any three out of the four observables $\{C_{mm},C_{ll}, A_{mm}, A_{ll}\}$, \ie, final- and initial-spin correlation in the scattering plane, the domain corresponds to a new type of  octahedron,  shown in Fig.~\ref{excl:fig:newo}.

But the most important message is that, whilst  the allowed domain of  pairs  is often the entire unit square, \emph{there is always  a constraint  for three observables}, that limits the allowed domain to a volume smaller than the unit cube, in agreement with the reasoning of Sec.~\ref{basic:sub:limited}.

\clearpage
\subsubsection{Understanding the observed domains}\label{excl:sub:underst}
Many of the inequalities detected by random simulations can be derived very simply. Others require somewhat the more sophisticated methods presented in Sec.\ref{basic:sub:methods}.

\paragraph{Anticommutation or commutation}
Many of the observed disk or sphere constraints are due  to anticommuting operators, see Sec.~\ref{basic:sub:antic}.
In particular, $(P_n,C_{ij})$ is in a disk for $i=l$ or $m$ %$i\in\{l,m\}$ 
and any $j\in\{l,m,n\}$, since the corresponding operators 
$\left\langle I\otimes I \otimes \sigma_n\otimes I\right\rangle$ and 
$\left\langle I\otimes I \otimes \sigma_i\otimes\sigma_j\right\rangle$ anticommute. Note that for $j=n$ the domain is restricted to the diameter $C_{in}=0$ of the disk, due to parity conservation.  
%$(A_n,K_{mi},D_{lj})$
Similarly $(P_n,C_{ij},K_{hi'})$ is in a sphere if $i\ne n$, $i'\ne i$ and $i'\ne n$, for any $h$ and $j$. 
Other examples are  
$K_{nn}^{2}+C_{ll}^{2}+O_{nlm}^2\leq 1$, and  $D_{nn}^{2}+C_{ll}^{2}+O_{nlm}^2\leq 1$.

When $\Ocal$ and $\Ocal'$ commute, the domain of $(\langle\Ocal\rangle,\langle\Ocal'\rangle)$ may reach a corner of the square $[-1,+1]^2$. This is the case, \eg, for $(P_n,C_{nn})$ (triangle) or $(D_{mm},C_{nml})$ (whole square). In the cases indicated by a $\oplus$ in Table~\ref{excl:tab:pimp}, however, the domain is a disk as if $\Ocal$ and $\Ocal'$ were anticommuting. The reason is that some symmetry imposes $\Ocal'=\Ocal''$ where $\Ocal''$ anticommutes with $\Ocal$. For instance $(D_{ml},C_{ml})$ is a commuting pair, but charge conjugation imposes $C_{ml}=C_{lm}$ and $C_{lm}$ anticommutes with $D_{ml}$. 
\paragraph{Final-state density matrix} 
Inequalities for the $\ppLL$ reaction relevant to a polarised target have been obtained in \cite{Elchikh:2004ex} using the positivity of the final state $4\times4$ density matrix for a given target polarisation $\Sv_\p$ (see also \cite{Richard:1996bb,Elchikh:1999ir}), 
\be\label{excl:eq:rholbarl1}
\rho_{\overline{\Lambda}\Lambda}(\Sv_\p)= 
{1\over{\cal I}(\Sv_\p)}\mathcal{M}\rho_\ppb{\cal M}^\dagger
\,,\qquad{\rm with}\quad\rho_\ppb=
\sum\limits_\lambda S_\p^\lambda{\mathbbm{1}\otimes\sigma^\lambda\over4}~.
\ee
${\cal I}(\Sv_\p)=\Tr\{\mathcal{M}\,\rho_\ppb\,\mathcal{M}^\dagger\}$ is the cross section when only the target is polarised. In terms of the Cartesian reaction parameter,
 \begin{equation}
\label{excl:eq:rholbarl2}
 \rho_\LLb(\Sv_\p) =
\frac{{\cal I}(0)}{4{\cal I}(\Sv_\p)}\sum\limits_{ \lambda,\mu,\nu=0,l,m,n}\ S_\p^{\lambda}\ C_{\lambda\mu\nu}\ (\sigma^\mu\otimes\sigma^\nu)~.      
\end{equation}
For transverse $\Sv_\p$, one can also write 
\begin{equation}\label{excl:eq:rhomat}
\rho_\LLb(\Sv_\p)=\frac{{\cal I}(0)}{4{\cal I}(\Sv_\p)}\,
(\Gamma_0+|\Sv_\p|\cos\phi\,\Gamma_n+|\Sv_\p|\sin\phi\,\Gamma_m)~,
\end{equation}
where $\phi$ is the azimuthal angle of $\Sv_\p$ in the $\{\vec{l}_\p,\vec{m}_\p,\vec{n}\}$ frame. For instance, in the helicity basis,
\be\label{dens:eq:Czmat}
   \Gamma_{0}=\begin{pmatrix}
1-C_{ll} & -C_{ml}-iP_{n} & -C_{ml}-iP_{n} & -C_{nn}-C_{mm} \\
-C_{ml}+iP_{n} & 1+C_{ll} & C_{nn}-C_{mm} & C_{ml}-iP_{n} \\
-C_{ml}+iP_{n} & C_{nn}-C_{mm} & 1+C_{ll} & C_{ml}-iP_{n} \\
-C_{nn}-C_{mm} & C_{ml}+iP_{n} & C_{ml}+iP_{n} & 1-C_{ll}
\end{pmatrix}~.
\ee
The quadratic inequalities of the type $\rho_{11} \ \rho_{22}\geq |\rho_{12}|^{2} $, applied to the above $\Gamma_0$ gives for instance
\be\label{dens:eq:r1}\begin{aligned}
C_{ml}^{2}+P_{n}^{2}+C_{ll}^{2}&\leq 1~,\\
 (C_{nn}-C_{mm})^2&\leq(1+C_{ll})^2~.
\end{aligned}\ee
Similarly, from
\be\label{dens:eq:Cnmat}
   \Gamma_{n}=\begin{pmatrix}
A_{n}+O_{nmm} & -O_{nlm}-iD_{nn} &
-O_{nml}-iK_{nn} & -A_{n}-O_{nmm} \\
-O_{nlm}+iD_{nn} & A_{n}-O_{nmm}&
A_{n}-O_{nmm} & O_{nml}-iK_{nn} \\
-O_{nml}+iK_{nn} & A_{n}-O_{nmm}
& A_{n}-O_{nmm} & O_{nlm}-iD_{nn} \\
-A_{n}-O_{nmm} & O_{nml}+iK_{nn}
 & O_{nlm}+iD_{nn} & A_{n}+O_{nmm}
\end{pmatrix}~,
\ee
corresponding to the proton spin along $\vec{n}$, one can derive  
\be \label{dens:eq:ineq2}\begin{aligned}
O_{nmm}^{2}+O_{nlm}^{2}+C_{ml}^{2}+P_{n}^{2}+D_{nn}^{2}+C_{ll}^{2}&\leq
1+A_{n}^{2}~,\cr
K_{nn}^{2}+C_{ll}^{2}+O_{nlm}^2&\leq 1~.
\end{aligned}\ee

\paragraph{Cross section matrix (CSM)}
The positivity of $\rho_\LLb(S_\p)$ does not give all possible inequalities for three polarised particles. In particular, it misses the positivity constraints required by ``entangled states in the $t$- or $u$-channel''. The CSM gives the complete set of inequalities. Let us apply it here, following the method of \cite{Artru:2004jx}, the CSM  of \ppLL\ can be defined as 
\be
\left\langle\ s'_\pb,s'_{\vphantom{pb}\p},s'_{\Lb},s'_{\vphantom{\Lb}\L}
 |  \R  |  s_\pb,s_{\vphantom{pb}\p},s_{\Lb},s_{\vphantom{\Lb}}\right\rangle
=  \langle\ s'_\pb,s'_{\vphantom{pb}\p} | \mathcal{M}^\dagger | s'_{\Lb},s'_{\vphantom{\Lb}\L} \rangle \langle\ s_{\Lb},s_{\vphantom{\Lb}\L} |  \M  | s_\pb,s_{\vphantom{pb}\p}\rangle~.
\label{excl:eq:RYbarY)}
\ee
 By construction $R$ is semi-positive definite and of rank 1. $\R$ is given terms of the correlation parameters, as per Eq.~(\ref{basic:eq:csdm}).

Let us use the transversity basis where $\sigma_y=\sigma_3$ is diagonal, and put the $|s_\pb,s_\p,s_{\overline{\Lambda}} s_\Lambda\rangle$ states in the following order: $|++++\rangle$,  $|+++-\rangle$, $|++-+\rangle$, $|++--\rangle$, etc., like numbers in base~2, but with 0 and 1 being replaced by $+$ and $-$. Due to parity conservation, $\mathcal{M}$ is zero when there is one + sign and three $-$ signs or vice-versa. Therefore $\R$ has the block-diagonal form $\R^+\oplus \boldsymbol{0}$, where $R^+$ is the $8\!\times\!8$ upper-left submatrix of $\R$, and $ \boldsymbol{0}$ the null matrix. One can also consider Eq.~(\ref{basic:eq:P-quant-R}). Using $\Pi_F=i\xi_F\,\sigma_F^3$ and $\xi_F\xi_{\bar F}=-1$ for each fermion $F$, it becomes
\be
\sigma_3\otimes\sigma_3\otimes\sigma_3\otimes\sigma_3\,\R=\R\,\sigma_3\otimes\sigma_3\otimes\sigma_3\otimes\sigma_3=\R~,
\label{excl:eq:equivalence-a}
\ee
which is the Bohr identity \cite{Bohr:1959}. From this one concludes that the $(\lambda,\mu|\nu,\tau)$'s are invariant under the  substitutions: $0\leftrightarrow3$, $1\to i2$, 2$\to -i1$ for $\lambda$ and $\mu$ and $0\leftrightarrow3$, 1$\to -i2$, 2$\to i1$ for $\nu$ and $\tau$, where, for instance $(\lambda,\mu|\nu,i\tau)\equiv i(\lambda,\mu|\nu,\tau)$.

The explicit form of $\R^+$ in terms of the $(\lambda\mu|\nu\tau)$'s will be needed when both beam and target will be polarised. We shall not write it here. Instead we consider the reduced CSM, $\R_3=\Tr_{\pb}\R$ describing the 3-particle observables between $\p$, $\Lambda$ and $\overline{\Lambda}$. It can be decomposed in the block-diagonal form $\R_3=\R_3^+\oplus \R_3^-$ where $\R_3^\pm=\Tr_{\pb} \R\,(\mathbbm{1}\pm\sigma_3)_{\pb}/2$ comes from antiprotons of + or $-$ transversity. $\R_3^+$ and $\R_3^-$ are the $4\!\times\!4$ upper-left and lower-right submatrices of $\R_3$. Table \ref{excl:tab:R3+} gives the matrix elements of $\R_3^+$, normalised to $\Tr R_3^+=\Tr \mathbbm{1}$.

\begin{table}[!hb]
\caption{\label{excl:tab:R3+}Submatrix $\R^+_3$ of the \CSdM{} of  $\bar{\rm p}+\mathrm{p}\!\uparrow\to\overline{\Lambda}\!\uparrow+\Lambda\!\uparrow$.}
\vspace*{-.2cm}
\centerline{
\scalebox{.9}{$
\begin{array}{c|cccc}
 & +++ & +-- & -+- & --+ \\   \hline 
+\!+\!+ & ( 0+\ii3\,|\,0+\ii3\,,0+\ii3 ) & ( 0+\ii3\,|\,1+i2\,,1+i2 ) & ( 1-i2\,|\,0+\ii3\,,1+i2 ) & ( 1-i2\,|\,1+i2\,,0+\ii3 )\\ 
+\!-\!- & ( 0+\ii3\,|\,1-i2\,,1-i2 ) & ( 0+\ii3\,|\,0-\ii3\,,0-\ii3 ) & ( 1-i2\,|\,1-i2\,,0-\ii3 ) & ( 1-i2\,|\,0-\ii3\,,1-i 2 )\\ 
-\!+\!- & ( 1+i2\,|\,0+\ii3\,,1-i2 ) & ( 1+i2\,|\,1+i2\,,0-\ii3 ) & ( 0-\ii3\,|\,0+\ii3\,,0-\ii3 ) & ( 0-\ii3\,|\,1+i2\,,1-i2 )\\ 
-\!-\!+ & ( 1+i2\,|\,1-i2\,,0+\ii3 ) & ( 1+i2\,|\,0-\ii3\,,1+i2 ) & ( 0-\ii3\,|\,1-i2\,,1+i2 ) & ( 0-\ii 3\,|\,0-\ii3\,,0+\ii3 )\\ 
\end{array}$}}
\end{table}
In this table, a compact notation is adopted. For instance 
\be\label{compact}\begin{aligned}
&(0+3|1-i2,1+i2 )=\left\langle
(\sigma_\p^0+\sigma_\p^3)\otimes(\sigma_\Lambda^1-i\sigma_\Lambda^2) \otimes (\sigma_{\bar\Lambda}^1+i\sigma_{\bar\Lambda}^2)\right\rangle=\cr
&( 00|11 )+ ( 03|11 ) -i ( 00|21 ) -i ( 03|21 )+i (0 0|12 ) +i ( 03|12 ) + (00|22 ) + (0 3|22 )~.
\end{aligned}\ee
$\R^-_3$ is obtained from $\R^+_3$ by changing the sign before the last digit in each box of Table \ref{excl:tab:R3+}. It can be also obtained by replacing each box by its symmetrical one about the centre of the matrix, then changing $\pm z$ into $\mp z$ and $\pm i2$ into $\mp i2$ everywhere. Both submatrices are of rank one, since each is associated to one spin state of the antiproton. Therefore all their $2\!\times\!2$ subdeterminants are zero. 
This gives a number of quadratic equations. A complete set of inequalities can - in principle - be derived from these relations and from the positivity of the diagonal elements. In particular, 
Eqs.~(\ref{dens:eq:r1}-\ref{dens:eq:ineq2}) are recovered, although less quickly than using the helicity basis.  

\emph{If the target is unpolarised}, the CSM is further reduced to a $4\!\times\!4$ matrix $\R_2$, which after transposition and normalisation becomes 
$\rho_{\overline{\Lambda}\Lambda}(0)=\Gamma_0/4$
(see Eqs.\ref{excl:eq:rhomat}-\ref{dens:eq:Czmat}). Let us rewrite $\Gamma_0$ in the transversity basis: 
\be\label{rhoYYbar}
%\rho_{\overline{\Lambda}\Lambda} = {1\over4} 
\Gamma_0=\begin{pmatrix}
1+C_{nn}+2P_n & 0 & 0 & C_{ll}-C_{mm} - 2i C_{lm} \\
0 & 1-C_{nn} & C_{ll}+C_{mm} & 0 \\  0 & C_{ll}+C_{mm} & 1-C_{nn} & 0 \\
C_{ll}-C_{mm} + 2i C_{lm} & 0 & 0 & 1+C_{nn} - 2P_n \end{pmatrix}~,       
\ee
%
%where we have put the $ |\bar\Lambda \Lambda\rangle $ basis in the order $\{|\uparrow\uparrow\rangle,\,|\uparrow\downarrow \rangle,\,|\downarrow\uparrow\rangle,\,|\downarrow\downarrow\rangle\}$ 	and 
where charge conjugation $C_{lm}=C_{ml}$, $P_n(\Lambda)=P_n(\overline{\Lambda})$ has been taken into account. 
As expected from the symmetry about the scattering plane, $\rho_{\overline{\Lambda}\Lambda}$ is block-diagonal in two rank-2 submatrices. This greatly simplifies the derivation of the positivity. One obtains the \emph{minimal complete set of constraints}
\be\label{YYbar}\begin{aligned}%
(1+C_{nn})^{2} &\ge 4 P_n^{2}  + 
( C_{ll} - C_{mm} )^{2}  + 4 C_{lm}^2 ~,\cr
(1-C_{nn})^{2} &\ge ( C_{ll} + C_{mm} )^{2} ~,\cr
|C_{nn}| &\le 1~,\end{aligned}\ee
which, in particular, recovers Eqs.(\ref{dens:eq:r1},\ref{dens:eq:ineq2}).

\paragraph{Projection on a subset of observables.}
Most often, only a subset of observables is measured and it is useful to draw the allowed domain for this subset. If, for instance $P_n$ and $C_{lm}$ are not measured, the answer is immediate, since the planes $P_n=0$ and $C_{lm}=0$ are {\it symmetry planes} for $\mathcal{D}$: the allowed domain $\mathcal{D}_3\{C_{ll},C_{mm},C_{nn}\}$ is obtained by setting $P_n=C_{lm}=0$ in Eqs.~(\ref{YYbar}), which gives 
\be
|1+C_{nn}|\ge|C_{ll}-C_{mm}|\,,\quad |1-C_{nn}|\ge|C_{ll}+C_{mm}|\,,\quad 1\ge|C_{nn}|\,. 
\ee
This is the tetrahedron of vertices  $(-1,-1,-1)$, $(-1,+1,+1)$, $(+1,-1,+1)$, $(+1,+1,-1)$ seen in Fig.\ref{2-3-5-4-9-10} (right). 

Let us now consider the case where at least one of the non-measured variables does not correspond to a symmetry plane of $\mathcal{D}$. For illustration we assume that $C_{lm}$ and $C_{ll}$ are not measured and we look for $\mathcal{D}_3\{P_n,C_{mm},C_{nn}\}$. First of all, since $C_{lm}=0$ is a symmetry plane of $\mathcal{D}$, one can immediately set $C_{lm}=0$ in (\ref{YYbar}). The projection on the non-symmetry plane $C_{ll}=0$ may be found with the two following different methods:
%Here we will compare two systematic methods, the \emph{apparent contour method} and the \emph{reciprocal polar method}, applied to $\mathcal{D}_3\{P_n,C_{mm},C_{nn}\}$ when $C_{lm}$ and $C_{ll}$ are not measured. 

\subparagraph{The apparent contour method.}
Let us rewrite the minimal complete set of constraints (\ref{YYbar}), with $C_{lm}=0$, as 
\be\label{exl:eq:minimal}\begin{aligned}
f(v,x,y,z) &= (1+z)^2 - ( x - y )^2 - 4 v^2 \ge 0\,,\\
g(x,y,z) &= (1-z)^2 - ( x + y )^2  \ge 0\,,\\
h(z) &= 1-z^2 \ge 0~\,,\end{aligned}
\ee
% $\mathcal{D}_4\{v,x,y,z\}$
where $v=P_n$ , $x=C_{ll}$ , $y=C_{mm}$ , $z=C_{nn}$. The projection $\mathcal{D}_3$ of $\mathcal{D}_4$ defined by (\ref{exl:eq:minimal}) on the plane $x=0$ can be obtained from the following rule: 

The boundary or {\it apparent contour} of $\mathcal{D}_3$ is the projection of a curve $\H$ lying on the boundary of $\mathcal{D}_4$ that we call {\it horizon}. A systematic method to find the horizon is the following:

- If $\mathcal{D}_4$ were defined by only one inequality, say $f \ge 0$,
a point of the horizon would satisfy
\be
f=0,\quad f'_x=0,\quad f''_x\le0~.
\ee
>From the two equalities one would eliminate $x$ and obtain the wanted equation in $v,y,z$ for the boundary. The inequality $f''_x\le0$ is the requirement that the projection line does not go inside $\mathcal{D}_4$ but is externally tangent to it. 
%(in the case $f''_x=0$ the first non-zero derivative has to be of even order and positive). 

- Actually, in our example, $\mathcal{D}$ is the intersection of 3 domains $\mathcal{D}_f$, $\mathcal{D}_g$, $\mathcal{D}_h$ respectively defined by $f\ge 0$, $g\ge 0$, $h\ge 0$. The projection of $\mathcal{D}$ is generally smaller than the intersection of the projections of $\mathcal{D}_f$, $\mathcal{D}_g$ and $\mathcal{D}_h$.  
A point of $\H$ can in fact be

\begin{itemize}
	\item[ A)] a point of the horizon $\H_f$ of $\mathcal{D}_f$, while belonging to $\mathcal{D}_g$ and $\mathcal{D}_h$:
\be
f=0 , \quad f'_x = 0, \quad f''_x \le 0, \quad g \ge 0, \quad h \ge 0~,
\ee	
or idem with permutation of the roles of $f$, $g$ and $h$. 
	\item[ B)] a point common to the boundaries of $\mathcal{D}_f$ and $\mathcal{D}_g$ and belonging to $\mathcal{D}_h$: 
\be\label{ProjB}
f=g=0, \quad f'_x \, g'_x \le 0, \quad h\ge 0
\ee
or idem with permutation of the roles of $f$, $g$ and $h$. 
The first inequality of (\ref{ProjB}) is the requirement that the projection line does not go inside $\mathcal{D}_f \bigcap \mathcal{D}_g$.
\end{itemize}

\ni
Let us apply these principles to (\ref{exl:eq:minimal}).
%Projection along $x$ gives the following boundaries for $\{v,y,z\}$:

\begin{itemize}\item[a)] $f=f'_x=0$, ~$f''_x\le0$, ~$g\ge0$, ~$h\ge0$ gives the two triangles 
\be
|z|\le1, \quad 2|y|\le1-z, \quad 2|v|\pm1+z~,
\ee
\item[b)] $g=g'_x=0$, ~$g''_x\le0$, ~$f\ge0$, ~$h\ge0$ gives the disk 
\be
z=1~, \quad v^2+y^2\le1 \qquad(h=0)~,
\ee
\item[c)] $h=h'_x=0$, ~$h''_x\le0$, ~$f\ge0$, ~$g\ge0$ gives the same disk, plus the edge 
\be
z=-1~,\ v=0, \quad |y|\le1  \qquad(f=0)~,
\ee
\item[d)] $g=h=0$, ~$g'_x\,h'_x\le0$, ~$f\ge0$ gives the disk again, plus the end points $A$ and $B$ of the edge,
\be
z=-1~,\ v=0, \quad y=\pm1  \qquad(f=0)~,
\ee
\item[e)] $f=h=0$, ~$f'_x\,h'_x\le0$, ~$g\ge0$ gives the edge again plus the disk perimeter.
\item[f)] $f=g=0$, ~$f'_x\,g'_x\le0$, ~$h\ge0$ gives two pieces of cones:
\be
|z|\le1,\qquad \pm2y=1-z+\sqrt{(1+z)^2-4v^2}~,
\ee
\end{itemize}
${\cal D}_3$ is then made of the convex completion of two cones, within the cube $[-1,+1]^{\otimes3}$. The cones, of equations $(1-z\pm2y)^2+4v^2=(1+z)^2$, have tips $(0,\mp 1, -1)$ and the disk (b) as a common basis. The walls looks like a paper coffee filter, see Fig.~\ref{excl:fig:filter-octo}, left. The interior of the domain is given by
\be
|z|\le1,\qquad 2|v|\le1+z,\qquad |2y|+z+\sqrt{(1+z)^2-4v^2}\le1~.
\ee
A section by a plane $z$ = constant is a stadium.

\subparagraph{The method of the reciprocal polar transform  \cite{Minnaert:1971}.}
Let us now try the second method outlined in Sec.~\ref{basic:sub:reciprocal-method}. 
First of all, we recall that due to parity and charge conjugation the effectively allowed domain $\mathcal{D}_{\rm eff}$ for the parameters $C_{\lambda\mu\nu\tau}$ is not the full domain of positivity $\mathcal{D}$ but the intersection of $\mathcal{D}$ with a hyperplane $H$. Then the vector $\vec C_\perp$ defined in Sec.~\ref{basic:sub:reciprocal-method} has redundant coordinates, for instance $C_{00n0}=P_n$ and $C_{000n}$, which are equal. Nevertheless, both coefficients contribute to the scalar product $\vec C_\perp\cdot\vec C'_\perp$ in (\ref{basic:eq:polar-recip'}), so that their contributions add to $2P_nP'_n$. An equivalent statement is that the corresponding Euclidean coordinate in $H$ is $:\hat P_n=(C_{00n0}+C_{000n})/\sqrt{2}=P_n\sqrt{2}$. One must keep this fact in mind when applying the reciprocal polar transform. 

We consider the intersection $\mathcal{I}_3\{P_n,C_{mm},C_{nn}\}$ of $\mathcal{D}_{\rm eff}$ with the hyperplane where all $C_{\lambda\mu\nu\tau}$ are vanishing except for $C_{0000}\equiv1$, $C_{00n0}=C_{000n}\equiv P_n$, $C_{00mm}\equiv C_{mm}$ and $C_{00nn}\equiv C_{nn}$. Using the notations of (\ref{exl:eq:minimal}) it is equivalent to the intersection $\mathcal{I}_3$ of $\mathcal{D}_4$ with the hyperplane $x=0$ (up to the rescaling $\hat P_n=\hat v=v\sqrt{2}$). From (\ref{exl:eq:minimal}), $\mathcal{I}_3$ is given by
\be\label{exl:eq:inter}\begin{aligned}
f(v,0,y,z) &= (1+z)^2 - y ^2 - 4 v^2 \ge 0~,\cr
g(0,y,z) &= (1-z)^2 - y^2  \ge 0~,\cr
h(z) &= 1-z^2 \ge 0~.\end{aligned}
\ee
The first and last equations define the interior of a cone whose tip is the centre of the bottom face of the $[-1,1]^{\otimes3}$ cube and whose intersection with the top face plane ($z=1$) is the ellipse $v^2+(y/2)^2=1$. Due to the second equation, this cone is amputated by a ``roof" made of the two planes $z=1\pm y$. 

According to Sec.\ref{basic:sub:reciprocal-method} the domain $\mathcal{D}_3$ we are looking for is the set of points $\vec C_\perp=(v,y,z)$ which satisfy
\be\label{excl:eq:polar-recip'} 
%(\ref{excl:eq:polar-recip'}) = from (\ref{basic:eq:recipro-proj-inter})
\vec C_\perp\cdot\vec C'_\perp \ge-1\,
\ee
for any $C'_\perp\in\mathcal{I}_3$, the scalar product in $\mathcal{D}_3$ being given by 
\be\label{Euclidean'}
\vec C_\perp\cdot\vec C'_\perp=2v.v'+y.y'+z.z' 
\quad{\rm or}\quad\hat v.\hat v'+y.y'+z.z'\,.
\ee
The boundary $\partial\mathcal{D}_3$ is the reciprocal polar transform, of power $-1$, of the boundary $\partial\mathcal{I}_3$ of $\mathcal{I}_3$. 

\begin{figure}[!!hbtc]
\begin{center}
\includegraphics[width=.8\textwidth]{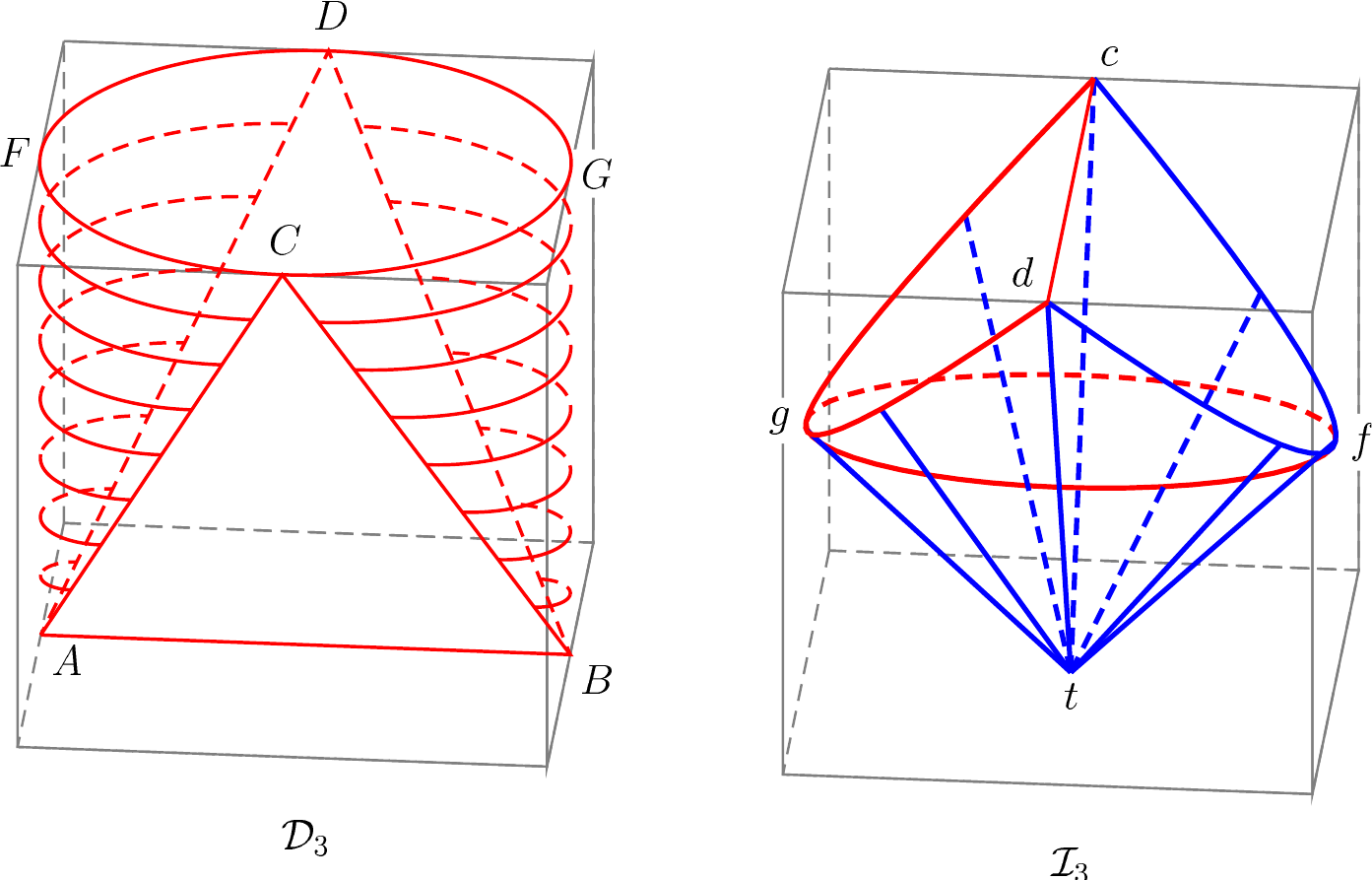}
\end{center}
\caption{\label{excl:fig:polar2} Polar transform (right) of the funnel-shape filter  (left)  encountered as the allowed domain of some triples of observables, \eg, $\{P_n, C_{mm}, C_{nn}\}$.}
\end{figure}

Instead of using (\ref{Euclidean'}) we may keep the usual form $\vec C_\perp\cdot\vec C'_\perp=v.v'+y.y'+z.z'$ but dilate $\mathcal{I}_3$ alone by a factor 2 in the $v$ dimension. It has the advantage of transforming the cone of $\mathcal{I}_3$ into a cone of revolution. Then $\partial\mathcal{D}_3$ is the reciprocal polar transform of $\partial\mathcal{I}'_3$, the boundary of the dilated $\mathcal{I}_3$, with the usual scalar product. Using the correspondences listed in (\ref{edge-edge}), one recovers the various pieces of the ``coffee filter": 

\be\label{excl:eq:edge-edge}
%\left(
\begin{array}{cccc}
\hbox{pieces of }\partial\mathcal{I}'_3: &  & \hbox{pieces of }\partial\mathcal{D}_3: \\
\hbox{cone tip t}  &  \rightarrow & \hbox{disk-face CFDG} \\
\hbox{trihedron-like tips c and d}&  \rightarrow & \hbox{triangular faces ABC and ABD} \\
\hbox{top edge cd}  & \rightarrow & \hbox{bottom edge AB} \\
\hbox{edges cfd and cgd (parabolas)}  & \rightarrow  & \hbox{conical faces CAD and CBD} \\
\hbox{planar faces cfd and cgd}  & \rightarrow  & \hbox{tips A and B} \\
\hbox{cone}  & \rightarrow  & \hbox{top circular edge}
\end{array}
%\right)
\ee
For this example at least, the reciprocal transform method involves less calculations but more geometrical reasoning  than the apparent contour method. The ``filter'' and its transform are shown in Fig.~\ref{excl:fig:polar2}.
\subsubsection{Special cases}
The inequalities relating the spin observables of $\ppLL$ have been derived without restriction on the underlying dynamics. As the early LEAR data indicated a clear dominance of spin-triplet transitions, the case where the spin-singlet is set to zero has been considered in \cite{Elchikh:1999ir}.

Now, the formalism of \ppLL\ applies without restriction to the reaction \eepp, which is measured, \eg, at Frascati \cite{Antonelli:1994kq}, and  to the inverse reaction, which has been measured for instance at LEAR \cite{Bardin:1991bz}, and which is on the agenda of future antiproton facilities.

However, \eepp\ is dominated by the one-photon intermediate state in the $s$-channel. For this mechanism, in the more general case of the $\overline{B}_1 B_1\to \overline{B}_2B_2$ reaction, the non-vanishing  helicity amplitudes read (see, \eg, \cite{Buttimore:2007cv} for a recent compilation)
\be\begin{aligned}
\mathcal{H}_{++;++}&= -{2\alpha m_1 m_2\over s} \cos\vartheta\, G^{\rm E}_1 G^{\rm E}_2~,\\
\mathcal{H}_{+-;+-}&= -{\alpha\over 2} (1+\cos\vartheta)\, G^{\rm M}_1 G^{\rm M}_2~,\\
\mathcal{H}_{+-;-+}&= -{\alpha\over 2} (1-\cos\vartheta)\, G^{\rm M}_1 G^{\rm M}_2~,\\
\mathcal{H}_{++;+-}&= -{\alpha m_1\over \sqrt{s}} \sin\vartheta\, G^{\rm E}_1 G^{\rm M}_2~,\\
\mathcal{H}_{+-;++}&= +{\alpha m_2\over \sqrt{s}} \sin\vartheta\, G^{\rm M}_1 G^{\rm E}_2~,
\end{aligned}\ee
where $\alpha$ is the fine-structure constant, $m_1$ and $m_2$ the mass of the baryon in the initial and final state, and $G^{\rm E,M}_i$ their electric and magnetic form factors, which are complex in the time-like region, due to elastic and annihilation channels. Here, the $1/s$ factor comes from the photon propagator, and each photon vertex is described as usual by a coupling
\be 
\bar{u} \left[ \gamma^\mu F(s) - {i\over 2m} \sigma^{\mu\nu} q_\nu F'(s)\right] u~,
\ee
the vector and tensor form factors being related to the usual electric and magnetic ones by
\be
G_{\rm M}=F+F'~,\qquad G_{\rm E}=F+{s\over 4 m^2} F'~.
\ee
In the case of an electron, $F=1$ and $F'=0$.

The amplitudes $a$, \dots, $g$ can be expressed in terms of these helicity amplitudes, 
\be\begin{aligned}
  a &=\cos\vartheta\left(\Hcal_{++;++} + \Hcal_{++;--} + \Hcal_{+-;+-} - \Hcal_{+-;-+}\right)/2 +\sin\vartheta \left(-\Hcal_{+;+-} + \Hcal_{+-;++}\right)~,\\
b&=\left(\Hcal_{++;++} - \Hcal_{++;--} + \Hcal_{+-;+-} + \Hcal_{+-;-+}\right)/2~,\\
 c &= \left(-\Hcal_{++;++} + \Hcal_{++;--} + \Hcal_{+-;+-} + \Hcal_{+-;-+}\right)/2~,\\
  d &=\cos\vartheta\left(\Hcal_{++;++} + \Hcal_{++;--} - \Hcal_{+-;+-} + \Hcal_{+-;-+}\right)/2 + \sin\vartheta\left(\Hcal_{++;+-} + \Hcal_{+-;++}\right)~,\\
  e &= -\cos\vartheta\left(\Hcal_{++;+-} - \Hcal_{+-;++}\right) - \sin\vartheta\left(\Hcal_{++;++} + \Hcal_{++;--} + \Hcal_{+-;+-} - \Hcal_{+-;-+}\right)/2~,\\
  g &=\cos\vartheta\left(\Hcal_{++;+-} + \Hcal_{+-;++}\right) - \sin\vartheta\left(\Hcal_{++;++} + \Hcal_{++;--} - \Hcal_{+-;+-} + \Hcal_{+-;-+}\right)/2~.\end{aligned}
\ee
The spin observables of the reaction $\eepp$ are discussed in \cite{Bilenky:1993cd,Buttimore:2006mq,Barnes:2007ub}\,\footnote{A detailed correspondence with T.~Barnes and W.~Roberts is gratefully acknowledged}. It is particularly interesting to compare the values on and off the $\mathrm{J}/\psi$ peak.
\subsubsection{Proton-proton scattering and other similar reactions}
The results derived for \ppLL\ can be adapted to any $1/2+1/2\to 1/2+1/2$ reaction. 
As already mentioned, the low-energy nucleon--nucleon observables have been measured in great detail, enabling one to reconstruct the five amplitudes in the each isospin channel, up to an overall phase. Hence the data are compatible with any possible model-independent inequality that can be devised beforehand.

\begin{figure}[!!htbc]
\begin{minipage}{.55\textwidth}
\begin{flushleft}
\includegraphics[width=.95\textwidth]{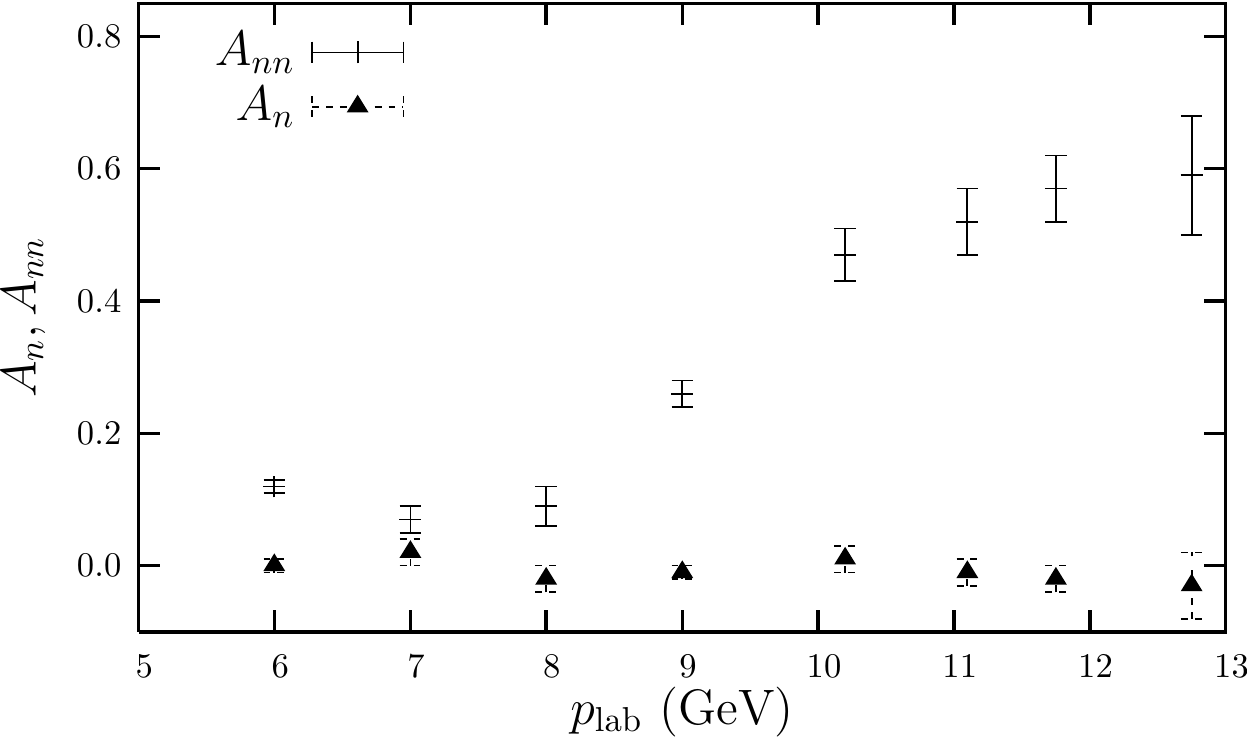}
\end{flushleft}
\end{minipage}
\begin{minipage}{.44\textwidth}
\caption{\label{excl:fig:pp90} Analysing power $A_n$ and spin-spin correlation $A_{nn}$ in proton-proton scattering at $90^\circ$, as a function of the beam momentum.}
\end{minipage}
\end{figure}

The proton--proton exclusive reaction has also been studied at high energy. With the advent  of QCD, the current wisdom based on perturbative contributions was that spin effects should generally be small as energy increases.  However, a measurement by  Krisch \etal\ \cite{O'Fallon:1977cp,Crabb:1978km,Crosbie:1980tb}, indicated a large value of the  spin--spin correlation $A_{nn}$ at $90^\circ$, as a function of the incoming momentum. Their results are reproduced in Fig.~\ref{excl:fig:pp90}.
Also shown is the analysing power $A_n$, which should vanish at $90^\circ$, due to permutation symmetry.

This striking result of \cite{O'Fallon:1977cp,Crabb:1978km,Crosbie:1980tb} motivated interesting studies on the underlying mechanisms. See, \eg, \cite{Brodsky:1979nc}. For our purpose, which is the allowed domain for observables, there is at $90^\circ$ a simplified situation, with only four independent helicity amplitudes, which, in the notation of \cite{Brodsky:1979nc}, can be chosen as
\be\label{excl:eq:pp90a}
\alpha=M_{++,++}~,\quad  \beta=M_{+-,+-}=M_{-+,+-}~,\quad \delta=M_{--,++}~,
\ee
leading to the angular distribution  $I_0=|\alpha|^2+2|\beta|^2+|\delta|^2$ and asymmetry coefficients given by
\be\label{excl:eq:pp90b}\begin{aligned}
I_0 A_{nn}&=2 \RE(\alpha^*\delta)+2|\beta|^2~,\\
I_0 A_{ll}&=|\alpha|^2+|\delta|^2-2|\beta|^2~,\\
I_0 A_{mm}&=-2 \RE(\alpha^*\delta)+2|\beta|^2~.\end{aligned}\ee
This implies  the identity 
\be\label{excl:eq:BCL}
A_{nn}+A_{ll}+A_{mm}=1~,
\ee
and  the (redundant) inequalities 
\be
A_{nn}+A_{ll}>0~,\quad A_{nn}+A_{mm}>0~,\quad A_{ll}+A_{mm}>0~.
\ee
Unfortunately, a very large $A_{N\! N}$ does not constrain much the other asymmetry coefficients, for instance 
$A_{nn}=0.8$ leaves the interval $[-0.8,1]$ available for both $A_{ll}$ and $A_{mm}$.

For completeness, let us make the link with the notation used for $\ppLL$.  In (\ref{excl:eq:amp-ppLL}), $d=0$ for the equal-mass case such as $\ppb\to\ppb$ or $\p\p\to\p\p$. Furthermore at $90^\circ$ for $\p\p$ elastic scattering,  $a=0$ and $b=c$.  If  these relations are plugged into (\ref{excl:eq:ppllobs}), 
\be
C_{nn}+C_{ll}+C_{mm}=1~,
\ee
which is the analogue of (\ref{excl:eq:BCL}), both already discussed in Sec.~\ref{se:basic}

Note that Gibbs and Loiseau \cite{Gibbs:2006ym} introduced a similar set of amplitudes, but following Ref.~\cite{Bystricky:1976jr}, they used unit vectors $\{\vec{\ell}',\,\vec{m}'\}$ which are rotated by $45^\circ$ as compared to  our  $\{\vec{\ell},\vec{m}\}$ for the scattered particle, so that their amplitudes are the same ($a'=a$, etc.), except for $c'=-c$, $d'=-g$ and $g'=-d$.
\subsection{Photoproduction of scalar or pseudoscalar mesons}\label{excl:sub:photps}
\subsubsection{Experimental results}
\begin{figure}
\centerline{\includegraphics[width=.7\textwidth]{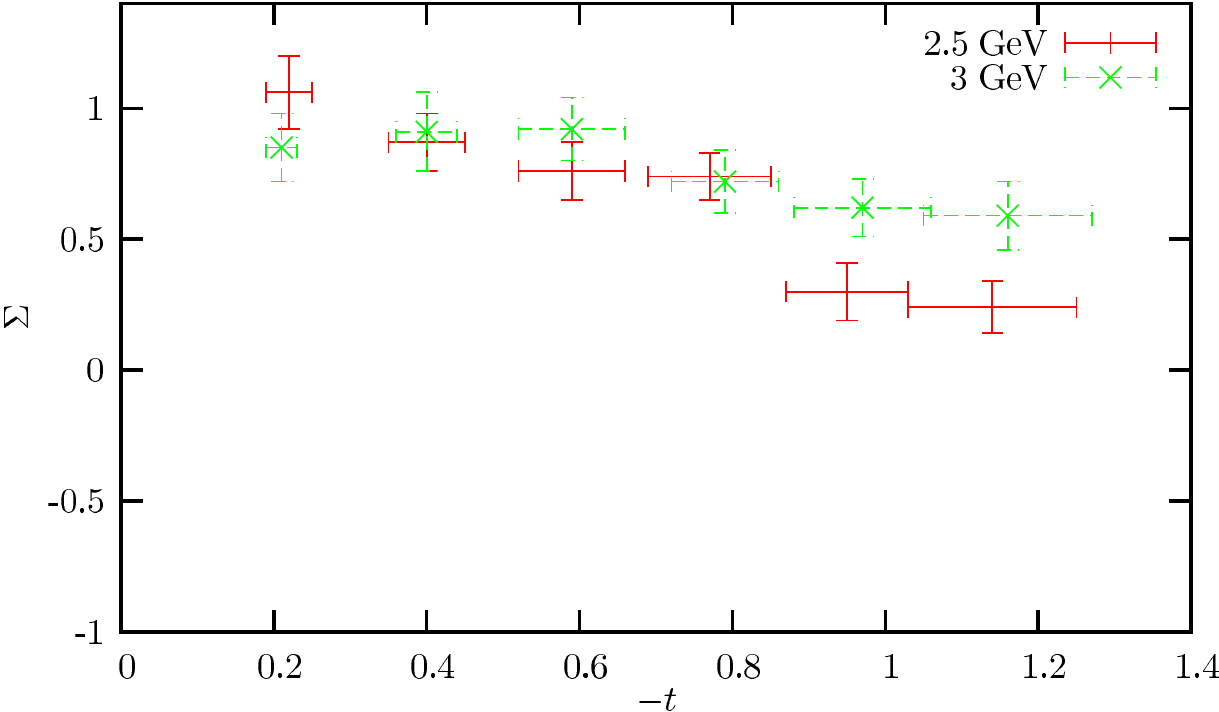}}
\caption{\label{excl:fig:eta}Beam asymmetry for $\gamma\mathrm{p}\to \eta \mathrm{p}$ at 2.5 and 3 GeV, as measured by 
Bussey \etal\ \protect\cite{Bussey:1976si}}
\end{figure}

There is an abundant literature on the reactions of photoproduction of a pseudoscalar meson, such as
\be\label{excl:eq:ph5}
\gamma + \N \to \pi+\N~,\qquad
\gamma + \N \to \mathrm{K}+\L~,
\ee
and many data are available  through the Durham data base \cite{Durham}.
Some interesting spin measurements have been performed many years ago, in particular with the aim of determining the spin and parity of new mesons, or to study the reaction mechanism. For instance, Bussey \etal\ \cite{Bussey:1976si} measured the beam asymmetry $\Sigma$ (to be defined shortly) of the reaction $\gamma\mathrm{p}\to \eta \mathrm{p}$ at 2.5 and 3 GeV, and found a value close to the maximum allowed,  $+1$, see Fig.~\ref{excl:fig:eta}. Large values of this observable $\Sigma$ were also found by the GRAAL collaboration \cite{Bartalini:2007fg}.

\begin{figure}
\centerline{\includegraphics[width=.7\textwidth]{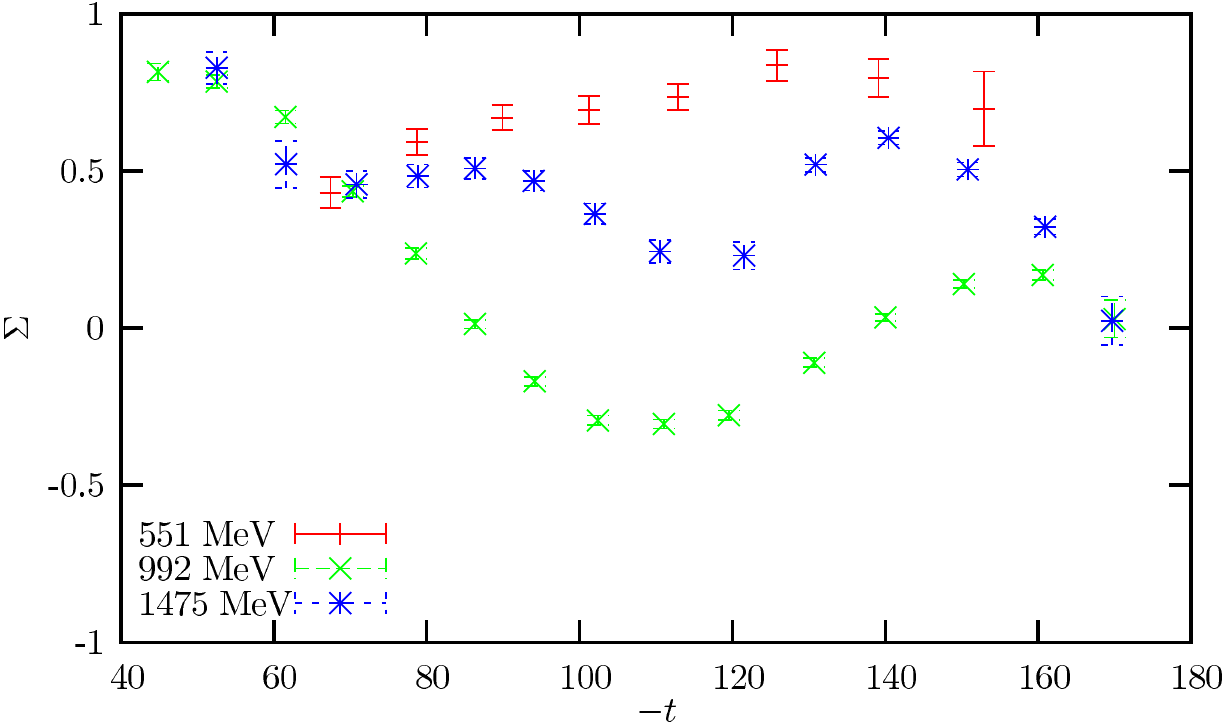}}
\caption{\label{excl:fig:piz}Beam asymmetry for $\gamma\mathrm{p}\to \pi^0 \mathrm{p}$ at several photon energies, as measured by 
GRAAL~\protect\cite{Bartalini:2005wx}.}
\end{figure}

Among the recent experiments are the spin measurements at LEPS \cite{Muramatsu:2006bk},
SAPHIR \cite{Klein:2006bj}, GRAAL \cite{Schaerf:2005ej,Bartalini:2007fg}  and CLAS \cite{Garcon:2006dy}.  
In particular , the analogue of Fig.~\ref{excl:fig:eta} for $\pi^0$, as measured by GRAAL \cite{Bartalini:2005wx}, is reproduced in Fig.~\ref{excl:fig:piz}, for three of their energies (see \cite{Bartalini:2005wx} for the other data), showing again a large asymmetry at some angles.

\begin{figure}[!hbt]
\centerline{\includegraphics[width=.95\textwidth]{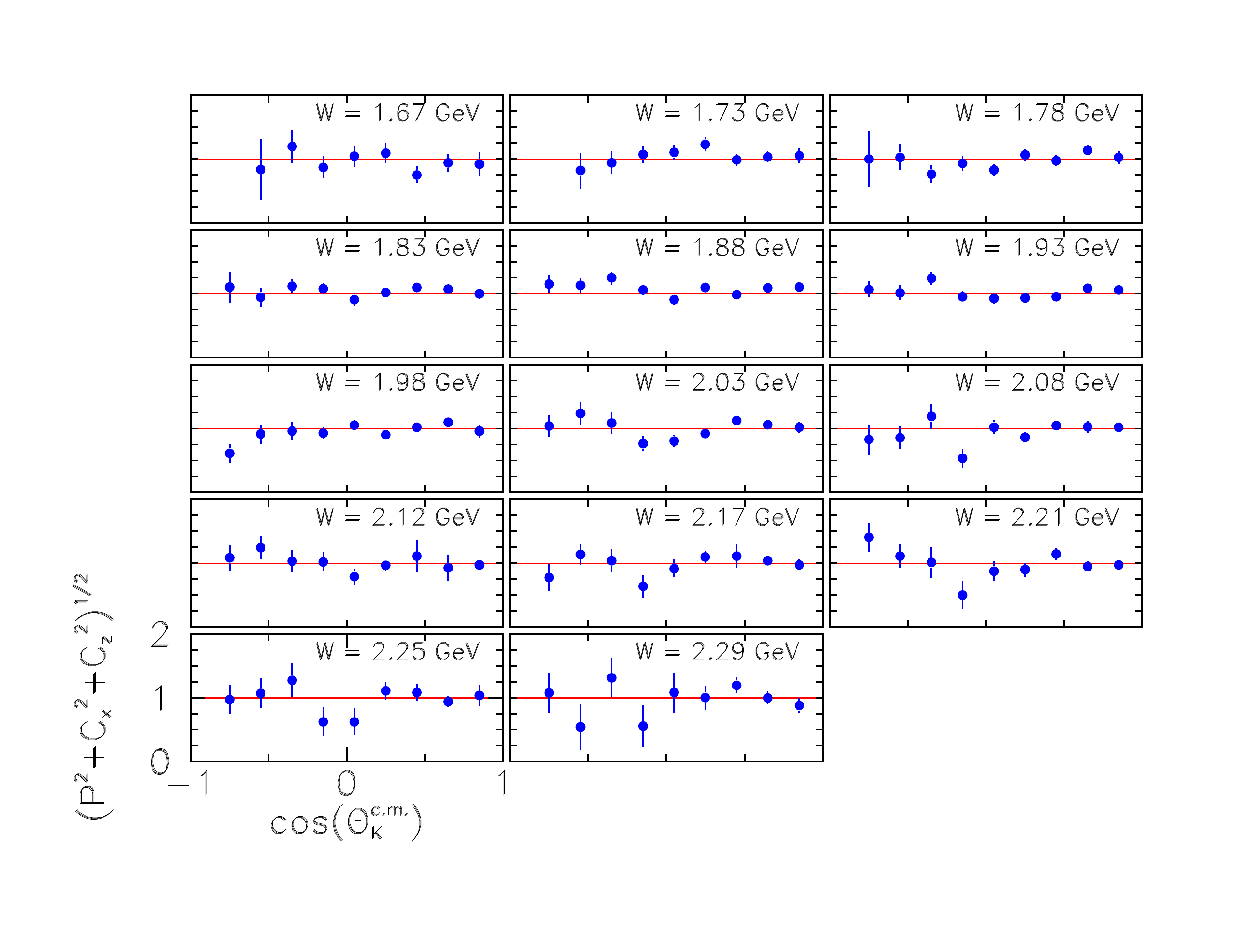}}
\caption{\label{excl:fig:sch}
The magnitude of the $\Lambda$ hyperon polarisation observable vector 
$\sqrt{P^2 + C_x^2 + C_z^2}$ 
from recent CLAS data \protect\cite{Bradford:2006ba}.}
\end{figure}

Much effort has also been devoted to the spin observables
for $\gamma +\N \to \K+\L$ at GRAAL and CLAS. The analysis of the CLAS data \cite{Bradford:2006ba} has revealed the following approximate relation (see Fig.~\ref{excl:fig:sch}):
\be\label{excl:eq:sch1}
P^2 + C_x^2 + C_z^2\simeq 1~.
\ee
It means that when photon has pure circular polarisation, the $\Lambda$ hyperon is fully polarised. This relation will be discussed later in this section.
\subsubsection{Amplitudes}\label{excl:sub:photpsamp}
In the pioneering paper by Chew \etal~\cite{Chew:1957tf}, the amplitude of the photoproduction of pseudoscalar mesons is written as a $2\!\times\!2$ matrix in the baryon spin space \cite{Chew:1957tf}
\be\label{excl:eq:ph6}
	\Mcal = i \vsigma.\ev \ F_1 + 
	{\vsigma.\qv \ \vsigma.(\kv\times\ev)\over qk} \ F_2 +
	i {\vsigma.\kv \ \qv.\ev\over qk} \ F_3 +
	i {\vsigma.\qv \ \qv.\ev\over q^2} \ F_4~,
\ee
where $\ev$ is the vector amplitude of the polarised photon, $\kv$ and $\qv$ are the initial and final relative momenta in the centre-of-mass system. 

Since parity is conserved in this reaction, it is convenient to express the transition matrix $\mathcal{M}$ %of reaction $\gamma+N \to K+Y$ 
in the \emph{transversity basis}: $|\boldsymbol{\pi},\pm\rangle$ and $|\boldsymbol{n},\pm\rangle$  for the initial state and $|\pm\rangle$ for the final state, where $\pm$ denotes the transversity of the initial nucleon and of the final-state nucleon or hyperon, \ie, the  projection $\pm1/2$ of its spin along  the normal to the scattering plane, and $\boldsymbol{\pi}$ (resp.\ $\boldsymbol{n}$) a photon state with linear polarisation parallel (resp.\ normal)  to the scattering plane.  These states are eigenstates of $\Pi$, the operator of reflection about the scattering plane. As explained in Sec.~\ref{basic:sub:parity}, one may assign $\Pi= + 1$ for  $ +1/2$ transversity fermions and $\vec\pi$-polarised photons, and $\Pi= - 1$ for pseudoscalar mesons, $ - 1/2$ transversity fermions and $\nv$-polarised photons. 
The four non-vanishing transversity amplitudes, which are independent, are
\begin{equation}
\begin{array}{ll}
 a_1=\langle + | \mathcal{M}| \boldsymbol{n}+\rangle~,\quad
 a_2=\langle - | \mathcal{M}| \boldsymbol{n}-\rangle~,\\[2pt]
 a_3=\langle + | \mathcal{M}| \boldsymbol{\pi}-\rangle~,\quad
 a_4=\langle - | \mathcal{M}| \boldsymbol{\pi}+\rangle~,
\end{array}
\label{transvampli}
\end{equation}
while $\langle + | \mathcal{M}| \boldsymbol{n}-\rangle=\langle - | \mathcal{M}| \boldsymbol{n}+\rangle
=\langle + | \mathcal{M}| \boldsymbol{\pi}+\rangle=\langle - | \mathcal{M}| \boldsymbol{\pi}-\rangle=0$.
If a scalar meson was produced instead of a kaon, the sets of vanishing and non-vanishing transversity amplitudes would be interchanged. 
%$F$ would have to be even under reflection $\Pi$ and the roles of $\ev_\cdot\vpi$ and $\ev\cdot\nv$ would be interchanged. 

The transversity amplitudes $a_i$ are related to the helicity ones \cite{Barker:1975bp}, and can be expressed in terms of the CGLN amplitudes $F_i$ \cite{Walker:1968xu}.

Several authors \cite{GRG,Barker:1975bp,Chiang:1996em} have considered the mathematical problem of fully reconstructing the amplitudes, up to an overall phase but without discrete ambiguities, from a well-chosen set of polarisation observables. Seven independent real functions can thus be determined. From all the possible experiments, one can extract \emph{sixteen} different quantities, which are the bilinear products of the four amplitudes, \ie, of the seven real functions. These quantities are therefore not independent. However the relations between them are non-linear, so that measuring only seven of them leaves discrete ambiguities. The measurement of at least an eight quantity is necessary to resolves these ambiguities. Here we will not review this problem, but the linear and nonlinear equalities constraints coming from the positivity and rank one of the \CSdM, as in Ref.~\cite{Artru:2006xf}.

\subsubsection{The observables}
The various observables are traditionally classified and denoted as follows: 
\begin{itemize}\itemsep-2pt
\item the unpolarised differential cross section $I_0$,
\item the linearly-polarised photon-beam  asymmetry $\Sigma$ (sometimes denoted $\Sigma^\gamma$ or $\Sigma_x$),
\item the polarised-target asymmetry $A$ (also denoted $A^\N$),
\item the  polarisation $P$ of the recoiling baryon (also denoted $P^\Y$ for $\gamma+\N\to\K +\Y$),
\item the baryon depolarisation  coefficients $T_i$ and $L_i$ expressing the correlation between the longitudinal or transverse (in the scattering plane) target polarisation and the spin of the recoil baryon,
\item the coefficients describing the  transfer of polarisation  from a photon beam   to the recoil baryon, in particular $O_i$ for oblique polarisation and $C_i$ for  circular polarisation,
\item the coefficients $G$, $H$, $E$ and $F$ of
double spin correlations between the photon beam and the nucleon target,
\item
triple correlations coefficients if both the beam and the target are polarised and the hyperon polarisation analysed.
\end{itemize}
The index $i$ refers to the component in a frame $\{\hat{x}, \hat{y}, \hat{z}\}$ attached to each baryon:
$\hat{y}$, the normal to the scattering plane, is the same for both, and $\hat{z}$ can be chosen along the centre-of-mass momentum $\vec{p}$, \ie, $\hat{z}=\vec{p}/p$ (in photoproduction experiments, $\hat{z}=-\vec{p}/p$ is usually chosen for the baryons). For the baryons, $S_y$ is the transversity. 
For the explicit calculations, it is convenient to choose a representation where $S_y$, instead of $S_z$, is diagonal. This leads to redefine the labelling of the Pauli matrices such that
\be\label{excl:eq:Pauli-trans}
\sigma_1=\begin{pmatrix} 0&1\\ 1&0 \end{pmatrix}=\sigma_z~,\quad
\sigma_2=\begin{pmatrix} 0&-i\\   i&0 \end{pmatrix}=\sigma_x~,\quad
\sigma_3=\begin{pmatrix} 1&0\\   0&-1 \end{pmatrix} =\sigma_y~,
\ee
with $\sigma_1$ corresponding to the helicity and $\sigma_3$ to the transversity of the baryon.
For the photon, we use the Stokes parameters $(S^1, S^2, S^3)$ defined in Sec.~\ref{basic:sub:single}: $S^3\equiv S_{\ominus}$ (``planarity'') is the polarisation along $\hat{x}$, $S^1\equiv S_{\oslash}$ (``obliquity'') is the polarisation along $(\hat{x}+\hat{y})/\sqrt{2}$ and $S^2\equiv S_{\odot}$ is the circular polarisation, or helicity.

Using the  general notation of Sec.~\ref{se:basic}, the fully polarised ($|\vec{S}|=1$) differential cross section is expressed as
\be
{d\sigma \over d\Omega} \left( \vec{S}_\gamma,\vec{S}_\N,\vec{S}_\Y\right)
= I_0 \,(\lambda\mu|\nu)\,S_\gamma^\lambda \, S_\N^\mu \, S_\Y^\nu~.
\ee
Here $\lambda, \mu, \nu$, run from 0 to 3, the summation is understood over repeated indices and the polarisations have been promoted to four-vectors with $S^{0} = 1$. The coefficients $(\lambda\mu|\nu)$ are related to the traditional ones by
\be\label{excl:eq:obs-def}\begin{aligned}
 ( 00|0 ) &= - ( 33|3 ) = 1~,\\
 \langle \ominus \rangle = - \langle yy' \rangle = ( 30|0 ) &= - ( 03|3 ) = -\Sigma~,\\
 \langle y \rangle = - \langle \ominus y' \rangle = ( 03|0 ) &= - ( 30|3 ) = + A ~,\\
 \langle y' \rangle = - \langle \ominus y \rangle = ( 00|3 ) &= - ( 33|0 ) = + P ~,\\
 \langle zz' \rangle = ( 01|1 ) &= - ( 32|2 ) = + L_z ~,\\
 \langle zx' \rangle = ( 01|2 ) &= + ( 32|1 ) = + L_x ~,\\
 \langle xz' \rangle = ( 02|1 ) &= + ( 31|2 ) = + T_z ~,\\
 \langle xx' \rangle = ( 02|2 ) &= - ( 31|1 ) = + T_x ~,\\
 \langle \oslash z \rangle = ( 11|0 ) &= + ( 22|3 ) = - G ~,\\
 \langle \oslash x \rangle = ( 12|0 ) &= - ( 21|3 ) = - H ~,\\
 \langle \odot z \rangle = ( 21|0 ) &= - ( 12|3 ) = + E ~,\\
 \langle \odot x \rangle = ( 22|0 ) &= + ( 11|3 ) = + F ~,\\
 \langle \oslash z' \rangle = ( 10|1 ) &= - ( 23|2 ) = - O_z~,\\
 \langle \oslash x' \rangle = ( 10|2 ) &= + ( 23|1 ) = - O_x ~,\\
 \langle \odot z' \rangle = ( 20|1 ) &= + ( 13|2 ) = + C_z ~,\\
 \langle \odot x' \rangle = ( 20|2 ) &= - ( 13|1 ) = + C_x~.
\end{aligned}\ee
The symbol $\langle\oslash x'\rangle$, for instance, is an intuitive notation for the correlation between the oblique polarisation of the photon (at $+\pi/4$) and the polarisation towards $\hat{x}$ of the final baryon.
The definition of $\Sigma$, $A$, ... $C_x$ is taken from \cite{Fasano:1992es}. They differ in sign with \cite{Barker:1975bp} concerning $L_x$, $G$, $E$, $C_x$ and $C_z$.  
\subsubsection{The \CSM}
The \CSM\ of $\gamma+\N+\overline{\L} \to \K$ is defined as 
\be\label{excl:eq:Rphotprod}
\left\langle \boldsymbol{e}', s'_\N, s'_\Y \left| \R \right| \boldsymbol{e}, s_\N, s_\Y \right\rangle
= \langle  \boldsymbol{e}', s'_\N | \mathcal{M}^\dagger | s'_\Y  \rangle
\langle s_\Y   | \mathcal{M} | \boldsymbol{e}, s_\N \rangle~.
\ee
By construction $\R$ is semi-positive definite and of rank 1. In terms of the observables for $\gamma+\N\to \K + \L$, the renormalised CSM reads
\be\label{excl:eq:decompose-R}
\Rh=
(2^3/ \Tr \R)\,\R=(\lambda\mu|\nu) \ \sigma^\lambda_\gamma \otimes 
\sigma^\mu_\N \otimes \left[\sigma^\nu_\Y\right]^t~.
\ee
Note in (\ref{excl:eq:Rphotprod}) the crossing $|s_Y\rangle \leftrightarrow \langle s_Y|$  of the hyperon and the corresponding transposition of $\sigma^\nu_Y$ in (\ref{excl:eq:decompose-R}).
%Table \ref{tab-app} of Appendix \ref{appa} gives the matrix elements of R. 
%

In the transversity basis, owing to (\ref{transvampli}), $\Rh=\Rh^+ \oplus \Rh^-$, where $\Rh^-$ acts in the subspace spanned by
\begin{equation}\label{excl:eq:oddPsubspace}
\left|\boldsymbol{\pi}+- \right\rangle~,\quad
\left|\boldsymbol{\pi}-+  \right\rangle~,\quad
\left|\boldsymbol{n}++\right\rangle \quad\hbox{and}\quad
\left|\boldsymbol{n}--\right\rangle~,
\end{equation}
and $\Rh^+$ acts on the complementary subspace. % vanishes identically. 
These matrices $\R^-$ are given by Table~\ref{tab-app}, where $( 0+3\,,\,1-i2\,|\,1+i2 )$, for instance, is a compact notation for 
\begin{equation}
( 01|1 ) + ( 31|1 ) -i ( 02|1 ) -i ( 32|1 )+i ( 01|2 ) +i ( 31|2 ) + ( 02|2 ) + ( 32|2 )~.
\label{excl:eq:compact}
\end{equation}

\begin{table}[here]
\caption{\label{tab-app}Submatrix $\Rh^\mp$ of the renormalised cross-section matrix $\Rh$. The upper sign refers to pseudoscalar photoproduction, the lower one to scalar.}
\vspace{-.6cm}
\begin{small}
$$
\hspace{-.5cm}\begin{array}{c|cccc}
%\hline 
 & \pi+\mp & \pi-\pm & n+\pm & n-\mp\\     \hline 
\pi+\mp & ( 0+\ii3\,,\,0+\ii3\,|\,0\mp\ii3 ) & ( 0+\ii3\,,\,1-i2\,|\,1\mp i2 ) & ( 1-i2\,,\,0+\ii3\,|\,1\mp i2 ) & ( 1-i2\,,\,1-i2\,|\,0\mp \ii3 )
\\ %\hline 
\pi-\pm & ( 0+\ii3\,,\,1+i2\,|\,1\pm i2 ) & ( 0+\ii3\,,\,0-\ii3\,|\,0\pm \ii3 ) & ( 1-i2\,,\,1+i2\,|\,0\pm \ii3 ) & ( 1-i2\,,\,0-\ii3\,|\,1\pm i2 )
\\ %\hline 
n+\pm & ( 1+i2\,,\,0+\ii3\,|\,1\pm i 2 ) & ( 1+i2\,,\,1-i2\,|\,0\pm \ii3 ) & ( 0-\ii3\,,\,0+\ii3\,|\,0\pm \ii3 ) & ( 0-\ii3\,,\,1-i2\,|\,1\pm i2 )
\\ %\hline 		
n-\mp & ( 1+i2\,,\,1+i2\,|\,0\mp \ii3 ) & ( 1+i2\,,\,0-\ii3\,|\,1\mp i2 ) & ( 0-\ii3\,,\,1+i2\,|\,1\mp i2 ) & ( 0-\ii3\,,\,0-\ii3\,|\,0\mp \ii3 )
\\ %\hline 		
\end{array}$$\end{small}
\end{table}

\paragraph{Parity conservation.}
The 4 states of (\ref{excl:eq:oddPsubspace}) correspond to the 4 helicity amplitudes which are allowed by parity for the photoproduction of a pseudoscalar meson. The other states correspond to vanishing amplitude, therefore $\R^+$ vanishes identically. The reverse situation would hold for a \emph{scalar} meson. 

The 16 equalities in Eqs.~(\ref{excl:eq:obs-def}) can be derived from $\R^+=0$. They correspond to the invariance of $(\lambda,\mu|\nu)$  under the  substitutions: $0 \to -3$, $3 \to -0$, $1\to i2$, 2$\to -i1$ for $\lambda$ and $\mu$ and 0$\to -3$, 3$\to -0$, 1$\to -i2$, 2$\to i1$ for $\nu$ and, in this way, they can be derived more quickly from 
\be\label{excl:eq:equivalence-b}
\sigma_3 \otimes \sigma_3 \otimes \sigma_3 \,\R = \R\, \sigma_3 \otimes \sigma_3 \otimes \sigma_3 = -\R~
\ee
(cf.~Eq.~(\ref{basic:eq:P-quant-O-bis})). This Bohr identity comes from the vanishing of amplitudes between an even number of particles with negative $\sigma_3$.

Using these equalities, one can simplify $\hat\R$ by replacing $k\pm k'$ (with $k$ = 0 or 1, $k'$ = 3 or 2) by $k$ \emph{once} in each box of Table \ref{tab-app}. The result is equal to $\hat\R/2$. 
%The vanishing of $\R^+$ gives the 16 \emph{linear} relations (\ref{excl:eq:obs-def}).

\subsubsection{Inequalities}
The positivity of the cross sections for normal polarisation of the baryon and $\vec\pi$ or $\nv$-polarised photons (diagonal elements of Table \ref{tab-app}) yields~\cite{GRG,Artru:2006xf}
\begin{equation}\label{excl:eq:tetra}
|A-P|  \le 1-\Sigma~,\quad
|A+P| \le 1+\Sigma~.
\end{equation}
\begin{figure}[here]
\begin{minipage}{.50\textwidth}
\centerline{
\includegraphics[width=.85\textwidth]{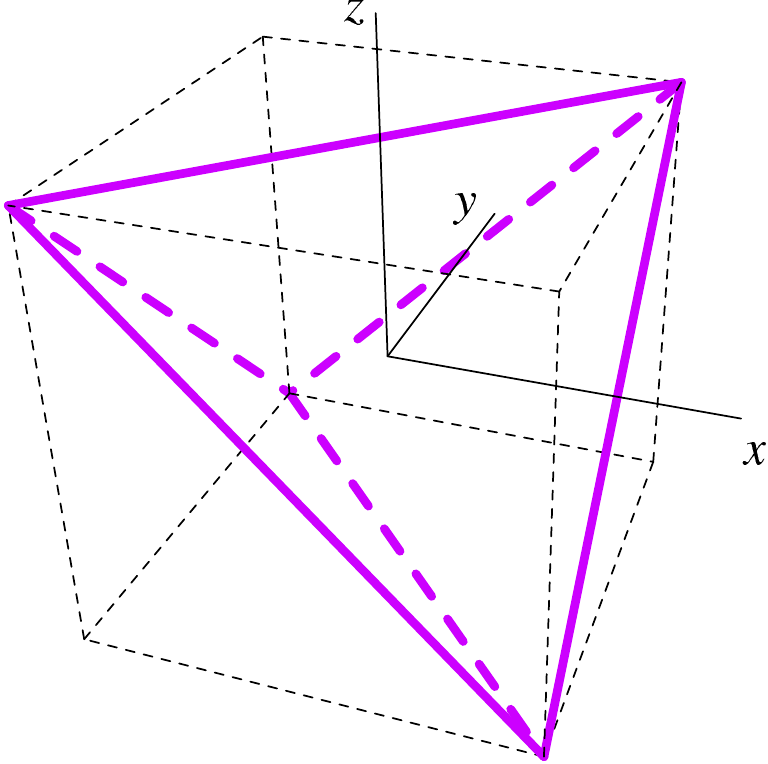}}
\end{minipage}\
\begin{minipage}{.49\textwidth}
\caption{Tetrahedron domain limited by the inequalities (\protect\ref{excl:eq:tetra}) for the observables  $x=A$, $y=P$ and $z=\Sigma$.\label{excl:fig:tetra}}
\end{minipage}
\end{figure}
The domain corresponds to a tetrahedron schematically drawn in Fig.~\ref{excl:fig:tetra}. Notice that its volume is only 1/3 of the volume of the entire cube, while its projection is the entire $[-1,+1]^2$ square on any face.
Unit-ball domains are obtained for many triples like
\be\label{eq:sph}
\{C_x,P,C_z\}~,\quad\{O_x,P, O_z\}~,\quad\{P, C_x, O_x\}~,\quad\{P, C_z, O_z\}~.
\ee 
\noindent
They can be deduced from the anticommutation of the observables operators (see Sec.~\ref{basic:sub:antic}). But they have also classical meaning: For an unpolarised target and a photon Stokes vector $(S_1,S_2,S_3)$ the polarisation vector of the recoil baryon is $(S_1O_x+S_2C_x,P,S_1O_z+S_2C_z)$. The first triple of (\ref{eq:sph}) contains the three components of that vector, when the photon is, say, right-handed ($S_2=1$). Eq.(\ref{excl:eq:sch1}) tells that the corresponding bound is experimentally saturated. Similarly for the second triple when the photon is oblique. For the third triple, let us introduce $a=[(O_x)^2+(C_x)^2]^{1/2}$. When the photon Stokes vector takes the particular value $(O_x/a,C_x/a,0)$, the hyperon polarisation vector becomes $(a,P,O_x O_z/a+C_x C_z/a)$. Ignoring the last component we deduce $P^2+a^2\le1$, \ie, the third triple is in the unit ball. A similar demonstration holds for the fourth triple. 
\subsubsection{Non-linear identities}
Many quadratic identities between observables can be derived from their expressions in terms of helicity or transversity amplitudes. A particularly elegant method \cite{Chiang:1996em} uses the Fierz identities of the Dirac algebra. Here, following the general method described in Sec.~\ref{basic:sub:iqr} (see also \cite{Artru:2006xf}), we will mainly use the fact that $\R$ (or $R^-$) has rank one, which implies the vanishing of all the $2\!\times\!2$ sub-determinants. We will also present more physical arguments, like the purity of the final polarisation when the initial particles are in pure states. 

From the vanishing of the principal minors, one gets
\begin{subequations}\label{excl:eq:quadratic}
\begin{eqnarray}
(1\pm A)^2 &=& (P\pm \Sigma)^2 + (O_z \pm  C_x)^2 + (C_z \mp O_x)^2
\label{excl:eq:quadratic-1}~,	\\
(1\pm P)^2 &=& (A\pm \Sigma)^2 + (G \mp F)^2 + (E \pm H)^2
\label{excl:eq:quadratic-2}~, \\
(1\pm \Sigma)^2 &=& (P\pm A)^2 + (L_z \pm T_x)^2 + (L_x\mp T_z)^2~.
\label{excl:eq:quadratic-3}
\end{eqnarray}
\end{subequations}
These equations can also be obtained by simple physical arguments. For instance, if the target has pure transversity $+$ and the photon in pure helicity $+$, then the hyperon spin components are
\be
S_Y^j=(0+2,0+3|j)/(0+2,0+3|0)=(C_z-O_x,\,O_z+C_x,\,P+\Sigma)/(1+A) \,.
\ee
Since the meson carries no spin information, the hyperon state must be pure, \ie, $|\Sv_Y|=1$. This gives (\ref{excl:eq:quadratic-1}) with the sign $+$. One get the same identity if the photon is oblique. 

From Eqs.~(\ref{excl:eq:quadratic}), one gets
\begin{subequations}
\begin{eqnarray}
1+ A^2 &=& P^2+\Sigma^2+O_z^2+C_x^2 +C_z^2+O_x^2
\label{excl:eq:quadratic-6}~,	\\
1+P^2 &=& A^2+\Sigma^2+G^2+F^2+E^2+H^2
\label{excl:eq:quadratic-7}~, \\
1+\Sigma^2&=&P^2+A^2+L_z^2+T_x^2+L_x^2+T_z^2
\label{excl:eq:quadratic-7a}~.
\end{eqnarray}
\end{subequations}

Complex identities, \ie, pairs of real identities, which do not contain $P$, $A$ nor $\Sigma$ can be obtained from the $2\!\times\!2$ determinants  made only of non-diagonal elements of Table 1, for instance
\begin{equation}
(E+iG)^2+(F+iH)^2+(O_z-iC_z)^2+(O_x-iC_x)^2=0~,
\label{excl:eq:quadratic-5}
\end{equation}
from which one deduces
\be
0 = E^2-G^2+F^2-H^2+O_z^2-C_z^2+O_x^2-C_x^2~.
\label{excl:eq:quadratic-8}\ee
From  the combinations (\ref{excl:eq:quadratic-6})-(\ref{excl:eq:quadratic-7})$\pm$(\ref{excl:eq:quadratic-8}), 
%and (\ref{excl:eq:quadratic-6})-(\ref{excl:eq:quadratic-7})+(\ref{excl:eq:quadratic-8}),
%
\begin{subequations}\label{excl:eq:quadratic-11-12}
\begin{eqnarray}
A^2+E^2+F^2 &=& P^2+C_x^2 +C_z^2
\label{excl:eq:quadratic-11}~,	\\
A^2+G^2+H^2 &=& P^2+O_x^2+O_z^2 
\label{excl:eq:quadratic-12}~.
\end{eqnarray}
\end{subequations}
One can derive Eqs.~(\ref{excl:eq:quadratic-11-12}) using crossing and basic results of quantum information theory. Consider the crossed reaction $\gamma+\bar K\to\bar N+\Lambda$ in which the $\gamma$ is fully polarised, \eg, in the +1 helicity state. Then the final $\bar N+\Lambda$ system is in a pure state. The individual density matrices $\rho_{Nb}= \Tr_{\Lambda} \left(\rho_{\Nb+\Lambda}\right)$ and $\rho_\Lambda=\Tr_{\Nb} \left(\rho_{\bar N+\Lambda}\right)$ should have equal entropies, that is to say, the $\Nb$ and $\Lambda$ polarisations are equal in magnitude:
\be%(|)
(0|30)^2+(2|10)^2+(2|20)^2=(0|03)^2+(2|01)^2+(2|02)^2~.
\ee
Crossing back to $\gamma+\mathrm{N}\to\Lambda+\mathrm{K}$, one changes $(\lambda|\mu\nu)^2$ into $(\lambda\mu|\nu)^2$ and obtains Eq.~(\ref{excl:eq:quadratic-11}). Equation (\ref{excl:eq:quadratic-12}) is obtained in a similar way, considering a photon of oblique polarisation.

Inequalities can be recovered from the identities. Equation (\ref{excl:eq:quadratic-1}), for instance, implies the eight spherical constraints
\be\label{excl:eq:sph-from-q}
1\ge(\Sigma^2 \ \mbox{or}\  P^2) +(O_z^2 \ \mbox{or}\ C_x^2)+(C_z^2\ \mbox{or}\  O_x^2)~,
\ee
% such as $P^2+O_z^2+O_x^2\le 1$. 
four of which are already listed in (\ref{eq:sph}). This can be seen from the inequality $(\vec{V}_1-\vec{V}_2)^2\le (|\vec{V}_1|+|\vec{V}_2|)^2$ applied to the vectors 
$\vec{V}_1=\{P+\Sigma, O_z+C_x,O_x+C_z\}$ and $\vec{V}_2=\{P-\Sigma, O_z-C_x,O_x-C_z\}$ with normalization  $|V_1|=1+A$ and $|V_2|=1-A$.  A slight variant consists of reading 
(\ref{excl:eq:quadratic-1}) as the four-vectors $(1\pm A;\,P\pm \Sigma, O_z\pm C_x,O_x\pm C_z))$ being light-like, and hence, since their time components  $1\pm A$ are positive, stating that their sum is time-like, in the same way as two photons combine into a massive state in elementary relativistic kinematics.

Very recently, Schumacher \cite{Schumacher:2006ii} stressed that in the CLAS data
\be\label{CyCzP}
P^2+C_x^2+C_z^2\simeq 1~.
\ee
This saturates the inequality $P^2+C_x^2+C_z^2\le1$ already encountered in (\ref{eq:sph}) and (\ref{excl:eq:sph-from-q}). Then, using (\ref{excl:eq:quadratic-6}), (\ref{excl:eq:quadratic-11}) and (\ref{excl:eq:quadratic-7}), one can predict
\begin{subequations}
\begin{eqnarray}
	\Sigma^2+O_x ^2 + O_z^2 &\simeq& A^2~,\label{AEFa}\\
	A^2+E^2+F^2 &\simeq& 1~,\label{AEFb}\\
		\Sigma^2+G^2+H^2 &\simeq& P^2~.\label{AEFc}
\end{eqnarray}
\end{subequations}
The relation (\ref{AEFb}) also means the saturation of an inequality. 

\subsubsection{Symmetry rules}
Many other identities can be listed \cite{Chiang:1996em}, but they are related by symmetry rules, in particular
\begin{itemize}
\item
rotation in the $(1,2)$ plane, in particular the substitution $1\to 2$, $2\to -1$ for any particle. For a baryon, it corresponds to a change of $\{z,x\}$ axes in the scattering plane. 
\item
permutation of the $\lambda,\mu,\nu$ indices, \ie, of the three particles with spin, except for the transposition of 
$\sigma^\nu_Y$. For instance, Eqs.~(\ref{excl:eq:quadratic-1}-\ref{excl:eq:quadratic-3}) are related by such a transformation.
%Also, from the triples (\ref{eq:sph}) one obtains unit balls for the triples
%\be\label{excl:eq:sph'}
%\{C_x,\Sigma,C_z\}~,\quad\{O_x,\Sigma, O_z\}~,\quad\{\Sigma, C_x, O_x\}~,\quad\{\Sigma, C_z, %O_z\}~.\ee%
\item
$\pi/2$ rotation in the (1,3) plane: 1$\to$3, 3$\to -1$ for the three particles simultaneously. This transformation violates the parity rules such as $(12|2)=0$ and the equalities  of Eqs.~(\ref{excl:eq:obs-def}). However, relaxing parity temporarily, we may apply it to a subdeterminant of $\R^+$ in Table~\ref{tab-app}, then enforce parity conservation. A practical recipe does not need this table but uses the fact that any observable $\langle{\cal O}\rangle$  of (\ref{excl:eq:obs-def}) corresponds to two $(\lambda\mu|\nu)$'s, one of which at least can be rotated without violating the parity rules. Applying this recipe to (\ref{excl:eq:quadratic-3}), for instance, yields 
\begin{equation}
(1\pm L_z)^2 = (E\pm C_z)^2 + (\Sigma\pm T_x)^2 + (G\pm O_z)^2~,
\label{excl:eq:rot1-3-ex}
\end{equation}
One can obtain this relations by considering the crossed reaction $\Lambda+\bar{\rm N}\to\gamma+\bar{\rm K}$, with $\Lambda$ and $\bar{\rm N}$ in pure helicity states and writing that the photon is fully polarised.
\item
substitution $0 \to i1$, $1 \to i0$ for all the particles, corresponding to an imaginary Lorentz transformation of the $\sigma_\mu$'s in the (0,1) plane. Equation (\ref{excl:eq:quadratic-3}), for instance, is transformed into
\begin{equation}
(1\pm T_x)^2 = (F\pm C_x)^2 + (\Sigma\pm L_z)^2 + (H\pm O_x)^2~.
\label{rot1-0-ex}
\end{equation}
One can interpret this relation with $\Lambda+\bar{\rm N}\to\gamma+\bar{\rm K}$ again, $\Lambda$ and $\bar{\rm N}$ having pure sideways polarisations. 
\end{itemize}
\subsubsection{Future applications}
At the time of the completion of this review, data are analysed by the CLAS and GRAAL collaboration, and for the first time,  several spin observables will be accurately measured. 
An ongoing analysis \cite{Lleres:2007} by Lleres \etal, based on the identities and inequalities listed in \cite{Artru:2006xf} shows that the results on different observables for the same reaction $\gamma \N\to \K\Lambda$ are compatible.

A remarkable feature of our analysis is that, as for $\ppLL$, there is no triple of observable whose allowed domain is the entire $[-1,+1]^3$ cube, as shown in Sec.~\ref{basic:sub:limited}. Every third observable is constrained by any pair of previously measured observable.
%
%%%%%%%%%%%%%%%%
\subsection{Photoproduction of vector mesons}\label{excl:sub:photvec}
\subsubsection{Experimental situation}
There are data on the  reactions of the type 
\be\label{excl:eq:phvec}
\gamma+\mathrm{N}\to V+ B~,
\ee
with $V$ being a vector meson such as $\rho$, $\omega$  or  $\phi$,  and $B=\mathrm{N'}$ a nucleon, or with $V$ being $\K^*$ and $B=Y$ an hyperon. The tensor polarisation is easily obtained, see (\ref{basic:eq:angul-distrib}-\ref{basic:eq:mom-Tij}), but there is no access to the vector polarisation of the meson.
Note that the crossed reaction, $\gamma V\to \pb\p$, has the same spin structure as the photodisintegration of the deuteron, $\gamma\mathrm{d}\to \mathrm{pn}$, which is extensively studied at Jlab (see, \eg, \cite{Jiang:2007ge}), and for which a formalism has been developed 
\cite{Dmitrasinovic:1989bf}.

\begin{figure}
\centerline{\includegraphics[width=.7\textwidth]{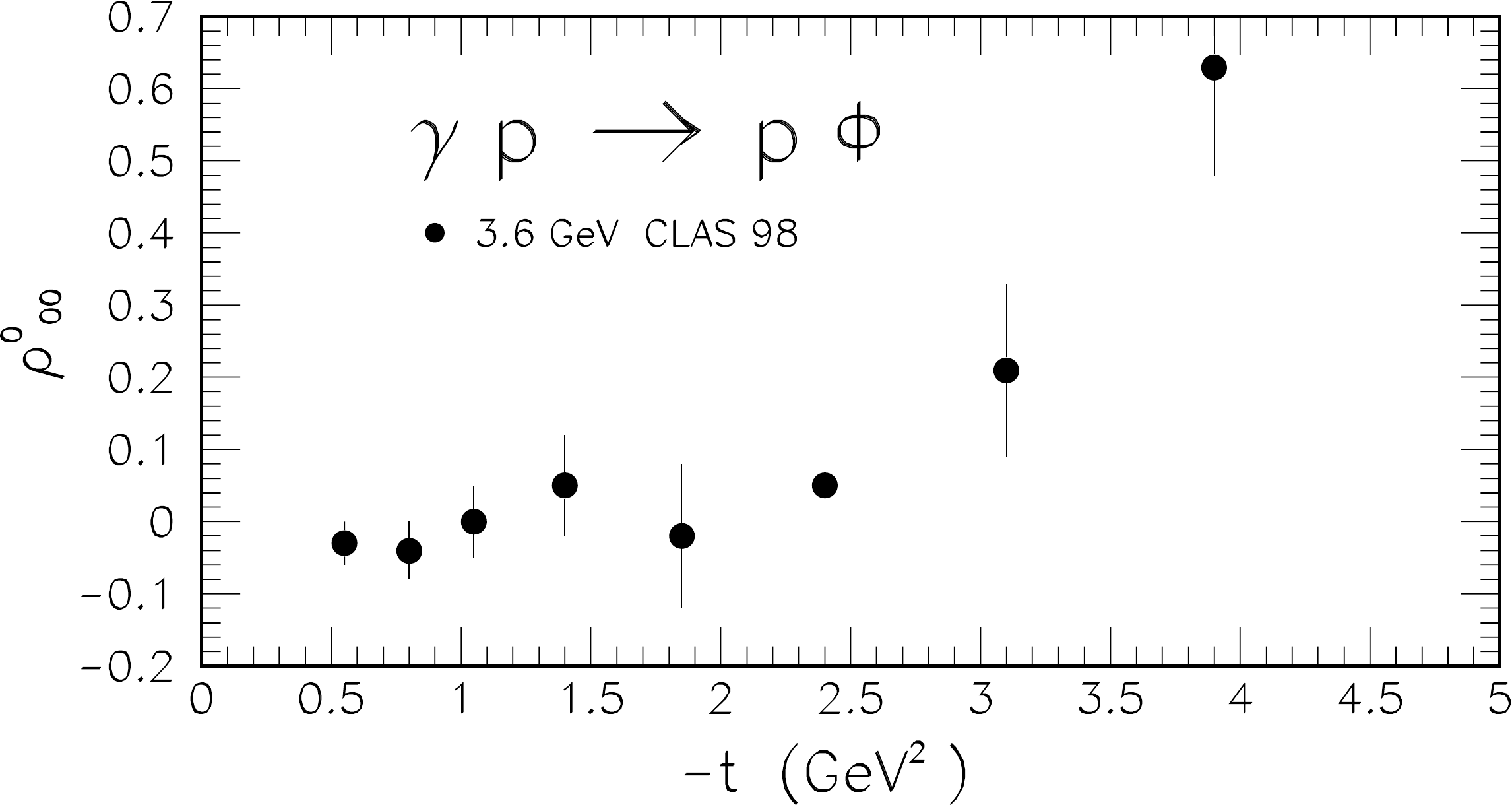}}
\vskip -.2cm
\caption{\label{excl:fig:phiprod} Tensor polarisation coefficient $\rho_{00}^0$  in $\gamma+\p\to \phi+\p$, as measured at CLAS \cite{McCormick:2003bb}.}
\end{figure}

A summary of early experiments can be found in \cite{Gilman:1972yb}.
The photoproduction of $\rho$, $\omega$ and $\Phi$ mesons with a linearly-polarised photon beam has been studied by Ballam \etal\ \cite{Ballam:1972eq} at 2.8, 4.7 and  9.3~GeV beam energy.
The reaction $\gamma+\N\to \K^*+\Y$ is studied at CLAS \cite{Hleiqawi:2007ad} for unpolarised photons.
This collaboration also studied $\phi$ production \cite{McCormick:2003bb}. Their results for the tensor-polarisation coefficient $\rho_{00}^0$ is shown in Fig.~\ref{excl:fig:phiprod}, with a large value at the highest transfer.  The smallness of  $\rho_{00}^0$ at low momentum transfer is also remarkable, and probably due to $s$-channel helicity conservation (an unpolarised vector meson  has $\rho_{00}^0=1/3$).
The density matrix $\rho_{\lambda\lambda'}$ of the vector meson is expanded as
a function of the possible photon polarisation $P_k$
\be
\rho_{\lambda\lambda'}\propto\sum_{k=0}^3 \rho_{\lambda\lambda'}^k P_k~,
\ee
where $k$ is the Stockes index, and $P_0=1$. The normalisation is specified later, see Eq.~(\ref{excl:eq:rho-norm}).
For this reaction $\gamma+\p\to\phi+\p$, Halpern \etal\ \cite{Halpern:1972ab} already noticed that the beam asymmetry  is sometimes close to saturation, with a value $\Sigma=0.985\pm0.12$ for a photon energy $8.14\;$GeV and a square transfer $t=-0.2\;(\mathrm{GeV}/c)^2$.  The photoproduction of $\omega$ has been measured, \eg, by  Ajaka \etal\ \cite{Ajaka:2006bn}. 

\subsubsection{Mechanisms}
In the review article \cite{Gilman:1972yb}, the data on $\rho$ photoproduction are interpreted in terms of Pomeron exchange, natural \vs\ unnatural-parity exchanges and $s$-channel helicity conservation. A comparison was also attempted between the photoproduction reaction (\ref{excl:eq:phvec}) and  the
$\rho+\mathrm{N}\to V+ B$ or $\omega+\mathrm{N}\to V+ B$ reactions in the spirit of the vector-dominance model.

The language has of course evolved and data are now discussed in terms of quark dynamics. In particular, a comparison of spin effects between  $\gamma+\p\to\phi+\p$ and  $\gamma+\p\to\omega +\p$ is interesting to understand the mechanisms of violation of the OZI rule.  It  for instance could reveal a signature of a possible strangeness content of the nucleon \cite{Titov:1998bw}.

\subsubsection{Formalism}
\begin{figure}[!hbtp]
\begin{minipage}{.55\textwidth}
\begin{flushleft}
\includegraphics[width=.99\textwidth]{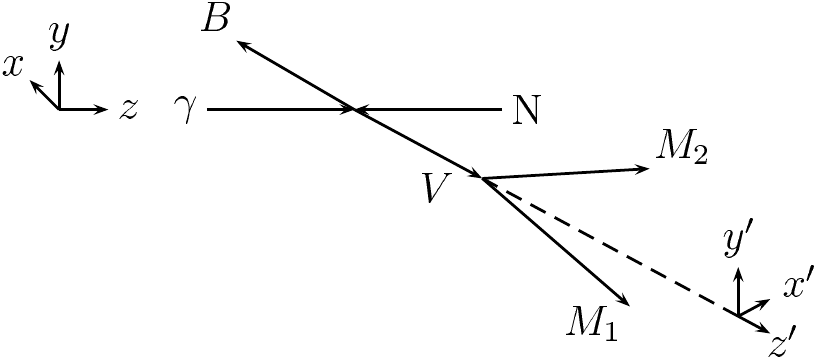}
\end{flushleft}
\end{minipage}
\ \begin{minipage}{.43\textwidth}
\caption{Kinematics for the $\gamma + \mathrm{N}\to V + B'$ and
$V\rightarrow M_1 M_2$ reactions.}\label{excl:fig:VecPhoto1}
\end{minipage}
\end{figure}

\paragraph{Kinematics}
The kinematics is shown in Fig.~\ref{excl:fig:VecPhoto1}, with an explicit two-body decay mode $V\to M_1+M_2$, which is the most current way of detecting a vector meson.
\paragraph{Helicity basis}
There are 24 helicity amplitudes, but this number reduces to 12 if parity is conserved
 \be \label{The helicity:parity} 
 \langle\lambda,\lambda_{B}|{\cal M}|\lambda_{\gamma}, \lambda_{\rm N}\rangle
 =(-1)^{\lambda-\lambda_B-\lambda_\gamma +\lambda_N}
 \langle-\lambda_{V} -\lambda_{B}|{\cal M}|-\lambda_{\gamma},-\lambda_{\rm N}\rangle~.
 \ee
It is possible to chose $\lambda_\gamma=+1$ and use
\be\label{The helicity:Ha} 
%H_{a, \lambda}\equiv\langle\lambda, \lambda_{B}|{\cal M}|1,\lambda_{N}\rangle~, 
a_i=\langle1, \lambda_{B}|{\cal M}|1,\lambda_{N}\rangle~, \quad
b_i=\langle0, \lambda_{B}|{\cal M}|1,\lambda_{N}\rangle~,\quad
c_i=\langle-1, \lambda_{B}|{\cal M}|1,\lambda_{N}\rangle~, 
 \ee 
for the 
$\lambda=\pm 1,0$ helicities the vector meson, and for the baryons the index $i=1,...4$ defined as follows:
\be
 \begin{tabular}{c|cccc}
  $i$ & 1 & 2 & 3 & 4 \\
  \hline
   $\lambda_{B}$ & $+1/2$ & $+1/2$ & $-1/2$ & $-1/2$ \\
   $\lambda_{\rm N}$ & $-1/2$ & $+1/2$ & $-1/2$ & $+1/2$\\
\end{tabular}
\ee
The various spin observables can be expressed in terms of these amplitudes $H_{a\lambda}$. 
For instance,
\be\label{excl:eq:obs-vec}\begin{aligned}
I_0&= |a_1|^2+|b_1|^2+ \cdots + |b_4|^2+|c_4|^2~,\\
I_0 \Sigma&=-2\RE\left(c_1a_4^*-b_1b_4^*+a_1c_4^*-c_2a_3^*+b_2b_3^*-a_2c_3^*\right)~,\\
I_0 T&=-2\IM\left(c_3c_4^*+b_3 b_4^*+a_3a_4^*+c_1 c_2^*+b_1 b_2^*+a_1 a_2^*\right)~,
\end{aligned}\ee
for the angular distribution, beam asymmetry and target analysing power. See, \eg, \cite{Pichowsky:1994gh,Titov:1998bw}.  In \cite{Titov:1998bw}, all single and double polarisation observables are listed, and expressed in terms of the above helicity amplitudes.

\paragraph{Transversity and linear polarisation bases}
For the initial and final baryons, and for the photon, we use the same notation $\pm$ and $\{\boldsymbol{\pi}, \, \boldsymbol{n}\}$ for the transversity states as in Sec.~\ref{excl:sub:photpsamp}.
For the vector meson, we will take the linearly polarised basis $\{|x\rangle,|y\rangle,|z\rangle\}$.
Assuming that ${\rm V}$ has the natural parity $P=-1$, the states $|x\rangle$ and $|z\rangle$ polarised in the scattering plane have $\Pi_{\rm V}=+1$ whereas $|z\rangle$ has $\Pi_{\rm V}=-1$. 
%We also choose the operators $\Sigma_1,\cdots\Sigma_9$ defined in (2.28!!!).
The conservation of $\Pi$, which reads 
\be
\Pi_\gamma \ \Pi_{\rm N} = \Pi_{\rm B} \ \Pi_{\rm V} 
\ee
selects 12 non-vanishing transversity amplitudes (out of 24), which can be chosen as:
\be \label{a1-12}\begin{aligned}
\langle x ,+ |\Mcal|\boldsymbol{\pi}, + \rangle &= f_1 ~,\quad
&\langle x ,- |\Mcal|\boldsymbol{n}, + \rangle &= g_1 ~,
\\
\langle x ,- |\Mcal| \boldsymbol{\pi},- \rangle &= f_2~,\quad
&\langle x ,+ |\Mcal|\boldsymbol{n}, - \rangle &= g_2 ~,
\\
\langle z ,+ |\Mcal|\boldsymbol{\pi}, + \rangle &= f_3 ~,\quad
&\langle z ,- |\Mcal|\boldsymbol{n},+ \rangle &= g_3 ~,
\\
\langle z ,- |\Mcal|\boldsymbol{\pi}, - \rangle &= f_4 ~,\quad
&\langle z ,+ |\Mcal|\boldsymbol{n},- \rangle &= g_4~,
\\
\langle y ,+ |\Mcal|\boldsymbol{n},+ \rangle &= f_5 ~,\quad
&\langle y ,- |\Mcal|\boldsymbol{\pi},+ \rangle &= g_5 ~,
\\
\langle y ,- |\Mcal|\boldsymbol{n}, - \rangle &= f_6 ~,\quad
&\langle y ,+ |\Mcal|\boldsymbol{\pi},- \rangle &= g_6 ~.
\end{aligned}\ee
The $f_i$ (resp.\ $g_i$) amplitudes correspond to $\Pi_{\rm V}=\Pi_\gamma$, 
$\ \Pi_{\rm B} = \Pi_{\rm N}$ (resp. $\Pi_{\rm V}=-\Pi_\gamma$, $\ \Pi_{\rm B} = -\Pi_{\rm N}$), that is to say, natural (resp.\ unnatural) parity exchange in the $t$-channel. 
\paragraph{The cross section matrix.}
The full spin dependence of the reaction is encoded in the cross section matrix $\R$ or its partial transpose $\Rt$ defined by (\ref{basic:eq:CSdM}). Equations (\ref{basic:eq:dpcsR}) and (\ref{basic:eq:csdm}) are generalised as
\ba\label{excl:eq:recompose-Rv}
(\lambda\mu|\alpha\nu)&=&\Tr\left\{
\Rt\,\left(
\sigma_\lambda^{\gamma}\otimes\sigma_\mu^{\N}\otimes
\Sigma_\alpha %(|i'\rangle\langle i|)
\otimes\sigma_\nu^{B}\right)\right\}~,\\
\label{excl:eq:decompose-Rv}
\Rt/(\Tr\Rt)&=& (\lambda\mu|\alpha\nu) \ 
2^{-3}\,g_\alpha^{-1}\ 
\sigma_\lambda^{\gamma}\otimes\sigma_\mu^{\N}\otimes
\Sigma_\alpha^\dagger %(|i\rangle\langle i'|)
\otimes\sigma_\nu^{B}~,
\ea
where $\lambda,\mu,\nu\in\{0,1,2,3\}$, $\ \Sigma_\alpha$ are nine basic spin-1 operators\,\footnote{%
The appearance of $\Sigma^\dagger$ in (\ref{excl:eq:decompose-Rv}) is analogous to what happens between (\ref{basic:eq:compact}) and (\ref{basic:eq:expectation}) using the orthogonality condition ( \ref{basic:eq:orth-spin1})}
% (see Sec.~\ref{basic:sub:massvec})
%$\ \alpha\in\{1,\cdots9\}$ 
and $g_\alpha=\Tr(\Sigma_\alpha\Sigma_\alpha^\dagger)$. 
Here we will simply take  the basis described by (\ref{basic:eq:SimpleBasis}).
%
%$\Sigma_\alpha=\Sigma_{\{ii'\}}=|i'\rangle\langle i|$, where $\alpha=\{ii'\}$ is a double index and $i,\,i'\in\{x,y,z\}$. For instance, $\Sigma_{\{xy\}}=|y\rangle\langle x|$,  $\ \left\langle\Sigma_{\{xy\}}\right\rangle =\rho_{xy}$.

$\R$ has rank 1 and effective dimension $12\!\times\!12$. Hence the spin observables are expressed in terms of 144 linearly independent (but non-linearly dependent) parameters. Only $2\!\times\!12-1=23$ parameters are  linearly and quadratically independent.

Among the $4\!\times\!4\!\times\!4\!\times\!9=576$ observables, half of them vanish due to the \emph{classical}  constraints of parity (\ref{basic:eq:P-class-R}-\ref{basic:eq:P-class-O}). They involve an odd number of $\Pi$-odd single particle operators  in (\ref{excl:eq:recompose-Rv}). The $\Pi$-odd operators are $\sigma_1$ and $\sigma_2$ for the photon and the baryons, $|x \rangle\langle y|$, $|y \rangle\langle x|$, $|z \rangle\langle y|$ and $|y \rangle\langle z|$ for the vector meson.  
Each vector-meson operator $\Sigma_{\alpha}$ is present in $4\!\times\!4\!\times\!2=32$ parity-allowed observables. 

The \emph{non-classical} constraint  (\ref{basic:eq:P-quant-O}) relates the nonvanishing observables two-by-two (in our basis of observables), up to a sign. For instance $(1,0|yy,1)=-(2,3|yy,2)$, $\ (3,1|yz,3)=+(0,2|yz,0)$, $\ (3,1|zy,3)=-(0,2|zy,0)$. This divides by 2 the number of linearly independent observables, which are therefore 144, as already predicted. 
For the $\rho$, $\omega$, $\Phi$ and $\mathrm{K}^*$ mesons, however, only the tensor polarisation can be measured through the angular distribution of the decay products and the axial polarisation $\IM\rho_{ii'}$ is hidden. This reduces to 128 the number of linearly independent accessible observables.
In what follows we consider successively the cases where one, two and the three spin-doublet particles (\ie, the photon, the nucleon and the recoil baryon) are unpolarised or non-analysed. 

\paragraph{One unpolarised or non-analysed spin-doublet particle.}
Let us assume for instance that the final baryon is not analysed: only observables involving $\sigma_0^{\rm (B)}$ are measured. They are all linearly independent since the identity (\ref{basic:eq:P-quant-O}) relates them to non-measured observables involving $\sigma_3^{\rm (B)}$. For each vector-meson operator $\Sigma_\alpha$ there is $4\times2=8$ measured observables. This gives $6\times8=48$ linearly independent accessible observables ($9\!\times\!8=72$ if the axial polarisation of $V$ was not hidden). The reduced cross section matrix for this experiment is 
% density matrix of the subsystem $\{N,\bar N',\bar V\}$,
%
%
\be\label{excl:eq:decompose-R3}
\R_{\gamma{\rm N,V}}\propto(\lambda\mu|\alpha0)\ g_\alpha^{-1}\ 
\,\sigma_\lambda^{\gamma} \otimes
\sigma_\mu^{\N} 
%\sigma_\nu^{t{\rm (N)}}  \left[\sigma^\nu_{N'}\right]^t
\otimes\left(\Sigma_\alpha^\dagger\right)^t~.
\ee
It is block-diagonal in two $6\!\times\!6$ matrices. The blocks $\R_{\gamma{\rm N,V}}^+$ and $\R_{\gamma{\rm N,V}}^-$ correspond to the final baryon having transversity $+1/2$ and $-{1/2}$ respectively. With the convention $\eta_{\rm B}=i$ we have 
\be
\Pi\,\R_{\gamma{\rm N,V}}^\pm = \R_{\gamma{\rm N,V}}^\pm \,\Pi = \pm \R_{\gamma{\rm N,V}}^\pm ~.
\label{excl:eq:P-quantique3}
\ee
$\R_{\gamma{\rm N,V}}^\pm$ has 36 linearly independent parameters. It is of rank one, since it corresponds to one photon state. So, in a non-linear way, it depends only on $2\!\times\!6-1=11$ parameters.   
\paragraph{Two unpolarised or non-analysed spin-doublet particles}
Let us now study the case where $\N$ is unpolarised, $B$ not analysed. 
For each vector-meson operator $\Sigma_\alpha$ there is 2 measured observables. This gives $6\!\times\!2=12$ linearly independent accessible observables ($9\!\times\!2=18$ if the axial polarisation of $V$ were not hidden). The reduced density matrix  is
\be
\R_{\gamma,V} \propto (\lambda0|\alpha0)\,g_\alpha^{-1}\,\sigma_\lambda^{\gamma}\otimes
\left(\Sigma_\alpha^\dagger\right)^t~.
\label{excl:eq:decompose-R2}
\ee
It is block-diagonal in two $3\!\times\!3$ matrices $\R_{\gamma,V}^+$ and $\R_{\gamma,V}^-$ obeying equations analogous to (\ref{excl:eq:P-quantique3}) and corresponding respectively to $\N$ and $B$ having equal or opposite transversities. In each case, the $(\N\overline{B})$ system in the $t$-channel is two-dimensional, so that $\R_{\gamma,V}^+$ and $\R_{\gamma,V}^-$ have rank 2. Each of them has 9 linearly independent parameters, but related by $\det(\R_{\gamma,V}^\pm)=0$, therefore 8 really independent parameters. We will come back to this case for more details in Sec.~\ref{excl:sub:ineq-ub}. 
\paragraph{No polarised or analysed spin-doublet particle.}
Repeating the reduction once again we arrive at the $3\!\times\!3$ density matrix of the vector meson, when all other spins are summed over:
\be
\Rt_{V} = \R^t_{V} \propto \rho = (00|\alpha0)\ g_\alpha^{-1}\ \Sigma_\alpha~.
\label{excl:eq:decompose-R1}
\ee
$\rho$ is the direct sum of a $2\!\times\!2$ matrix $\rho^+$ acting on the $\Pi_{\rm V}=+1$ subspace $\{a|x\rangle+b|z\rangle\}$ (linear polarisation in the scattering plane) 
and the simple number $\rho_{yy}$ acting on the $\Pi_{\rm V}=-1$ subspace colinear to $|y\rangle$: 
\be
\left(\rho_{ij}\right)=
\begin{pmatrix}\rho_{xx}&0&\rho_{xz}\\ 0&\rho_{yy}&0\\ \rho_{zx}&0&\rho_{zz}\end{pmatrix}.
\label{excl:eq:rhoSpin1}
\ee
The dimension 4 of the $(\gamma \N\overline{B})$ channel for $\Pi=+1$ does not
restricts the rank of $\rho^+$, so that $\rho_{xx}$, $\rho_{zz}$,  $\RE\rho_{xz}$ and $\IM\rho_{xz}$ are fully independent parameters. Note that the latter is generally not accessible. The positivity conditions are
 \be
  \rho_{xx} \ge 0, \quad \rho_{yy} \ge 0, \quad \rho_{zz} \ge 0, \quad (\RE\rho_{xz})^2 + (\IM\rho_{xz}))^2 \le \rho_{xx} \, \rho_{zz}~. 
\label{excl:eq:positiV}
\ee
\subsubsection{Inequalities for the case of unpolarised baryons}
\label{excl:sub:ineq-ub}
We return to the case $\N$ unpolarised, $B$ not analysed, where experimental data exist \cite{Ballam:1972eq}. Writing (\ref{excl:eq:decompose-R2}) 
%with explicit normalisation 
with the $\Sigma_{\{ii'\}}$ operator basis, we have
\be
\Rh\equiv 2\,\R_{\gamma,V} /\left(\Tr \R_{\gamma,V}\right) = (\lambda|{ii'})\ \sigma_\lambda^{\gamma} \otimes (|i'\rangle\langle i|)
~,
%\label{excl:eq:decompose-R2}
\ee
where $(\lambda,0|{\alpha},0)=(\lambda,0|{\{ii'\}},0)$ has been abbreviated as $(\lambda|{ii'})$. The matrix $\Rh$ is normalised such that $(0|{ii'})$ is equal to the density matrix $\rho^0_{ii'}$ of the vector meson when the photon is unpolarised.  Its explicit expression is
\be\label{excl:eq:Rgamma-V}
\begin{pmatrix}
 (0+\ii3|xx) & 0 & (0+\ii3|zx) & 0 & (1-i2|yx) & 0 \\
 0 & (0+\ii3|yy) & 0 & (1-i2|xy) & 0 & (1-i2|zy) \\
 (0+\ii3|xz) & 0 & (0+\ii3|zz) & 0 & (1-i2|yz) & 0 \\
 0 & (1+i2|yx) & 0 & (0-\ii3|xx) & 0 & (0-\ii3|zx) \\
 (1+i2|xy) & 0 & (1+i2|zy) & 0 & (0-\ii3|yy) & 0 \\
 0 & (1+i2|yz) & 0 & (0-\ii3|xz) & 0 & (0-\ii3|zz) 
\end{pmatrix} % \,.
\ee
The  linear polarisation basis for the photon + vector meson states has been ordered as 
$|x\rangle\otimes|x\rangle$, $|x\rangle \otimes|y\rangle$, $|x\rangle \otimes|z\rangle$, $|y\rangle \otimes|x\rangle$, $|y\rangle \otimes|y\rangle$, $|y\rangle \otimes|z\rangle$.
% \otimes $(x,z)$, $(y,x)$, $(y,y)$, $(y,z)$. 
The notation $(1-i2|xy)$, for instance, stands for $(1|xy)-i(2|xy)$.  Adopting a notation similar to that of Schilling \etal\ \cite{Schilling:1969um} we rewrite $(0|ij)$, $(1|ij)$, $(2|ij)$ and $(3|ij)$ as $\rho^0_{ij}$, $\rho^\oslash_{ij}$, $\rho^\odot_{ij}$ and $\rho^\ominus_{ij}$. The beam asymmetry is 
 \be
 \Sigma={\Tr\left\{\R \left(\sigma_\ominus^{\gamma}\otimes\mathbbm{1}^{V}\right)\right\}\over \Tr\R} = \rho^\ominus_{xx}+\rho^\ominus_{yy}+\rho^\ominus_{zz}~.
 \ee 
 The meson density matrix takes the form
 \be\label{excl:eq:rho-norm}
 \rho={1\over1+S_\ominus\Sigma}\sum_{\lambda=0,\oslash,\odot,\ominus} \rho^\lambda\,
 S_\lambda~.
 \ee
 Note that $\rho$ depends nonlinearly on the photon Stokes parameter $S_\ominus$, due to the denominator which is inversely proportional to the cross section with polarised photons. Schilling \etal~\cite{Schilling:1969um} use the helicity basis and their Stokes parameters,  $\{P_1,P_2,P_3\}=\{-S_\ominus,-S_\oslash,S_\odot\}$ are different from ours,$\{S_1,S_2,S_3\}=\{S_\oslash,S_\odot,S_\ominus\}$. We give here the correspondence between their notations (left-hand side) and ours (right-hand side) for the accessible observables:
 \goodbreak\goodbreak\goodbreak
 %%%
%
\be\begin{aligned}
(1-\rho^0_{00})/2=\rho^0_{11}=\rho^0_{-1-1}
&\leftrightarrow
\left(\rho^0_{xx}+\rho^0_{yy}\right)/2=(1-\rho^0_{zz})/2\\
\rho^0_{1,-1}=\rho^0_{-1,1}&\leftrightarrow\left(-\rho^0_{xx}+\rho^0_{yy}\right)/2\\
\RE\rho^0_{10}=-\RE\rho^0_{0,-1}&\leftrightarrow - \RE\rho^0_{zx} /\sqrt{2}\\
\rho^1_{00} &\leftrightarrow - \rho^\ominus_{zz}\\
\rho^1_{11}=\rho^1_{-1-1}&\leftrightarrow - \left(\rho^\ominus_{xx}+\rho^\ominus_{yy}\right)/2\\
\rho^1_{1,-1}=\rho^1_{-1,1}&\leftrightarrow \left(\rho^\ominus_{xx}-\rho^\ominus_{yy}\right)/2\\
\RE\rho^1_{1,0}= - \RE\rho^1_{-1,0}&\leftrightarrow \RE\rho^\ominus_{zx} /\sqrt{2}\\
\rho^2_{1,-1} &\leftrightarrow -i\, \RE\rho^\oslash_{xy}\\
\IM\rho^2_{10} = \IM\rho^2_{-1,0} &\leftrightarrow  - \RE\rho^\oslash_{yz} /\sqrt{2}\\
\rho^3_{1,-1} &\leftrightarrow i\, \RE\rho^\odot_{xy}\\
\IM\rho^3_{10} = \IM\rho^3_{-1,0}  &\leftrightarrow  \RE\rho^\odot_{yz} /\sqrt{2}~.
\end{aligned}\ee
As predicted above, the matrix $\Rh_{\gamma,V}$ given by Eq.~(\ref{excl:eq:Rgamma-V}) is block-diagonal in two semi-positive $3\!\times\!3$ matrices of rank 2: 
$\Rh_{\gamma,V}^+$, made of the odd lines and columns, is associated to equal transversities of the initial and final baryons and to natural parity exchange in the $t$-channel. $\Rh_{\gamma,V}^-$, made of the even lines and columns, is associated to a baryon transversity flip and unnatural parity exchange.  For instance
\be\label{excl:eq:R+gamma-V}
R_{\gamma,V}^+ = \begin{pmatrix}
 (0|xx) + \ii(3|xx)  & (0|zx) + \ii(3|zx) & (1|yx) -i(2|yx)  \\
 (0|xz) + \ii(3|xz) & (0|zz) + \ii(3|zz) & (1|yz) -i(2|yz)  \\
  (1|xy) +i (2|xy) & (1|zy) +i (2|zy) & (0|yy) - \ii(3|yy)  \\
\end{pmatrix}\,.
\ee
The vanishing of the determinants of $\R_{\gamma,V}^\pm$  yield two cubic relations between the observables. We do not write them here, since they involve the axial polarisation of the vector meson, which is not usually accessible. The positivity of the upper-left $2\!\times\!2$ minor of (\ref{excl:eq:R+gamma-V}), and a similar condition for $\Rh_{\gamma,V}^-$, read
\be\label{excl:eq:upper-left} \left(\RE [ \rho^0_{zx}  \pm \rho^\ominus_{zx} ]\right)^2 +  \left(\IM [ \rho^0_{zx}  \pm \rho^\ominus_{zx} ]\right)^2  \le  \left( \rho^0_{xx}  \pm  \rho^\ominus_{xx}  \right) \,    \left(  \rho^0_{zz}   \pm \rho^\ominus_{zz} \right) ~.\ee
These two  inequalities are necessary conditions for the positivity of the meson density matrix when the photon is polarised in the scattering plane. They are therefore \emph{classical} positivity conditions. The positivity of the other  $2\!\times\!2$ principal minors of $\R_{\gamma,V}^\pm$ yield the \emph{non-classical} conditions
\ba\label{excl:eq:big}
[ \RE \rho^\oslash_{yx} +\IM  \rho^\odot_{yx} ]^2 + [ \IM  \rho^\oslash_{yx} - \RE  \rho^\odot_{yx} ]^2 &\le&
  \left( \rho^0_{xx} +  \rho^\ominus_{xx}  \right) \left(  \rho^0_{yy}  -  \rho^\ominus_{yy} \right)~,\\
\label{excl:eq:lower-right}[ \RE \rho^\oslash_{yz} +\IM  \rho^\odot_{yz} ]^2 + [ \IM  \rho^\oslash_{yz} - \RE  \rho^\odot_{yz} ]^2 &\le&  \left( \rho^0_{zz} +  \rho^\ominus_{zz}  \right) \left(  \rho^0_{yy}  -  \rho^\ominus_{yy} \right)~,\\
\label{excl:eq:big'}[ \RE \rho^\oslash_{xy} +\IM  \rho^\odot_{xy} ]^2 + [ \IM  \rho^\oslash_{xy} - \RE  \rho^\odot_{xy} ]^2 &\le&   \left(  \rho^0_{xx}  -  \rho^\ominus_{xx} \right) \left( \rho^0_{yy} +  \rho^\ominus_{yy}  \right)~,\\
\label{excl:eq:lower-right'}[ \RE \rho^\oslash_{zy} +\IM  \rho^\odot_{zy} ]^2 + [ \IM  \rho^\oslash_{zy} - \RE  \rho^\odot_{zy} ]^2 &\le&  \left(  \rho^0_{zz}  -  \rho^\ominus_{zz} \right) \left( \rho^0_{yy} +  \rho^\ominus_{yy}  \right)~.\ea
Combining (\ref{excl:eq:big}) and (\ref{excl:eq:big'}),  (\ref{excl:eq:lower-right}) and  (\ref{excl:eq:lower-right'}) one gets an inequality without the inaccessible quantities $\IM\rho_{ii'}$:
\ba\label{excl:eq:big''}
[ \RE \rho^\oslash_{xy} ]^2 + [ \RE  \rho^\odot_{xy} ]^2 &\le& \rho^0_{yy}  \,   \rho^0_{xx} ~, \\
\label{excl:eq:lower-right''}
[ \RE \rho^\oslash_{yz} ]^2 + [ \RE  \rho^\odot_{yz} ]^2 &\le& \rho^0_{yy}  \,   \rho^0_{zz}  ~.
\ea
In the notations of \cite{Schilling:1969um}  (\ref{excl:eq:big''}) and (\ref{excl:eq:lower-right''}) read
\ba\label{excl:eq:big'''}
[ \rho^0_{1,-1} ]^2 + [ \IM \rho^2_{1,-1} ]^2 + [ \IM \rho^3_{1,-1} ]^2  &\le&[ \rho^0_{11} ]^2 \,,
\\
\label{excl:eq:lower-right'''}
2\, [ \IM \rho^2_{10} ]^2 + 2\, [ \IM \rho^3_{10} ]^2 
&\le&
\rho^0_{00} \, [\rho^0_{11} + \rho^0_{1,-1}] \,.
\ea
The inequality (\ref{excl:eq:big'''}) is stronger than (4) and (10) of Table 2 in \cite{Schilling:1969um}.
The inequality (\ref{excl:eq:lower-right'''}) is stronger than (5) and (11) of the same table.%
\,\footnote{The $12^{\rm th}$ inequality of this table of Ref.~\cite{Schilling:1969um} seems erroneous}

The inequalities (\ref{excl:eq:big''}) and (\ref{excl:eq:lower-right''}) are \emph{classical}. For instance(\ref{excl:eq:big''}) can also be derived from the positivity of the meson density matrix when the photon has the polarisation $(S_{\oslash},S_{\odot},S_{\ominus})=(\cos\psi,\sin\psi,0)$ with $\psi=\arg(\RE\rho^\oslash_{xy}+i\RE\rho^\odot_{xy})$. The non-classical character of (\ref{excl:eq:big}-\ref{excl:eq:lower-right'}) has been lost when throwing away the information of $\IM\rho$. 
%%%%%%%%%%%%%
\subsubsection{Inequalities for polarised baryons}
The generalisation to the case of target polarisation, or the case where the spin of the recoiling baryon is analysed, is tedious but straightforward.
Kloet \etal \cite{Kloet:1998js,Kloet:1999wc} have listed the observables that are accessible when the vector meson is identified through its strong-decay products.

From the expression given, \eg, by Titov \etal\ \cite{Titov:1998bw}, it is possible to compute fictitious observables from randomly-generated amplitudes, and to get a first view at the allowed domain, before studying the constraints rigorously, as done for $\ppLL$. However, there are twelve amplitudes, and the method becomes more time-consuming, and more delicate, as the points populate preferentially the centre of each domain, and the contour is more elusive. Some refinements are needed.

\subsubsection{Production of two pions}
A natural extension of $\pi+\N\to\pi+ \N$ is the two-pion production $\pi+\N\to \pi+\pi+\N$ and,
similarly, the single-pion production $\gamma+\N\to\pi+ \N$
can be studied in parallel to $\gamma+\N\to\pi+ \pi+\N$. 
This is the point of view of Ref.~\cite{Roberts:2004mn}, where a rather comprehensive formalism is presented, and identities and inequalities among observables are listed.

Note that with three or more particles in the final states, the reaction is not planar any more.
In particular, the constraints induced by parity conservation cannot be formulated as mirror
symmetry with respect to the scattering plane, as discussed in Sec.~\ref{basic:sub:parity}. An
exhaustive analysis of such symmetry constraints is carried out in
\cite{Roberts:2004mn}. Note, however, that all the parity constraints are classical, \ie, can be
implemented directly at the level of the observables without writing down explicitly the amplitudes or the cross section matrix.

The two-pion production can also be thought as a generalisation of $\rho$ production followed by
$\rho\to \pi+\pi$, without the restriction of the $\pi\pi$ system to the mass and quantum numbers
of the vector meson resonance. This is the point of view adopted in most experimental analyses, with quantities measured on and off the $\rho$ peak of the $\pi\pi$ invariant mass. The aim is to
distinguish the mechanism of $\rho$ production from that responsible for background $\pi\pi$
production.
\clearpage
\newpage\setcounter{equation}{0}\section{Inclusive reactions}\label{se:incl}
\subsection{Spin observables in hadronic inclusive reactions}
The positivity constraints have been widely used in hadron physics to reduce the
allowed domain of spin observables and, as an illustration, we will now envisage
some specific examples in inclusive reactions.
\subsubsection{Total cross sections}
Let us consider the spin-dependent total cross sections for the scattering of
two spin-1/2 particles
\be \label{incl:eq:1}
a (\hbox{spin}\, 1/2) + b  (\hbox{spin}\, 1/2) \to \hbox{anything} ~,
\ee
which are, for example, $\p\p, \pbp, \p\Lambda, \p\mathrm{n}$, etc.
The reaction (\ref{incl:eq:1}) can be described in terms of three independent observables
which are the unpolarised cross section
\be \label{incl:eq:2}
\sigma_{\rm tot} =[\sigma_{\rm tot}(++)+\sigma_{\rm tot}(+-)]/ 2=[\sigma_{\rm tot}(\uparrow
\uparrow)+ \sigma_{\rm tot}(\uparrow\downarrow)]/2 ~,
\ee
and the two asymmetries
\be \label{incl:eq:3}
\Delta\sigma_L=\sigma_{\rm tot}(++)-\sigma_{\rm tot}(+-)~,\quad \mbox{and} \quad \Delta\sigma_T=
\sigma_{\rm tot}(\uparrow\downarrow)- \sigma_{\rm tot}(\uparrow \uparrow) ~.
\ee
Here $+(-)$ denotes the longitudinally polarised (or helicity) states of $a$ and $b$
and $\uparrow(\downarrow)$ their transversely polarised states.

\begin{figure}
\begin{minipage}{.6\textwidth}
\centerline{\includegraphics[width=.85\textwidth]{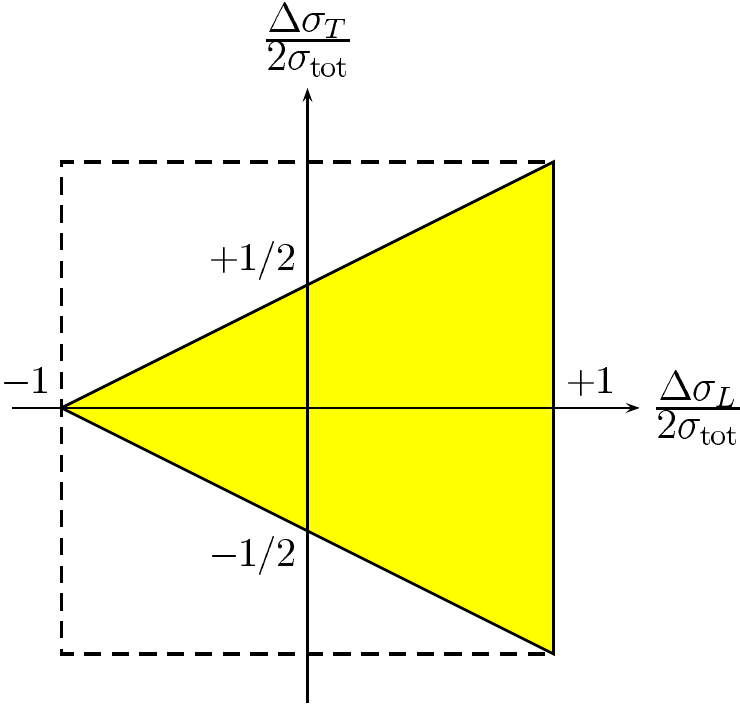}}
\end{minipage}\
\begin{minipage}{.39\textwidth}
\caption{\label{incl:fig:sectot} Domain allowed for $\Delta \sigma_L$ and $\Delta \sigma _T$}
\end{minipage}
\end{figure}

If $\vec{P}_a$ and $\vec{P}_b$ are the polarisation unit vectors of
$a$ and $b$, the polarised total cross sections corresponding to Eq.~(\ref{incl:eq:1}) are \cite{Bourrely:1980mr}
\be\label{incl:eq:4}
\sigma_{\rm tot}(\vec {P}_a,\vec {P}_b)= \Tr(\mathsf{M} \rho)~.
\ee
Here  $\rho$ is the $4\!\times\! 4$ density matrix $\rho = ( \mathbbm{1}_2+ \vec {P}_a .\vec{\sigma})
\otimes(\mathbbm{1}_2+ \vec{P}_b.\vec{\sigma})/4$, and 
$\mathsf{M}$ denotes the imaginary part of the forward scattering amplitude for the elastic reaction
\be\label{incl:eq:5}
a + b \to a + b  ~,
\ee
and  is the following $4\times4$ matrix\footnote{Using the same axes $x, y, z$ for $a$ and $b$} 
\be\label{incl:eq:tot}
\mathsf{M} =
2\sigma_{\rm tot} \mathbbm{1}_4 - \Delta \sigma_{L} \sigma_{az}\otimes\sigma_{bz} - \Delta\sigma_{T}(\sigma_{ax}\otimes
\sigma_{bx} + \sigma_{ay}\otimes\sigma_{by})~.
\ee
$\mathsf{M}$ must be Hermitian and \emph{positive}, which implies that all its principal minors, (\ie, subdeterminants of $
\mathsf{M}$ with diagonal elements) must be \emph{positive}, as reminded in Sec.~\ref{basic:sub:dens}, since $\mathsf{M}$ is  a particular case of CSM. In terms of
the three observables defined above, we get two trivial conditions, \ie,
$|\Delta \sigma_i | \leq 2 \sigma_{\rm tot},\  (i=L,T)$ and one non-trivial positivity bound \cite{Soffer:1973di},
namely
\be\label{incl:eq:6}
|\Delta \sigma_T| \leq \sigma_{\rm tot} + \Delta  \sigma_L/2 ~.
\ee
The corresponding domain is shown in Fig.~\ref{incl:fig:sectot}.

In the case of proton--proton scattering, this rigorous bound is easily fulfilled
for $p_{\rm lab}  \geq 1 \,\mbox{GeV/c}$, where $\Delta \sigma_L$ and $\Delta \sigma_T$ have been measured
\cite {Yokosawa:1980cx}, because  $\sigma_{\rm tot}$ is rather large. For $ p_{\rm lab} \leq 1\,\mbox{GeV/c}$, 
$\sigma_{\rm tot}$ can be as small as $25\,\mbox{mb}$ or so
and $\Delta \sigma_L$ can reach the value $-30\,\mbox{mb}$, so by using Eq.~(\ref{incl:eq:6})
one gets $|\Delta \sigma_T| \leq 10 \,\mbox{mb}$, with                                                                                                     
some errors, an interesting limit which is, indeed, satisfied by the data 
\cite{Stanley:1983sb,Auer:1984qh}.
\subsubsection{Spin transfer observables}\label{incl:sub:spintrans}
For a parity conserving inclusive reaction of the type, 
\be
\overrightarrow{a}(\mbox{spin 1/2}) + \overrightarrow{b}(\mbox{unpolarised}) \to c (\mbox{spin 1/2}) + X~,
\label{incl:eq:increa}
\ee
one can define \emph{eight} independent observables, which depend on three kinematic variables, $\sqrt{s}$ the centre-of-mass energy and $x_F$, $p_T$ the scaled longitudinal momentum and the transverse momentum of the final particle $c$.
In order to define these observables, we adopt here the standard notation used in Ref.~\cite{Bourrely:1980mr} $(ab|c X)$, where the spin direction is specified 
 in one of the three possible directions
$L,N,S$ of the Argonne convention: $\vec{L}_a=\vec{L}_b=\pv_a$, $\vec{L}_c=\pv_c$, $\vec{N}_a=\vec{N}_b=\vec{N}_c$ is along $\vec{p}_a\times \vec{p}_c$, and $\vec{S}_i=\vec{N}_i\times\vec{L}_i$. Note that $\{\vec{L},\,\vec{N},\,\vec{S}\}=\{\vec{l},\,\vec{n}\,\vec{m}\}$ in the notation of Eq.~(\ref{basic:eq:frame})
for particles $a$ and $c$ and $\{\vec{L},\,\vec{N},\,\vec{S}\}=\{-\vec{l},\,\vec{n}\,-\vec{m}\}$ for particle $b$.
%
%Since only one initial and one final spin are observed, we have in fact $(a 0|c 0)$. 
%$\vec{L},\vec{N},\vec{S}$ are unit vectors, in the centre-of-mass
%system, along the particle momentum ($\vec{p}_a$ or $\vec{p}_c$), along the normal to the scattering plane which contains $a,b$ and $c$, and along $\vec{N}\times
%\vec {L}$, respectively. 
In addition to the unpolarised cross section $\sigma_0$, there
are \emph{seven} spin dependent observables, \emph{two} single transverse spin asymmetries
\be
A_{aN} = (N0|00)~,\quad\mbox{and} \quad P_{cN}= (00|N0)~,
\label{incl:eq:stasy}
\ee
and \emph{five} depolarisation parameters
\be\label{incl:eq:dpol}\begin{aligned}
D_{LL}&=(L0|L0)~,\quad  &D_{SS} &= (S0|S0)~,\quad &D_{NN} &= (N0|N0)~,\\
D_{LS} &= (L0|S0)~, & &\text{and}   &D_{SL} &= (S0|L0)~.\end{aligned}
\ee
%
%\begin{eqnarray}
%D_{LL}&=&(L0|L0)~,~ D_{SS} = (S0|S0)~,~ D_{NN} = (N0|N0)~,
%\nonumber \\
%&& D_{LS} = (L0|S0)~~ \mbox{and} ~~ D_{SL} = (S0|L0)~.
%\label{dpol}
%\end{eqnarray}

Several years ago, Doncel and M\'endez \cite{Doncel:1972vk} (see also \cite{Artru:2004jx})  have derived very
general inequalities constraining  these parameters, which read
 \be
(1 \pm D_{NN})^2 \geq ( A_{aN} \pm P_{cN})^2 + ( D_{LL} \pm D_{SS})^2 + ( D_{LS} \mp D_{SL})^2~.
\label{incl:eq:DM}
\ee

\begin{figure}[!htb]
\begin{center}
\includegraphics[width=1.05\textwidth]{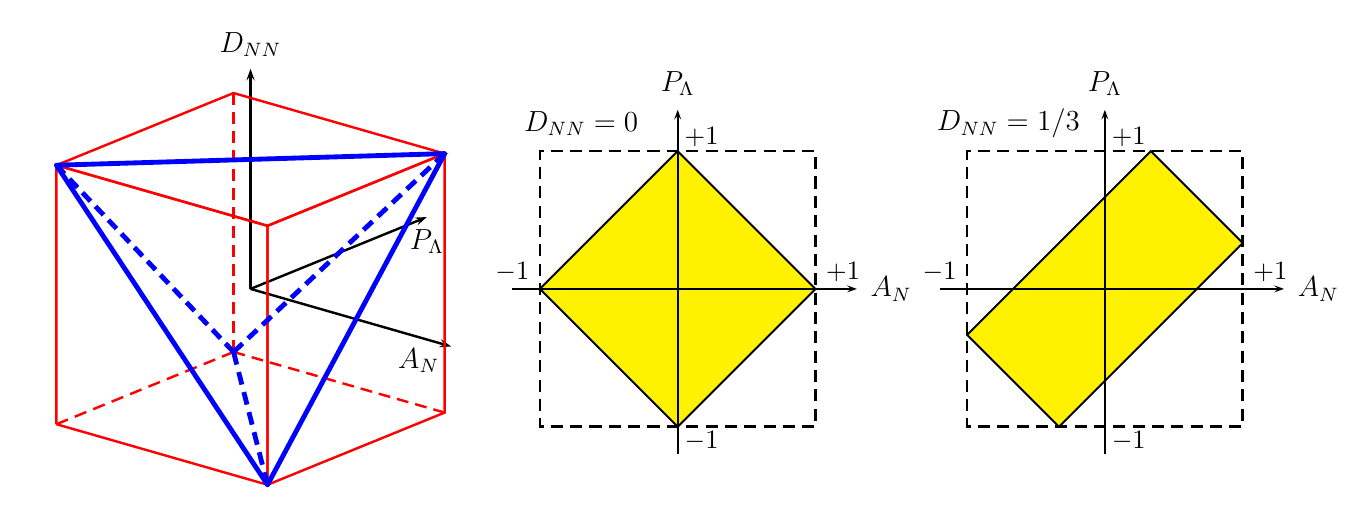}
\end{center}
\caption{The allowed domain corresponding to the constraints Eq.~(\ref{incl:eq:pos1}) (\emph{left}). 
The slice of the full domain for $D_{NN}=0$ (\emph{ middle}) and for $D_{NN}=1/3$ (\emph{right}).}
\label{incl:fig:posit2}
\end{figure}

 Let us now consider
the particular reaction $\p^{\uparrow}\p \to \Lambda^{\uparrow} X$, where the incoming
proton beam is polarised and the polarisation of the outgoing $\Lambda$ is
measured.
If we concentrate for the moment, on the
case where the particle spins are \emph{normal} to the scattering 
plane, the relevant observables are the $\Lambda$ polarisation $P_{\Lambda}$, the
analysing power $A_N$ (denoted before $P_{cN}$ and $A_{aN}$ respectively) and the spin transfer $D_{NN}$, between the 
proton and the $\Lambda$. As a special case of Eq.~(\ref{incl:eq:DM}), we obtain
\be
1 \pm D_{NN}  \geq  | P_{\Lambda} \pm A_N |~,  
\label{incl:eq:pos1}
\ee
two constraints which must be satisfied for any values of the kinematic variables
$\sqrt{s}, x_F, p_T$. These linear conditions are similar to those reported earlier
(see Sec.~\ref{excl:sub:photps}) in the case of exclusive photoproduction of pseudoscalar mesons. The allowed domain 
corresponds to the tetrahedron shown in Fig.~\ref{incl:fig:posit2}, which is identical to the tetrahedron encountered in \ppLL,  
see Fig.~\ref{2-3-5-4-9-10}, and in photoproduction, see Fig.~\ref{excl:fig:tetra}.

These inequalities constraints involve three spin parameters, once we fixe
the value of one parameter, the other two are restricted to lie in a certain
domain. For instance, Fig.~\ref{incl:fig:posit2} shows also that for $D_{NN} = 0$,
$P_{\Lambda}$ and $A_N$ are correlated within the shaded area of a square
with boundaries $(-1, +1)$. This domain is more restricted in the case $D_{NN} =1/3$. 
In the limit $D_{NN} =1$, we immediately deduce from
Eqs.~(\ref{incl:eq:pos1}) that $P_{\Lambda} = A_N $. In the case where
$D_{NN}$ is negative, one obtains the same regions, but $P_{\Lambda}~\mbox{and}
 ~A_N $ are interchanged with respect to their axis. In particular, for $D_{NN}=-1$
one should have $P_{\Lambda} = - A_N $.

Various applications can be envisaged with the above inequalities, one concerns
the analysis of experimental data and another one the validity of the
spin observables predicted by theoretical models \cite{Bourrely:2007gr}.
For example, at FNAL, the experiment E-704 \cite{Bravar:1995fw,Bravar:1997fb,Pondrom:1985aw} has performed the 
measurement of
$P_{\Lambda},~A_N~\mbox{and}~D_{NN}$ with transversely polarised proton beam
at $200 \,\mbox{GeV/c}$, in the kinematic range $0.2 \leq x_F \leq 1.0$ and
$0.1 \leq p_T \leq 1.5 \,\mbox{GeV/c}$. Their data indicate a negative $A_N$ and a value
of $D_{NN}$ up to 30\%. If we take for instance $p_T \sim 1 \,\mbox{GeV/c}$, 
$x_F \sim 0.8$, then $D_{NN} \sim 30\%$, for these kinematical values, they
have $A_N \sim -10\% $ and $P_{\Lambda} \sim -30\% $, the inequalities 
(\ref{incl:eq:pos1}) are well satisfied. These constraints can be very useful
to check the experimental data and errors, if they lie outside the allow
domain of positivity.

When the polarisation is not normal to the scattering plane, one has new depolarisation parameters.
In the case where both $p$ and $\Lambda$ are longitudinally polarised, one
can measure $D_{LL}$ and when each particle has a transverse polarisation 
in the scattering plane, one can measure $D_{SS}$. In this case, one gets, as a consequence
of Eq.~(\ref{incl:eq:DM}), the non-trivial, and in fact, \emph{non-classical} constraints
\be
 1 \pm D_{NN} \geq |D_{LL} \pm D_{SS}|~,
 \qquad  1 \pm D_{NN} \geq |D_{LS} \mp D_{SL}|~.
\label{incl:eq:pos3}
\ee
Note that the two depolarisations in the right
hand side Eq.~(\ref{incl:eq:pos3}) have not yet been measured. Thus from the positivity constraints and
the knowledge of some measured observables, 
 bounds on some unknown observables are deduced.

We would like to stress the importance of new measurements which are
needed to understand the spin mechanism in these reactions. The polarised $\p\p$ 
collider at BNL-RHIC is certainly the most appropriate machine for those 
experiments of single and double spin measurements \cite{Bunce:2000uv}
and positivity was used to determine the allowed domain for $D_{LL}$ \cite{deFlorian:1998ba}
and $D_{NN}$ \cite{deFlorian:1998am,Soffer:2002tf}.
\subsubsection{Initial spin observables}
Let us consider an inclusive reaction of the type
\be
a(\mbox {spin} 1/2)+ b(\mbox {spin} 1/2) \rightarrow c + X ~,
\label{incl:eq:reac}
\ee
where the spins of both initial spin 1/2 particles can be in any possible 
directions and no polarisation is observed in the final state. Clearly, this case is related to the previous one and here we will 
present more details in the derivation of the results. The observables of this
reaction, which are the spin-dependent differential cross sections with respect to the momentum of $c$, can  be expressed 
in terms of the discontinuities (with respect to the invariant mass
squared of $X$) of the amplitudes for the forward three-body scattering
\be
a + b + \bar {c} \rightarrow a + b + \bar {c}~,
\ee
as given by the generalised optical theorem. With parity conservation,  the 
knowledge of the dependence of this reaction upon the spins of $a$ and $b$ requires the determination of \emph{eight} real functions, \ie, 
the measurement of eight  independent spin observables \cite{Goldstein:1975ci}. 
%In order to define these observables, we use the standard notation  $(A_1 A_2|B X)$, by
%which the spin directions of $A_1,A_2,B$ and $X$ are specified in one of the three possible directions
%$L,N,S$, which were defined earlier. 
Since the final spins are not observed, we have in fact 
$(a b|0 0)$ and in addition to the unpolarised cross section $\sigma_0=(00|00)$, there
are \emph{seven} spin dependent observables, \emph{two} single transverse spin asymmetries
\be
A_{aN} = (N0|00)~,\quad \mbox{and}\quad A_{bN}= (0N|00)~,
\ee
and \emph{five} double-spin asymmetries
\be\begin{aligned}
A_{LL}&=(LL|00)~,\quad &A_{SS} &= (SS|00)~,\quad &A_{NN} &= (NN|00)~,\\
A_{LS} &= (LS|00)~, &&\text{and}   &A_{SL} &= (SL|00)~,
\end{aligned}\ee
which are in correspondence with Eqs.~(\ref{incl:eq:stasy}) and (\ref{incl:eq:dpol}).
The state of polarisation of the two spin 1/2 particles $a$ and $b$ is characterised by the $2 \times 2$  density matrices $\rho_{i}=(\mathbbm{1}_2  +\vec{P}_{i}. \vec{\sigma}_i)/2$, $i=a,b$. The state of  polarisation of the incoming system
in the reaction (\ref{incl:eq:reac}) is described by the $4 \times 4$ density matrix $\rho$, which is the direct product  $\rho =  \rho_a \otimes \rho_b$.
The spin-dependent cross section corresponding to this reaction is
\be\label{incl:eq:M-rho}
\sigma(\vec{P}_a,\vec{P}_b) =\Tr( \mathsf{M} \rho)~,
\ee
where $\mathcal{M}$ denotes the $4 \times 4$ cross section matrix
which we shall parametrise in the following way\,\footnote{A much simpler form was used in the case
of the $\p\p$ total cross section, in pure spin states, to derive positivity bounds as discussed
above (See Eq.~(\ref{incl:eq:tot})).}
\begin{multline}\label{incl:eq:M}
\mathsf{M} = \sigma_0[\mathbbm{1}_4 + A_{aN} \sigma_{ay}\otimes \mathbbm{1}_2+ 
A_{bN}\mathbbm{1}_2\otimes \sigma_{by} +
A_{NN}\sigma_{ay}\otimes\sigma_{by} + A_{LL}\sigma_{az}\otimes\sigma_{bz}\\
{}+ A_{SS}\sigma_{ax}\otimes\sigma_{bx} + A_{LS} \sigma_{az}\otimes\sigma_{bx} +
A_{SL}\sigma_{ax}\otimes\sigma_{bz}]~.
\end{multline}
Here $\sigma_0$ stands for the spin-averaged cross section.
This expression is fully justified, since we have explicitly
\begin{multline}
\sigma(\vec {P}_a,\vec{P}_b)= \sigma_0[1 + A_{aN} P_{ay} + A_{bN} P_{by} + A_{NN}P_{ay} P_{by} + A_{LL}P_{az} P_{bz} \\
{}+ A_{SS}P_{ax} P_{bx} + A_{LS}P_{az} P_{bx} + A_{SL}P_{ax} P_{bz}]~.
\end{multline}
The crucial point is that $\mathsf{M}$ is a Hermitian and \emph{positive} matrix and in order to derive 
the positivity conditions one should write the explicit
expression of $\mathsf{M}$ as given by Eq.~(\ref{incl:eq:M}). In the transversity basis, where $\sigma_y$ is diagonal, $
\mathcal{M}$ is diagonal, and   by permuting two rows and two columns, it reduces to the simple form 
\be
\left ( \begin {array}{c|c} \mathsf{M}_{+} & 0 \\ \hline  0 & \mathsf{M}_{-} \end {array} \right)~, 
\label{incl:eq:M1}
\ee
where $\mathcal{M}_{\pm}$ are 
$2 \times 2$ Hermitian matrices which must be positive, leading to the 
following \emph{two} strongest constraints \cite{Soffer:2003qj}\footnote{This article contains sign errors}
\be
(1 \pm A_{NN})^2 \geq ( A_{aN} \pm A_{bN})^2 + ( A_{LL} \mp  A_{SS})^2 + ( A_{LS} \pm A_{SL})^2~.
\label{incl:eq:constr}
\ee
These inequalities are related to (\ref{incl:eq:DM}) by crossing, as mentioned earlier (see Sec.~\ref{se:basic}).
As special cases of the above equation, we have the \emph{six} weaker constraints
\be\label{incl:eq:11}\begin{aligned}
1 \pm A_{NN} &\geq | A_{aN} \pm A_{bN} |~,\\
1 \pm A_{NN} &\geq | A_{LL} \mp  A_{SS} |~,\\
1 \pm A_{NN} &\geq | A_{LS} \pm  A_{SL} |~.
\end{aligned}\ee
The first inequality is \emph{classical} in the sense defined in Ref.~\cite{Artru:2008ax}
(see also Sec.~\ref{furth:sub:quantum}): it results from the positivity of the cross section when $a$ and $b$ are polarised along $\pm\yu$. The last two ones are non-classical. They forbid a spin dependence of the form $1+\Pv_a.\Pv_b$ or $1+\Pv_a.R\Pv_b$, where $R$ is any rotation.

These constraints are very general 
and must hold in any kinematical region and for many different situations
such as electron--proton scattering, electron--positron scattering or quark--quark scattering, but we now turn to a specific 
case, which is of direct relevance to the spin programme at the BNL-RHIC polarised $\p\p$ collider \cite{Bunce:2000uv}.
Now let us consider a proton--proton collision and let us call $y$ the rapidity of the outgoing particle $c$. In this case 
since the initial particles are identical, we have $A_{aN}(y)=-A_{bN}(-y)$ and
$A_{LS}(y)=-A_{SL}(-y)$.
In this case Eq.~(\ref{incl:eq:constr}), which becomes two constraints among five independent spin observables, reads
\be
(1 \pm A_{NN}(y))^2 \geq ( A_{aN}(y) \mp A_{aN}(-y))^2 + ( A_{LL}(y) \mp A_{SS}(y))^2 
+ ( A_{LS}(y) \mp A_{LS}(-y))^2~.
\ee
This implies in particular, for $y=0$,
\begin{equation}
1 - A_{NN}(0) \geq 2| A_{N}(0) |~,
\qquad
1 - A_{NN}(0) \geq 2| A_{SL}(0) |~,
\end{equation}
so that the allowed range of $A_N$ and $A_{SL}$ is strongly reduced 
if $A_{N\! N}$ turns out to be large and positive. Conversely
if $A_{N\! N} \simeq -1$, these constraints are useless. 
Note that, in the kinematical region accessible to the $\p\p$ polarised collider,
a calculation of $A_{NN}$ for direct photon production and jet 
production has been performed \cite {Soffer:2002tf,Mukherjee:2003pf}; it was found that $|A_{NN}|$ is of the order of 1 or~2 percent.
Similarly, based on Ref.~\cite{Soffer:1994ww}, this double transverse spin asymmetry for lepton pair 
production was estimated to be a few percents \cite{Martin:1999mg}. The direct consequence of these
estimates is that $| A_{N} |$ and $| A_{SL} |$, for these processes\,\footnote{It is amusing to recall
that, using a phenomenological approach for lepton-pair production, 
bounds on $| A_{N} |$ larger than 50\% were obtained in Ref.~\cite{Soffer:1979kg}, but at that time it was
not known that $A_{NN}$ is small.}, are essentially bounded by 1/2. In 
addition, from Eq.~(\ref{incl:eq:11}), there are two other non-trivial constraints: $1 \geq | A_{LL} \pm A_{SS}|$ when $A_{NN}$ vanishes.

Single transverse spin asymmetries in inclusive reactions at high energies are now considered to be directly related to the 
transverse momentum of the fundamental partons involved in the process. This new viewpoint, which has been advocated 
to explain the existing data in semi-inclusive deep inelastic scattering, will have to be more firmly established also by 
means of future data from BNL-RHIC. On the theoretical side several possible leading-twist QCD mechanisms \cite{Sivers:
1989cc,Sivers:1990fh,Collins:1992kk} have been proposed to generate these asymmetries  in leptoproduction \cite{Brodsky:2002rv}, but also in 
$\p\p$ collisions 
%\cite{Anselmino:2004ky,Schmidt:2005gv,Artru:1993ad,Artru:1995bh}
\cite{Anselmino:1994tv,Anselmino:2005sh,Schmidt:2005gv,Artru:1993ad,Artru:1995bh}. We believe that these new  positivity constraints on spin observables for a wide class of reactions will be of interest for model builders as well as for  future measurements.
\subsection{Spin observables in deep inelastic scattering and structure functions}
\boldmath\subsubsection{Positivity constraint for the $A_2$ transverse asymmetry}\unboldmath
\label{incl:sub:pca2}
It is well known that Deep Inelastic Scattering (DIS), in the one-photon exchange approximation, is
described as forward Compton scattering, where a very off-shell photon is probing the nucleon. In 
polarised DIS one has four cross sections, $\sigma_T^{m}$, where $m=|J_z^\gamma+ J_z^\N|=1/2$ or $3/2$, for photons and nucleons of definite helicities,  $\sigma_L$ for a zero-helicity photon, and $\sigma_{LT}$ for the interference between the two cases. $L$ or $T$ indicates the longitudinal or transverse nature of the virtual photon. $\sigma_{LT}$ occurs for both the nucleon and the photon carrying transverse spin.
Then one can define two asymmetries
\be
A_1 ={ \sigma_T^{1/2} - \sigma_T^{3/2}\over \sigma_T^{1/2} + \sigma_T^{3/2}}~,
\quad \mbox{and}\quad A_2 = {\sigma_{LT}\over \sigma_T^{1/2} + \sigma_T^{3/2}}~.
\ee
There is a well-known condition established long time ago \cite{PhysRev.143.1310}  and re-derived on a recent extensive 
study \cite{Doncel:1971qb}, written in the form
\be
\label{incl:eq:DDR}
|A_2| \leq \sqrt{R}~,
\ee
where $R=\sigma_L/\sigma_T$ and $\sigma_T = \sigma_T^{1/2}  +  \sigma_T^{3/2}$. 
It reflects a non-trivial positivity condition on the photon--nucleon helicity amplitudes.
We will see that it is possible to establish a  bound  stronger than Eq.~(\ref{incl:eq:DDR}), namely \cite{Soffer:2000zd} 
\be
\label{incl:eq:DDRn}
|A_2| \leq \sqrt{R(1+A_1)/2}~,
\ee
which is also implicit from Eq.~(2.32) of \cite{Doncel:1971qb}, once their notation is translated into the modern one.

Let us start with the following expressions for 
these four photon-nucleon cross sections in terms of the 
matrix elements describing the transition from the state $|H,h\rangle$ of a
nucleon with helicity $H$ and a photon with helicity $h$, 
to the unobserved state  $|X\rangle$
\be\label{incl:eq:def-sig}\begin{aligned}
\sigma^{1/2}_T&=\sum_{X} |\langle + 1/2,+ 1|X\rangle |^2~,
\qquad\sigma^{3/2}_T=\sum_{X} |\langle - 1/2,+ 1|X\rangle |^2\\
\sigma_L&=\sum_{X} |\langle+1/2, 0|X\rangle|^2=\sum_{X}|\langle-1/2,0|X\rangle|^2~, \\
\sigma_{LT}&=\RE\sum_{X} \langle +1/2,+1|X\rangle\langle-1/2, 0|X\rangle^*~.
\end{aligned}\ee
%
%Note that while longitudinal and transverse cross sections
%are symmetric with 
%respect to the reverse of the nucleon and photon helicities, this
%is not the case for the interference term. The reason is very simple:
%the opposite helicities of photon and nucleon correspond to their spins 
%parallel, so that the angular momentum of the state  $|X\rangle$ has its maximum value 
%$3/2$. The amplitude, which could  possibly interfere with it
%to produce the transverse asymmetry, should have the same total
%angular momentum of the state  $|X\rangle$.
%This is however impossible, as 
%the flip of the one of the helicities
%would require another one to exceed its maximal possible value,
%in order to keep the angular momentum of $|X\rangle$ the same. 
%Therefore the interference, responsible for $A_2$, does not occur. 
%This is quite a general reason,
%for the occurrence of the $+$ helicity configurations 
%in all the cases considered above.
Note the different behaviour of $\sigma^{\{1,3\}/2}_T$, $\sigma_L$ and $\sigma_{LT}$ 
when one or two helicities flip sign, due to symmetries and conservation of angular momentum. 
 We are now ready to write down the Cauchy--Schwarz inequality as
\be
\label{incl:eq:CSa}
\sum_{X}|\langle+1/2,+1|X\rangle  \pm a \langle-1/2,0|X\rangle |^2 \geq 0~,
\ee
where $a$ is a positive real number. 
By making use of the definitions (\ref{incl:eq:def-sig}) and
after the standard minimisation with respect to the choice of
$a$, one immediately arrives at
\be
\label{incl:eq:DDRs}
|\sigma_{LT}| \leq \sqrt{\sigma_L\,  \sigma^{1/2}_T}~, 
\ee
leading directly to Eq.~(\ref{incl:eq:DDRn}). The value of this new bound is clearly
related to the size of $A_1$ and it becomes very useful for negative $A_1$, like in
the neutron case.

%The next step is to apply this method for each quark flavour separately.
%This may be achieved by considering a fictitious ``photon'' 
%coupled to only one flavour. In other words, this is just the consideration 
%of positivity  for each flavour contribution to the structure function $W^{\mu \nu}$, \ie,
%

One may apply this method for each quark flavour separately.
%\be
%\label{incl:eq:wf}
%W^{\mu \nu}_f \epsilon_\mu \epsilon^{*}_\nu > 0~.
%\ee
%
Decomposing the cross-sections as
\be
\label{incl:eq:defq}
\sigma_{i} = \sum_f e^2_i \sigma_{i}^f \quad (i=L,T,LT)~,
\ee
positivity implies that
\be
\label{incl:eq:DDRq}
|\sigma_{LT}^f| \leq \sqrt{\sigma_L^f \,\sigma^{1/2,f}_T}~, \qquad
|A_2^f| \leq \sqrt{R^f(1+A_1^f)/2}~.
\ee
%
%or analogously to Eq.~(\ref{incl:eq:DDRn})
%
%\be
%\label{incl:eq:DDRnf}
%|A_2^f| \leq \sqrt{R^f(1+A_1^f)/2}~.
%\ee
%
Some consequences of these positivity constraints in connection
with higher twists contributions have been discussed in Ref.~\cite{Soffer:2000zd}. 
\subsubsection{Positivity  Constraints for Structure Functions}\label{incl:sub:pcsf}
In the previous Sec.~\ref{incl:sub:pca2}, we derived constraints for DIS
observables related to the photon--nucleon forward helicity amplitudes\,\footnote{
Positivity constraints for non-forward Compton amplitudes were derived in 
Ref.~ \cite{DeRujula:1973iw}.}. However, DIS
is fully described by the hadronic tensor, so
the positivity constraints in DIS may be also derived from
the positivity properties of this tensor. We will present their
derivation in the most general case, when the nucleon is polarised and
parity violation (PV) is taken into account.

In PV lepton--nucleon DIS all the observables 
arise form the contraction
\be \label{Incl:eq:DIS}
\ell^{\mu \nu} W_{\mu\nu}~,\ee
which involves the leptonic tensor $\ell^{\mu\nu}$ and 
 the hadronic
tensor $W_{\mu\nu}(p,\mathsf{s},q)$ of the nucleon, where $p$ and $\mathsf{s}$ 
are the energy-momentum and polarisation vectors of the nucleon, respectively 
($p^2=M^2, \mathsf{s}^2=-1$ and $p . \mathsf{s}=0$); 
$k$ and $k'$ are the four momenta of the initial and final leptons,
%$\nu_{\tau}$($\bar {\nu}_{\tau}$) and produced $\tau^-$
%($\tau^+$)
respectively, and $q = k - k'$ is the momentum
transfer. Using Lorentz invariance and time reversal invariance,
one can express $W_{\mu\nu}(p,\mathsf{s},q)$ in terms of 14 real structure
functions as follows \cite{Ji:1993ey}\,\footnote{This expression  slightly differs from Eq.~(3) of Ref.~\cite{Ji:1993ey}.},
\be \label{incl:eq:htensor}\begin{aligned}
W_{\mu\nu}(p,\mathsf{s},q)=&-g_{\mu\nu}(W_{1}(\nu,q^{2})+X_{1}(\nu,q^{2})\frac{\mathsf{s} . q}{M})\\
&{}+\frac{p_{\mu}p_{\nu}}{M^{2}}\,(W_{2}(\nu,q^{2})+X_{2}(\nu,q^{2})\frac{\mathsf{s} . q}{M})r\\ 
&{}-i\epsilon_{\mu\nu\alpha\beta}[\frac{p^{\alpha}}
{2M^{2}}\,(q^{\beta}W_{3}(\nu,q^{2}) - 2 M
{\mathsf{s}^\beta}G_{3}(\nu,q^{2}))\\
&{}+ \frac{q^{\alpha}} {M}\,(\mathsf{s}^{\beta}G_1(\nu,q^{2}) + ( \nu
\mathsf{s}^\beta - \frac{(\mathsf{s} . q)}{M} p^\beta)\frac{G_{2}}{M}(\nu,q^{2}))]
\\
&{}+\frac{q_{\mu}q_{\nu}}{M^{2}}\,(W_{4}(\nu,q^{2})+X_{4}(\nu,q^{2})\frac{\mathsf{s} . q}{M})\\
&{}+\frac{p_{\mu}q_{\nu}+
q_{\mu}p_{\nu}}{2M^{2}}\,(W_{5}(\nu,q^{2})+X_{5}(\nu,q^{2})\frac{\mathsf{s} . q}{M})\\
&{}+\frac{p_{\mu}\mathsf{s}_{\nu}+
\mathsf{s}_{\mu}p_{\nu}}{2M}\,Z_{1}(\nu,q^{2})+\frac{q_{\mu}\mathsf{s}_{\nu}+
\mathsf{s}_{\mu}q_{\nu}}{2M}\,Z_{2}(\nu,q^{2})~.
\end{aligned}\ee
Here $\epsilon_{\mu\nu\alpha\beta}$ is the total antisymmetric
tensor with $\epsilon_{0123} = +1$.
All structure functions, which are made dimensionless by including
appropriate mass factors, depend on two Lorentz scalars, $\nu =
p. q/M$ and $Q^2 = -q^2 > 0$, where $M$ is the nucleon mass. Seven
of them, $W_1$, $W_2$, $W_4$, $W_5$, $G_1$, $G_2$ and $G_3$ appear in parity-conserving
processes and the other seven $W_3$, $X_1$, $X_2$, $X_3$, $X_4$, $Z_1$ and $Z_2$, are
related to PV processes involving weak interactions. Note that only the $W_i$'s are independent
of the polarisation vector $\mathsf{s}$ and they characterise the scattering of an unpolarised nucleon.

Clearly the hadronic tensor $W_{\mu\nu}(p,\mathsf{s},q)$ is Hermitian
\be
W_{\mu\nu}(p,\mathsf{s},q) = W_{\nu\mu}^*(p,\mathsf{s},q)~, \label{incl:eq:her}
\ee
and semi-positive. This last property follows from its definition as a
bilinear expression of quarks currents and implies that
\be
a^*_{\mu}W_{\mu\nu}(p,\mathsf{s},q)a_{\nu} \geq 0 ~, \label{incl:eq:gpos}
\ee
for \emph{any} complex 4-vector $a_{\mu}$.
The resulting constraints are valid for the most
general case of PV and spin-dependent scattering, mentioned above.
Moreover, they can be generalised to Semi-Inclusive Deep Inelastic Scattering (SIDIS), which involves 18 structure functions, if
parity is conserved  \cite{DeRujula:1973uk,Dmitrasinovic:1995qe},
and they may also be applied
to the Drell-Yan process, related to it by crossing.

Let us now further discuss the complexity arising when the nucleon target polarisation is taken into account. Following Refs.~\cite{Doncel:1971qb,DeRujula:1973uk}, one should consider 
instead of Eq.~(\ref{incl:eq:gpos}), the more general condition
\be
\sum_{\lambda.\lambda'}a^*_{\mu,\lambda}W_{\mu\nu}(p,q,\lambda,\lambda')a_{\nu,\lambda'}
\geq 0 ~, \label{incl:eq:gpos'}
\ee
in 8-dimensional space including, besides Lorentz indices, the spinor indices $\lambda$ and $\lambda'$. $a_{\mu\lambda}$ 
is a compound photon--nucleon spin wave function, which might be entangled%
\footnote{$W_{\mu\nu}(p,q,\lambda,\lambda')$ is a particular type of cross section matrix  which could be written $\mathcal{R}_{\mu,\lambda,\nu,\lambda'}$ in the notation of Sec.~\ref{se:basic}.}.
In particular, the inequalities (\ref{incl:eq:DDRn})  and (\ref{incl:eq:DDRq}) on $A_2$  require entangled states. 
The  expressions resulting from (\ref{incl:eq:gpos'}) can be very complicated and we will restrict ourselves to the
positivity constraints in the case of longitudinal polarisation of degree $\xi$, so we have
$\mathsf{s}=(0,\xi,0,0)$. In that case $W_{\mu\nu}(p,\mathsf{s},q)$ in the laboratory frame has the block-diagonal form
\be\begin{pmatrix}N_1&0\\  0&N_0\end{pmatrix}~,\ \ 
N_\alpha=M_\alpha + \xi \Delta M_\alpha = \left(\begin{array}{cc}
M_{\alpha,11}+\xi \Delta M_{\alpha,11} &
M_{\alpha,12}+\xi \Delta M_{\alpha,12}\\
M^{*}_{\alpha,12}+\xi \Delta M^{*}_{\alpha, 12} &
M_{\alpha,22}+\xi \Delta M_{\alpha, 22}
\end {array} \right)~,
\ee
where $\alpha=0,1$ and $M_{\alpha,ij}$ (resp. $\Delta M_{\alpha,ij}$) stands for the $ij$
matrix element of $M_{\alpha}$ (resp. $\Delta M_{\alpha}$). The
target spin-averaged part is described by 
$M_1$ and $M_0$ which are the following $2\times 2$ Hermitian matrices
\be \label{incl:eq:M10}
M_1 =\begin{pmatrix}%
\displaystyle -W_1+W_2+\nu'^2W_4+\nu' W_5 &
\displaystyle Q'\left(\nu' W_4+W_5/2\right) \\
\displaystyle Q' \left(\nu' W_4+W_5/2\right) &
\displaystyle W_1+Q'^2 W_4                 \end{pmatrix}~,
\ee
and
\be\label{incl:eq:M13}
M_0 =\begin{pmatrix}%
\displaystyle W_1 & \displaystyle -iQ'W_3 /2\\
\displaystyle + i Q' W_3/2  &\displaystyle  W_1  \end{pmatrix}~,
\ee
where $\nu'=\nu/M$ and $Q'=\sqrt{\nu^2+Q^2}/M$.
The dependence on target polarisation is 
described by two similar $2\times2$ Hermitian matrices $\Delta M_1$ and $\Delta M_0$
with the following entries:
\be\begin{aligned}
\Delta M_{1,11}&=Q'\left(X_2-X_1+ \nu'^2 X_4 + \nu' X_5\right)~,\\
\Delta M_{1,12}&=Z_1/2+ \nu' Z_2/2+Q'^2 \left(X_5+\nu' X_4\right)~,\\
\Delta M_{1,22}&=Q' \left(-X_1+Q'^2 X_4 + Z_2\right)~,\\
\Delta M_{0,11}&=\Delta M_{0,22}=Q' X_1~,\\
\Delta M_{0,12}&=-i\left(G_3+\nu' G_1 -Q^2G_2/M^2\right)~.\end{aligned}\ee
Note that $N_0$ is diagonal in the circular-polarisation basis, because of the conservation of $J_z$.

The \emph{necessary and sufficient conditions} for
$W_{\mu\nu}(p,\mathsf{s},q)$ to satisfy the inequality~(\ref{incl:eq:gpos}) are,
according to Sec.~\ref{basic:sub:dens}, that all
the principal minors of $N_1$ and $N_0$ should be positive
definite for any $|\xi| <1$. For the diagonal elements, since two inequalities for the matrix $N_0$ coincide, we have three inequalities linear in the $W_i$
\be
M_{\alpha,ii}- |\Delta M_{\alpha,ii}| \geq 0~, \label{incl:eq:genlin}
\ee
where $i=1,2$. The positivity
of the  $2\!\times\!2$ determinants of $N_0$ and $N_1$ reads
\be\label{incl:eq:pos-a-b-c}
  a_\alpha + (b^{(1)}_\alpha +b^{(2)}_\alpha)\,\xi +c_\alpha\,\xi^2 \geq 0~,\ee
where
\be\begin{aligned}
& a_{\alpha} = \det( M_\alpha)~,\qquad &c_{\alpha} &= \det(\Delta M_\alpha)~, \\
& b^{(1)}_\alpha  =\begin{vmatrix}
M_{\alpha,11} &  M_{\alpha,12} \\
\Delta M^{*}_{\alpha, 12} & \Delta M_{\alpha, 22}
\end{vmatrix}~,\qquad
& b^{(2)}_\alpha  &= \begin{vmatrix}
 \Delta M_{\alpha,11}  & \Delta M_{\alpha,12}\\
M^{*}_{\alpha, 12} & M_{\alpha, 22}\end{vmatrix}~.\end{aligned}
\ee

Due to convexity, it is sufficient that (\ref{incl:eq:pos-a-b-c}) is satisfied for $\xi=+1$ and $\xi=-1$. This gives the following inequalities quadratic in the $W_i$'s
\be
|b_{\alpha}^{(1)}+b_{\alpha}^{(2)}| \leq a_{\alpha}+c_{\alpha}~. \label{incl:eq:abc}
\ee
%while for $c \geq a \geq 0$ the global constraint
%$(b^{(1)}+b^{(2)})^2 \leq 4 ac$ should hold.

Note that $a_\alpha$ is positive because of the positivity in the spin-averaged case $\xi=0$. The condition $|c_{\alpha}| \leq a_{\alpha}$ follows from Eq.~(\ref{incl:eq:genlin}). This
is most conveniently seen by performing a unitary
transformation such  that $M_\alpha$ becomes diagonal. 
As the moduli of diagonal entries of $\Delta M_\alpha$ cannot exceed
those of $M_\alpha$, $\det(\Delta M_\alpha) \leq
\det(M_\alpha)$. 
%
%The only possible way of violation of relation
%$|c_{\alpha}| \leq a_{\alpha}$, when $c_{\alpha} \leq -a_{\alpha}$, is excluded, because in that case
%Eq.~(\ref{incl:eq:pos-a-b-c}) cannot hold for any $b$'s. This completes the
%proof, that Eq.~(\ref{incl:eq:abc}) provides the quadratic positivity
%constraints for parity-violating DIS on longitudinally polarised target.
%
\subsubsection{Neutrino Deep Inelastic Scattering}
We are now turning to an application\,\footnote{For other applications to DIS or SIDIS, see
Refs.~\cite{Korthals-Altes:1974wh, Korthals-Altes:1974nk}.} of positivity inequalities for 
neutrino deep inelastic scattering \cite{Bourrely:2004iy}.
We consider the case of unpolarised nucleons which is
technically simpler and is
of more practical importance. We keep the
possibility of the detecting the polarisation of the final-state
lepton, to better exhibit the
power of positivity constraints. The corresponding application concerns
the polarisation of a produced $\tau$ lepton which is of major
importance for neutrino physics.

We should now consider the inequalities of Sec.~\ref{incl:sub:pcsf} in the special case $\xi =0$.
So for the diagonal elements we have from Eq.~(\ref{incl:eq:genlin}) the following
three inequalities linear in the $W_i'$s:
\begin{align}
W_1 \geq 0 ~, \label{incl:eq:W1}\\
-W_1+W_2+\frac{\nu^2}{M^2}W_4+\frac{\nu}{M}W_5 \geq 0 ~,\label{incl:eq:M11}\\
W_1+\frac{\nu^2+Q^2}{M^2}W_4 \geq 0 ~. \label{incl:eq:M22}
\end{align}
The quadratic inequalities following from  from the $2\times 2$ determinants of $M_0$ and $M_1$
are of the form:
\be\label{incl:eq:W132}
W_1^2 \geq \frac{\nu^2+Q^2}{4M^2}W_3^2 ~, 
%\ee
%
\quad \hbox{or equivalently}\quad
%
%\be
W_1 \geq \frac{\sqrt{\nu^2+Q^2}}{2M}|W_3| ~, 
%\label{incl:eq:W13}
\ee
and
\be
\left(-W_1+W_2+\frac{\nu^2}{M^2}W_4+\frac{\nu}{M}W_5 \right)
\left( W_1+\frac{\nu^2+Q^2}{M^2}W_4 \right)
\geq \frac{\nu^2 + Q^2}{M^2} \left(
\frac{\nu}{M}W_4+\frac{1}{2}W_5 \right)^2~. \label{incl:eq:quad}
\ee
By imposing the last condition, only one of the two inequalities
Eq.~(\ref{incl:eq:M11})
 or Eq.~(\ref{incl:eq:M22}) is needed, which is a consequence of
 the existence of no more than 4 independent inequalities
 as discussed in Sec.~\ref{basic:sub:single}.

Let us pass to the applications of the inequalities.
In the laboratory frame, let us denote by $E_{\nu}$, $E_{\tau}$
and $p_{\tau}$ the neutrino energy, $\tau$ energy and momentum,
respectively and $\theta$ the scattering angle. We then have $\nu
= E_{\nu} - E_{\tau}$ and
$Q^2=2E_{\nu}[E_{\tau}-p_{\tau}\cos{\theta}]-m^2_{\tau}$, where
$m_{\tau} = 1.777\, \mbox{GeV}$ is the $\tau$ mass. Finally, the
Bjorken variable $x$ is defined as usual:
$x= Q^2/2p. q$
and the physical region is $x_{\rm min} \leq x \leq 1$, where $x_{\rm min}  =
m_{\tau}^2/2 M (E_{\nu} - m_{\tau})$. The unpolarised cross
sections for deep inelastic scattering  are expressed as
\be
\label{incl:eq:cs}
\frac{\d\sigma^{\pm}}{\d E_{\tau}\d\cos{\theta}}=\frac{G^2_F}{2\pi}
\frac{M_W^{4}p_{\tau}}{(Q^2+M_W^{2})^2}\,R_{\pm}~,
\ee
where $G_F$ is the Fermi constant and $M_W$ is the $W$-boson mass. Here
\begin{multline} \label{incl:eq:cross}
R_{\pm}= \frac{1}{M}\,\bigg\{
\Big(2W_{1}+\frac{m_{\tau}^{2}}{M^{2}}\,W_{4}\Big)\left(E_{\tau}-p_{\tau}\cos\theta\right)
+W_{2}\left(E_{\tau}+p_{\tau}\cos\theta\right)\\
{}\pm\frac{W_{3}}{M}\,\Big(E_{\nu}E_{\tau}+p_{\tau}^{2}
-(E_{\nu}+E_{\tau})p_{\tau}\cos\theta\Big)
-\frac{m_{\tau}^{2}}{M}\,W_{5}\bigg\}~,
\end{multline}
where the $\pm$ signs correspond to $\tau^{\mp}$ productions.

Because of time reversal invariance, the polarisation vector
$\overrightarrow P$ of the $\tau$ in its rest frame, lies in the
scattering plane defined by the momenta of the incident neutrino
and the produced $\tau$. It has a component $P_L$ along the
direction of $\overrightarrow {p_{\tau}}$ and a component $P_P$
perpendicular to $\overrightarrow {p_{\tau}}$, whose expressions are, in the laboratory
frame,
\be\label{incl:eq:polvec}\begin{aligned}
P_P&=\mp\,\frac{m_{\tau}\sin\theta}{MR_{\pm}}
\bigg(2W_{1}-W_{2}\pm\frac{E_{\nu}}{M}\,W_{3}
-\frac{m_{\tau}^{2}}{M^{2}}\,W_{4}+\frac{E_{\tau}}{M}\,W_{5}\bigg)~,\\
P_L&=\mp\,\frac{1}{MR_{\pm}}\bigg\{
\Big(2W_{1}-\frac{m_{\tau}^{2}}{M^{2}}\,W_{4}\Big)
\left(p_{\tau}-E_{\tau}\cos\theta\right)
+W_{2}\left(p_{\tau}+E_{\tau}\cos\theta\right)\hfill\\
&\hspace*{1.5cm}\pm\frac{W_{3}}{M}\,\Big((E_{\nu}+E_{\tau})p_{\tau}
-(E_{\nu}E_{\tau}+p_{\tau}^{2})\cos\theta\Big)
-\frac{m_{\tau}^{2}}{M}\,W_{5}\cos\theta\bigg\}~.
\end{aligned}\ee
 In addition, it is
convenient to introduce also the degree of polarisation defined as
$P = \sqrt{P_P^2+P_L^2}$. As previously the $\pm$ signs correspond
to $\tau^{\mp}$ productions and it is clear that if $W_3=0$, one
has $R_{+}=R_{-}$ and $\tau^+$ and $\tau^-$ have opposite
polarisations.
%We also note that if one can neglect the mass of the produced lepton
%($m_{\tau}=0$), $P_P=0$, so such a lepton is purely left-handed, if negatively
%charged, or purely right-handed, if positive.

\begin{figure}[!htb]
\centering{%
  \includegraphics[width=.32\textwidth]{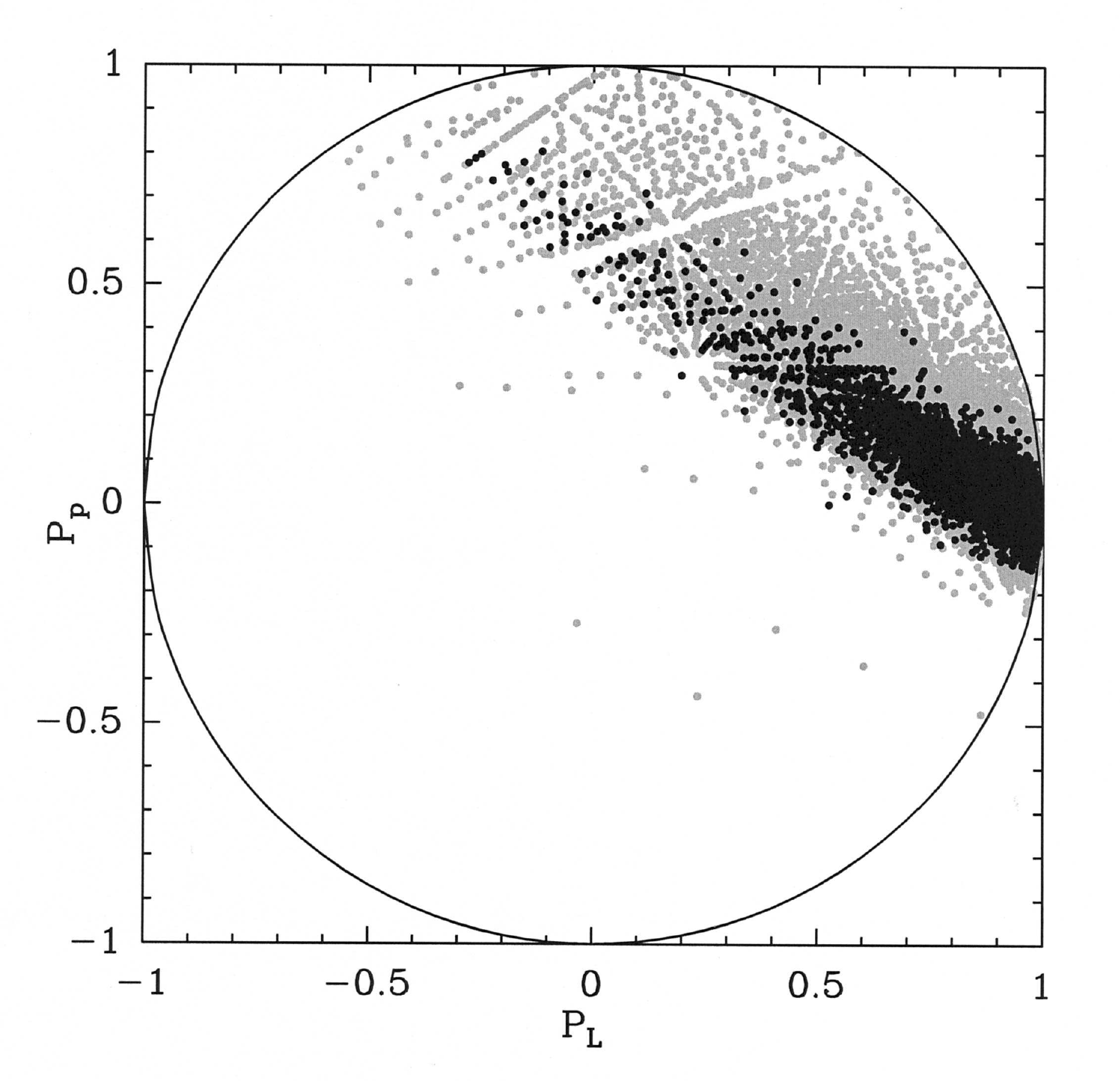}\
    \includegraphics[width=.32\textwidth]{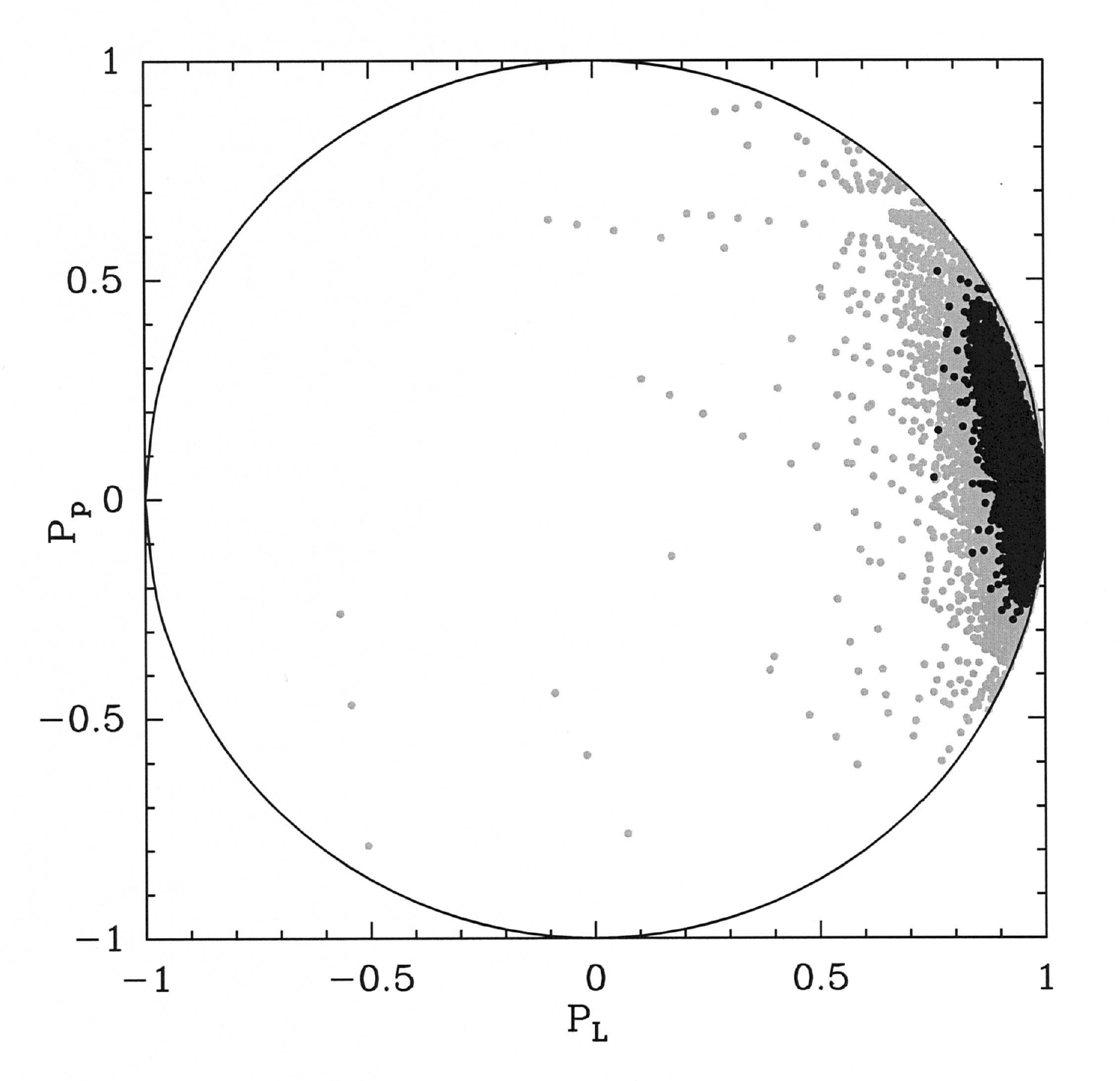}\
      \includegraphics[width=.32\textwidth]{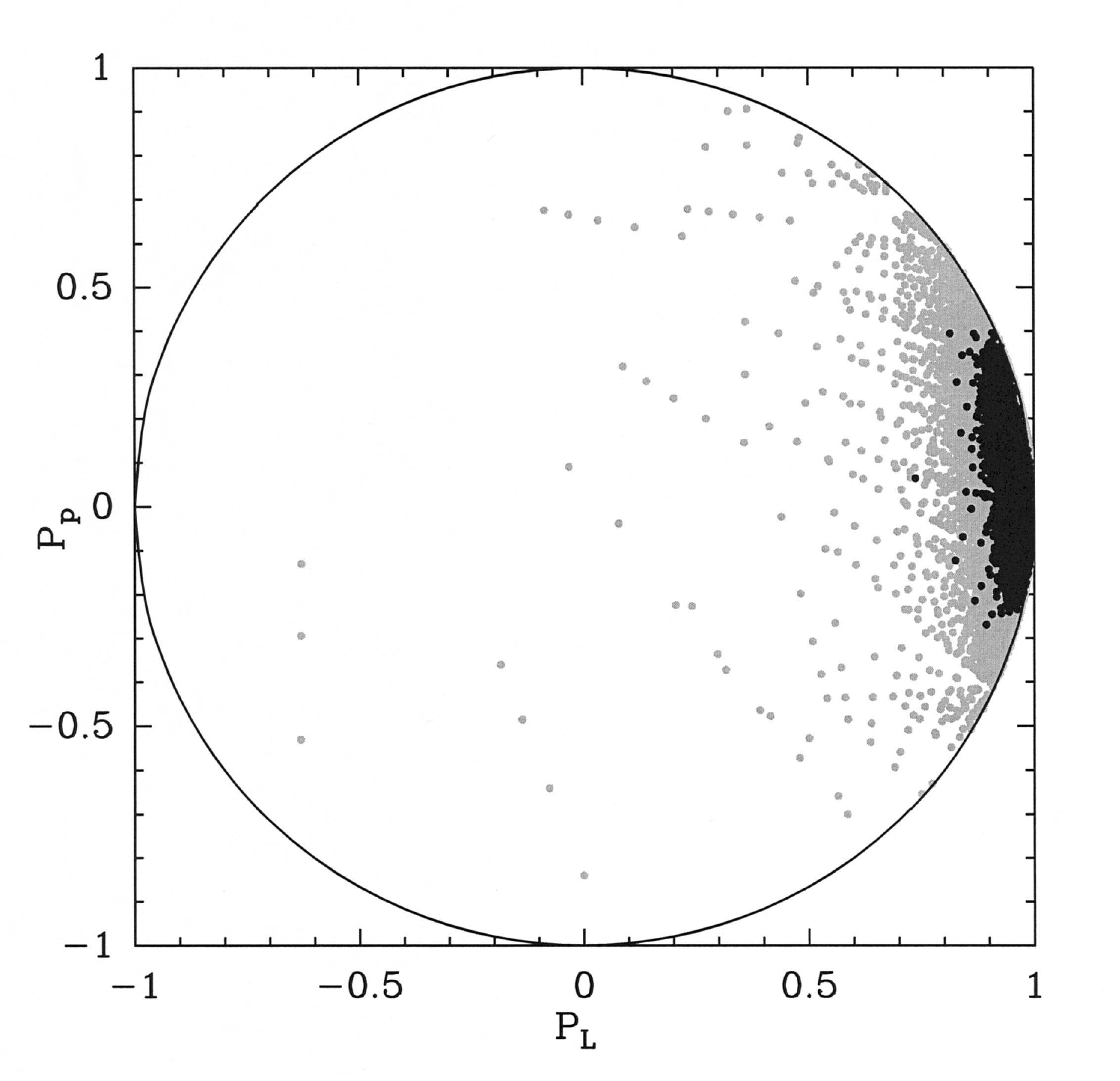}}\\[-4pt]
 \centering{%
  \includegraphics[width=.32\textwidth]{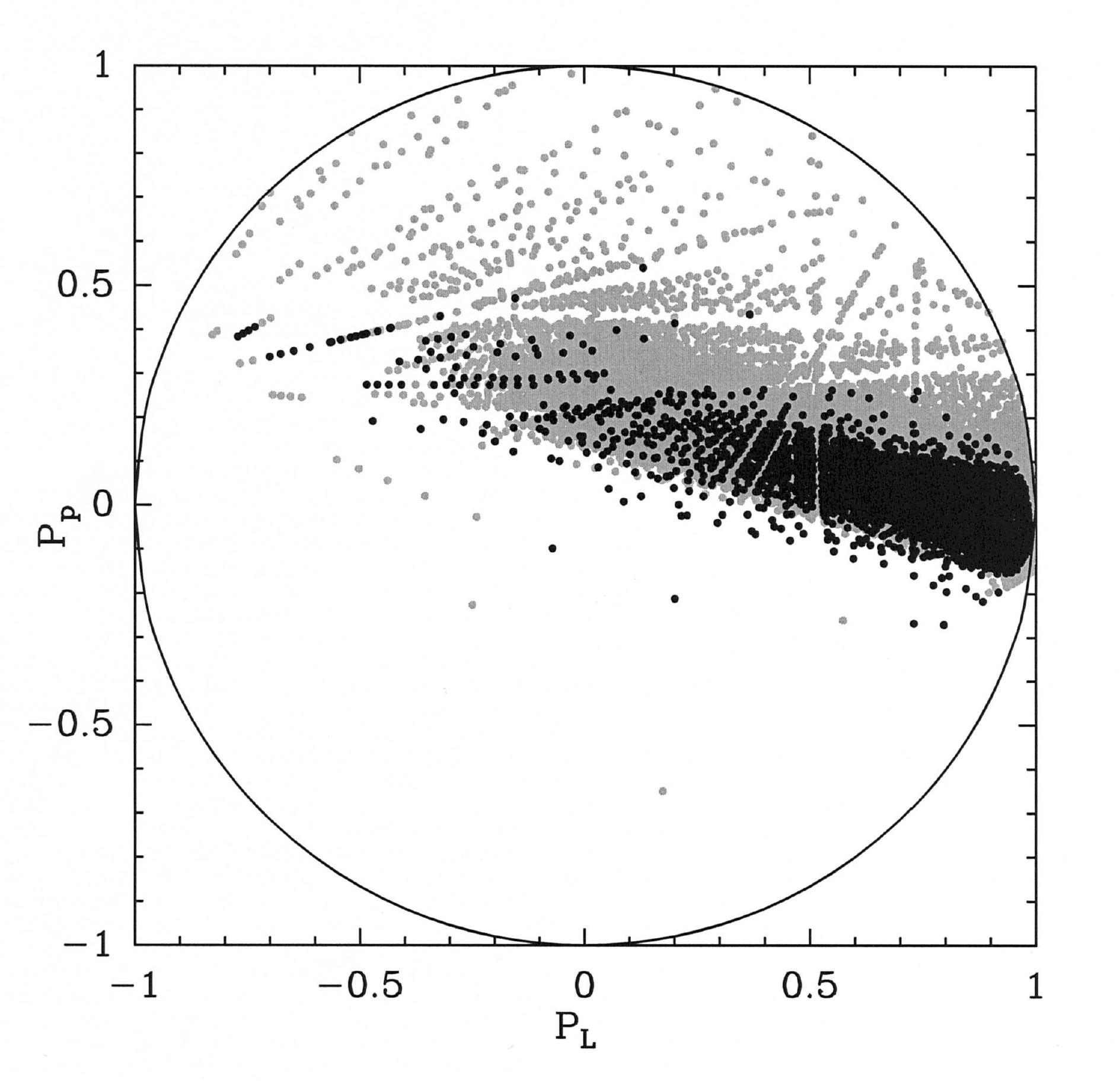}\
    \includegraphics[width=.32\textwidth]{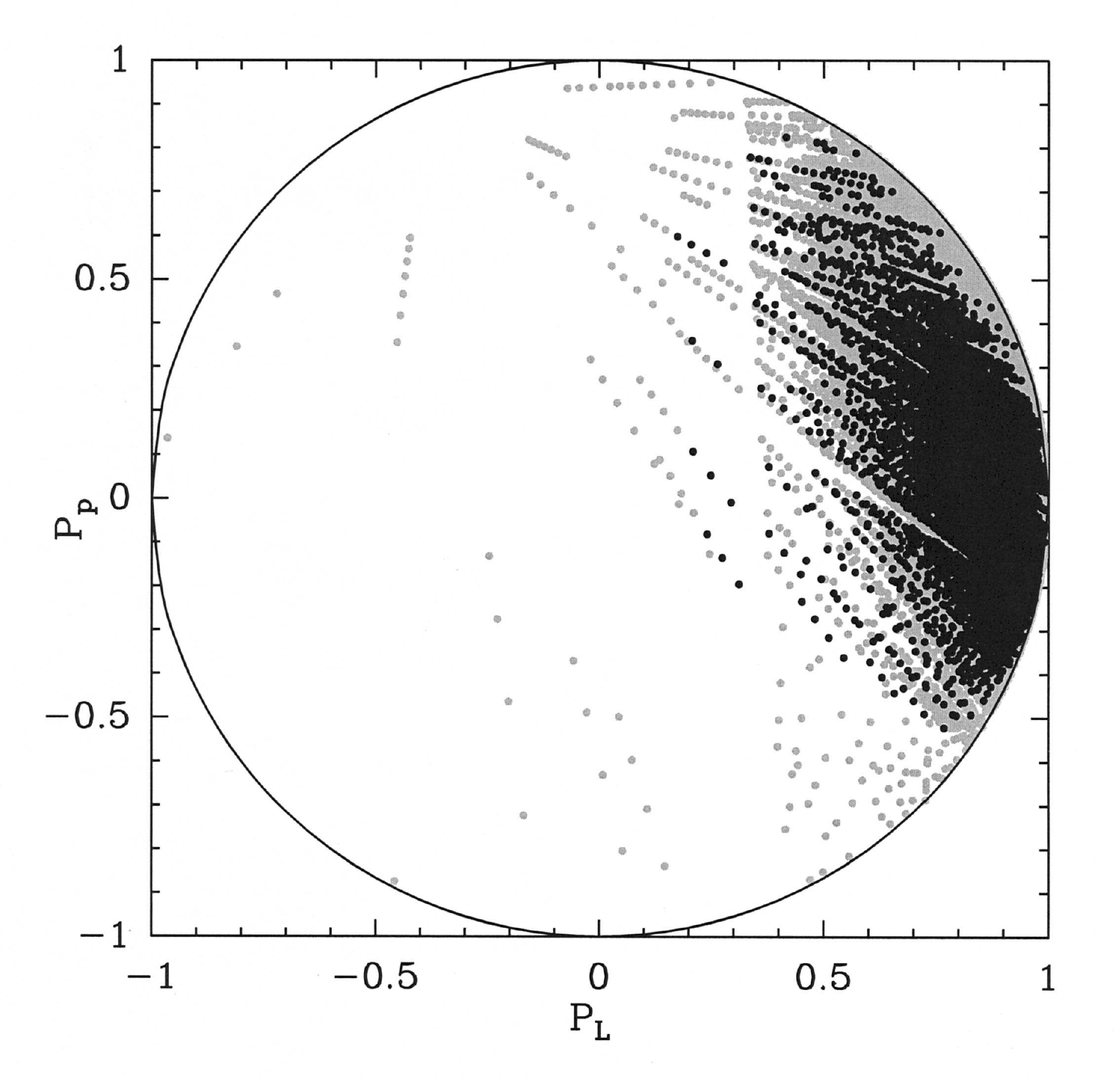}\
      \includegraphics[width=.32\textwidth]{FigIncl/f10-1-90}} \\[-4pt] 
  \centering{%
  \includegraphics[width=.32\textwidth]{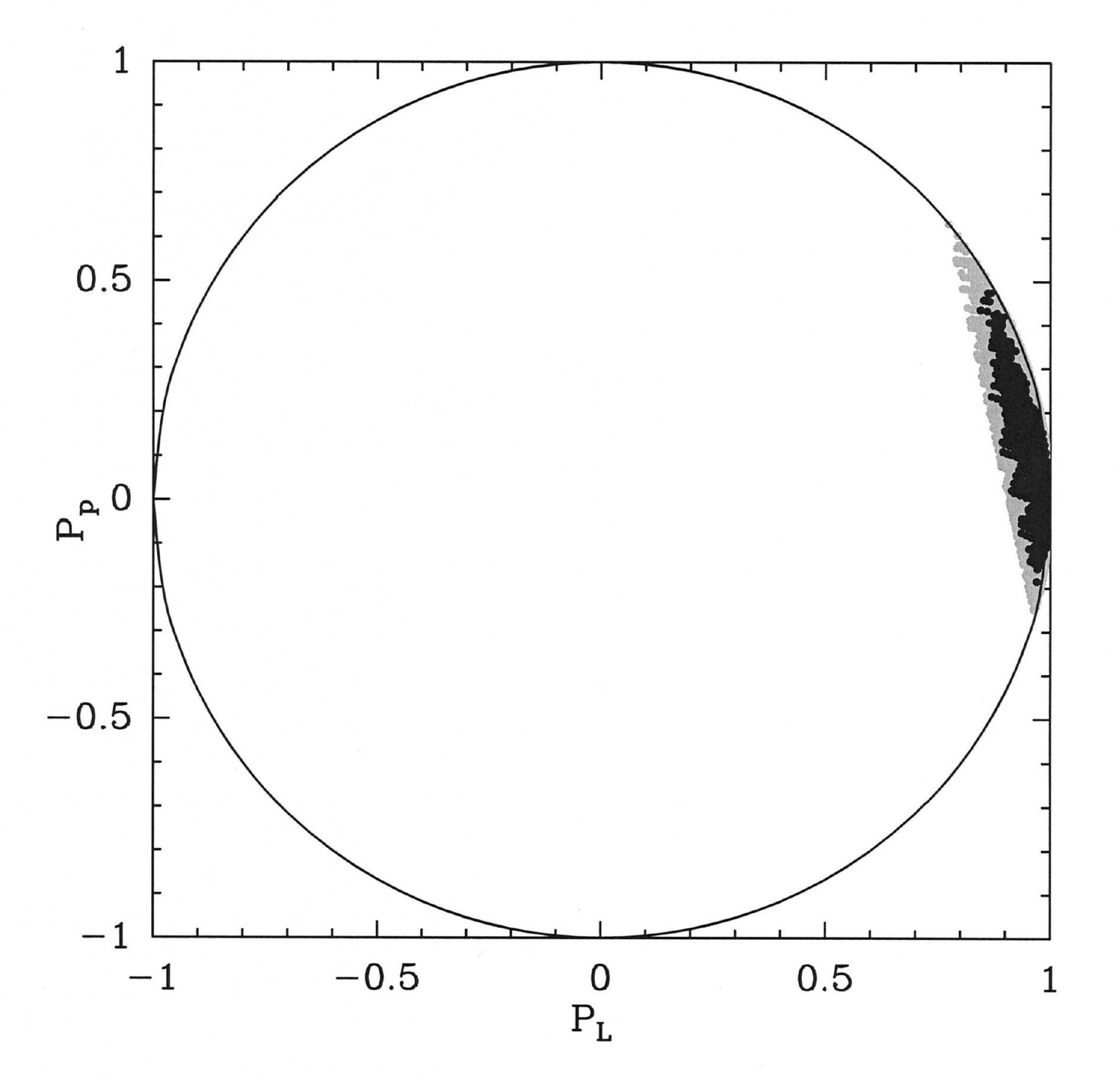}\
    \includegraphics[width=.32\textwidth]{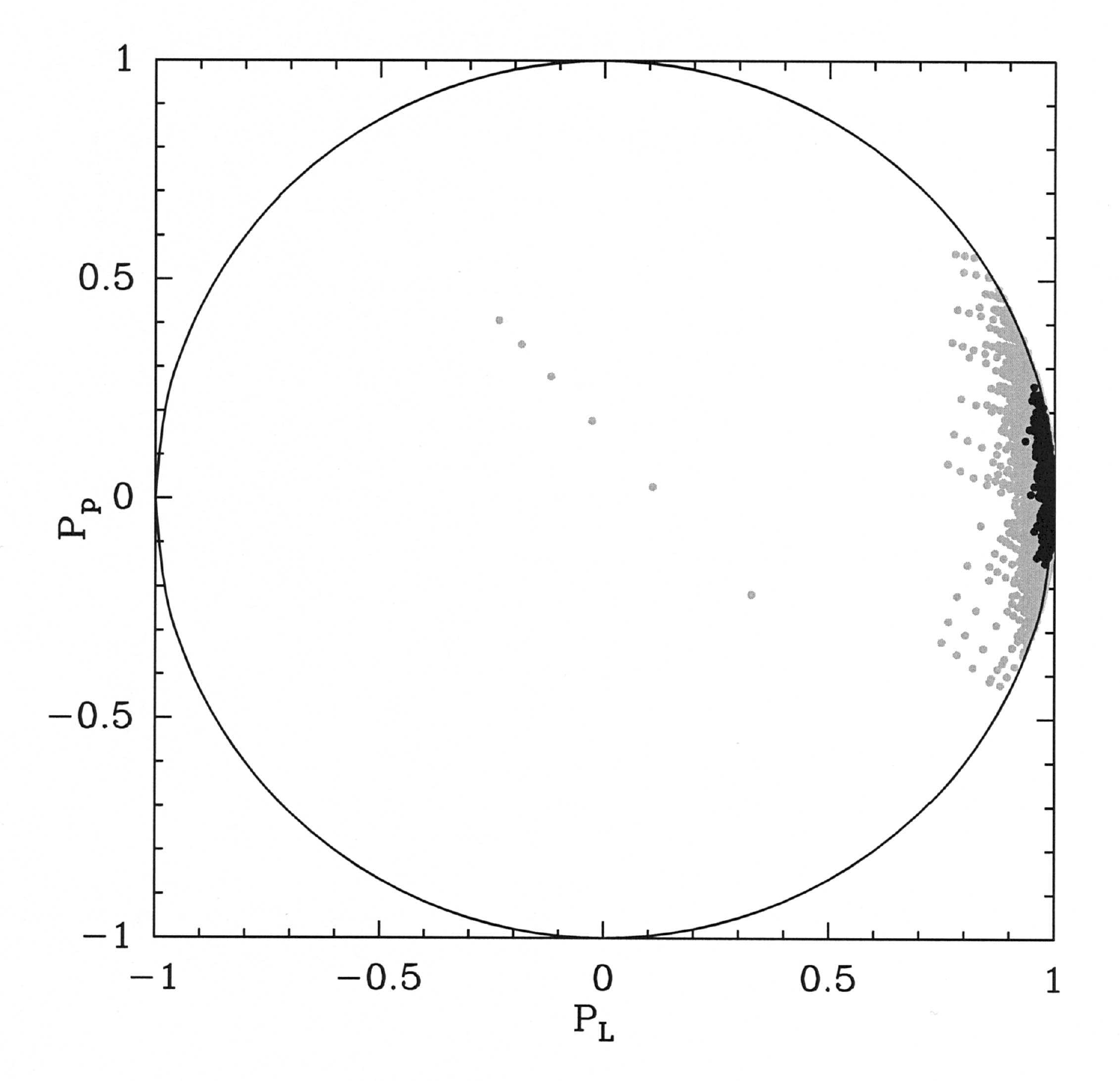}\
      \includegraphics[width=.32\textwidth]{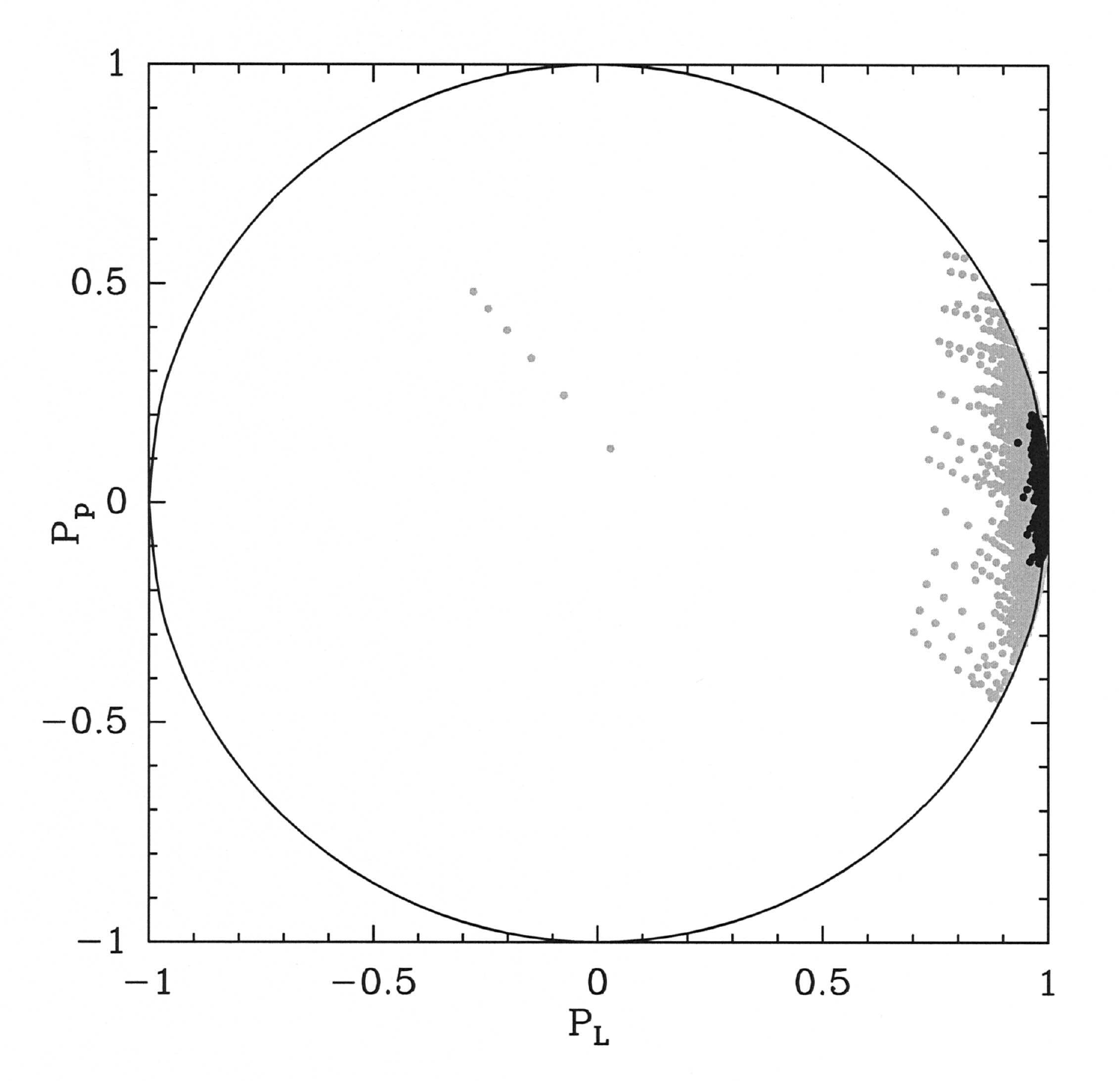}} \\[-4pt]       
  \centering{%
  \includegraphics[width=.32\textwidth]{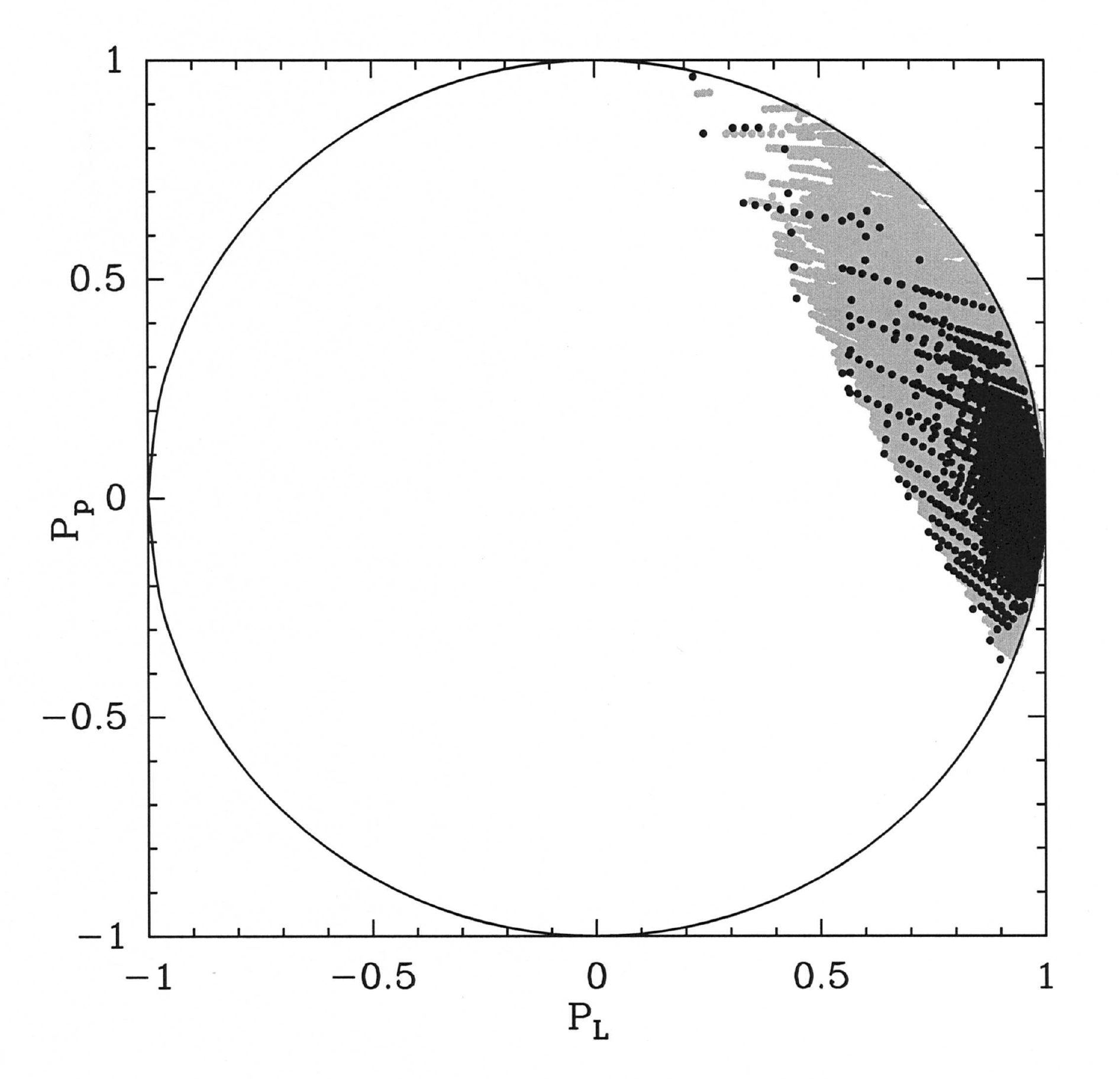}\
    \includegraphics[width=.32\textwidth]{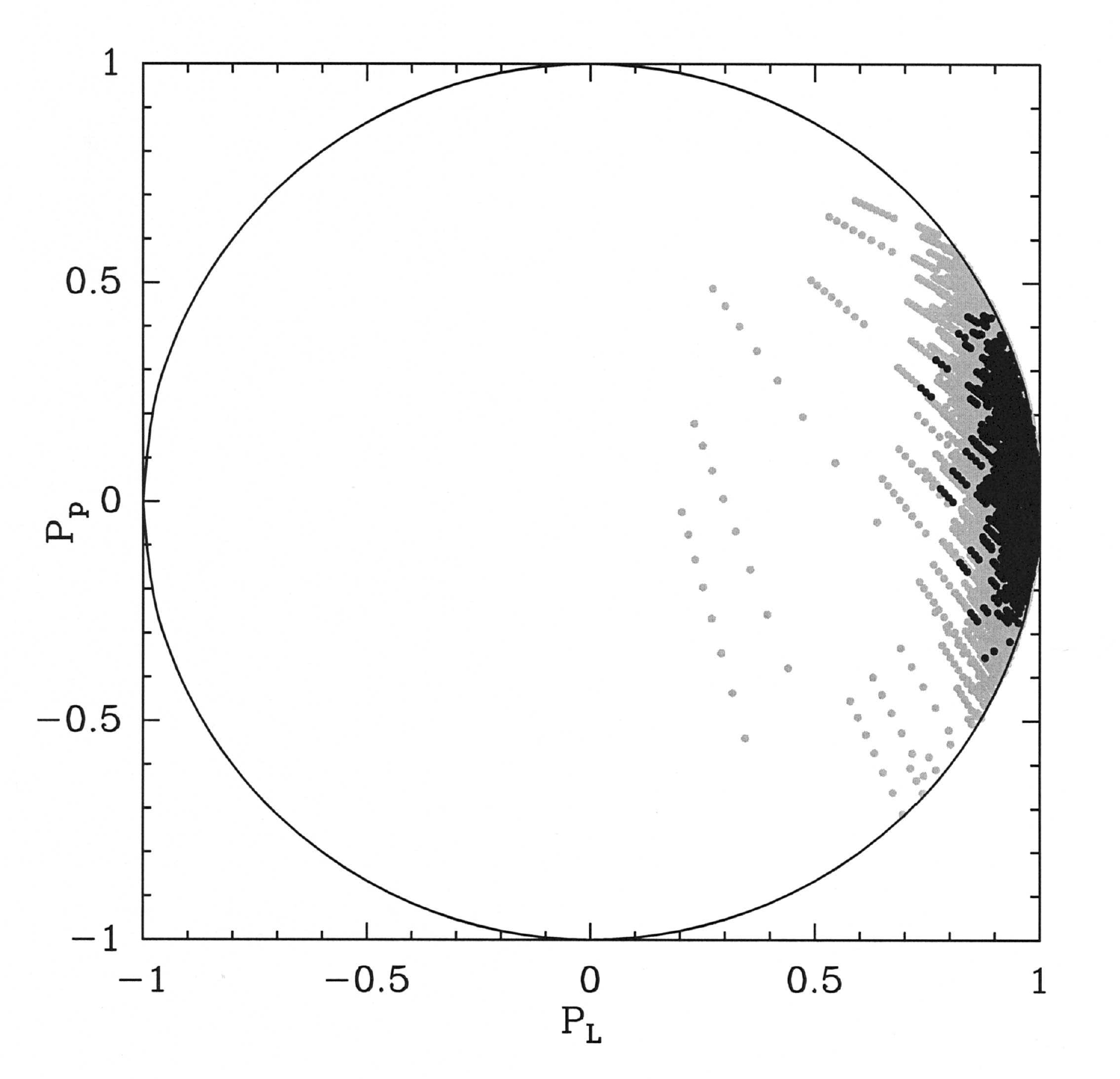}\
      \includegraphics[width=.32\textwidth]{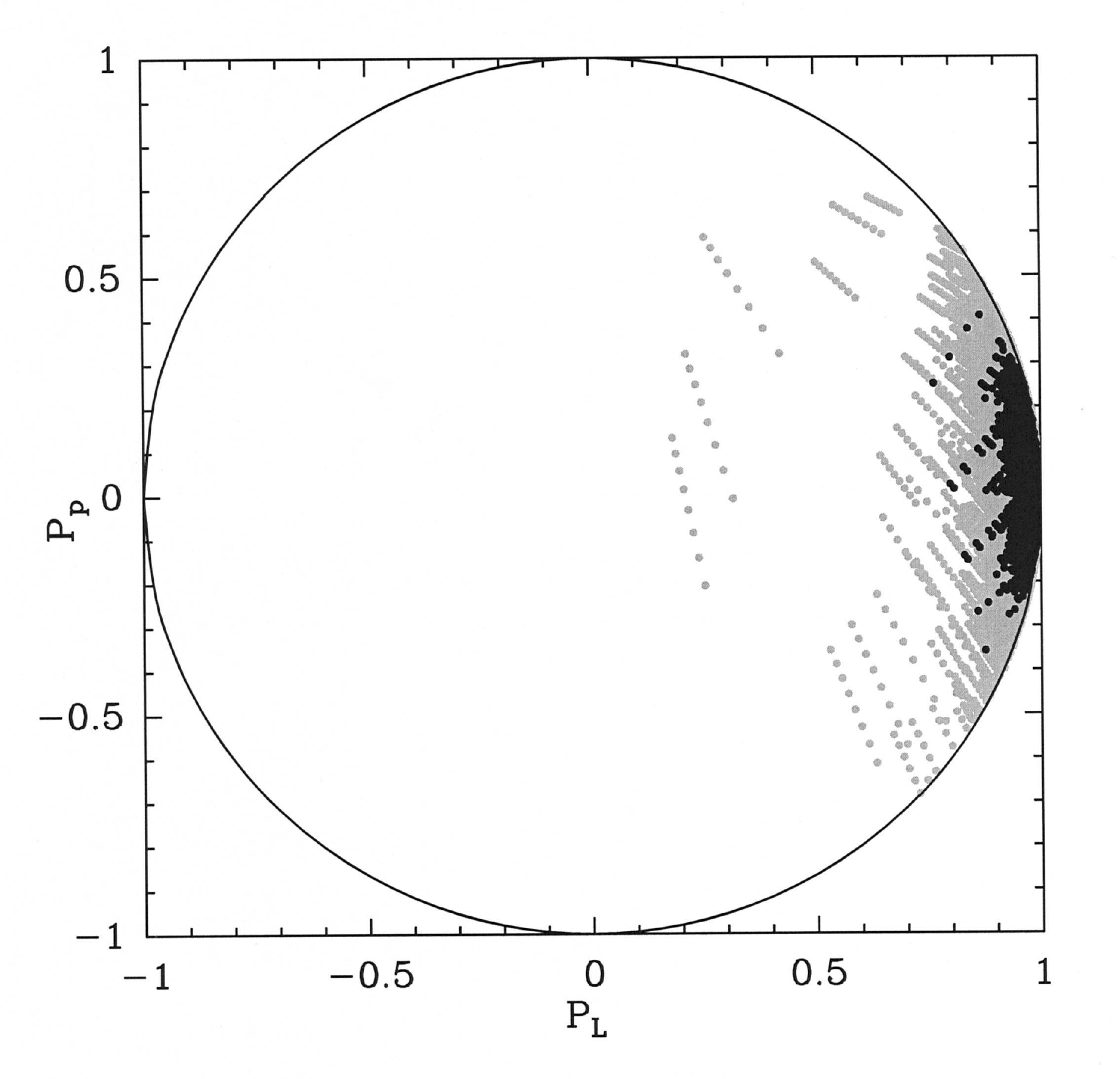}}
\caption{\label{incl:fig:1} For $\tau^+$ production, $P_P$ versus $P_L$ in a
domain limited by $R_{+} \geq 0$, $P \leq 1$ (grey dots inside disk)
plus non trivial positivity constraints (black dots inside disk).
From top to bottom and left to right,
$E_{\nu} = 10\,\mbox{GeV},~Q^2 = 1\,\mbox{GeV}^2,~x = 0.25, 0.6, 0.9$,
$E_{\nu} = 10\,\mbox{GeV},~Q^2 = 4\,\mbox{GeV}^2,~x = 0.4, 0.6, 0.9$,
$E_{\nu} = 20\,\mbox{GeV},~Q^2 = 1\,\mbox{GeV}^2,~x = 0.25, 0.6, 0.9$,
$E_{\nu} = 20\,\mbox{GeV},~Q^2 = 4\,\mbox{GeV}^2,~x = 0.25, 0.6, 0.9$.}
\end{figure}

In order to test the usefulness of positivity constraints to restrict
the allowed domains for $P_P$ and $P_L$, we proceed by the
following method, similar to that of Sec.~\ref{excl:sub:hyp-pair},
without referring to a specific model for the
$W_i$'s. We generate randomly the values of the $W_i$'s, in the
range $[0,+1]$ for $W_1$ and $W_2$, which are clearly positive and
$[-1,+1]$ for $i=3,4,5$. The most trivial positivity constraints are
$R_{\pm} \geq 0$, but in fact they are too weak and do not imply
the obvious requirements $|P_L| \leq 1$ and $|P_P| \leq 1$ or $P
\leq 1$\,\footnote{ Note that in the trivial case where
$W_3=W_4=W_5=0$, $R \geq 0$ implies $P \leq 1$.}. So we first
impose $R_{\pm} \geq 0$ and $P \leq 1$ for different values of
$E_{\nu}$, $Q^2$ and $x$ and as shown in Fig.~\ref{incl:fig:1}, for $\tau^+$
production, the points which satisfy these constraints are
represented by grey dots inside the disk, $P_{L}^2 + P_{P}^2 \leq
1$. If we now add the non trivial positivity constraints
Eqs.~(\ref{incl:eq:W1}-\ref{incl:eq:quad}), which also guarantee that $P\leq 1$, we get the black
dots, giving a much smaller area. In Fig.~\ref{incl:fig:1}, the top row
corresponds to $E_{\nu}= 10\,\mbox{GeV}$ and $Q^2 = 1\,\mbox{GeV}^2$,
the row below to $E_{\nu}= 10\,\mbox{GeV}$ and $Q^2 =
4\,\mbox{GeV}^2$ and the next two rows to $E_{\nu}= 20\,\mbox{GeV}$
and $Q^2 = 1$ or $4\,\mbox{GeV}^2$. Going from left to right $x$
increases from a value close to its minimum to 0.9. It is
interesting to note that the black allowed area increases with
$Q^2$ and becomes smaller for increasing incident energy and
increasing $x$. For $\tau^-$ production, the corresponding areas
are obtained by symmetry with respect to the centre of the disk.
For increasing $x$, since $P_L$ is more and more restricted to
values close to $+1$ for $\tau^+$ ($-1$ for $\tau^-$), it is striking
to observe that the non trivial positivity constraints lead to a
situation where the $\tau^+$ ( $\tau^-$) is almost purely
right-handed (left-handed), although it has a non zero mass.

\begin{figure}[thb]
\begin{minipage}{.5\textwidth}
\includegraphics[width=.9\textwidth]{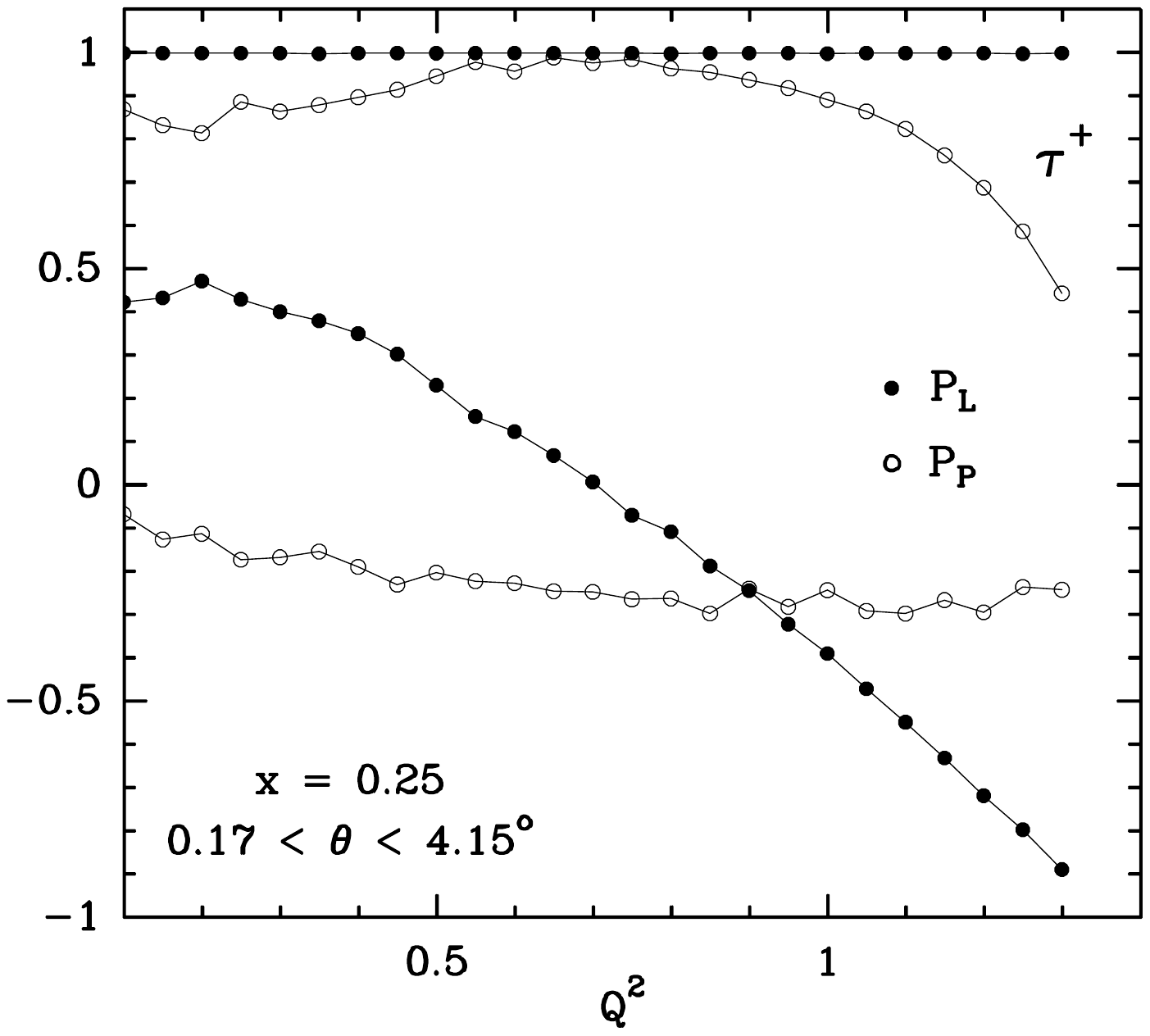}\\[5pt]
 \includegraphics[width=.9\textwidth]{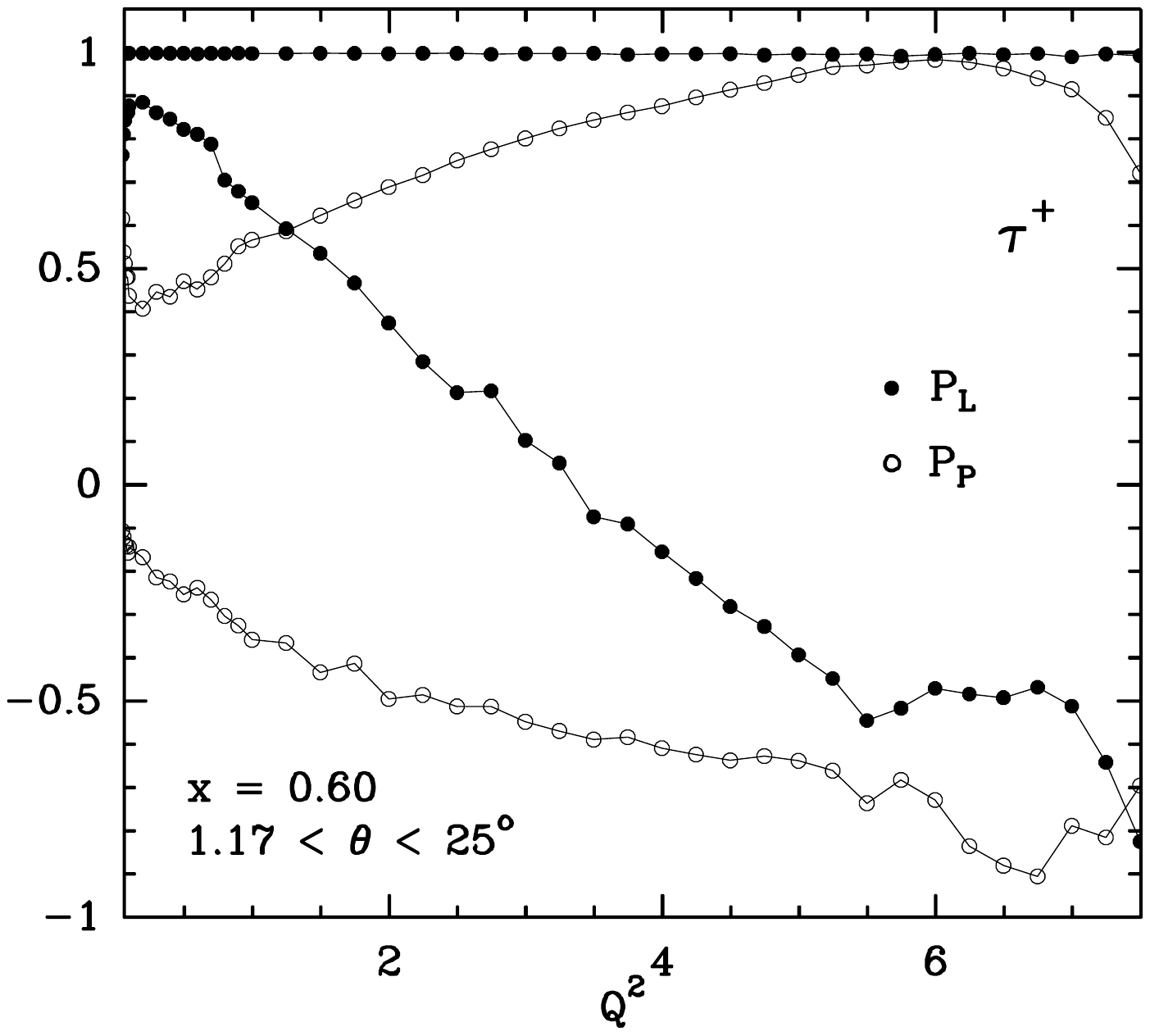}\\[5pt]
 \includegraphics[width=.9\textwidth]{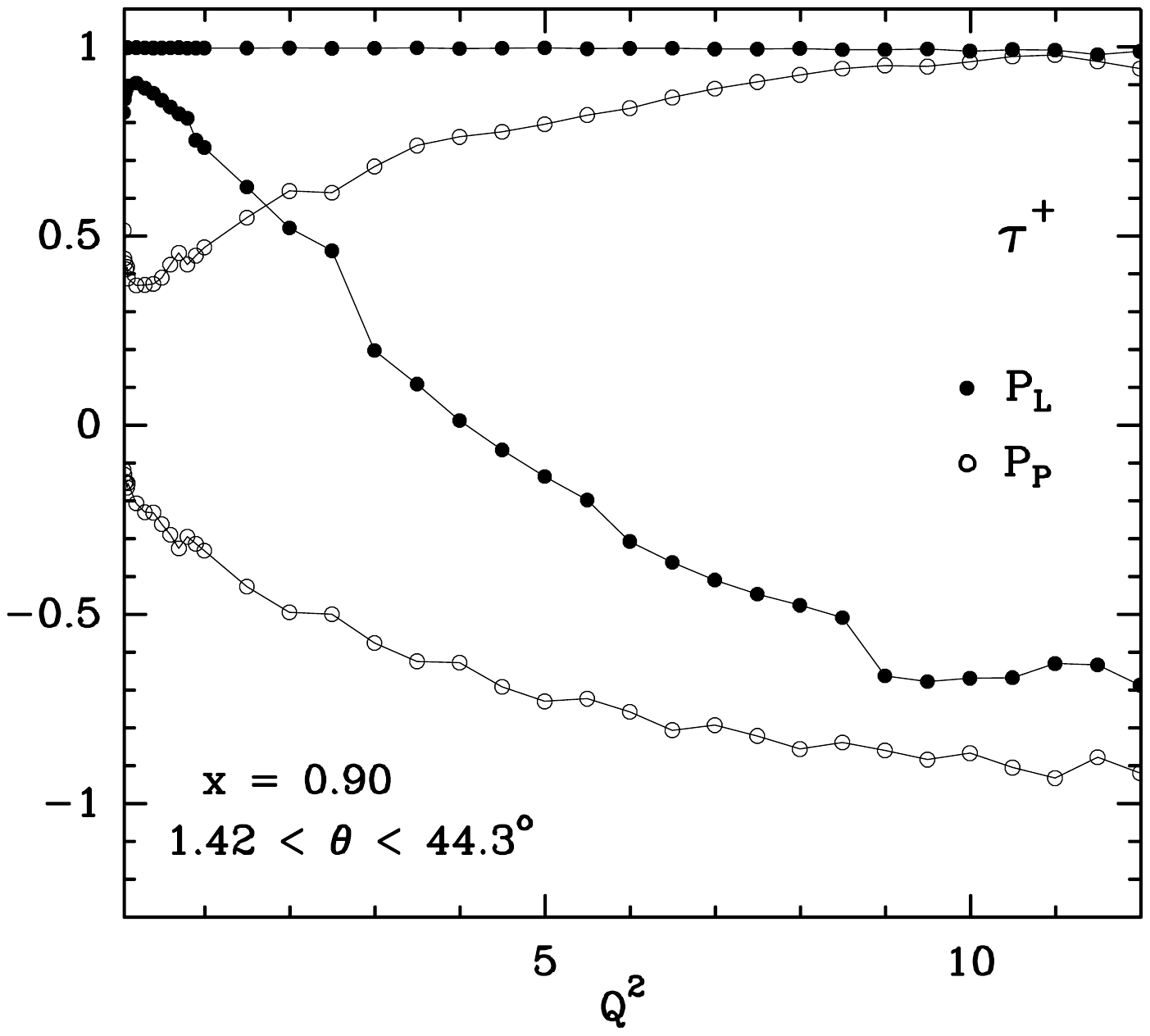}
\end{minipage}
\begin{minipage}{.49\textwidth}
\caption{\label{incl:fig:2}For $\tau^+$ production, upper and lower bounds 
on $P_P$ (open circles) and  $P_L$ (full circles) as a function of $Q^2$ for 
$E_{\nu} = 10\,\mbox{GeV}$ and $x = 0.25, 0.6, 0.9$.}
\end{minipage}
\end{figure}

Another way to present our results is seen in Fig.~\ref{incl:fig:2}, which shows the upper and
lower bounds from the non trivial positivity constraints for a given incident
energy and different $x$ values, versus $Q^2$.
These bounds are obtained by selecting the larger and smaller allowed values
of $P_L$ and $P_P$, when the $W_i$'s are varied for a fixed bin
of $E_{\nu}$ and $x$.
We also indicate the scattering angle which increases with $Q^2$ and we recall
that for $\theta=0$ we have $P_P=0$ (see Eq.~(\ref{incl:eq:polvec})).

\begin{figure}[!thb]
\begin{center}
\includegraphics[width=.35\textwidth]{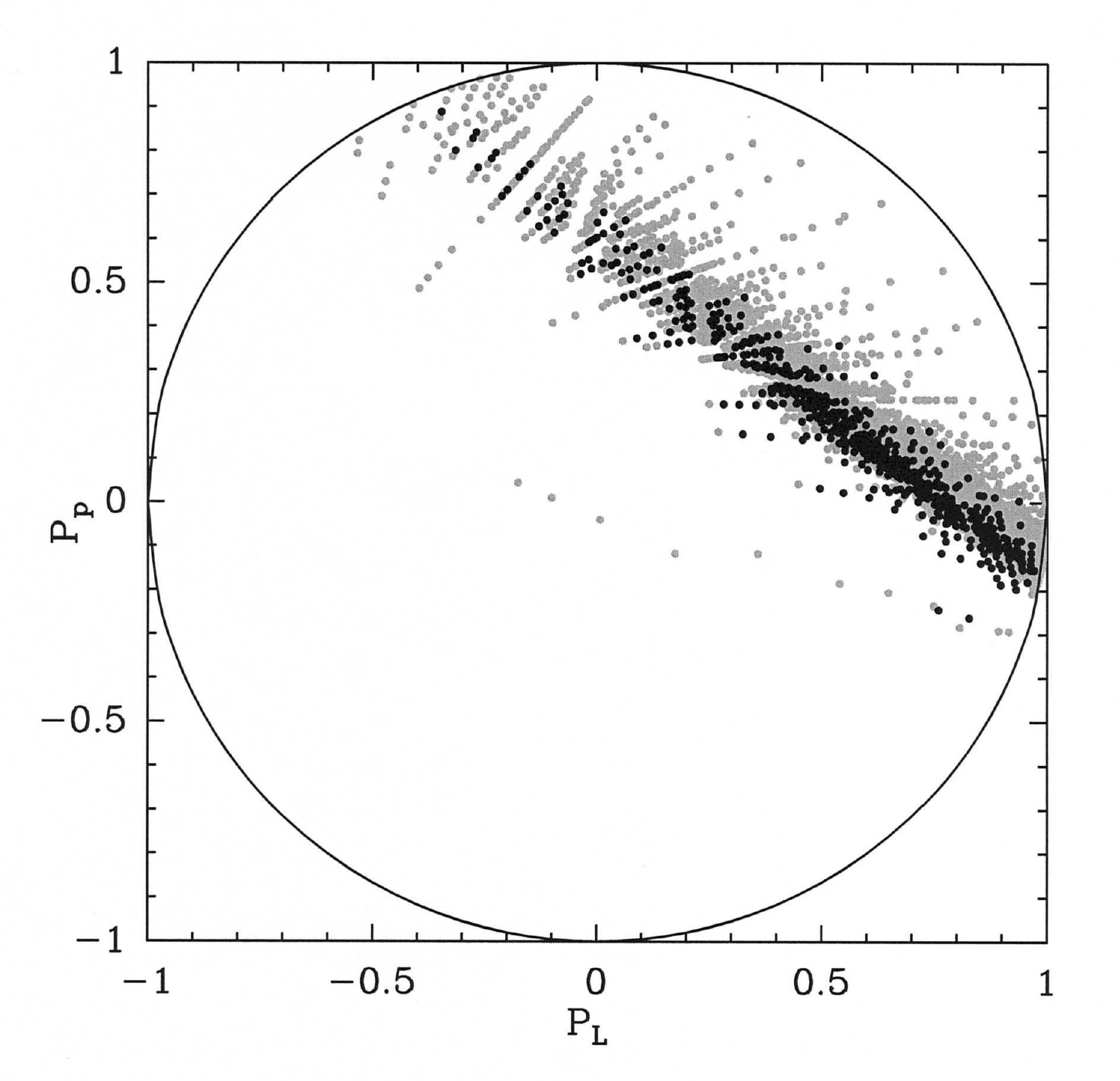}\hspace*{.2\textwidth}
  \includegraphics[width=.35\textwidth]{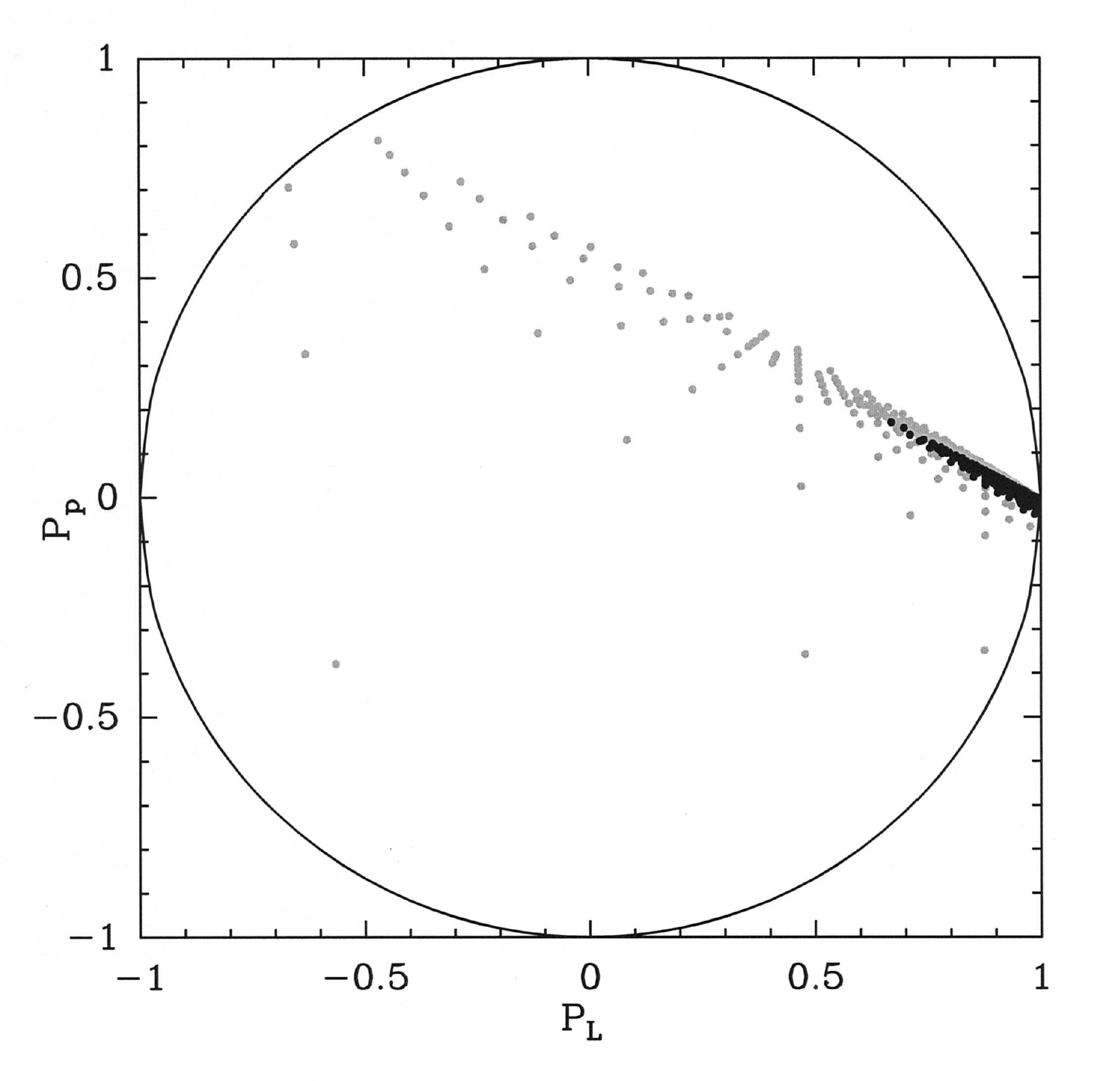}

  \includegraphics[width=.35\textwidth]{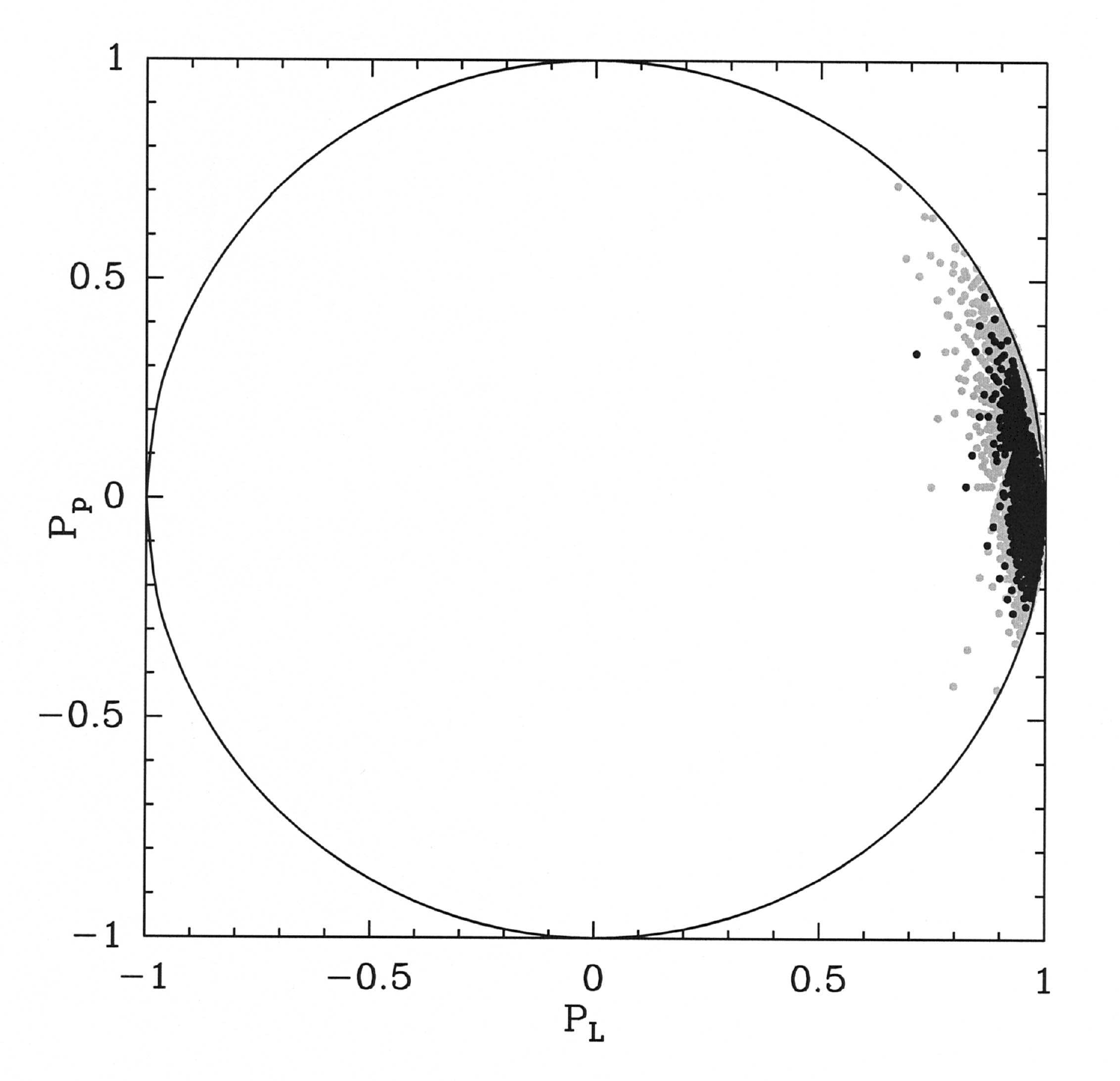}\hspace*{.2\textwidth}
  \includegraphics[width=.35\textwidth]{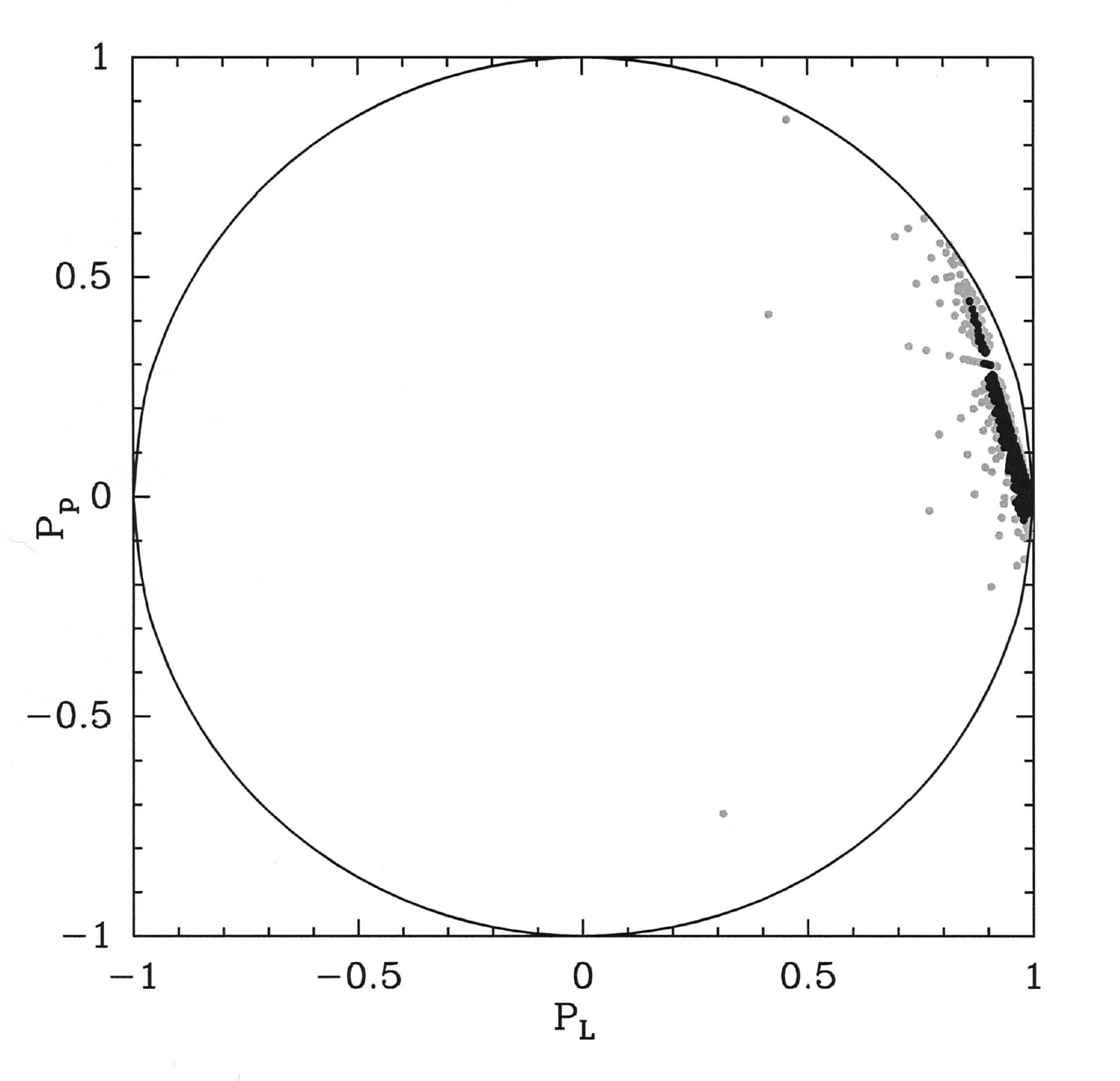}
  
  \includegraphics[width=.35\textwidth]{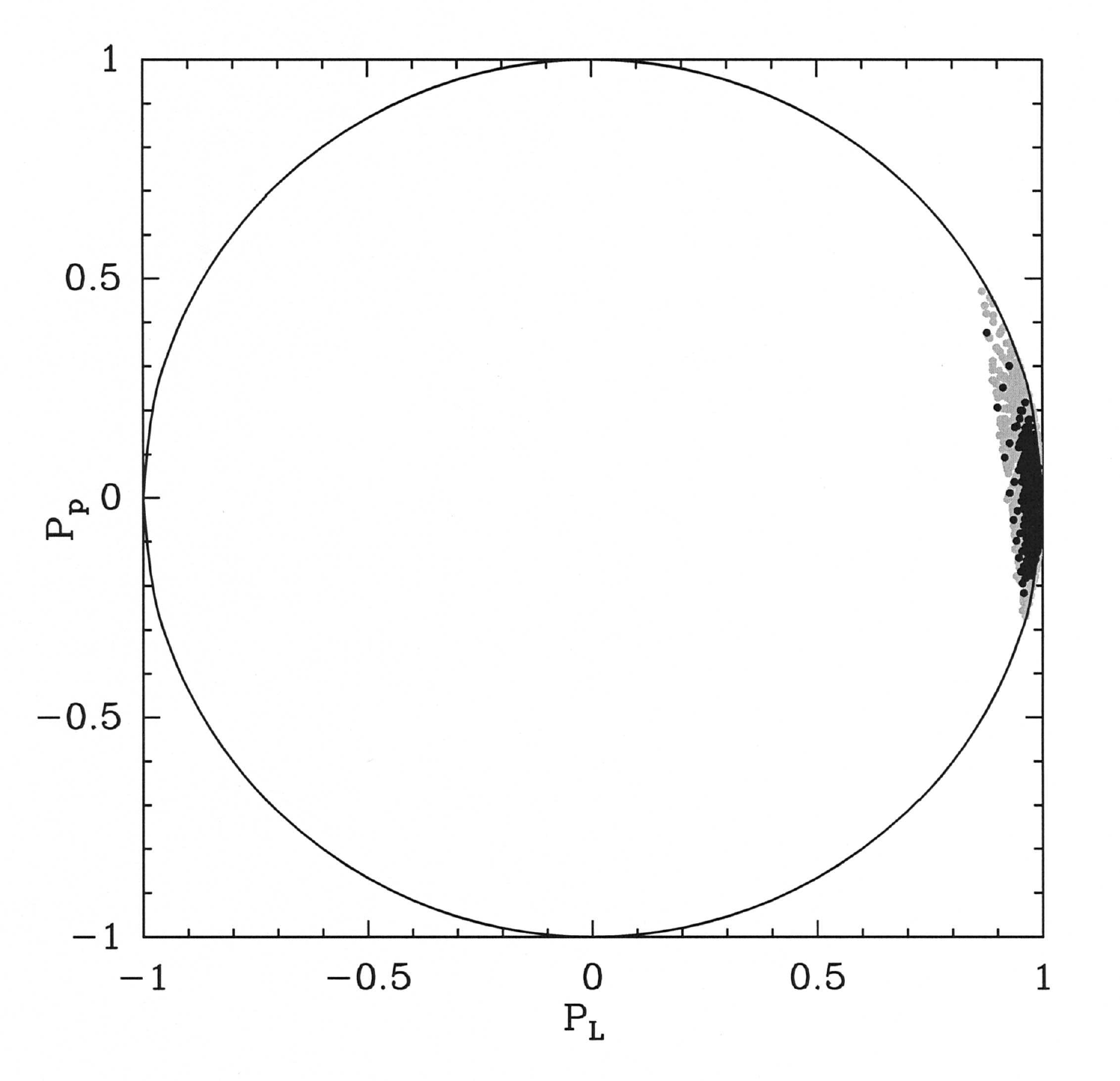}\hspace*{.2\textwidth}
  \includegraphics[width=.35\textwidth]{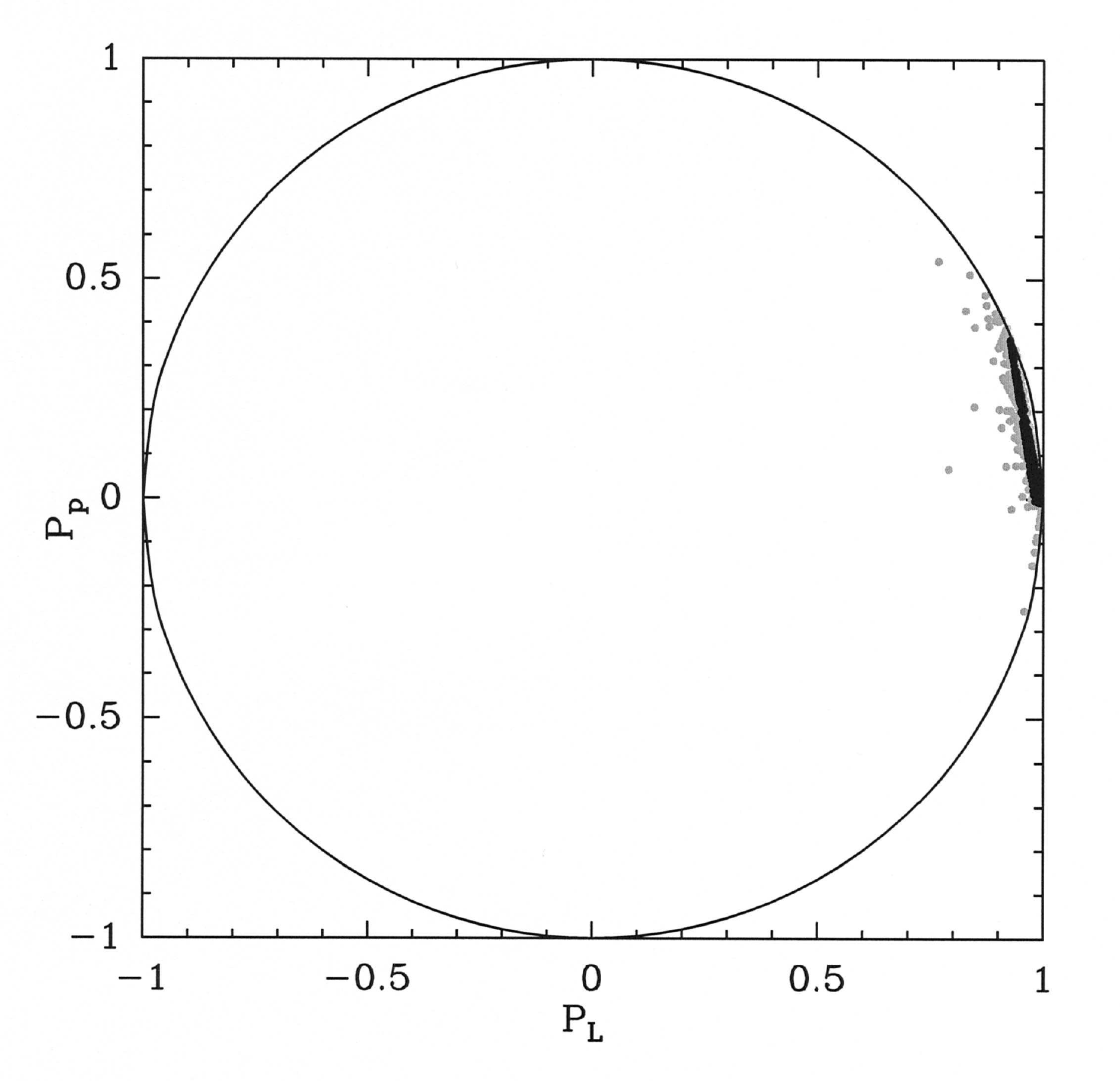}
  \end{center}
  \caption{\label{incl:fig:3}For $\tau^+$ production, $P_P$ versus $P_L$ in a domain 
  limited by $R_{+} \geq 0$, $P \leq 1$ assuming the Callan--Gross (left) or the Albright--Jarlskog (right)
  relations (grey dots inside disk) plus non trivial positivity constraints (black dots inside disk).
$E_{\nu} = 10\,\mbox{GeV},~Q^2 = 1\,\mbox{GeV}^2$, from top to bottom,
$x = 0.25, 0.6, 0.9$.}
  \end{figure}

Finally, as an example for a particular kinematic situation we show in Fig.~\ref{incl:fig:3}
the effect of imposing the Callan--Gross relation \cite{Callan:1969uq},
namely $Q^2W_1=\nu^2W_2$. It further reduces both the grey dots and the black
dots areas, since this has to be compared with the first row of Fig.~\ref{incl:fig:1}.
For the same kinematic situation we also show in Fig.~\ref{incl:fig:3}, the effect of the
Albright-Jarlskog relations \cite{Albright:1974sv}, namely $MW_1=\nu W_5$ and $W_4 =0$,
and we observe again that the allowed regions are much smaller.
This example illustrates the fact that a more precise knowledge of the
structure functions $W_i$'s, will certainly further restrict the domains
shown in Fig.~\ref{incl:fig:1}.

\subsubsection{Photon structure functions}
Let us now investigate the model-independent
constraints for the structure functions of virtual (off-shell) and
real (on-shell) photon target. In Refs.~\cite{Sasaki:2001pc, Sasaki:2002zd}, three positivity conditions for
the virtual photon case and one condition for the real photon were derived, the
latter of which relates the polarised and unpolarised structure functions.

\begin{figure}[!!htc]
\begin{center}
\includegraphics[width=6cm]{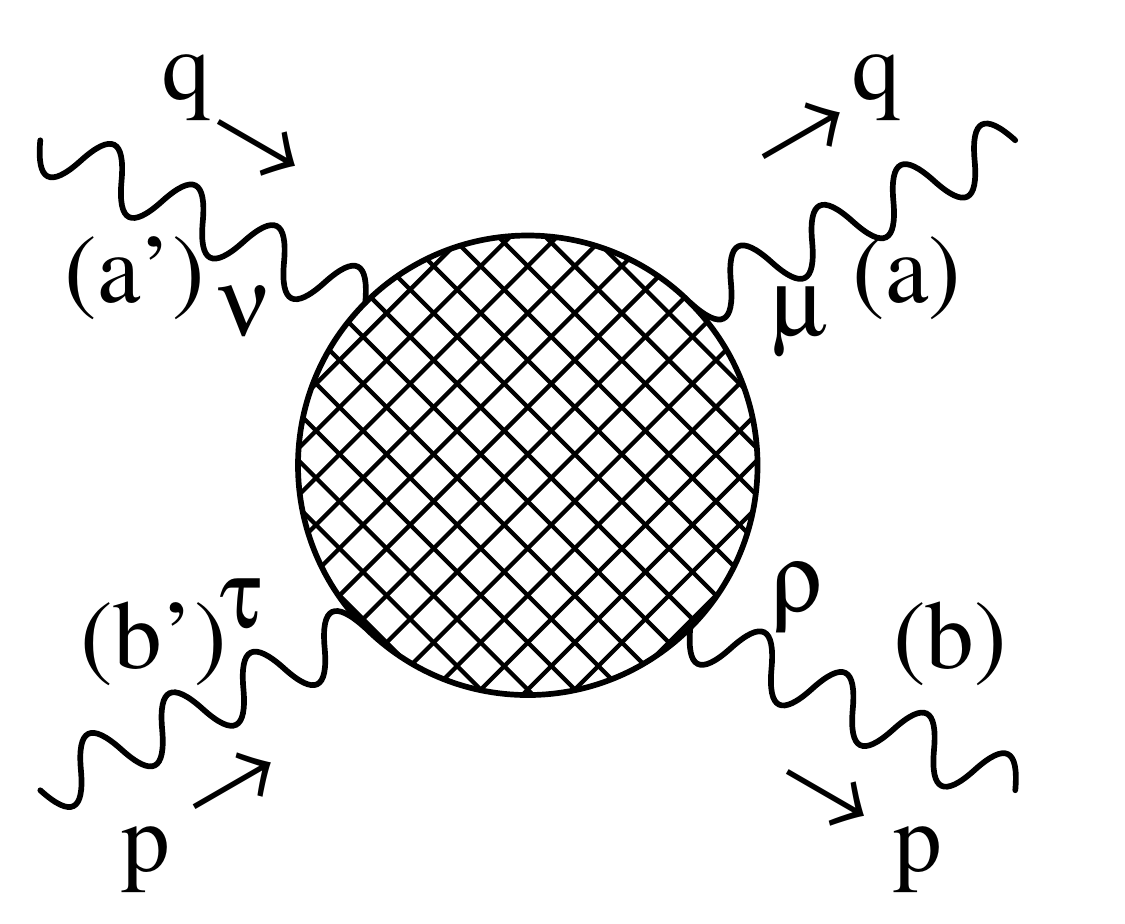}
\end{center}
\caption{\label{incl:fig:fig1photon}
Virtual photon-photon forward scattering with momenta $q(p)$
and helicities $a(b)$ and $a'(b')$.
}
\end{figure}

Consider the virtual photon-photon forward scattering
$\gamma(q)+\gamma(p)\rightarrow \gamma(q)+\gamma(p)$ illustrated in Fig.~\ref{incl:fig:fig1photon}.
The $s$-channel helicity amplitudes are given by
\be
W(a,b\vert a',b')=\epsilon^*_\mu(a)\epsilon^*_\rho(b)
W^{\mu\nu\rho\tau}\epsilon_\nu(a')\epsilon_\tau(b')~,
\ee
where $p$ and $q$ are the four-momenta of the incoming and outgoing photons and
$\epsilon_\mu (a)$ represents the photon polarisation
vector with helicity $a = 0,\pm1$. Similarly for the other polarisation vectors we have  and $a', b, b'=0, \pm1$.
Due to the angular momentum conservation, $W(a,b\vert a',b')$ vanishes unless it
satisfies the condition $a-b=a'-b'$. Parity conservation and time reversal
invariance lead to the following properties for
$W(a,b\vert a',b')$~\cite{Bourrely:1980mr}:
\be\begin{aligned}
W(a,b\vert a',b') &= W(-a,-b\vert -a',-b')\quad&&\text{Parity conservation}~,\\
W(a,b\vert a',b') &= W(a',b'\vert a,b) &&\text{Time reversal invariance}~.
\end{aligned}\ee
Thus in total we have eight independent $s$-channel helicity amplitudes,
which we may take as $W(1,1\vert 1,1)$,~
$W(1,-1\vert 1,-1)$, ~$W(1,0\vert 1,0)$, ~$W(0,1\vert 0,1)$, ~$W(0,0\vert 0,0)$,
~$W(1,1\vert -1,-1)$, ~$W(1,1\vert 0,0)$ and ~$W(1,0\vert 0,-1)$.
The first five amplitudes are helicity-nonflip  and the rest are
helicity-flip. We note that $s$-channel helicity-nonflip amplitudes
are semi-positive, but not the heli\-city-flip ones. Moreover
corresponding to these three helicity-flip amplitudes,
we will obtain three non-trivial positivity constraints.

The helicity amplitudes may be expressed in terms of the transition
matrix elements from the state $|a,b\rangle$ of two virtual photons with
helicities $a$ and $b$, to the unobserved state $|X\rangle$ as
\be\begin{aligned}
W(a,b\vert a,b) &= \sum_X |\langle X|a,b\rangle|^2,   \\
W(a,b\vert a',b') &= \RE\sum_X \langle X|a,b\rangle^{*} \langle X|a',b'\rangle\quad\mbox{for}\quad (a\neq a', b\neq b')~.
\end{aligned}\ee
Then, a Cauchy--Schwarz inequality~\cite{Soffer:1994ww,Soffer:2000zd}
\be
\sum_X \Bigl| \langle X\vert a,b\rangle +\alpha
\langle X\vert a',b'\rangle \Bigr|^2 \geq 0~,
\ee
which holds for an arbitrary real number $\alpha$, leads to
a positivity bound for the helicity amplitudes, \ie,
$\Bigl|W(a,b\vert a',b')  \Bigr|\leq \sqrt{W(a,b\vert a,b)
W(a',b'\vert a',b')}$~.
Writing down explicitly, we obtain
the following three positivity constraints:
\be\label{incl:eq:CS123}\begin{aligned}
\Bigl|W(1,1\vert -1,-1)  \Bigr|&\leq W(1,1\vert 1,1)~,\\
\Bigl|W(1,1\vert 0,0)  \Bigr|&\leq \sqrt{W(1,1\vert 1,1)
W(0,0\vert 0,0)}~,\\
\Bigl|W(1,0\vert 0,-1)  \Bigr|&\leq \sqrt{W(1,0\vert 1,0)W(0,1\vert 0,1)}~.
\end{aligned}\ee
In terms of the eight independent amplitudes introduced by Budnev, Chernyak
and Ginzburg \cite{Budnev:1971sz}, the $s$-channel helicity amplitudes are
related as follows
\be\begin{aligned}
W^{\tau}_{TT}&= W(1,1\vert -1,-1)~,\quad
&W_{TT}=&\frac{1}{2}\left[W(1,1\vert 1,1)+W(1,-1\vert 1,-1)   \right]~, \\
W_{ST}&= W(0,1\vert 0,1)~,\quad &W^a_{TT}=&\frac{1}{2}\left[W(1,1\vert 1,1)-W(1,-1\vert 1,-1)   \right]~, \\
W_{TS}&=W(1,0\vert 1,0)~, \quad &W^{\tau}_{TS}=& \frac{1}{2}\left[W(1,1\vert 0,0)-W(1,0\vert 0, -1)   \right]~,\\
W_{SS}&=W(0,0\vert 0,0)~,\quad
&W^{\tau a}_{TS}=&\frac{1}{2}\left[W(1,1\vert 0,0)+W(1,0\vert 0, -1)   \right] ~. 
\end{aligned}\ee
Since the helicity-nonflip amplitudes are non-negative, the first four structure
functions are  positive definite and the last four are not.
 
The above three conditions Eqs.~(\ref{incl:eq:CS123}) can be rewritten as
\be\label{incl:eq:BCG123}\begin{aligned}
\Bigl|W_{\rm TT}^\tau \Bigr|&\leq\left(W_{\rm TT}+W_{\rm TT}^a\right)~,\\
\Bigl|W_{\rm TS}^\tau +W_{\rm TS}^{\tau a}  \Bigr|&\leq\sqrt{(W_{\rm TT}+W_{\rm TT}^a)W_{\rm SS}}~,\\
\Bigl| W_{\rm TS}^\tau -W_{\rm TS}^{\tau a} \Bigr|&\leq\sqrt{W_{\rm TS}W_{\rm ST}}~,
\end{aligned}\ee
where T and S refer to the transverse and longitudinal
photon, respectively, and the superscripts "$\tau$" and "$a$" imply the 
relevance
to the helicity-flip  amplitudes and polarised ones, respectively.
For the real photon, $p^2=0$, the number of
independent helicity amplitudes reduces to four. They are
$W(1,1|1,1)$, $W(1,-1|1,-1)$,
$W(0,1|0,1)$, and $W(1,1|-1,-1)$~, which are related to four
structure functions $W_i^{\gamma}$ as follows
\cite{Sasaki:1980ss}:
\be\begin{aligned}
&W(1,1|1,1)+W(1,-1|1,-1)=2W_1^\gamma~,\\
&W(0,1|0,1)=-W_1^\gamma+(p. q)^2W_2^\gamma/Q^2~,\\
&W(1,1|-1,-1)=2W_3^\gamma~,\\
&W(1,1|1,1)-W(1,-1|1,-1)=2W_4^\gamma~,
\end{aligned}\ee
where the last one is the polarised structure function and
usually denoted by $g_1^\gamma$ with
$W_4^\gamma=g_1^\gamma/2$~. Also the first one, $W_1^\gamma$,
is often referred to as $F_1^\gamma$ with $W_1^\gamma=F_1^\gamma/2$~.

For the real photon case we have only one constraint, \ie,,
the first inequality of Eq.~(\ref{incl:eq:CS123}), which is rewritten as
\be
2|W_3^\gamma| \leq (W_1^\gamma +W_4^\gamma)~.\label{incl:eq:realbound}
\ee
It is interesting to recall that the polarised structure function $W_4^\gamma$ of
the real photon  satisfies a remarkable sum rule~\cite{Bass:1998bw}
\be
\int_0^1 W_4^\gamma (x, Q^2)\,\d x=0~.
\ee
The integral of $|W_3^\gamma|$ is, therefore, bounded from above by the first
moment of $W_1^\gamma$,
\be
\int_0^1 |W_3^\gamma (x, Q^2)| \,\d x\leq \frac{1}{2}\int_0^1 W_1^\gamma (x,Q^2)\,\d x~.
\ee
These constraints could be studied in future experiments on the photon structure, through the
two-photon processes in $\ep\e$ collisions as well as the resolved photon processes in a 
$\e\p$ collider. The validity of the inequality Eq.~(\ref{incl:eq:realbound}) has been checked in a simple
parton model, as discussed in Ref.~\cite{Sasaki:2001pc}.

In the case of virtual photon, $p^2=-P^2\neq 0$, there appear
eight structure functions (four of them are new) and we have derived three
positivity  constraints on these functions Eqs.~(\ref{incl:eq:BCG123}).
However, up to now very little
attention has been paid to the virtual photon case and therefore,
we have almost no knowledge of the new photon structure functions. In this
situation it is worthwhile to investigate these new structure functions in
the simple parton model to find out whether or not the three positivity constraints
actually hold. They were proven to be satisfied in Ref.~\cite{Sasaki:2002zd}, both in 
a massive and massless quark cases, although the two cases show quite different
behaviours.
\subsection{Positivity constraints for parton distributions and fragmentation functions}
\label{incl:sub:pdpd}
In hard QCD processes, the transition from hadrons to quarks (antiquarks) and gluons is
specified in terms of distribution functions and the transition from quarks (antiquarks)
and gluons to hadrons is specified in terms of fragmentation functions. For example, 
the cross section for inclusive electroproduction $\mathrm{e}H \to \mathrm{e'}X$,  is given as a charge squared
weighted sum over quark (and antiquark) distribution functions $q(x)$, describing the 
probability of finding quarks inside hadron $H$. Similarly, in electron-positron annihilation,
the one-particle inclusive cross section $\ep\e \to h X$ is given as a charge squared weighted
sum over quark (and antiquark) fragmentation functions $D_q(z)$, describing the decay
of the produced quarks (and antiquarks) into hadron $h$.
\subsubsection{Parton distributions and fragmentation functions (integrated over transverse momentum)}\label{incl:sub:PD}
The parton distribution functions (PDF) constitute the large-distance
non-perturbative QCD inputs entering the expressions for the DIS
structure functions, as well as for other hard QCD processes, in
particular Semi-Inclusive DIS (SIDIS), Drell-Yan (DY) process and
high $p_T$ hadron production in hadronic collisions.

For the derivation of positivity constraints for
the PDF, we will combine the general operator method \cite{Jaffe:1991kp}
with the approach based on
helicity amplitudes, closely related to the consideration of structure
functions, as used in Sec.~\ref{incl:sub:pcsf}.
\paragraph{i) - Unpolarised and polarised PDF}

Let us start the discussion with the simplest case of
unpolarised quark distribution, which is, by introducing the
light-cone decomposition of the quark fields
\cite{Jaffe:1991kp}, analogous to the scalar case. The quark forward distribution is just
\be\la{incl:eq:q(x)}
q(x)=\int{\d \lambda \over{4\pi}}\exp({i\lambda x})\langle p|\bar\psi(0) \slashed{n}% /\!\!\! n
 \psi(\lambda n)|p\rangle
=
{1 \over{\sqrt{2}\,p^+}}
\int{\d \lambda \over{2\pi}} \exp({i\lambda x})\langle p|\phi^\dagger(0)\phi(\lambda n)|p\rangle~,
\ee
where $\phi$ is the good component of the quark field and the light-cone
vectors, $p$ and $n$ are normalised such as $p^2=0$, $n^2=0$, $p^-=n^+=0$, $pn=p^+ n^-=1$. By inserting
a complete set of intermediate states $|X\rangle$ of momentum $p_X$
and making use of the generalised optical theorem
and the fact that the matrix elements may be replaced by their
imaginary parts \cite{Diehl:1998sm}, the forward
distribution can be written as
\be
q(x)=\sum_{X} {1 \over{\sqrt{2}\,p^+}} |\langle p|\phi^\dagger(0)|X\rangle |^2
\delta(x-(p-p_X).n)~.
\la{incl:eq:av}
\ee
A comment is in order here. The states $|X\rangle$ are in fact coloured
ones and therefore unphysical. One should therefore have in mind
that the full unobserved state contains also the jet in the
corresponding hard subprocess. The QCD factorisation implies the
decomposition $|X\rangle \to |X_h\rangle |X_s\rangle $ to the states in the hard and
soft jets. The soft colour neutralisation implies, in turn, the
possibility to consider them separately. Therefore, in what
follows these states should be understood as formal objects rather than
physical objects, which is convenient for studies of positivity
properties. Indeed, the expression (\ref{incl:eq:av}) guarantees immediately 
the positivity of the spin-averaged quark distribution
\be
q (x) \geq 0~.
\la{incl:eq:avp}
\ee
In order to match the general framework of density matrix positivity (Sec.~\ref{basic:sub:dens}), 
one may interpret the quantity $q(x)$ as the \emph{diagonal element} $q(x,x)$ of a generalised parton distribution matrix $q(x_1,x_2)$, which has infinite dimension, the continuous partonic variable
$x\in[0,1]$ replacing the discrete index $n\in[1,N]$.

Note that our analysis can be simply extended to the case of \emph{fragmentation
functions} $D_q(z)$, describing the process $q \to h X$, which correspond to 
the following modification of Eq.~(\ref{incl:eq:av})
\be
D_q (z)=\sum_{X} {1 \over{\sqrt{2}\,p^+}} |\langle0|\phi(0)|p,X\rangle|^2 \,\delta(1/z-(p-p_X).n)~.
\la{incl:eq:ad}
\ee
Here $z = p/k$, where $k$ and $p$ are the quark and hadron $h$ momenta, respectively.
As a result the spin averaged fragmentation function is also positive
\be
D_q (z) \geq 0~,
\la{incl:eq:adp}
\ee
and all other positivity constraints which we will be considered for distribution
functions, will have their counterparts for fragmentation functions (see, \eg, Ref.~\cite{deFlorian:1998am}).

The generalisation to the case of a longitudinally polarised
proton of spin $S_L$ may be easily achieved, by considering
the PDF with definite helicities $(\pm)$
\be\label{incl:eq:qpm(x)}\begin{aligned}
q_{\pm}(x)&=\int{\d \lambda \over{4\pi}}\exp({i\lambda x})\langle p,S_L|\bar\psi_{\pm}(0) \slashed{n}
 \psi_{\pm}(\lambda n)|p,S_L\rangle\\
&={1 \over{\sqrt{2}p^+}}
\int{\d \lambda \over{2\pi}} \exp({i\lambda x})\langle p,S_L|\phi_{\pm}^\dagger(0)\phi_{\pm}
(\lambda n)|p,S_L\rangle~,
\end{aligned}\ee
where $\psi_{\pm} = (1+\gamma^5) \psi /2$ and $\phi_{\pm}$ are the respective
good components, so that $q(x)=q_+(x)+q_-(x)$.
It is now straightforward to get,
in complete analogy to (\ref{incl:eq:av}), the positivity of spin-dependent
distributions which reads
\be
q_{\pm} (x)=\sum_{X} {1 \over{\sqrt{2}p^+}} |\langle p,S_L|\phi^\dagger_{\pm}(0)|X\rangle|^2
\delta(x-(p-p_X).n)~ \geq 0~.
\la{incl:eq:avpm}
\ee
It is clear that now the helicity distributions correspond to diagonal elements 
of the density matrix, in the $x\otimes$ helicity basis. By introducing the more commonly used
spin-dependent distribution $\Delta q(x)=q_+(x)-q_-(x)$,
one has
\be
q (x) \geq |\Delta q(x)|~.
\la{incl:eq:avp1}
\ee
\paragraph{ii) - Quark transversity distribution and Soffer inequality}

Let us introduce the new ingredient in our analysis,
corresponding to the possibility of non-diagonal elements of density matrices.
The best known example is represented by the case of quark helicity
flipping chiral-odd transversity distribution, defined as
\be\label{incl:eq:dq(x)}\begin{aligned}
\delta q(x)&=\int{\d \lambda \over{4\pi}}\exp({i\lambda x})\langle p,S_T|\bar\psi(0)
\slashed{n} S \gamma_5  \psi(\lambda n)|p,S_T\rangle\\
&= \RE \sum_{X} {1 \over {\sqrt{2 p^+}}}\langle p,+|\phi^\dagger_+(0)|X\rangle 
\langle p,-|\phi^\dagger_-(0)|X\rangle^* \delta(x-(p-p_X).n)~.
\end{aligned}\ee
where $S_T$ stands for the spin of the transversely polarised proton.
%
%The positivity bound in this case is provided by the Cauchy--Schwarz (CS) inequality
%which can be derived by considering the following positive quantity
%%
%\be\label{incl:eq:CSh}
%\sum_{X}|\langle p,+|\phi^\dagger_+(0)|X\rangle  \pm a\langle p,-|\phi^\dagger_-(0)|X\rangle|^2 
%\delta(x-(p-p_X).n) \geq 0~.
%\ee
%%
%Here $a$ is a positive number, which we put equal to 1 for the time being.
%While the interference term of (\ref{incl:eq:CSh}) is producing just the
%transversity distribution, both the  diagonal terms - the distribution $q_+(x)$. The equality of diagonal terms is the 
%immediate consequence of parity, as the simultaneous change of quark and hadron helicity
%(\ie, different quark helicities pick up different helicities from $S_T$, as  required by angular momentum conservation) does 
%not affect the result, if the summation over all the unobserved states is performed.
%Therefore the minimal value of the sum of diagonal terms is provided by $a=1$. The result is just the Soffer inequality \cite
%{Soffer:1994ww}
%%
%\be
%q_+ (x)= (q(x)+\Delta q (x))/2  \geq |\delta q(x)|~.
%\la{incl:eq:avp2}
%\ee
%%=
The Soffer inequality \cite{Soffer:1994ww}
\be
q_+ (x)= (q(x)+\Delta q (x))/2  \geq |\delta q(x)|~,
\la{incl:eq:avp2}
\ee
%
%This inequality 
is especially transparent, if the parton distributions are
considered as the following bilinears of quark--hadron helicity amplitudes, corresponding $N \to q + X$,
\be\la{incl:eq:def}\begin{aligned}
q(x) &=\sum_{X} |\langle+1/2|{+1/2},X\rangle|^2 + |\langle+1/2|{-1/2},X\rangle|^2~,\\
\Delta q(x) &=\sum_{X} |\langle+1/2|{+1/2},X\rangle|^2 - \sum_{X} |\langle+1/2|{-1/2},X\rangle|^2~, \\
\delta q(x)&=2\RE\sum_{X} \langle+1/2|{+1/2},X\rangle \langle-1/2|-1/2,X\rangle^*~,
\end{aligned}\ee
where the unobserved state $|X\rangle$ is the same as discussed above.
Now Soffer's inequality follows from the positivity of
\be
\sum_{X}|\langle +1/2|+1/2,X\rangle \pm a \langle -1/2|{-1/2},X\rangle |^2 \geq 0~.
\la{incl:eq:csb}
\ee
Note that only the amplitudes $|h,q\rangle$, with the 
same quark and hadron helicities enter (to) this equation
due to helicity conservation
resulting from angular momentum conservation in collinear kinematics.
As the total helicity
of both amplitudes is equal to the projection of angular momentum of the state $|X\rangle$,
the quark helicity flip in the interfering amplitudes should be accompanied by the
hadron helicity flip. This explains why only the distribution 
$q_+ =\sum_{X} |\langle+1/2|+1/2,X\rangle|^2=
\sum_{X} |\langle-1/2|-1/2,X\rangle|^2$ enters the Soffer inequality.

\begin{figure} 
\centerline{\includegraphics[width=.5\textwidth]{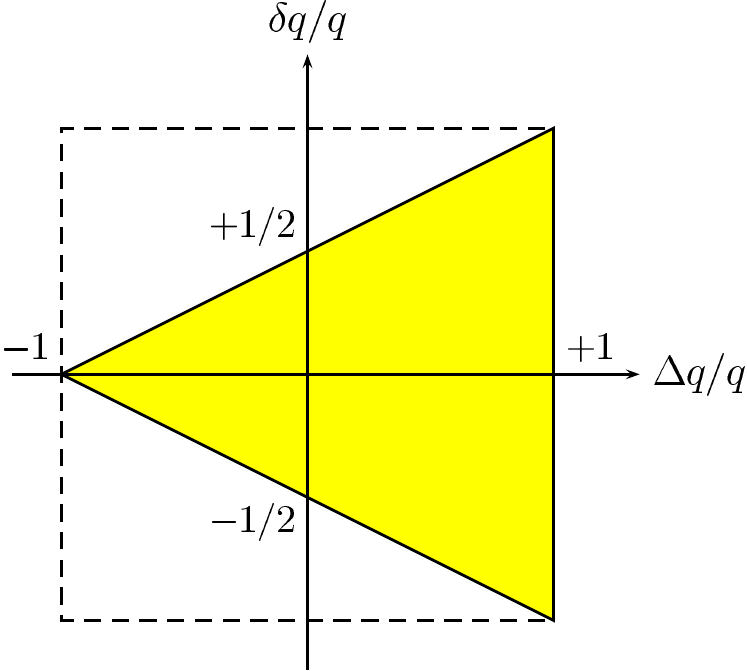}}
\caption{The allowed domain corresponding to the constraint Eq.~(\protect\ref{incl:eq:avp2}).}
\label{incl:fig:soffer}
\end{figure}

For completeness we show in Fig.~\ref{incl:fig:soffer} that the region restricted by Eq.~(\ref{incl:eq:avp2})
is only half of the entire square.
\paragraph{iii) - Bounds for different polarised gluon distributions}
One can easily generalise these considerations to
the case of offshell gluons, when respective distributions are also related to the helicity amplitudes, corresponding to $N \to G + X$,
\be\la{incl:eq:defG}\begin{aligned}
G_{\pm} (x) &=\sum_{X} |\langle+1/2|{\pm 1},X\rangle|^2~, \cr
G_L(x)&=\sum_{X} |\langle+1/2|0,X\rangle |^2=\sum_{X}|\langle-1/2|0,X\rangle|^2~,\cr
\Delta G_{T}&= \RE \sum_{X} \langle+1/2|{+1},X\rangle \langle-1/2|0,X\rangle^*~,
\end{aligned}\ee
where $G_{\pm} (x)=(G(x) \pm \Delta G(x))/2$ are combinations of the well known spin-averaged $G$
and longitudinally polarised $\Delta G$ gluon distributions, $\Delta G_T$ is the counterpart
of the latter for transverse polarised nucleon \cite{Soffer:1997zy}, while $G_L$ is the distribution
of longitudinal (scalar) gluons in a nucleon \cite{Gorsky:1990wv}. $\Delta G_T$ is related to the distributions $H_1$ and $H_2$ introduced in \cite{Ali:1992qj}. 
We can now write the CS inequality as
\be
\sum_{X}|\langle+1/2|{+1},X\rangle \pm a\langle-1/2|0,X\rangle|^2 \geq 0~.
\la{incl:eq:CSG}
\ee
By performing the minimisation with respect to the variation of $a$, which was unnecessary in the case of Soffer inequality, 
because the minimum was provided for $a=1$, we obtain the following bound
\be
|\Delta G_T(x)| \leq \sqrt{ G_+(x)G_L(x)}~.
\la{incl:eq:ineq}
\ee
This was first derived \cite{Soffer:1997kx} in a weaker form with $G$ instead of $G_+$,
whose appearance is again due to helicity conservation, since only the state
with parallel helicities of nucleon and gluon may interfere with
some other helicity state which is $|{-1/2},0\rangle$. This bound is completely similar to the bound
for $A_2$ mentioned in Sec.~\ref{incl:sub:pcsf}.

Note that because the scalar gluon polarisation is not enhanced by Lorentz boost
contrary to the transverse one, it produces the distributions of higher twist-three ($\Delta G_T$) and twist-four ($G_L$) 
which are related by positivity constraints, like in the case of $A_2$. It is most instructive to use it to bound the less known 
twist-four distribution from above \cite{Soffer:1997kx}
\be
G_L(x) \geq [\Delta G_T(x)]^2/G_+(x)~.
\la{incl:eq:bound}
\ee

Another application of similar reasoning may be used in the case of twist-three single spin asymmetries
(SSA), which are enhanced at large $x$, to deduce that twist-four corrections to unpolarised 
cross-sections should be also large in this region \cite{Teryaev:1998gn}.
This may explain the observation \cite{Bourrely:2003bw} that the NLO QCD calculation largely underestimates the
pion production in unpolarised hadronic collisions at lower energies, where large SSA are observed, whereas it
describes it well at higher energies, where SSA are smaller.
Indeed, the lower energies at the same $p_T$ mean the higher typical partonic $x$.
In that region enhanced twist-three leads to SSA while simultaneously
enhanced, because of positivity, twist-four,
spoils the NLO QCD applicability. It is known that the account for intrinsic partonic transverse momentum
improves the description of spin-averaged  data \cite{D'Alesio:2004up}. At the same time,
such intrinsic momentum 
may be considered \cite{Teryaev:2004df} as a model for infinite resummed tower of higher twists,
supporting the suggested picture. 

In order to avoid the contribution of scalar gluons polarisation and
to have gluonic helicity flip effects at leading twist level one should be able to
compensate the helicity flip two of the gluon, which is possible in the case of a spin one
target like the deuteron \cite{Jaffe:1989xy}. The {\it polarised} leading twist gluonic distributions in spin one hadron take the form
\be
\label{incl:eq:defG1}\begin{aligned}
G_{\pm}^1 (x)&= \sum_{X} |\langle+1|{\pm 1},X\rangle |^2~,\cr
\delta G_{T}^1(x)&= \RE\sum_{X} \langle+1|{+1},X\rangle \langle -1|{-1},X\rangle^*~,
\end{aligned}\ee
where $\delta G_{T}^1(x)= G_{x\!/\! x}(x)-G_{y\! /\! x}(x)$ and
$G_{n\! /\! x}(x)$ is the density of gluon linearly polarised along
$\vec{n}$, in a spin-one hadron linearly polarised along $\hat{\vec{x}}$. It can be
called the leading twist gluonic transversity; it has an evolution kernel close to
$\delta q(x)$ \cite{Artru:1989zv} and an analogue of the Soffer inequality
\be
G_+^1 (x)= (G^1(x)+\Delta G^1 (x))/2  \geq |\delta G_{T}^1(x)|~,
\label{incl:eq:avpG}
\ee
may be derived in complete similarity with its quark counterpart.
\subsubsection{Spin and transverse-momentum dependent quark distributions}\label{incl:sub:kt}
The spin- and $\vec k_T$-dependent quark distributions in the nucleon can be viewed as the probability of the virtual 
\emph{inclusive} process 
\be\label{incl:eq:N=q+X} % (\ref{eq:N=q+X})
{\rm nucleon}(\pv,\Sv) \to {\rm quark}(\kv,\Sv') + X \,,
\ee
with $\kv = x \pv + \kv_T$.
This reaction has the same spin content as the inclusive reaction (\ref{incl:eq:increa}). The $(\pv,\kv_T)$ plane is a 
symmetry plane and plays the same role as the scattering plane in (\ref{incl:eq:increa}). 
Therefore the spin observables of (\ref{incl:eq:N=q+X}) obey the same parity constraints and inequalities as (\ref
{incl:eq:increa}). The distribution of quarks of polarisation $\Sv'$ in a nucleon of polarisation $\Sv$ can be written in terms of 
the unpolarised quantity $q(x,|\kv_T|)$ and seven Cartesian parameters $D_{ij}$
\begin{multline}
{\d N\over \d x \, \d^2\kv_T}\equiv
q(x,\kv_T,\Sv,\Sv') = {q(x,|\kv_T|)\over2}\left[1+D_{0n} \ \Sv'.\nv +D_{n0} \ \Sv.\nv +D_{nn} \ \Sv.\nv\ \Sv'.\nv \right. \\
\left. {}+ D_{ll} \ \Sv.\lv\ \Sv'.\lv +D_{lm} \ \Sv.\lv\ \Sv'.\mv +D_{ml} \ \Sv.\mv\ \Sv'.\lv +D_{mm} \ \Sv.\mv\ \Sv'.\mv\right]~.
\label{densPol}
\end{multline}
The $D_{ij}$'s also are functions of $x$ and $|\kv_T|$. The dependence of the quark density with respect to the azimuth of 
$\kv_T$ at given $\Sv$ and $\Sv'$ is contained in the factors $\Sv.\mv$, $\Sv.\nv$, $\Sv'.\mv$ and $\Sv'.\nv$, where $\lv 
= \pv_T/|\pv| $, $\mv = \kv_T/|\kv_T| $ and $\nv = \lv \times \mv$.
The Cartesian parameters are related to the functions $f_1$, $f_{1T}^\perp$, etc., of the Amsterdam notation (see, \eg, \cite{Barone:2001sp,Bacchetta:1999kz}) by 
\be\label{incl:eq:Amsterdam}\begin{aligned}
f_{1} (x,k_T) &= q(x,k_T)\,,\\
g_{1} (x,k_T) &= q(x,k_T) \,  D_{ll}\,,\\
h_1(x,k_T) &= q(x,k_T)\,  (D_{mm}+D_{nn})/2\,,\\
(k_T^2/(2M^2))\, h_{1T}^\perp (x,k_T) &= q(x,k_T) \,  (D_{mm}-D_{nn})/2\,,\\
(k_T/M)\, g_{1T} (x,k_T) &= q(x,k_T) \,D_{ml}\,,\\
(k_T/M)\, h_{1L}^\perp (x,k_T) &= q(x,k_T) \, D_{lm}\,,\\
(k_T/M)\, h_{1}^\perp (x,k_T) &= -q(x,k_T) \,  D_{0n}\,,\\
(k_T/M)\, f_{1T}^\perp (x,k_T) &= q(x,k_T) \,  D_{n0}\,,
\end{aligned}\ee
where $k_T=|\kv_T|$. The integrals of the first three functions over $\kv_T$ give $q(x)$, $\Delta q(x)$ and $\delta q(x)$. 
The last function describes the Sivers effect \cite{Sivers:1989cc,Sivers:1990fh}. The $4^{\rm th}$ to $7^{\rm th}$ functions 
were introduced by Kotzinian \cite{Kotzinian:1994dv}, Mulders, Boer and Tangerman \cite{Boer:1997nt,Tangerman:1995hw}.  The $4^{\rm th}$ one, $h_{1T}^\perp$, has been given the name 
\emph{pretzelosity} \cite{Avakian:2008dz}.
The positivity constraints have been first derived in the Amsterdam notation by Bacchetta \etal~\cite{Bacchetta:1999kz}. 
Using the Cartesian parameters they take the simple form \cite{Artru:2004jx}
\begin{equation}\label{incl:eq:kT-ineg}
(1\pm D_{nn})^2\ge(D_{0n}\pm D_{n0})^2+(D_{ll}\pm D_{mm})^2+(D_{lm}\mp D_{ml})^2\,,
\end{equation}
together with the trivial bounds $|D_{ij}|\le1$. These inequalities are easily derived in  the transversity basis and have the same form as (\ref{incl:eq:DM}).

Similar formulas also apply to the quark fragmentation $q \to\hbox{baryon}+ X$, just by interchanging the quark and the nucleon. 

\paragraph{An example: the relativistic (quark + scalar diquark) model} 
See \cite{Artru:1989zv,Meyer:1990fr,Melnitchouk:1993nk,Suzuki:1997vu,Jakob:1997wg,Brodsky:2000ii}.
The simplest spectator system $X$ is a $0^+$ diquark, like the $(ud)$ pair of the $\Lambda$ baryon in the SU(6) model. 
Assuming a quark--baryon--diquark vertex of the form $g(k^2)\,\bar u(k,\Sv')\,u(p,\Sv)$, the quark polarisation $\Sv'$ is 
obtained from $\Sv$ by a rotation of angle $\theta=-2\arctan[k_T/ (m_q + x m_B)]$ in the $(\pv,\kv_T)$ plane, as seen from formula (43-44) of \cite{Artru:2002pu}. 
This model gives $D_{nn}=1$, $D_{ll}=D_{mm}=\cos\theta$, $D_{lm}=-D_{ml}=\sin\theta$ and predicts no Sivers or 
Boer--Mulders effect: $D_{n0}=D_{0n}=0$, thus satisfying Hermiticity and time reversal invariance (see Sec.~\ref{se:basic}). 
Note that this model exhibits a strong ``spin crisis" (small $D_{ll}$) at large $\langle k_T\rangle/M$. The positivity 
constraints (\ref{incl:eq:kT-ineg}) as well as the Soffer bound are saturated. A pure baryon spin state ($|\Sv|=1$) leads to a pure 
quark spin state ($|\Sv'|$=1). The reason is that a spin-zero spectator cannot take away any  spin information.
 This is no more the case when the model is enriched with $1^+$ spectator diquarks, as in
 \cite{Meyer:1990fr,Melnitchouk:1993nk,Suzuki:1997vu,Jakob:1997wg}, or in a more realistic model with gluons and sea quarks.  The corresponding kinematical domain corresponds to small $x$, and one expects Soffer's inequality far from saturation there.

Adding the gluon field also gives rise to \emph{non-intrinsic} Sivers  ($D_{n0}\ne0$) and Boer--Mulders 
($D_{0n}\ne0$) effects. In deep inelastic electron scattering, the Sivers effect is due to the final state 
interaction of the scattered quark with 
spectator partons \cite{Brodsky:2002rv,Ji:2002aa}.
In the Drell--Yan lepton pair production, the Sivers effect in the target is due to the initial state interaction 
between the active 
quark of the projectile with the spectators of the target, and its sign is opposite to the DIS one \cite
{Collins:2002kn}.
For references, see also \cite{Barone:2001sp}. A ``theoretical laboratory" \cite{Artru:2005rb,Artru:
2007xx}  to study these 
relativistic effects is the large-$Z$ hydrogen-like atom, where the electron plays the role of a parton. 
This model has many common features with the models describing the nucleon with a quark and a 
scalar diquark. It also possesses a photon cloud and 
an electron--positron sea, interpreted as the local deformation of the Dirac sea.

\paragraph{Role of the Sivers and Collins effects in single-spin asymmetries}
The various functions of (\ref{incl:eq:Amsterdam}) can be separated unambiguously by their different azimuthal dependences in the semi-inclusive deep inelastic scattering experiments being presently performed at HERMES \cite{Hasch:2007zz}, Jlab \cite{deJager:2006ia}, 
COMPASS \cite{Alexakhin:2005iw,Ageev:2006da,Bradamante:2006jw,Alekseev:2008dn}.
%\cite{Parsamyan:2007ju}. 
%%%
Their roles in the single-spin asymmetries like $\p\!\!
\uparrow+\p\to \pi+X$ at $p_T\sim 1-2\;$GeV is still controversial. A model based on the Collins effect and a large transversity distribution $\delta q(x)$ \cite{Artru:1993ad,Artru:1995bh} could reproduce the E704 experiment \cite{Adams:1991cs,Bravar:1996ki}, but was hardly compatible with the Soffer bound \cite{Boglione:1999dq}. An alternative model based on the Sivers effect gave reasonable results \cite{Anselmino:1994tv,D'Alesio:2004up}.  See, also, \cite{D'Alesio:2007jt}. On the other hand, the model of \cite{Artru:1995bh} 
can be improved by taking into account the azimuthal $k_T$-dependence of the polarised quark distribution \cite{artru:rikenbnl}.  
Let the proton spin points in the  $\yu$ direction and the pion be produced in the ($z,x$) plane. 
%the quark polarisation is $\Sv'=(D_{mm}-D_{nn}\sin\phi\cos\phi \xu+(D_{nn}\cos^2\phi+D_{mm}\sin^2\phi)\yu$. 
Allowing for a semi-hard collision of momentum transfer $q=k'- k$ and assuming that the quark fragments into a pion of 
transverse momentum $\tilde\pv_T$ relative to $k'$, one has $\pv_T\simeq(p/k) (\kv_T+\qv_T)+\tilde\pv_T$. Due to the 
trigger bias, $\kv_T$ points preferably towards  $\pv_T$, therefore is mainly in the $(z,x)$ plane. Accordingly the quark 
polarisation is $S_q\simeq D_{nn}(x,\kv_T)\,\yu$. Scattering does not change $S_q$ too much and one gets a Collins effect 
of strength $\langle\tilde\pv_T\rangle\propto D_{nn}(x,\kv_T)$ instead of $\delta q(x)/q(x)$. Contrary to $\delta q(x)/q(x)$, 
$D_{nn}(x,\kv_T)$ is not constrained by the Soffer bound, as illustrated by the quark--scalar diquark model where $D_{nn}
=1$. 
%$D_{mm}+D_{nn}\le1+D_{ll}$,  
Therefore the Collins effect combined with $k_T$-dependent polarised quark distribution  bypasses the Soffer bound and 
could well be responsible for a dominant part of the single-spin asymmetry.

\paragraph{T-odd distributions functions} Here it is worth mentioning the case of T-odd fragmentation functions\,%
\footnote{T-odd distribution functions, arising from initial state interactions, will be discussed in Sec.~\ref{incl:sub:kt}}%
,  which was given recently special attention. It is a new topic which belongs to the field one could call
{\it quark and gluon polarimetry}. This is a consequence of possible final
state interactions in the fragmentation process, giving rise to a non-trivial
phase. A particular example of such a function is
the Collins function \cite{Collins:1992kk}, which describes the following
process $q(\uparrow) \to h X$, namely the fragmentation of
a transversely polarised quark into an unpolarised hadron $h$ (\eg, a pion). In this case, provided
$h$ has a transverse momentum $\bf{p_T}$,         if $\vec{S}^{q}_{T}$ denotes the transverse 
spin of the quark, the event rate reads
\be
\d N(q(\uparrow) \to h X) = \d z\,\d{\vec{p}_T}\,D(z,\vec{p}_T)\left[1 + A_C(z,p_T)
\frac{(\vec{k} \times \vec{p}_T).\vec{S}^q_T}{|\vec{k} \times \vec{p}_T|}\right]~,
\ee
where $|A_C|\leq 1$ and $DA_C$ is the Collins function.
More generally, one introduces the concept of \emph{jet handedness} \cite{Efremov:1992pe},
an old idea first proposed by Nachtmann \cite{Nachtmann:1977ek}, which can be used
to determine the polarisation of a high energy quark or gluon. It is due to
the correlations between the spin of the fragmenting parton and the
momenta of the produced hadrons and for example the quark helicity can be obtained from the sign of
$\vec{p}^{(1)}.(\vec{p}^{(2)} \times \vec{p}^{(3)})$, where $\vec{p}^{(1)}$, $\vec{p}^{(2)}$ and $\vec{p}^{(3)}$ are the three 
leading meson momenta. 
Positivity requires that these functions cannot exceed
the corresponding spin-averaged functions.
\subsubsection{Generalised Parton Distributions}\label{incl:sub:GPD}
Let us finally consider the possibility of non-diagonality in
the momentum, rather than in helicity, which is
the case of generalised parton distributions (GPD) representing the most general
non-perturbative QCD inputs allowing to describe the amplitudes of
exclusive hard processes.
The review of the positivity bounds for these objects may be found in
Sec.~3.11 of \cite{Belitsky:2005qn}.
Here we simply describe the derivation of positivity bounds for GPD,
with the use of the operator method \cite{Jaffe:1991kp}
and we mention some new versions of these inequalities.

The unpolarised quark GPD reads
\be\la{incl:eq:qzeta(x)}
H_q (x,\xi)={1 \over \sqrt{1-\xi^2}}
\int {\d \lambda \over {4\pi}} \exp({i\lambda x})\langle p'|\bar\psi(0) \slashed{n}\psi(\lambda n)|p\rangle~,
\ee
where the factor $\sqrt{1-\xi^2}$ comes from the bilinear $\bar u(p')
u(p)$, in the definition of the GPD \cite{Goeke:2001tz,Belitsky:2005qn}.
We have assumed the strictly collinear kinematics, so that
$p=P(1+\xi)$ and $p'=P(1-\xi)$ with $P=(p + p')/2$.

By an analogous procedure as above, it becomes
\be
H_q (x,\xi)=\RE\sum_{X} {1 \over {\sqrt{2(1-\xi^2)}p^+}}\langle p|\phi(0)|X\rangle 
\langle p'|\phi(0)|X\rangle ^*\delta(x-(p-p_X).n)~,
\la{incl:eq:H}
\ee
where we used the hermiticity of the matrix element.

We are now ready to write down \cite{Pire:1998nw}
the Cauchy--Schwarz inequality as follows
\be
\sum_{X}|\langle p|\phi(0)|X\rangle  \pm a\langle p'|\phi(0)|X\rangle |^2 \delta(x-(p-p_X).n) \geq 0~,
\la{incl:eq:CS1}
\ee
where $a$ is a positive number.
%which we put equal to 1 for the time being.
While the non-diagonal term of (\ref{incl:eq:CS1}) is just producing the
GPD, the diagonal terms
%should be studied in more details
are related to the usual (diagonal) parton distributions. Performing
the  minimisation with respect to the variation of $a$, we finally get
\be\label{incl:eq:q(geo)}
|H_q(x, \xi)| \le \sqrt{{ q (x_1) q (x_2)} \over{1-\xi^2}}~,
\ee
where $x_{1,2}=(x \pm \xi)/(1 \pm \xi)$.
The positivity constraints should be generalised if one goes beyond the
collinear approximation for the initial and final momentum, when
the helicity-flip GPD, denoted by E, starts to contribute \cite{Diehl:2001pm,Pobylitsa:2001nt,Kirch:2005in}.
The general analysis of the parton-hadron eight-dimensional helicity matrix, which by use 
of symmetry properties, may be reduced to dimensions four and two, allows to get 
\cite{Pobylitsa:2002gw} the complete set of inequalities involving 
the full basis of GPD's.
Another general framework for the derivation of positivity constraints is
also provided by impact space representation~\cite{Pobylitsa:2002iu}.

The derivation for the gluons is analogous
\cite{Pire:1998nw,Radyushkin:1996nd,Radyushkin:1996ru,
Radyushkin:1997ki,Martin:1997wy,Pobylitsa:2002gw}.
The corresponding positivity conditions will be discussed in more detail in
Sec.~\ref{incl:sub:QCDevol}, when QCD evolution will be considered.

Finally, let us consider the possibility of transition GPD \cite{Goeke:2001tz},
when  not only momentum but also the type of the hadron can be different
\be\label{incl:eq:qAB(x)}
H_{AB} (x,\xi)={1\over \sqrt{1-\xi^2}}
\int {\d \lambda \over {4\pi}} \exp({i\lambda x}) \langle B,p'|\bar\psi(0) \slashed{n}\
\psi(\lambda n)|A,p\rangle~.
\ee
It is straightforward to bound them by the distributions for hadrons $A$ and $B$
\be\label{incl:eq:ab}
|H_{AB}(x, \xi)| \le \sqrt
{{ q_A (x_1) q_B (x_2)} \over{1-\xi^2}}~.
\ee
This bound was recently studied \cite{Tiburzi:2005nj}
for the case of pion--photon transition GPD \cite{Lansberg:2006fv} and the consideration  
of photonic GPD in the perturbative regime allows explicit tests of
these inequalities~\cite{Friot:2006mm}.
\subsubsection{QCD Evolution of Positivity  Constraints}\label{incl:sub:QCDevol}
 One may wonder, to what extent the $Q^2$ evolution of QCD is compatible with the 
 positivity constraints. In this section it is shown,
 that positivity becomes especially clear when the
 evolution equation is interpreted as a kinetic master equation.
In particular, one obtains a complete proof of
the corresponding inequalities, at
leading order (LO) and also at next-to-leading order (NLO), provided
a suitable factorisation scheme is chosen.
New positivity properties, which currently may be proved only
at the perturbative QCD level, will be discussed.
They include preservation of positivity by the Balitsky--Fadin--Kuraev--Lipatov (BFKL)
equation \cite{ Kuraev:1977fs,Balitsky:1978ic,Lipatov:1985uk} (see v) below) and the convexity of parton distributions (see 
vi) below).
The important property of positivity in QCD is that it is
preserved only in one direction of evolution, providing a sort of
``scale arrow".
\paragraph{i) -   Master equations for QCD evolution}
The crucial point in preserving the positivity property is that \emph{all}
 the known evolution equations may be represented
 \cite{Bourrely:1997bx,Bourrely:1997nc,Teryaev:1998gn,Teryaev:2005uf} in a form of master,
 or gain--loss equation, like
\be
 {\d q(x,t) \over{\d t}}=
 \int \d y \left[w(y \to x) q(y,t)-w(x \to y) q(x,t)\right]~.
 \label{incl:eq:gme0}\ee
Here $q$ is a generic parton distribution and the role of ``time" $t$ is played either by the longitudinal momentum
fraction for the BFKL equation \cite{Lipatov:2004sb} or
by the transverse momentum for the Dokshitzer--Gribov--Lipatov--Altarelli--Parisi (DGLAP) equations 
\cite{Gribov:1972ri,Altarelli:1977zs,Dokshitzer:1977sg}.
% Note also that for
% BFKL equation, besides the effect of diffusion $D$ and drift $v$, which are just the local counterparts
% of master equation, there is also the effect of exponential fission $\lambda$.
The probabilistic understanding of the DGLAP evolution equations
 came already from some pioneering papers
 \cite{Gribov:1972ri,Lipatov:1974qm,Bukhvostov:1974uu,Altarelli:1977zs,Dokshitzer:1977sg}.
 Actually, its standard form\,\footnote{For brevity, the
 argument $t$ will not be written down explicitly in the parton densities.}
 (for the time being, we shall confine ourselves to the non-singlet
 (NS) case)
\be
 {\d q(x) \over{\d t}}=
 {\alpha_s \over {2 \pi}} \int_x^1 \,\d y\, {{q(y)} \over y} P \left({x/ y}\right)~,
 \label{incl:eq:GLAP}\ee
may be interpreted as a ``time'' $t=\ln Q^2$  evolution  of the
 ``particle'' density $q$ in the one-dimensional space
 $0 \leq x \leq 1$, due to the flow from right to left,
 with the probability equal to the splitting kernel $P$.
 The key element of such interpretation is the problem of the
 infrared (IR) singular terms in $P$, which was considered in detail
 some years ago \cite{Collins:1988wj} (see also \cite{Durand:1986te}).
The kinetic interpretation is obviously preserved
 provided the ``+''  form of the kernel is presented in the following way
\be
P_+ (z) =P(z)-\delta(1-z)\int_0^1 P(y) \,\d y~,
 \label{incl:eq:+}\ee
leading to the corresponding expression for the evolution equation
 \be
 {\d q(x) \over{\d t}}=
 {\alpha_s \over {2 \pi}}\left[ \int_x^1 \,\d y \,{{q(y)} \over y}
 P ({x/y})- q(x) \int_0^1 P(z) dz\right]~.
 \label{incl:eq:+e}\ee
This form has already a kinetic interpretation because
 the second term in the square brackets describes the flow of
 the partons at the point $x$ \cite{Collins:1988wj}. It seems however
 instructive to make this similarity even clearer
 by the simple change of variables  $z=y/x$ in the second term, which
 leads to the following symmetric form
\be
 {\d q(x) \over{\d t}}=
 {\alpha_s \over {2 \pi}}\left[ \int_x^1 \,\d y \, {{q(y)} \over y}
 P ({x/y})- \int_0^x \,\d y \, {q(x) \over x} P({y/x}) \right]~.
 \label{incl:eq:b}\ee
It allows us to write it down in a form
of master equation (see Eq.~(\ref{incl:eq:gme0}))
with the 'transition probability'  defined as
\be
  w (y \to x) = {\alpha_s \over {2 \pi}} P ({x /y})
  {{\theta (y -x)} \over y}~.
 \label{incl:eq:sigma}\ee
As a result, the known properties of the evolution equation
 receive the following simple interpretation: the cancellation of
 the IR divergences between the contributions of the real and virtual
 gluons emission is coming from the equality of in- and out- flows, when
 in both terms of Eq.~(\ref{incl:eq:gme0}) one has $y \sim x$, following from
 the continuity condition for the ``particle''  number.
Also, the conservation of the vector current
\be
 \int_0^1 \,\d x \, {\d q(x) \over{\d t}}=
 \int\limits_0^1\hskip -5pt\int\limits_0^1 \d x \d y  \left[w (y \to x) q(y)- w(x \to y) q(x)\right]=0~,
 \label{incl:eq:c}
 \ee
comes from the integration of an antisymmetric function in a symmetric
 region.
%\paragraph{ii)- Drift and Diffusion for DGLAP equation}
 
 As the transition probability is peaked for $x \sim y$,
 which is the reflection of the IR divergence, it is natural to
 apply the Kramers--Moyal (KM) expansion \cite{Cafarella:2003up}.
 However, the natural expansion variable is $Y=\ln x$ \cite{Teryaev:2005uf}, rather than $x$ as in Ref.~\cite{Cafarella:
2003up}. 
 The reason is that the transition probability
 acquires than the translation invariant form.
The master equation holds for function  $f(Y)=xq(x)$, with $x=\exp(Y)$, as its integral over $Y$ conserves
\be
 \int_0^1 \d x\, q(x) = \int_{-\infty}^0 \d Y\, f(Y)~.
 \label{incl:eq:t0}\ee
 By rewriting Eq.~(\ref{incl:eq:c}) in terms of the new variable $Y$, one may expand the first term in power series in $Y$,
 which is just the KM expansion,
 and extend the integration to the whole $Y$ axis $(-\infty <Y< \infty)$.
 If the initial distribution remains non-zero only for $0<x<1$ $(-\infty <Y<0)$, the directional property
 of the DGLAP equation, resulting from the $\theta$-function in Eq.~(\ref{incl:eq:sigma}), guarantees that the evolved 
function  also remains non-zero only for $0<x<1$ $ (-\infty <Y<0)$.
 As a result, the moments of transition probability appearing in the KM expansion, coincide with the derivatives of 
anomalous dimension at $n=1$
 \be
 \int_0^1 \,\d x\, \ln^k x\, P(x) =-{2\pi\over\alpha_s} \left.{\d^k \over \d n^k} \gamma(n)\right |_{n=1}~,\quad\text{with}
 \quad \gamma(n)=-{\alpha_s\over 2\pi} \int_0^1 P(x) \,x^{n-1} \,\d x ~,
 \label{incl:eq:t}\ee
 which leads to the differential form of DGLAP equation
 \be
 {\partial q(x) \over{ \partial t}}=\left.-{1 \over x} \exp\left[{\partial \over \partial \ln(1/x)} {\d \over
 \d n}\right]x\, q(x) \gamma(n)\right |_{n=1}~.
 \label{incl:eq:dif}
 \ee

The Fokker--Planck (diffusion) approximation corresponds to keeping
 only the two first terms in this equation. Substituting 
 the expression for LO non-singlet kernel
 to (\ref{incl:eq:t}) with $n=1,2$  
 one gets
\be \label{incl:eq:FP}\begin{aligned}
 &{\partial q(x) \over{\partial t}}={1 \over x} \left( v {\partial (x q(x))
 \over \partial \ln(1/x)} + D {\partial^2 (x q(x))\over \partial \ln^2 (1/x)}
 \right)~,\\
\mbox{with}\quad& v= {\alpha_s \over {2 \pi}}( 5/4 - \pi^2/3)~,\quad
\mbox{and}\quad D=  {\alpha_s \over {4 \pi}} ( -9/8 + 2\zeta(3))~.
\end{aligned} \ee
 Here $v$ stands for the drift velocity. Note that the positivity of the diffusion coefficient $D$ 
 naturally explains the convexity of the curve of anomalous dimension, first
 pointed by Nachtmann \cite{Nachtmann:1973mr,LlewellynSmith:1978me}.
 It is also possible to observe the convexity of the anomalous dimension curve in the points different
 from $n=1$. In this case, one should consider the evolution of the function $x^k q(x)$
 with arbitrary positive $k$ and perform a similar KM expansion.
 Note that besides the diffusion and drift effects one will get also the  
 ``decay" effect, since the corresponding moment is not conserved.

  \paragraph{ii) - Preservation of positivity constraints for spin-dependent distributions by QCD evolution}
 Let us start with obvious remarks about the unpolarised distributions.
 As soon as the master equation has a clear probabilistic meaning, and the probabilities 
 of the corresponding Markovian process are
 always positive, the preservation of positivity becomes clear.
 
 For further analysis it is, however, instructive, to 
 describe in some detail the formal reasons of this preservation. 
 The negative second term in Eq.~(\ref{incl:eq:gme0}) cannot change the sign of the
 distribution
 because it is 'diagonal' in $x$ (recall that $x$ may be considered as a
 label of density matrix), which means that it is proportional
 to the function at the same point $x$, as in the l.h.s..
 When the distribution gets too close to zero,
 its stops decreasing. This is true for both $'+'$ and $\delta(1-z)$
 terms, for any value of their coefficient (if it is positive,
 it will reinforce the positivity of the distribution).

Let us consider now the spin-dependent case. For simplicity, we
 postpone the discussion of quark-gluon mixing for a moment, but allow
 the spin-dependent and spin-independent kernels to be different,
 as they are at NLO.
 It is most convenient to write down the equations for definite parton
 helicities, which was actually the starting point in deriving
 the equations for the spin-dependent quantities \cite{Altarelli:1977zs}.
 Although the form, which we shall use, mixes the contributions of
 different helicities, it makes the positivity properties especially
 clear. So we have
\be
 {\d q_{\pm}(x) \over{\d t}}=
 {\alpha_s \over {2 \pi}} \left[P_{+\pm} ({x/y}) \otimes q_+(y)+
 P_{+\mp} ({x/ y}) \otimes q_-(y)\right]~.
 \label{incl:eq:h}\ee
Here $P_{+\pm}(z)=(P(z) \pm \Delta P(z))/2$, 
 are the evolution kernels for definite helicities, and the shorthand
 notation for the convolution is adopted.
 As soon as $x < y$, the positivity of the initial distributions
 ($q_{\pm}(x, Q_0^2) \geq 0$, or
 $|\Delta q (x, Q_0^2)| \leq q (x,Q_0^2)$)
 is preserved, if both kernels
 $P_{+\pm}$ are positive, which is true, if
\be
 |\Delta P (z)| \leq P (z)~,\quad  z < 1~.
 \label{incl:eq:ineq1}\ee

The singular terms at $z=1$  are not altering positivity,
 because they appear only in the diagonal (now in helicities)
 kernel $P_{++}$ (only forward scattering is
 IR dangerous). From the kinetic interpretation again
 the distributions $q_+,~q_-$ stop decreasing, as soon as they
 are close to changing sign.
From the general point of view, the quantities in the l.h.s. of
Eq.~(\ref{incl:eq:h}) may be considered as diagonal elements of density matrix remaining positive 
in the course of the QCD evolution.

Now to extend the proof to the quark gluon mixing is trivial.
 One should write down the expressions for the evolutions of quark
 and gluon distributions of each helicity
\be \label{incl:eq:hs}\begin{aligned}
{\d q_{\pm}(x) \over{\d t}}&=
 {\alpha_s \over {2 \pi}} \left[ P_{+\pm}^{qq} ({x/ y}) \otimes q_+(y)+
 P_{+\mp}^{qq} ({x/ y}) \otimes q_-(y)\right]\cr
 &\hspace*{1.5cm} {}+P_{+\pm}^{qG} ({x /  y}) \otimes G_+(y)+
 P_{+\mp}^{qG} ({x / y}) \otimes G_-(y)~,\cr
 {\d G_\pm(x) \over{\d t}}&=
 {\alpha_s \over {2 \pi}} \bigl[P_{+\pm}^{Gq} ({x / y}) \otimes q_+(y)+
 P_{+\mp}^{Gq} ({x /  y}) \otimes q_-(y)\cr
 &\hspace*{1.5cm} {}+P_{+\pm}^{GG} ({x / y}) \otimes G_+(y)+
 P_{+\mp}^{GG} ({x / y}) \otimes G_-(y)\bigr]~.
 \end{aligned}\ee
 If the inequality (\ref{incl:eq:ineq1}) is valid for each type of partons \cite{Bourrely:1997nc},
\be
 |\Delta P^{ij} (z)| \leq P^{ij} (z), \quad z < 1; \quad i,j = q,G~,
 \label{incl:eq:ineqs}\ee
all the kernels, appearing in the r.h.s. of such a system, are
 positive. Concerning the singular terms, they are again diagonal,
 now in parton type, and do not affect positivity.
 The validity of these equations at LO comes just from the
 way they were derived, as the (positive) helicity-dependent
 kernels were in fact first calculated in Ref.~\cite{Altarelli:1977zs}.
 At  NLO, the situation is more controversial \cite{Bourrely:1997nc}
 and positivity can be used for the choice of factorisation scheme.
The positivity is, generally speaking, not preserved order by order,
but is typically preserved for the sum of LO and NLO terms.

To conclude, the stability of positivity under $Q^2$ evolution
 comes from two sources:
\textsl{i)} 
 the inequalities (\ref{incl:eq:ineqs}), leading to
 the increasing of distributions,
\textsl{ii)} the kinetic interpretation
 of the decreasing terms.
For the latter it is crucially important, that they
 are diagonal in $x$, in helicity and in parton type, which is
 related to their IR nature.
Such property of
 virtual correction makes preservation of positivity especially simple when
 the evolution in $x$ space is considered, while in the space of the moments 
 of parton distributions it
 requires a more elaborate analysis \cite{Nachtmann:1973mr,LlewellynSmith:1978me}.

\paragraph{iii) - Preservation of Soffer inequality by QCD evolution}

Let us now come to the evolution of Soffer inequality.
This is really a crucial problem as the very different evolution
of transversity lead to the suggestion that
it cannot be stable against QCD evolution
\cite{Goldstein:1995ek}, However, this evolution, albeit different, is very
precisely correlated with the evolution of chiral-even distributions
leading to the stability of Soffer inequality.

According to
 the previous analysis it is straightforward \cite{Bourrely:1997bx} 
to define the following
 'super'-distributions, corresponding to the eigenvalues
 of density matrix:
\be  Q_\pm(x) =   q_+(x) \pm  h_1(x)~.\label{incl:eq:Q}\ee

Due to Soffer inequality, both these distributions are positive
 at some point $Q^2_0$, and the evolution equations for the NS case
 take the form
\be
 {\d Q_\pm(x) \over{\d t}}=
 {\alpha_s \over {2 \pi}} (P^Q_{+\pm} ({x/ y}) \otimes Q_+(y)+
 P^Q_{+\mp} ({x/ y}) \otimes Q_-(y))~,
 \label{incl:eq:eQ}\ee
where the 'super'-kernels at LO are just
 \be \label{ineq:eq:PQ}\begin{aligned}
 &P^Q_{++}(z)  \equiv  {P_{qq}^{(0)}(z) + P_h^{(0)}(z) \over 2}
={C_F \over 2} \left[{{(1+z)^2} \over{(1-z)_+}} + 3\delta (1-z)\right]~,\\
& P^Q_{+-}(z) \equiv {P_{qq}^{(0)}(z) - P_h^{(0)}(z)\over 2}
={C_F \over 2} (1-z)~.
\end{aligned}\ee

 One can easily see \cite{Bourrely:1997bx}, that the inequalities analogous to Eq.~(\ref{incl:eq:ineqs})
 are satisfied, so that both $P^Q_{++}(z)$ and $P^Q_{+-}(z)$ \cite{Barone:1997fh}
 are positive for $z<1$, while the singular term does appear
 only in the diagonal kernel.
 So, both requirements are valid and Soffer inequality is preserved
 under LO evolution. The extension to the singlet case is trivial,
 as the chiral-odd transversity distributions does not mix with gluons.
 Therefore, they affect only the evolution of quarks, and lead to the
 presence in the r.h.s. of the same extra terms as in Eq.~(\ref{incl:eq:hs}). We have now
 \begin{multline} \label{ineg:eq:PQ1}
 {\d Q_{\pm}(x) \over{\d t}}=
 {\alpha_s \over {2 \pi}} \biggl[P^Q_{+{\pm}} ({x/y}) \otimes Q_+(y)+
 P^Q_{+{\mp}} ({x/ y}) \otimes Q_-(y)\\
{} +P_{+-}^{qG} ({x/y}) \otimes G_+(y)+
 P_{++}^{qG} ({x/ y}) \otimes G_-(y)\biggr]~,
\end{multline}
where the non-diagonal kernels are all positive and free from singular terms, so positivity is preserved. This completes the proof of the positivity at leading order, while the NLO case is considered in detail in  \cite{Bourrely:1997bx} (see also \cite{Vogelsang:1997ak, Kumano:1997qp}).  

\paragraph{iv)- Preservation of positivity constraints for Generalized Parton Distributions by QCD evolution}

We consider the evolution of the constraints for gluonic GPD, thus completing our discussion
of quark GPD's in Sec.~\ref{incl:sub:GPD}.
We restrict ourselves to the pure gluodynamics case, by means of the non-forward
gluon distribution, or generalised momentum density, $M(x_1,x_2)=M(x_2,x_1)$ \cite{Teryaev:1998gn}, where $x_1,x_2$, defined as in Sec.~\ref{incl:sub:GPD}, are suitable for expressing the simplest symmetry properties, following from T-invariance. 
The $Q^2$ evolution in the DGLAP region $x_1, x_2 > 0$, which is the only region 
where positivity constraints are applicable, is given by the following equation 
\begin{multline} \label{incl:eq:evm}
 {\d M(x_1,x_2) \over{\d t}}={\alpha_s \over {2 \pi}}\biggl [\int_{x_1}^1 {\d z\over{z(1-z)}} \tilde P(z,z')
 M (x_1/z,x_2/z') \\
{}-{{M (x_1,x_2)}\over 2}
\left(\int_0^1 {\d z\over{1-z}} \tilde P(z)+\int_0^1 {\d z'\over{1-z'}}
 \tilde P(z')\right)\biggr]~,
\end{multline}
where $t=\ln Q^2$, $\tilde P(z,z')$ and $\tilde P(z), \tilde P(z')$ are
related to the off-diagonal and diagonal splitting functions $ P(z,z')$, $P(z)$, $ P(z')$ 
in the following ways: $\tilde P(z,z')= \tilde P(z',z)= z'(1-z) P(z,z')$ implying
a similar relation for diagonal kernels $\tilde P(z,z)= \tilde P(z)= z(1-z) P(z)$. The above evolution equation preserves the symmetry with respect to the interchange of $x_1$ and $x_2$, because of the simple relation between the integration variables $z$ and $z'$ implied by the $t$-channel momentum conservation
\be
{1- z\over z}\,{1-x_2\over x_2}= {1- z'\over z'}\,{1-x_1\over x_1} ~.
\label{sim:z:z'}
\ee

 The above evolution equation allows one to prove the stability
 of the positivity constraint against $Q^2$ evolution, similarly to 
 the quark case,   
 following the general line of \cite{Bourrely:1997nc,Bourrely:1997bx}.
 To do so, one
 may consider the positive quantities (at some initial scale $Q_0$)
 $M_{\pm}(x_1,x_2)=a M (x_1) + M (x_2 )/a \pm 2 M (x_1,x_2)$,
 where $a$ is an arbitrary positive number.

The inequality 
 \be\label{incl:eq:g(geo1)}
 |M (x_1,x_2)| \le \sqrt {M (x_1) M (x_2)}~,
 \ee
is than the result of the minimization with
 respect to the variation of $a$, where $M(x)=xg(x)$
 is a diagonal momentum distribution. 
 One may write the evolution equations as
\begin{multline}
 {\d M_{\pm}(x_1,x_2) \over{\d t}}={{\alpha_s} \over {2 \pi}}
 \left[\int_{x_1}^1 {\d z \over{z(1-z)}} (a \tilde P(z) M(x_1/z)+\tilde
 P(z') M(x_2/z')/a \right.\\
\left. \pm 2 \tilde P(z,z') M (x_1/z,x_2/z')) - M_{\pm} (x_1,x_2)
 \int_0^1 {\d z\over{1-z}} \tilde P(z)\right]~.
 \label{evm+}\end{multline}

It is very important that the virtual contributions are {\it diagonal}
 in the index $\pm$, so that they cannot change the positivity of the
 distribution.

To prove positivity of the real term
 it is sufficient to consider the minimisation with respect to the
 variation of $z$-dependent (positive) $a(z)$,
 which can only make the sum of two positive
 diagonal terms smaller, than in the actual case of minimisation with respect
 to constant $a$:
\begin{multline} \label{incl:eq:az}
 \min_a \int\limits_{x_1}^1 {\d z \over{z(1-z)}}
\biggl[a \tilde P(z) M({x_1/ z})+{\tilde P(z')\over a} M({x_2/ z'})
 \pm 2 \tilde P(z,z') M ({x_1/z},{x_2/ z'})\biggr] \geq \\
\int\limits_{x_1}^1 {\d z\over{z(1-z)}} \min_{a(z)}\biggl[a(z) \tilde P(z) M({x_1/z})+
 {\tilde P(z')\over a(z)} M({x_2/z'}) \pm 2 \tilde P(z,z') M ({x_1/z},{x_2/z'})\biggr]\\
\hspace*{0.5cm}{}= 2 \int\limits_{x_1}^1 {\d z\over{z(1-z)}} \biggr[\sqrt {\tilde P(z) M({x_1/z})
 \tilde P(z') M({x_2/z'})}
 \pm \tilde P(z,z') M ({x_1/z},{x_2/z'})\biggl]~.
\end{multline}

Writing down Eq.~(\ref{incl:eq:g(geo1)}) for $x_1 \to x_1/z, x_2 \to x_2/z'$,
 the sufficient condition of positivity of (\ref{incl:eq:az}) can be
 easily found \cite{Pire:1998nw}, namely
\be\label{incl:eq:g(p)}
 |\tilde P (z,z')| \le \sqrt{\tilde P (z) \tilde P (z')}~.
 \ee

Such inequality is really valid \cite{Pire:1998nw} for all $z,z'$,
 completing the proof of positivity. Although we have considered here a
 pure gluodynamics, the mixing is improving the situation
 with positivity, providing extra positive terms, like in the forward case
 \cite{Bourrely:1997nc}.

\paragraph{v) - Preservation of positivity for the BFKL equation}

Let us now move to the positivity properties which can be currently established only at the perturbative level. Since the non-perturbative counterpart of BFKL evolution, namely, the hadron impact factor,
 is not expressed in the form of a matrix element, the proof of positivity by the methods of Sec.~\ref{incl:sub:pdpd}
 is not possible. The BFKL evolution, as it will be discussed,
  may be presented in a form of generalised master equation, so that the gain and loss probabilities differ,
 \be
 {df(x,t) \over{dt}}=
 \int \d y \left[w_+(y \to x) f(y,t)-w_-(x \to y) f(x,t)\right]~.
 \label{incl:eq:exp}
 \ee
 It will preserve positivity as long as one considers
 the evolution towards larger longitudinal ``time'' $t=\ln(1/x)$, that is towards 
 smaller partonic momentum fractions $x$. Note that in the subsequent expressions of this section, $x=\ln k_T$ is used to 
denote the variable which is the transverse coordinate in the BFKL equation \footnote{ Note the difference with the DGLAP definitions of $t$ and $x$. These notations will be used only throughout this paragraph.}.

 Moreover, the extra non-linear negative term \cite{Jalilian-Marian:1997jx}, responsible for gluons
 fusion will not violate this property. The specific property of generalised  master equation
is the different values of gain $w_+$ and loss $w_-$
probabilities. Therefore, it describes not only transitions
of "particles" but also their fission (multiplication), when
the gain probability prevails, as in the case of the BFKL equation, or
fusion in the opposite case.
However, this does not affect the proof of positivity,
which in fact is making use only of the positivity of
$w_+$ and locality of loss term due to $w_-$.
The non linearity of the loss term also does not spoil
the proof of positivity, as soon as it remains local.

 In order to separate the effects of diffusion and fission, one may consider
 \cite{Teryaev:2005uf} the master equation for the weighted function $f_\sigma(x,t)=f(x,t)\sigma(x)$, so that
 $w_+(x,y) \to w_\sigma(x,y)= w_+(x,y)\sigma(x)/\sigma(y)$, while $w_-$ remains unchanged. For the BFKL case, the
 function $\sigma(x)$ can be simply $(k_T)^{\alpha}$ with $0<\alpha<1$, in order to avoid divergences.
 Let us also define the ``relative" function as
 $\bar f_\sigma(x,t) = f_\sigma(x,t)/\langle f_\sigma(t)\rangle$, with the $k_T$-integrated
 function $\langle f_\sigma(t)\rangle =\int dx f_\sigma(x,t)$.
 It satisfies the standard master equation
\be
 {\d  \bar f_\sigma (x,t) \over{\d t}}=
 \int \d y \left[ w_\sigma(y \to x) f_\sigma (y)-w_\sigma (x \to y) \bar f_\sigma (x)\right]~.
 \label{incl:eq:rel}
 \ee

If $\langle f_\sigma(t)\rangle$ evolves exponentially, as $\d \langle f_\sigma(t)\rangle /{\d t} = \lambda_\sigma \langle f_\sigma 
(t)\rangle$, $\lambda_\sigma$ is solution of an eigenvalue problem defining simultaneously the eigenfunction $\sigma$
 \be
 \int \d x\, w_+(y \to x)\sigma(x) = \left[\lambda_\sigma  +  \int \d x\,  w_- (y \to x)\right] \sigma(y)~,
 \label{incl:eq:gme}\ee
 and we have
 \be
 \lambda_\sigma= \int \d x \left[ w_\sigma (y \to x)-w_- (y \to x)\right]~,
 \label{incl:eq:ev}
\ee
which is independent of $y$.

It turns out that the separation of these effects for the BFKL equation is ambiguous, due to the freedom in the choice of the function $\sigma$, reflecting the scale invariance of gain and loss probabilities,  while, when they are not separated, their combination is invariant \cite{Teryaev:2005uf}. This invariance in the short-range approximation leads to
\be
 \lambda_0=\lambda_\sigma-\frac{v_\sigma^2}{4 D_\sigma}~.
 \label{incl:eq:B1}
 \ee
where $\lambda_0$ is independent of $\sigma$.
In fact, it is just this combination which gives the
famous BFKL value for the Pomeron intercept, proportional to $4\ln 2$,
corresponding to the minimal fission and diffusion and
to the absence of drift  in this approach.
\paragraph{vi) - Convexity properties of parton distributions}
Another property which can be established at the perturbative level is the
stability in $t$ of the signs of the derivative with respect to $x$ of the parton distribution. It follows \cite{OT07} from 
 the differential form (\ref{incl:eq:FP}) in which the operator in the r.h.s. commutes with the
(logarithmic) derivative in $x$. This immediately means that the derivatives obey the same evolution equation.
Therefore, their sign is also preserved by the evolution. One has the pattern of alternate
signs for the derivatives of increasing order. 

This may explain the success of a simple concave parametrisation
of $x^{-a}(1-x)^{b}$ type. It is interesting, that if one uses the  
parton distributions as an input to construct a model for GPD, 
the concave behaviour guarantees  \cite{Radyushkin:1998es}
the validity of the positivity constraints for the GPD (see Sec.~\ref{incl:sub:GPD}).

This concave parametrisation explains why the results which are,  strictly
speaking, valid at very small $x$, still holds at much larger $x$, if  the factor $(1-x)^{b}$ is still  close to 1. Such a  
situation  happened \cite{Soffer:1996ft} when the behaviour of neutron spin-dependent structure function $g_1$ at SLAC 
energies was found to be compatible with the perturbative low-$x$ asymptotics (see \cite{Ermolaev:2005ny} and references 
therein). 
%Note that this latter approach corresponds, in fact, to another definition of non-perturbative matrix
%elements with the singular in $x$ terms being subtracted. Therefore, there is no contradiction with DGLAP approach,
%providing the preservation of concavity and possibility to apply the results of \cite{Ermolaev:2005ny} at moderate $x$.  
Repeating this
argumentation for spin-averaged case one may expect the applicability of BFKL approach at larger $x$ as well, because 
the BFKL equation (as well as its non-linear generalisations) also preserves the concavity of (unintegrated) gluon 
distribution in $x$, 
as soon as the gain and loss probabilities in the generalised master equation do not depend on the longitudinal momentum 
fraction $x$.
The observation \cite{Haidt:2004ck}  
%\footnote{I am indebted to L.N. Lipatov for reference to this work and valuable comments}
on its violation may be considered as a signal of a different kind of evolution at very small $x$.
At the same time, another recent analysis \cite{Gluck:2006pm} shows an agreement with the predicted concave behaviour.
%It is instructive to compare the positivity of (dominating) gluon distribution and its curvature at very small $x$.
%The positivity of both may be violated in DGLAP due to the negative kernel.
%At the same time, both stay positive in the BFKL approach. There is, however,  essential physical difference
%$f_2$ should stay positive, contrary to its curvature.
%One may also use the similar arguments to apply the very large $x$ asymptotics \cite{Gardi:2002bk} for
%the study of Bloom-Gilman duality \cite{Teryaev:2005hy}.

 \paragraph{vii) - Irreversibility:  longitudinal and transverse scale arrows}
The  positivity property is preserved when the 
respective ``time" variable in the master equation is increasing, but it 
may be violated for the backward evolution. 
 This defines a  ``scale arrow" analogous to the famous time's arrow.
 The reason for such irreversibility in positivity preservation is the following.
The master equation clearly distinguishes the directions of time, in
other words time reflection makes it  ``antiprobabilistic''.
To understand the general origin of this distinction, let us start, 
following G.M. Zaslavsky \cite{Zaslavsky:1981aw}, from the reversible Kolmogorov--Chapman equation
\be
 W(x_1,t_1|x_2,t_2)=
 \int \d y\, W(x_1,t_1|y,\tau) W(y,\tau |x_2,t_2)~,
 \label{incl:eq:KC}\ee
with a probability depending only on $t_1-t_2$\,\footnote{It corresponds in QCD to the choice of the variable $t=\ln k_T$
for the DGLAP equation and of $t=-\ln x$  for the BFKL equation.}
\be
 W(x_1,x_2,t)= \int \d y\, W(x_1,|y,t-\tau) W(x_2,y,\tau)~,
 \label{incl:eq:KC1}\ee
 $\tau \in [t_1, t_2]$ being any intermediate time.
Let us write the following expression for the difference
 \be
W(x,\xi,t+\delta t)-W(x,\xi,t)
=\int \d y\left[W(x,y,\delta t) W(y,\xi,t)- W(y,x, \delta t) W(x,\xi, t)\right]~,
 \label{incl:eq:meW}\ee
where we use the completeness condition
\be
 \int \d y\, W(y,x,\delta t)=1~.  
 \label{incl:eq:B}
 \ee

Note that Eq.~(\ref{incl:eq:meW})  is \emph{already}  a master equation, in a more general form, with finite difference 
instead of derivative corresponding to the special case $\delta t\to 0$.
 The probability $ W(y,\xi,t)$ depending on the initial value $\xi$
 should be considered as a distribution function, while  $ W(x,y,\delta t)$  is the transition probability.
It is not \emph{ necessary}, although possible, to integrate over the  initial value $\xi$ or to "forget" it.
 Its very existence in \emph{past} is \emph{sufficient}. Choosing the value $\xi$ in \emph{future}, one immediately comes
 to the ``antiprobabilistic" master equation. The short-range limits for these two cases correspond to two Kolmogorov 
equations, the first one coinciding with the Fokker--Planck equation.
%Note that the infinitesimal value of $\delta t$ is also unnecessary.

These two points mark the appearance of the time ordering at the earlier stage of comparison with Ref.~\cite{Zaslavsky:
1981aw}.
 It is possible, following this reference, to consider the limit of small $\delta t$ and to define the differential probability rate 
(which can be also fractional \cite{Zaslavsky:1981aw}), resulting in the standard differential form of master equation.
Note finally, that to manifest irreversibility, the transition probability should also not be reversible in the sense 
that the transition from  different points to a single one should be allowed. This is an analogue of the coarse-graining 
procedure  which is a well-known ingredient of the emergence of irreversibility. This property is obviously valid in the case of QCD  evolution equations, where a corresponding scale arrow arises.

 The striking difference between the BFKL and DGLAP 
 equations is that the ``time" (whose role is played by scale) direction  is pointed to the infrared (IR) in the BFKL case, and  to the ultraviolet (UV) in the DGLAP one.
 The IR direction is quite natural for the Wilson renormalisation group (RG) and, indeed, there are
 studies \cite{Jalilian-Marian:1997jx} where BFKL dynamics
 emerges within the Wilson RG approach.

Note also that for the DGLAP equation, the Wilson RG may be considered
in momentum, rather than in coordinate space.
 Indeed, the parton distributions may be described as
 integrated over transverse momentum up to the actual energy scale $Q^2$
\be
F(x,Q^2)= \int_0^{Q^2} \d k_T^2\, f(x, k_T^2)~.
 \label{W}
 \ee

Note that this equation is compatible with the probabilistic properties
of the integrand only when the l.h.s is increasing with $Q^2$, \ie, for small enough $x$.
The different directions of longitudinal and transverse scale arrows
 may be unified by assuming the fundamental role of an angular ordering,
 related to QCD coherence equations \cite{Ciafaloni:1987ur, Catani:1989yc, Catani:1989sg, Marchesini:1994wr}. If so,
 its projection  onto longitudinal direction provides the IR longitudinal scale arrow,
 while its projection to transverse direction results in UV transverse scale arrow.

Let us compare the role of irreversibility for distribution and fragmentation functions. 
The evolution of the latter is also described by the DGLAP equation. The time
development of the hard scattering process  
corresponds to the probabilistic evolution of distributions and antiprobabilistic evolution
of fragmentation functions. The latter property may be important for the explanation 
of the strong dependence of the final state characteristics on the initial conditions,
and for the possible emergence of turbulence-like phenomena.  

\newpage\setcounter{equation}{0}\section{Further developments}\label{se:furth}
\subsection{Polarised cascades}

\subsubsection{Dual types of density matrices}

In Eq.~(\ref{basic:eq:reac-rhoCD}) we have introduced the \emph{acceptance} matrix $\acc$ to characterise a detector. For a spin one-half,
\be\label{furth:eq:acc}
\acc = \epsilon_0\ (1+\Av.\vec\sigma)~.
\ee
$\epsilon_0$ is the unpolarised efficiency and $\Av$ is the vector 
analysing power. $\acc$ has the same properties as a density matrix, 
except for the trace condition. More generally an acceptance matrix is 
attached to any process that the particle may undergo in the 
\emph{future}, for instance scattering, decay, absorption or 
detection. The probability that a particle of density matrix $\emit$ 
undergoes such a process is given by
\be\label{dual}
w\propto\Tr(\emit\,\acc)\, .
\ee
For spin one-half, $\emit=(\mathbbm{1}+\Sv.\vec\sigma)/2$, then
\be
w=\epsilon_0\ (1+\Sv.\vec\Av)\,.
\ee
As opposed to $\acc$, the density matrix $\emit$ is the result of the 
\emph{past} interactions and is an \emph{emittance} 
matrix. More generally, emittance matrices can describe 
\emph{intensity-like} quantities, like the spin-dependent quark 
distribution, in which case their traces are not normalised to unity.
An acceptance matrix can be seen as an emittance matrix in the 
time-reversed process. The \emph{cross section matrix} is ambivalent: 
it acts as an acceptance matrix for the initial particles and as an 
emittance matrix for the final ones.
An example where emittance and acceptance matrices are implicitly 
used is in the expression $l^{\mu\nu}\,W_{\mu\nu}$ for the cross 
section of deep inelastic electron scattering. It is of the form
$\Tr(\emit\,\acc)$ with $\emit=l^{\mu\nu}$ and $\acc=W^{\mu\nu}$ (or 
vice-versa, depending on the chosen arrow of the virtual photon line).
\subsubsection{Monte-Carlo simulations of entangled cascades}
Suppose, for instance, that one needs a Monte-Carlo simulation of the cascade of reactions
\be\label{furth:eq:cascade}\begin{aligned}
&(1)\qquad & \ep\e&\to\tau^+\tau^-~,\\
&(2)\qquad &\tau^+&\to\mu^+\nu_\mu\bar\nu_\tau~,\\
&(3)\qquad & \tau^-&\to \mu^-\bar\nu_\mu\nu_\tau~.
\end{aligned}\ee
Can the event generator treat the different reactions (\ref{furth:eq:cascade}) successively?
Even %if the $e^+$ and $e^-$ are unpolarised and
if we are not interested in the muon polarisations, the 
generator has to include the spin degree of freedom of the $\tau$ 
leptons. Polarised $\tau$ have, indeed, anisotropic decays of the form 
$1+ A\,\Sv.\pu$ and $1- A\,\Sv'.\pu'$,
%$1\pm A\,\Sv^\pm.{\hat\pv^\pm}$,$1\pm A\,\Sv^\pm\cdot\pu(\mu^\pm)$, 
where $\Sv$ and $\Sv'$ are the $\tau^\pm$ polarisations and $\pu$ and $\pu'$ 
%\pu_\mu^\pm$ %$\hat\pv^\pm$
the muon directions in the $\tau^\pm$ rest frames. A simple Monte-Carlo 
program would first generate the $\tau^+$ and $\tau^-$ momenta, 
calculate their individual polarisations $\Sv$ and $\Sv'$ (which 
are nonzero due to the contribution of $Z^0$ exchange), then 
simulate the $\tau^+$ and $\tau^-$ anisotropic decays independently. 
Such a program would correctly give the individual muon spectra, but 
would miss the influence of the $\tau^+\tau^-$ spin correlations on 
the joint momentum distribution of the two muons.
As a remedy, one might try to generate \emph{correlated} polarisation vectors 
$\Sv$ and $\Sv'$ at the $\tau^+\tau^-$ production vertex, with a 
joint distribution $C(\Sv,\Sv')$. But whichever the latter may be, 
it would only describe a \emph{classical} correlation, corresponding 
to the \emph{separable} $\tau^+\tau^-$ density matrix
\be
\rho_{\rm sep}=\int 
{\d^2\Sv\over4\pi}\,{\d^2\Sv'\over4\pi}\,C(\Sv,\Sv')\,{\mathbbm{1}+\Sv.\vec\sigma\over2} 
\otimes{\mathbbm{1}+\Sv'.\vec\sigma'\over2}\,,
\ee
whereas the actual $\tau^+\tau^-$ spin correlation is often of 
the \emph{entangled} type. For instance, at $90^\circ$, one-photon 
exchange produces the $\tau^+\tau^-$ pair in the spin-triplet state corresponding to  $\langle\vec\sigma+\vec\sigma'\rangle.\vec{n}=0$  and described by the density matrix $\rho=(\mathbbm{1}-\vec\sigma.R\vec\sigma')/4$, where $R$ is the symmetry about the $\nv$ axis. With the convention of Fig.\ref{basic:fig:frame}, it means $C_{ll}=C_{mm}=C_{nn}=-C_{00}\equiv-1$ and $C_{\mu\nu}=0$ for $\mu\ne\nu$. % ($\mu,\nu=0,l,m,n$).
This is a maximally entangled state. The strongest classical correlation of the form $C_{ij}\propto\delta_{ij}$,~ $C_{i0}=C_{0i}=0$, that is to say, $\Sv=-R\Sv'$, is given by 
$C(\Sv,\Sv')=\delta^2(\Sv+R\Sv')$. It leads to $\rho_{\rm sep}=
[\mathbbm{1}-(\vec\sigma.R\vec\sigma')/3]/4$, which represents a correlation three time too weak. 

\begin{figure}[here]
{}\centerline{\hfill\includegraphics[height=.4\textwidth]{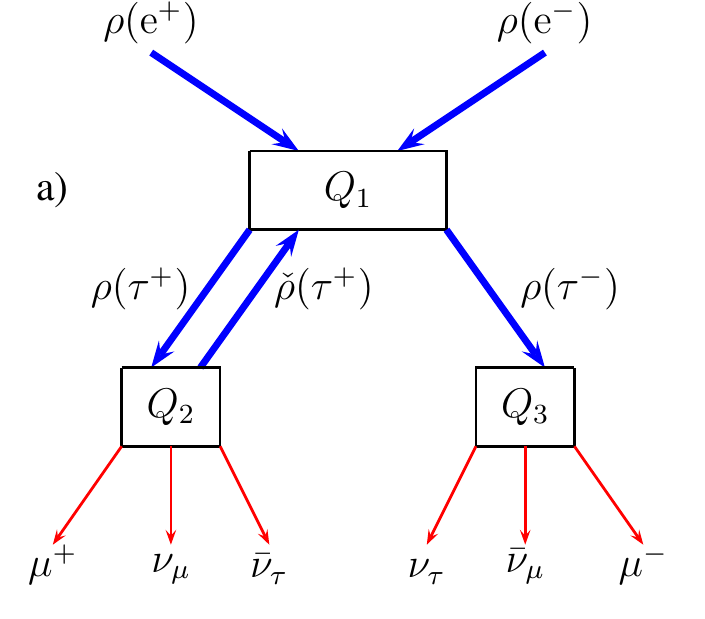}
{}\hfill \includegraphics[height=.4\textwidth]{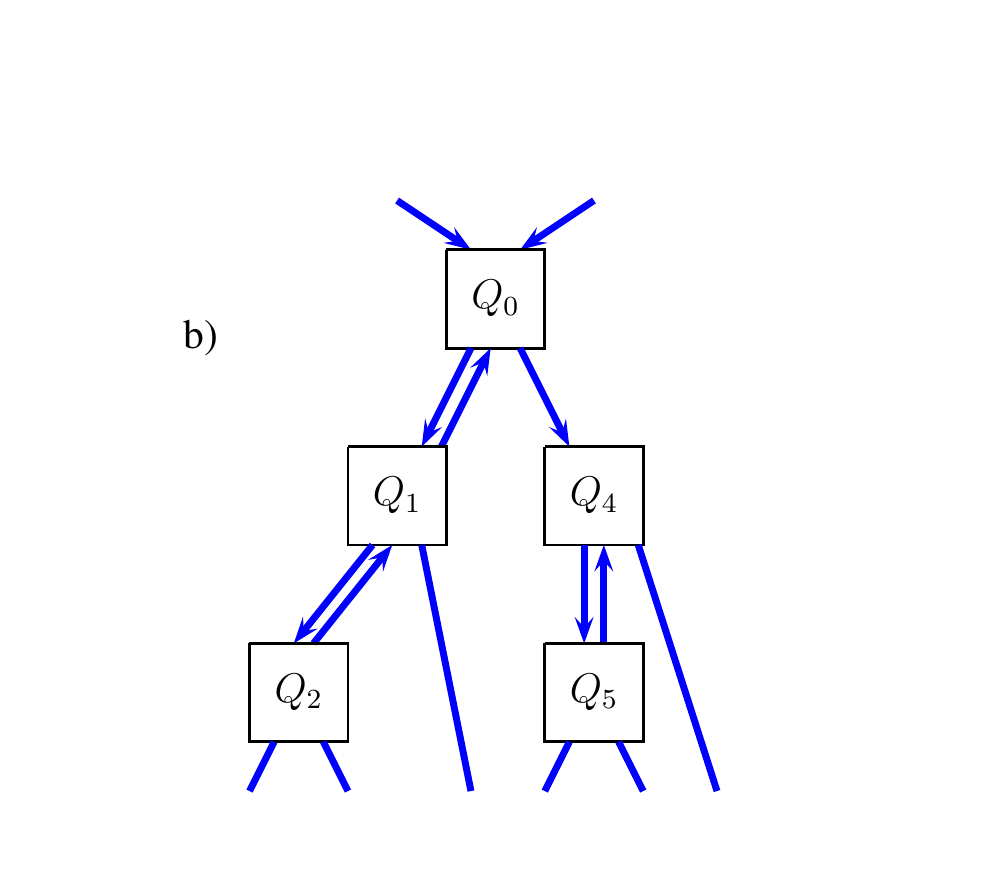}\hfill}
\caption{\label{furth:fig:XXX}
Successive steps of the Monte-Carlo algorithm generating momentum 
correlations in cascade reactions. a) reactions 
$\ep\e\to\tau^+\tau^-$, $\ \tau^+\to\mu^+\nu_\mu\bar\nu_\tau$ and 
$\tau^-\to \mu^-\bar\nu_\mu\nu_\tau$; b) a more complicated tree 
diagram. Downward arrows represent density matrices, upward arrows 
represent acceptance matrices. }
\end{figure}

A correct Monte-Carlo algorithm, taking entanglement into account, has been proposed by Collins and Knowles \cite{Collins:1987cp,Knowles:1987cu} (see also \cite{Richardson:2001df}). Here we show how it applies to the reaction (\ref{furth:eq:cascade}).
To treat the spin in the various subprocesses of (\ref{furth:eq:cascade}), it is convenient to use the partially transposed cross section matrix $\Rt$ defined in (\ref{basic:eq:CSdM}) or a reduced form obtained by partial trace, like in (\ref{basic:eq:reducedAC}).
%Equations (\ref{basic:eq:dpcsR}) 
More particularly, one needs the expressions of the matrices
$\Rt_1(\theta,\varphi)$, $\Rt_2(p_\tau,p_\mu)$, and 
$\Rt_3(p_\tau^-,p_\mu^-)$ of the subprocesses (1-3) in function of 
their kinematical variables. Since one is only interested in the muon 
momenta, the summation is made over the final polarisations in $\Rt_2$ 
and $\Rt_3$ and the integration is made over the (unobservable) 
$\nu_\mu-\bar\nu_\tau$ and $\bar\nu_\mu-\nu_\tau$ relative momenta. The initial electrons may be polarised, described by density matrices $\rho(\ep)$ and $\rho(\e)$.
To lighten the equations,the density matrices will not be explicitly normalised. 
The event generation for the full reaction (\ref{furth:eq:cascade}) 
proceeds in the following steps:
\begin{itemize}
%\item[$-$]
%
\item[$-$]
generate the momenta $p_\tau^+$ and $p_\tau^-$ (\ie, $\theta$ and $\varphi$) according to the 
differential cross section
\be
{d\sigma\over d\Omega}(\overrightarrow{\ep}\,\overrightarrow{\e}\to\tau^+\tau^-) =
\langle a|\rho(\ep)|a'\rangle\,\langle b|\rho(\e)|b'\rangle\,
\langle a',b';c,d|\Rt_1(\theta,\varphi)| a,b,c,d \rangle~
%\Rt(\vec e\vec e^-\to\vec \tau^+\vec \tau^-)
\ee
(summation over repeated indices is understood),
\item[$-$]
keeping these momenta, calculate the (emittance-type) density matrix 
of $\tau^+$
\be
\langle c|\rho( \tau^+)|c'\rangle \propto
\langle a|\rho(\ep)|a'\rangle\,\langle b|\rho(\e)|b'\rangle\,
\langle a',b';c,d|\Rt_1(\theta,\varphi)| a,b,c',d \rangle~,
\ee
\item[$-$]
generate the $\mu^+$ momentum according to the decay distribution
\be
\langle c |\rho( \tau^+)| c' \rangle
\langle c'|\Rt_2(p_\tau^+,p_\mu^+)|c \rangle ~,
%\Rt(\vec\tau\to\mu\nu_\mu\bar\nu_\tau)
\ee
\item[$-$]
keeping this momentum, calculate the $\tau^+$ \emph{acceptance} matrix
\be
\langle c' |\check\rho( \tau^+)| c \rangle\propto\langle 
c'|\Rt_2(p_\tau^+,p_\mu^+)|c\rangle~,
\ee
\item[$-$]
going back to the first reaction, calculate the (emittance-type) 
density matrix of $\tau^-$
\be
\langle d|\rho( \tau^-)| d' \rangle\propto
\langle a|\rho(\ep)|a'\rangle\,\langle b|\rho(\e)|b'\rangle
\langle c' |\check\rho( \tau^+)| c \rangle
\langle a',b';c,d|\Rt_1(\theta,\varphi)| a,b,c',d' \rangle~,
\ee
\item[$-$]
generate the  $\mu^-$ momentum according to the decay distribution
\be
\langle d |\rho( \tau^-)| d' \rangle
\langle d'|\Rt_3(p_\tau^-,p_\mu^-)|d \rangle\,.
\ee
\end{itemize}
The different steps of the simulation are schematised in 
Fig.~\ref{furth:fig:XXX}-a).

This method can be generalised to any process described by tree 
graphs: cascade decays, QCD jets, electromagnetic showers, etc.,
when one needs the momentum or spin correlations between particles 
which are on different branches of the tree. This is illustrated in 
Fig.~\ref{furth:fig:XXX}-b).
\subsection{Quantum information aspects}
The present report brings out the role of \emph{vector of information} played by the spin. Polarised scattering experiments make use of this vector to extract the information on
\begin{itemize}
	\item a physical state, \eg, the internal structure of 
nucleons or the spin-parity of a new particle. The information is 
encoded in a density matrix $\rho$.
	\item an interaction mechanism, \eg, a Reggeon exchange or a 
non-minimal coupling. The information is encoded in a cross section 
matrix $\R$.
\end{itemize}
In this subsection we give some qualitative and quantitive laws 
governing the efficiency of the polarised  experiment in extracting 
the desired information, taking into account its quantum aspects 
which have been briefly presented in Secs.~\ref{basic:sub:dens} and \ref{basic:sub:several}.
\subsubsection{Various kinds of entropy}
We recall that a \emph{pure state} of a quantum system is a 
statistical state on which one has the maximum information. Its 
density matrix is of the form $\rho=|\psi\rangle\langle \psi|$ and 
has rank 1. The opposite case (no information at all) is given by 
the matrix $\rho_0 = \mathbbm{1}_N /N$ of rank $N$, where 
$\mathbbm{1}_N$ is the unit $N\times N$ matrix and $N$ the number of 
basic states. Between these two cases, one can quantify the lack of 
information about the system by various kinds of \emph{entropy} (see, 
for instance, \cite{Wehrl:1978}), among which the Shannon entropy
\be\label{furth:eq:Shannon}     E^{(1)}(\rho) = - \Tr \{\rho \log \rho\} ~, \ee
and the R\'enyi entropy of order $q$
\be\label{furth:eq:Renyi}  E^{(q)} (\rho) = \frac{\ln \Tr 
\rho^q}{1-q} \qquad (0\le q\ne 1)~, \ee
which converges to the Shannon entropy for $q\to1$. In the limiting 
case $q=0$ and $q=+\infty$, we have respectively $E^{(0)}(\rho) = 
\ln[{\rm rank}(\rho)]$ and $E^{(\infty)}(\rho) %-\ln({\rm Sup}_i,\{w_i\})$,
=-\ln(w_{max})$, where $w_{max}$ is the largest eigenvalue of $\rho$. 
Roughly speaking, for small $q$, $E^{(q)}(\rho)$ is an ``egalitarian'' 
measure of entropy (all nonzero eigenvalues count more or less 
equally), whereas for large $q$, $E^{(q)}(\rho)$ is an ``elitist'' 
measure (only the largest eigenvalues count).

The Shannon entropy has the following properties
\begin{enumerate}
	\item
	$0\le E(\rho) \le \ln N$. The left equality is obtained  for 
pure states and the right one for $\rho=\rho_0$.
	\item
	For a mixing $\rho=\lambda\rho_1+\mu\rho_2$ with 
$0\le\lambda=1-\mu\le1$, one has\par
	$\lambda E_1+\mu E_2\le E\le \lambda E_1+\mu E_2+E_{\rm mixing}$, 
where $E_i\equiv E(\rho_i)$ and 
$\rho_{\rm mixing}=\begin{pmatrix}\lambda&0\\   0&\mu \end{pmatrix}$.
	\item
	Removing the non-diagonal elements of a density matrix 
increases its entropy.
	\item
	For a composite system $C=A+B$, one has the ``triangle 
inequality'' $|E_A-E_B|\le E_C\le E_A+E_B$, where $E_i\equiv 
E(\rho_i)$ and $\rho_A$ and $\rho_B$ are the individual density 
matrices of $A$ and $B$, given by the partial traces 
$\rho_A=\Tr_B(\rho_C)$ and $\rho_B=\Tr_A(\rho_C)$.
	\item
	$E_C=E_A+E_B$ if $A$ and $B$ are uncorrelated, \ie, $\rho_C 
= \rho_A \otimes \rho_B$.
   \item
   If $C$ is a \emph{separable pure} state, \ie, $ 
\rho_C=|\psi\rangle\langle\psi|$ with 
$|\psi\rangle=|\psi_A\rangle\otimes|\psi_B\rangle$, then $E_A=E_B=0$.
   \item
   If $C$ is an \emph{entangled pure} state, $E_A=E_B>0$.
\end{enumerate}
For R\'enyi entropies with $q\ne1$, properties 1,5,6,7 are still valid. The mixing property 2 is replaced by the weaker one
\be\label{furth:eq:convex-entrop}
E\ge\inf\{E_1,E_2\}\,,
\ee
which implies that the density matrices of entropy greater than some 
fixed value $E_m$ form a convex subset, containing $\rho_0$,  of the 
positivity domain. More properties are given in \cite{633557}.

Other types of entropies can be defined. We will generalise the name of 
entropy to any symmetric functions of the eigenvalues of $\rho$ 
satisfying (\ref{furth:eq:convex-entrop}), \ie, which can only 
increase upon mixing. This is the case for $\Delta_2$, $\Delta_3$,... 
$\Delta_n$ of \ref{basic:eq:gendens1} and for the distance to the boundary of the positivity 
domain. The various entropies are not independent, since they are 
functions of the $n$ quantities $\Delta_i$'s.

\subsubsection{Entropy of the cross section matrix}
Let us consider, for definiteness, the scattering experiment $A+B\to C+D\ (+X)$, where $X$ is the set of non-analysed or undetected particles. The initial spins are generally uncorrelated: $\rho_{A+B}=\rho_A\otimes\rho_B$, therefore the initial entropy is 
$E_{A+B}=E_A+E_B$ and its ranks is $r_{A+B}=r_A\,r_B$.
%(with $\rho(A)\equiv\rho_A$, $E(A)\equiv E_A$, $r(A)\equiv r_A$, etc.).

Let us first treat the completely polarised exclusive experiment (no $X$). If the initial particles are fully polarised, both initial and final states are pure: 
$\rho_{A+B}=|\Psi_{A+B}\rangle \langle \Psi_{A+B}|$ with $|\Psi_{A+B}\rangle=|\Psi_{A}\rangle\otimes|\Psi_{B}\rangle$ and
$\rho_{C+D}=|\Psi_{C+D}\rangle \langle \Psi_{C+D}|$
with $|\Psi_{C+D}\rangle =\Mcal|\Psi_{A+B}\rangle $ and $r_{A+B}=r_{C+D}=1$. However, $C$ and $D$, taken separately, are only partially polarised if 
$|\Psi_{C+D}\rangle $ is an \emph{entangled} state, which is the most general case. If $A$ and $B$
are partially polarised (mixed $A+B$\/ state, $r_{A+B}>1$), the $C+D$ state can be pure or mixed with $r_{C+D}\le \min\{r_{A+B},r_\Mcal\}$, $r_\Mcal$ being the rank of the transition matrix. Examples of $r_{C+D}<r_{A+B}$ are $\ep\e\to\nu\bar\nu$ and $\gamma\gamma\to\pi^0\to \ep\e$. In these two cases $r_\Mcal=1$ and $r_{C+D}=1$, even if the initial particles are unpolarised ($r_{A+B}=4$).

Let us now consider the inclusive case ($X\ne\O$). 
Instead of $\Mcal$ we use the inclusive cross section matrix $\R$ defined in Sec.~\ref{basic:sub:redCSM} or its partial transpose $\Rt$.   %(\ref{basic:eq:CSdM}),(\ref{basic:eq:reducedAC}), 
%Equations (\ref{basic:eq:dpcsR}) 
The density matrix of $C+D$ as a function of $\rho_{A+B}$ is given, according to \ref{basic:eq:dpcsR}, by
\be\label{furth:eq:rhoB=Tr}
\rho_{C+D}(\rho_{A+B})=\Tr_{A,B}(\Rt\,\rho_{A+B})/\Tr(\Rt\,\rho_{A+B})\,.
\ee
One can introduce the entropy $E(\R)$ of the CSM, replacing $\rho$ in (\ref{furth:eq:Shannon}-\ref{furth:eq:Renyi}) by the unit trace matrix $\R/\Tr\R$. This entropy comes from the leakage of information toward $X$. We may call it \emph{entropy of inclusiveness}%
\footnote{For a \emph{reduced} CSM like $\R_{A,C+D}$, part of this entropy also comes from the unpolarised $B$ state.
}. 
The rank $r(\R)$ of $\R$ cannot exceed the dimension $d(X)$ of the Hilbert space of $X$ (which is infinite if $X$ contains several particles, due to their continuous relative momenta). The former case of completely polarised exclusive reaction can be treated as a special case with $X=\O$, $\ r(\R)=d(X)=1$ and $E(\R)=0$.
It is to be expected that the CSM of an inclusive reaction with high 
missing mass has a large entropy, manifested in spin observables being well 
below the positivity bounds. This should apply to the Soffer bound at small $x$, corresponding to a high mass of the spectator system, as mentionned in Sec.~\ref{incl:sub:kt}.
%At medium or large $x$ one cannot at present exclude a $\delta q(x)$ close to the Soffer bound. In fact this bound is saturated when the spectator partons form a spin-zero diquark \cite{Artru:2002pu}.

\subsubsection{Factorisation, separability or entanglement of the cross section matrix}
The CSM may be \emph{factorised} in an initial state acceptance matrix and a final state emittance matrix (transposed): $\R=\acc_{A+B}\otimes\rho^t_{C+D}$. This is the case when the reaction is strongly dominated by a spin-zero exchange in the $s$-channel, as in $\gamma\gamma\to\pi^0\to \ep\e$. The factorisation of $\R$ comes from the factorisation of the amplitude $ {\cal M}=|\Psi\rangle\langle\Phi|$. No spin information is transmitted between the initial and the final state. More generally, the CSM may be ``initial-final separable'':
%An example of separable CSM is provided by the incoherent superposition of several exchanges of this type. otherwise \emph{entangled}.
%
\be\label{R-sep} %\acc^{(k)}_A\otimes\left(\rho_B^{(k)}\right)^t\,, 
\R=\sum_n w_n\,\acc_{n,A+B}\otimes\rho^t_{n,C+D} \,,\quad w_n>0 \,,
\ee
in analogy with Eq.~(\ref{basic:eq:sep}).
%, but with $w_n$ absorbed in $\acc_{n,A+B}$ or $\rho_{n,C+D}$. 
Then $\Rt=\sum w_n\,\acc_{n,A+B}\otimes\rho_{n,C+D}$ is also positive and separable and the final density matrix is given in terms of the initial one by
\be
\rho_{C+D}(\rho_{A+B})=\sum w_n\,\Tr(\rho_{n,A+B}\,\acc_{n,A+B})\,\rho_{n,C+D} \,.
\ee
%
%$\rho_B(\rho_A)=\sum_n \Tr(\acc_{A,n}\rho_A)\linebreak[2]{\rho_{B,n}}$.  
If the form (\ref{R-sep}) is not possible, the CSM is ``initial-final entangled''. For exclusive reactions, if the amplitude is a sum of factorised terms, ${\cal M}= \sum|\Psi_{n}\rangle\langle\Phi_{n}|$, the interference between these terms gives rise to entanglement between initial and final particles. For inclusive reactions the interference may be washed out when summing over the $X$ states, giving an ``initial-final separable'' CSM.

Factorisation, separability or entanglement of the CSM may also refer to partitions of the external particles other than initial versus final ones. Many  \emph{peripheral} reactions, like single or double diffraction dissociation, are dominated by the exchange of a particle, Reggeon or Pomeron, here denoted $Q$, in the $t$-channel. They can be decomposed, for instance, in the subreactions $A+Q\to C$, $X'$ or $C+X'$ and $B+\bar Q\to D$, $X''$ or $D+X''$. 
%or in $A+Q\to C(+X')$ and $B+\bar Q\to X''$. 
Besides, even if $Q$ has nonzero spin, it may be dominantly exchanged in only one spin state, therefore effectively carry no spin information. Examples are the electromagnetic dissociation of relativistic nuclei, which mainly occurs through $S_z=0$ photon exchange, and the reaction $\ep\e\to W^+W^-$ at $s\gg t, m_W^2$, where neutrino exchange is dominating. For these reactions the amplitude and the CSM are factorised as follows
\be
{\cal M} = {\cal M}_A \, F(Q^2)\, {\cal M}_B \,,\qquad \R=|F(Q^2)|^2\ \R_{A,C} \otimes \R_{B,D}\,.
\ee
It means no $A-B$ or $C-D$ spin correlation, no $A\to D$ or $B\to C$ spin transfer and  positivity yields separate conditions for the two sub-reactions. If $X'$ is empty, the inequalities for $A+Q\to C$ are saturated. 

%
%(the $Q$ propagator being absorbed in ${\cal M}_1$ or ${\cal M}_2$). 

% \cite{Peres:1996dw} 
%
\subsubsection{Entropy of the observed final state}
The entropy $E_{C+D}$ of the analysed particles comes partly from $A$ and $B$,
partly from $\R$. One can draw the more or less qualitative laws
\begin{itemize}
\item
  $E_{C+D}$  is a non-decreasing function of $E_A$, $E_B$ and $E_\R$.
%$\ \partial E_B /\partial E_{A}\ge0$ and $\partial E_B /\partial E_{R}\ge0$.
\item
if $E_A=E_B=E_\R=0$, then $E_{C+D}=0$ (case of an exclusive reaction with 
perfect initial polarisation).
\item
  if $\R$ is ``initial-final factorised'' (Eq.~(\ref{R-sep}) with only one term)  
%$R=R_A\otimes R_B $\rho_B=R_B^t/\Tr R_B$,
then $\rho_{C+D}$ is independent of $\rho_A$ and $\rho_B$ and one has $E_{C+D}=E_\R$. An example is $\ppb\to\eta_c\to \K^{*0}+\Kb{}^{*0}+X$. 
%and $\partial E_B /\partial E_{A}=0$. This is the case in $\gamma\gamma\to\pi^0\to \ep\e$; the $\pi^0$ cannot transmit any spin information.
\item
if $R$ is \emph{pure} (no $X$) and ``initial-final factorised'': $R\propto(|\Phi\rangle\langle\Phi|)_i\otimes\left(|\Psi\rangle\langle\Psi||\right)_f^t $, then $\rho_{C+D}=|\Psi\rangle\langle\Psi|$ is pure. This is the case of $\ppb \to\eta_c\to \K^{*}+\Kb{}^{*}$.
\item
if $R$ is \emph{pure} and ``initial-final entangled'', $E_{C+D}$ may be nonzero. For 
unpolarised $A$ and $B$ % \ie, $\rho_{A+B}=\mathbbm{1}/(n_An_B)$, 
$E_{C+D}$ is a measure of the entanglement of $R$.
\end{itemize}
\subsubsection{Transfer of information from initial to final state} 
%\bar{\cal B}
%
Let us focus on the sensitivity of $\rho_{C+D}$ to $\rho_{A+B}$. To simplify, we will ignore the spins of $B$ and $D$. % and here $\R$ stands for the reduced CSM $\R_{A,C}$. 
We will also use the shorter notations $\R$ for the reduced CSM $\R_{A,C}$, $\mathcal{A}\equiv\rho_A$ and $\mathcal{C}=\Tr_A\{\Rt A\}$ (wherefrom $\rho_C=\mathcal{C}/\Tr \mathcal{C}$). The matrix $\mathcal{C}$ contains the same spin 
information as $\rho_C$ and in addition it gives the initially polarised cross section 
$\sigma(\rho_A)\propto\Tr\mathcal{C}$. To express the linear dependence of 
$\mathcal{C}$ on $\mathcal{A}$, we unwind these matrices into column vectors of dimension $n_A^2$ and $n_C^2$:
\be\label{unwind}
  \mathcal{A}\to\vec{A}=(A_{11},A_{12}, ... A_{1n_A},A_{21},A_{22},...)^t \,
  \ee
and similarly for $C\to\mathcal{C}$. One thus has two equivalent writings: 
\be\label{equivalent writing}\begin{aligned}
\mathcal{C}&=\Tr_A(\Rt\mathcal{A}) &&\rightarrow& \langle j|\mathcal{C}|j'\rangle &=        \langle i'j|\Rt|ij'\rangle \ \langle i|\mathcal{A}|i'\rangle \,, \\
{\vec C}&=J\vec{A} &&\rightarrow& \langle jj'|{\vec C}\rangle &=\langle jj'|J|ii'\rangle \	\langle ii'|\vec{A}\rangle \,,	
\end{aligned}\ee
%	{\cal B}^*=K\mathcal{A} \quad&\rightarrow& \langle j'j|{\cal 
%B}^*\rangle &=\langle j'j|K|ii'\rangle \ %	\langle ii'|\mathcal{A}\rangle \,, \cr
%\mathcal{B}_{j'j} = K_{jj';ii'} \ \mathcal{A}_{ii'}
% %\be\langle i|B |i' \rangle  = J_{ii';kk'}\langle k|A|k' \rangle\ee
where we have used the Dirac notation both for the original matrices 
and the unwound ones, \eg,
$\langle ii'|\vec{A}\rangle \equiv A_{ii'}\equiv\langle i|\mathcal{A}|i'\rangle $.
The matrix $J$ is obtained from $\Rt$ by the circular permutation 
$\{i',j,j'\}\to\{j,j',i'\}$. 
%$J=p(\Rt)$. Similarly $B=J\mathcal{A}$, 
%with $\langle jj'|J|ii'\rangle =\langle j'j|K|ii'\rangle $.
%The matrices $J$ and $K$ only differ by permutations of lines and one 
%has $\det(J)=(-1)^{n_B(n_B-1)/2}\,\det(K)$.

The Jacobian $|\partial \vec C/\partial \vec A|=|\det(J)|$ is one possible 
measure of the efficiency of the spin information transfer from $A$ 
to $C$. If it vanishes, $\vec A$ can vary along in some directions without 
changing $\vec C$: this is a loss of information. If $J=0$, the cross 
section vanishes and obviously no information at all can be 
transmitted. If $J$ is a pure rotation, $\det(J)=+1$. If it is a 
rotation times a spin reversal, $\det(J)=-1$ (however this is forbidden, as shown in 2.5.3).

In the case $n_A=n_C=2$, for instance in %${1\over2}\to{1\over2}\ (+X)$
$\vold\p+\p\to\vold\Lambda+X$, one can also use the Cartesian components $X^\mu$ and $Y^\mu$ of $\mathcal{A}$ and $\mathcal{C}$, defined by $\mathcal{A}\equiv\rho_A={1\over2}X^\mu\sigma^\mu_A$,~ $\mathcal{C}\equiv Y^0\rho_C={1\over2}Y^\mu\sigma^\mu_B$. $\vec X$ and $\vec Y/Y^0$ are the polarisation 3-vectors of $A$ and $C$.
The column vector
$\vec A$ introduced in (\ref{unwind}) is related to $X^\mu$ by $\vec A={1\over2}(X^0+X^3,X^1-iX^2,X^1+iX^2,X^0-X^3)^t$, and similarly for $Y^\mu$ and $\vec C$. A third form of (\ref{equivalent writing}) is 
\be
Y^\nu=(\Tr\R/2)\ X^\mu D_{\mu\nu}\,,\quad\hbox{with}\quad
\det \left(D_{\mu\nu}\right)=\left(\Tr R/2\right)^{-4}\,\det J\,.
\ee
$D_{\mu\nu}=(\mu 0|\nu 0)$ generalises  the ``depolarisation parameters'' of  Sec.~\ref{incl:sub:spintrans}. An interesting relation,
\be\label{furth:eq:3determinants}
\det(\R)-\det(J)-\det(\Rt)=0\,,
\ee
links three different information parameters of the reaction:
\begin{itemize}
\item
$\det(\R)$, which has to be positive, is a measure of the  ``entropy 
of inclusiveness",
\item
$\det(J)$ is an algebraic measure of the spin transfer efficiency,
\item %=-\det(p(R))
$-\det(\Rt)$ is a measure of the ``entanglement in the $A\bar C$ channel".
\end{itemize} %=\det(p\circ p(R))
For instance $\det(\Rt)<0$ implies that $\R$ is entangled and that 
$\det(J)>0$.  A simple example is 
the scattering by a spin-independent potential: 
$\langle j|\Mcal|i\rangle =a\delta_{ij}$. Then
\be
{\R\over|a|^2}=\begin{pmatrix}  1&0&0&1 \\   0&0&0&0 \\  0&0&0&0 \\  1&0&0&1
\end{pmatrix}\,,\quad 
{J\over|a|^2}=\mathbbm{1}_4=D\,,\quad
{\Rt\over|a|^2}={\mathbbm{1}+\vec\sigma_A.\vec\sigma_B\over2}=
\begin{pmatrix}1&0&0&0 \\  0&0&1&0 \\ 0&1&0&0 \\  0&0&0&1\end{pmatrix}\,,
\ee
which gives $\det(\R)=0$, $\det(J)=|a|^8$, $\det(\Rt)=-|a|^8$. The 
negative value of $\det(\Rt)$ means that $\R$ is entangled. In fact 
$\langle j|\Mcal|i\rangle\propto\delta_{ij}$ corresponds, in the $t$-channel, to the 
exchange of the spin-singlet, which is an entangled state. A relation similar to (\ref{furth:eq:3determinants}) can be written for the correlations between two initial or two final spin one-half:
\be
\det\rho_{AB}+\left[\Tr(\rho_{AB})/2\right]^4\,\det \left(C_{\mu\nu}\right)-\det \rho^{\rm pt}_{AB}=0~.
\ee
 It would be interesting to find an analogous relation for higher spins. These equations, indeed, suggest some general relation between the entanglement of $\R$, the entropy of inclusiveness and the strength of the correlation between $A$ and $B$ or $C$. 
\subsection{Domains of quantum positivity, classical positivity and 
separability}\label{furth:sub:quantum}
\begin{figure}[here]
\centerline{\includegraphics[width=.6\textwidth]{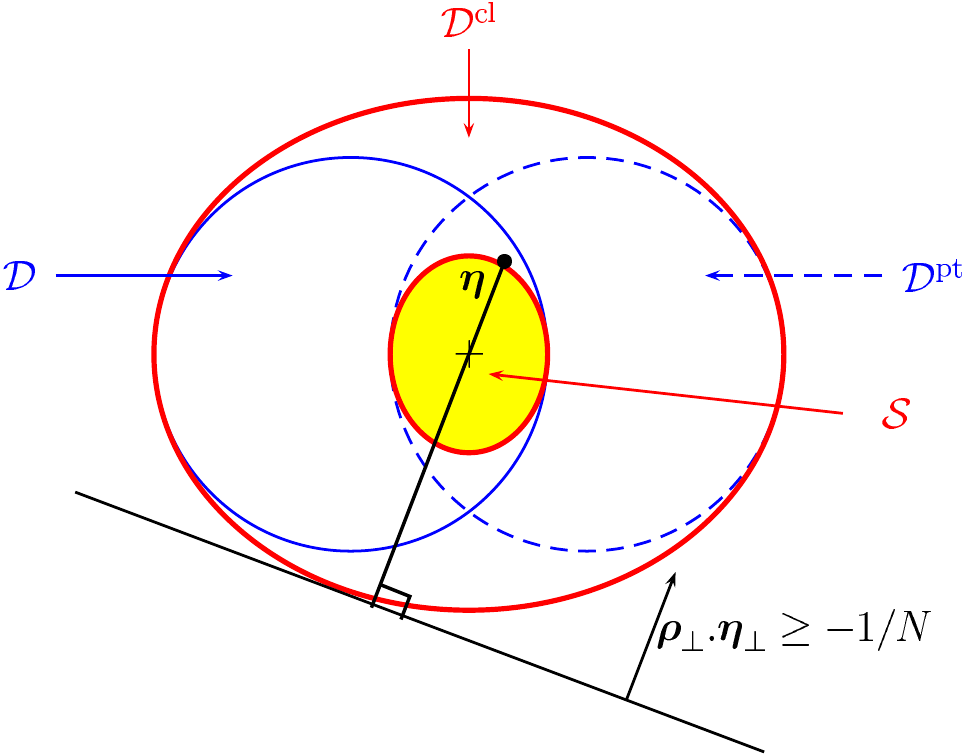}}
\caption{\label{furth:fig:YYY}%
Relative positions of the classical positivity domain $\mathcal{D}^{\rm cl}$: 
the separability domain  $\mathcal{S}$, 
the true positivity domain $\mathcal{D}$ 
and the domain of positive partial transform 
$\mathcal{D}^{\rm pt}$ (dashed contour). 
$\mathcal{D}\cap\mathcal{D}^{\rm pt}$ is generally larger than 
$\mathcal{S}$, but for two spin one-half, one has 
$\mathcal{D}\cap\mathcal{D}^{\rm pt}=\mathcal{S}$.
A matrix $\eta$ of the boundary $\partial\mathcal{S}$ is represented 
together with its reciprocal polar line, which is tangent to 
$\partial\mathcal{D}^{\rm cl}$. }
\end{figure}

As we have seen in Sec.~\ref{se:basic}, the cross section 
(\ref{basic:eq:cs4pol}) of the reaction (\ref{basic:eq:reac}) has to 
be positive for arbitrary \emph{independent} $\Sv_A$, $\Sv_B$, 
$\Sv_C$ and $\Sv_D$ in the unit ball $\Sv^2\le1$, but this 
\emph{classical} condition is not a sufficient condition for the 
positivity of the cross section matrix. On can thus distinguish a 
\emph{classical positivity domain} $\mathcal{D}^{\rm cl}$, which is larger than the true or 
\emph{quantum} positivity domain $\mathcal{D}$. To illustrate this fact, let us 
consider the final density matrix $\rho_{C+D}$ of a reaction $A+B\to 
C+D+X_f$. Classical positivity reads $\Tr\{\rho_{C+D}\,(\acc_C\otimes\acc_D)\}\ge0$ for any acceptance matrices $\acc_C$ and $\acc_D$ of the detectors. More generally
\be\label{furth:eq:eq:classic+}
\Tr \{\rho_{C+D}\, \acc_{C+D}\}\ge0 \quad \hbox{for any 
\emph{separable}} \ \acc_{C+D}\,,
\ee
whereas quantum positivity requires
\be\label{furth:eq:quant+}
\Tr \{\rho_{C+D}\,\acc_{C+D}\}\ge0 \quad \hbox{for any separable 
\emph{or entangled}} \ \acc_{C+D}\,.
\ee
One can say that the classical positivity domain $\mathcal{D}^{\rm cl}$ of $\rho_{C+D}$ is 
\emph{dual} to the separability domain $\mathcal{S}$ of $\acc_{C+D}$, in the sense that $\Tr \{\rho\,\eta\}\ge0$ for any pair $\{\rho\in\mathcal{D}^{\rm cl}\,,\eta\in\mathcal{S}\}$. 
We recall that the quantum positivity domain $\mathcal{D}$ is dual to itself, according to (\ref{basic:eq:polar-recip}). 
%\cite{Minnaert:1971}. 
We have
\be\label{furth:eq:SDC}
\mathcal{S}\subseteq\mathcal{D}\subseteq\mathcal{D}^{\rm cl}~,
\qquad
\mathcal{S}\subseteq\mathcal{D}^{\rm pt}\,\subseteq\mathcal{D}^{\rm cl}~,
\ee
where $\mathcal{D}^{\rm pt}$ contains the matrices $\rho$ whose partial 
transform $\rho^{\rm pt}$ are semi-positive. Separability and classical 
positivity are preserved under partial transposition:  $\mathcal{S}^{\rm pt}=\mathcal{S}$,
$(\mathcal{D}^{\rm cl})^{\rm pt}=\mathcal{D}^{\rm cl}$. All the domains 
in (\ref{furth:eq:SDC}) are convex. Figure \ref{furth:fig:YYY} schematises their respective extensions.
%----------FIGURE:

\begin{figure}[here]
{}\centerline{\includegraphics[width=.6\textwidth]{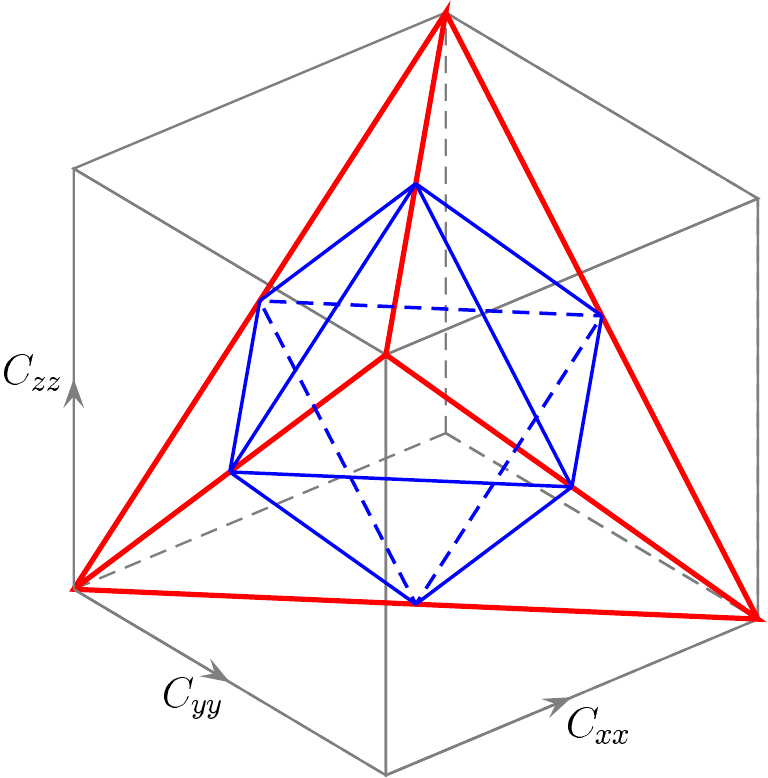}}
\caption{\label{furth:fig:ZZZ}%
Classical positivity domain (cube), true positivity domain 
(tetrahedron) and separability domain (octahedron) for the triple 
$\{C_{xx}\,,C_{yy}\,,C_{zz}\}$ of observables. }
\end{figure}

%Using the decomposition $\rho=\rho_\parallel+\rho_\perp$ and generalising (\ref{basic:eq:rho-dec}) as $\vec\rho_\perp.\vec\eta_\perp=\Tr\{\rho_\perp\,\eta_\perp\}$ 
In analogy with (\ref{basic:eq:polar-recip}), we can write the duality between $\mathcal{D}^{\rm cl}$ and $\mathcal{S}$ as
\be\label{furth:eq:polar-recip} %(\ref{eq:polar-recip})
\vec\rho_\perp.\vec\eta_\perp\ge-1/N\quad\hbox{for all}\quad\rho\in\mathcal{D}^{\rm cl}\,,\ 
\eta\in\mathcal{S}\,,
\ee
after normalisation of $\rho$ and $\eta$ to unit trace.
This implies that the boundaries $\partial\mathcal{D}^{\rm cl}$ and 
$\partial\mathcal{S}$ of the two domains are \emph{polar reciprocal} 
of each other, 
% when $\vec\eta_\perp$ moves on $\partial\mathcal{S}$, the reciprocal plane in $\vec\rho_\perp$ space defined by $\vec\rho_\perp.\vec\eta_\perp=-1/N$ envelops $\partial\mathcal{D}^{\rm cl}$, 
as shown in Fig.~\ref{furth:fig:YYY}. 
This reciprocity is also visible on a subset of observables, in the following form: let $P$ be the hyperplane defined by the vanishing of all other observables, 
% $P\cap\mathcal{D}^{\rm cl}$, $\ P\cap\mathcal{S}$ its intersections with 
$P(\mathcal{D}^{\rm cl})$ and $P(\mathcal{S})$ denote the projections of $\mathcal{D}^{\rm cl}$ and $\mathcal{S}$ on $P$. In analogy with (\ref{basic:eq:recipro-proj-inter}) we have
% P\cap\mathcal{D} contains the centre $\vec\rho_\perp=0$. 
%
\be\label{furth:eq:recipro-projS-interC} %(\ref{eq:polar-recip})
\vec\rho_\perp.\vec\eta_\perp\ge-1/N
\quad\hbox{for all}\quad %\,,\quad
\rho\in P\cap\mathcal{D}^{\rm cl}\,,\ \eta\in P(\mathcal{S}) 
% P\cap\mathcal{D}^{\rm cl}
\quad{\rm or}\quad
\eta\in P(\mathcal{D}^{\rm cl})\,,\ \eta\in P\cap\mathcal{S}\,.
%\ \eta\in P(\mathcal{D}^{\rm cl})\,.
\ee
For instance, for a two-fermion system of density matrix 
$\rho_{C+D}={1\over4}C_{\mu\nu}\,\sigma_\mu(C)\otimes\sigma_\nu(D)$, 
the classical positivity domain of the triple 
$\{C_{xx}\,,C_{yy}\,,C_{zz}\}$ is the whole cube $[-1,+1]^2$, the 
quantum positivity domain is the tetrahedron defined by
\be\label{furth:eq:xxx}
C_{xx}-C_{yy}-C_{zz}\le1\quad\hbox{and circular permutations,}\quad 
C_{xx}+C_{yy}+C_{zz}\le1\,,
\ee
and the separability domain, an octahedron, is the intersection of 
the tetrahedron with its mirror figure. One can see on 
Fig.~\ref{furth:fig:ZZZ} the polar reciprocity (edge 
$\leftrightarrow$ edge) and (summit $\leftrightarrow$ face) between 
the cube and the octahedron. Here %In the case of Fig.~\ref{furth:fig:ZZZ} 
the full domains $\mathcal{S}$, $\mathcal{D}$ and $\mathcal{D}^{\rm cl}$ are symmetrical about $P$, therefore their intersections with $P$ coincide with their projections on $P$.
%projections coincide with their intersections with $P$. 
%This explains why the dualities $\mathcal{S}\leftrightarrow\mathcal{D}^{\rm cl}$ and $\mathcal{D}\leftrightarrow\mathcal{D}$ are still visible after projection.

%
\clearpage\setcounter{equation}{0}\section{Conclusions and outlook}\label{se:outl}
\paragraph{Spin dynamics and inequalities}
In this review, we have discussed several aspects of spin physics, as studied in exclusive hadronic reactions, lepton-induced reactions, inclusive reactions and in particular spin-dependent structure functions and analogous aspects in generalised parton distributions and BFKL evolution.
%It is unfortunately not possible to cover all aspects. Our aim was not to study how to reconstruct the amplitudes from a selected set of observables

The main focus of this survey is on the inequalities relating  spin observables. When suitably projected, these inequalities delimit the domain allowed for two, three or more observables.
In the case of a pair, the domain is a disk or a triangle inside the unit square, but their are many cases where the entire square is accessible.  For a triple of observables of a reaction involving less than eight amplitudes, there is always a constraint, and the geometry of the allowed domain can assume a variety of shapes: sphere, cone, pyramid, tetrahedron, octahedron, intersection  of  cylinders, etc.

To list the inequalities, different methods can be used. For a preliminary investigation,  it is possible to generate randomly fictitious amplitudes and to plot an observable against the other ones. Once the domain is identified, its boundary can be derived algebraically from the explicit expression of the observables in terms of a set of independent amplitudes, an exercise that is not always easy.
The positivity of the density matrix of the initial or final state, or of a state relevant for a crossed reaction, gives some physics insight into the origin of these inequalities.

The inequalities reviewed in this article are not relevant for the cases where only one spin observable is measured, for instance  one polarisation or one analysing power. They are also not needed in the rare cases (for instance elastic nucleon--nucleon scattering at low energy), where a reliable amplitude analysis has been carried out which fits well the data points: the inequalities are then automatically satisfied. The optimal circumstances for using these inequalities are those where a few observables are measured,  and some further ones are foreseen: it is possible to check the compatibility of the existing data, the range allowed for the future ones, although a complete amplitude analysis is yet out of reach.

Examples have been chosen to illustrate the formalism. For exclusive reactions, the cases of $\pi \N\to\pi \N$, $\ppb\to \pi\pi$, $\ppLL$, $\gamma \p\to \K \Lambda$ and $\gamma \p\to \rho \N$ have been described in details.  The domains for the spin observables are of course identical for reactions with exactly the same spin structure, such as $\K \N\to \K \N$, whose detailed investigation has been proposed in relation with the tentative pentaquark, and $\pi\N\to\pi\N$. For 
$\pb\p\to \ep\e$, as compared to $\ppLL$, further restrictions applied if the leading mechanism is one-photon exchange in the $s$-channel. Some reactions are related by crossing, and hence are constrained by similar inequalities, once crossing is applied to the observables. This is the case, for instance, for the photoproduction of vector mesons and the photodisintegration of the deuteron.
%For instance, we discuss in Sec.~\ref{excl:sub:photvec} the observables for the photoproduction of vector mesons, and could have chosen as well the photodesintegration of deuteron, which  involves the same types of particles: a photon, two nucleons and a massive vector particle. This latter reaction is studied at Jlab \cite{Jiang:2007ge}.
%Similarly the spin structure of the reaction $\mathrm{e}^+\mathrm{e}^-\to \bar{\mathrm{p}}+\mathrm{p}$ is identical to that or $\ppLL$ which is discussed at lenght in our review. The former reaction gives access to the relative phase between the electric and magnetic form factors of the nucleon \cite{Buttimore:2006mq}.

For inclusive reactions, the constraint on the spin-dependent integrated cross-section has been reminded, \ie,  the inequality relating $\sigma_{\rm tot}$, $\Delta \sigma_T$ and $\Delta\sigma_L$. For reactions $a+b\to c+X$, with $a$, $b$ or $c$ polarised or analysed, there are inequalities for the spin-correlation or spin-transfer observables. Inequalities are also for the structure functions and parton distributions. A king of universality is encountered, with the same shape observed for different exclusive reactions and for the inclusive reactions which have been studied, for instance the triangle $2|y|\le 1+x$ and the tetrahedron $x\pm y\le 1\mp z$ (or its mirror image).

% On the theoretical side, a number of questions mentioned along this review would deserve some deeper investigation, or perhaps, an explanation  somewhat simpler than these available in the literature, including the present review. For instance, what is the maximal degree of the limiting curves and surfaces? In our investigations, this degree is always either 1 or~2. Why are the allowed domain convex? What is the minimal number $n$ of amplitudes needed to envisage that an hypercube $[-1,+1]^m$ becomes fully allowed for $m$ observables? How to distinguish genuinely quantum effects from inequalities  which can be understood at the classical level? 
 
  On the theoretical side, a number of questions have been discussed along this review and clarified, although,  perhaps, alternative  explanations could be envisaged. For instance, what is the maximal degree of the limiting curves and surfaces? In our investigations, this degree is always either 1 or~2. Why are the allowed domain convex? What is the minimal number $n$ of amplitudes needed to envisage that an hypercube $[-1,+1]^m$ becomes fully allowed for $m$ observables? How to distinguish genuinely quantum effects from inequalities  which can be understood at the classical level? 
\paragraph{New polarised beams}
In the near future, new experiments are expected to take data, either at existing accelerators  with improved polarisation devices, or at new facilities designed with polarisation equipment.
Much progress is observed in the art of building and maintaining a high degree of polarisation in targets, see, \eg, \cite{Jeffries:1990dt} for the history and \cite{0034-4885-66-11-R02} for the recent developments. 

Polarised beams of electrons, protons, neutrons or photons are routinely produced, and 
a hot issue at the time of completion of this review is whether intensive polarised beams of positrons or antiprotons can be built. The physics case of polarised positrons at the future ILC (International Linear Collider) is discussed, \eg, in \cite{MoortgatPick:2006qp}. Several devices have been proposed, for instance, the production of high-energy polarised photons by back scattering of an intense electron beam on a polarised laser and pair creation by these photons \cite{Sakaue:2007zr}.

 The problem of polarising antiprotons was debated in the 80's when LEAR came into operation at CERN, but was not given very high priority in the CERN agenda.
Debates were also organized in view of implementing polarisation at the Fermilab collider
\cite{Krisch:1985dg}. The discussion is now
resumed for  studying antiproton-induced reactions at FAIR or JParc.

As polarised protons have been produced and used in the E704 experiment at Fermilab from the weak decay $\Lambda\to\mathrm{p}\pi$, polarised antiprotons have be  obtained in a similar way  \cite{Grosnick:1989qv,Bravar:1996ki}. This method could be further used,  provided a large enough number $\overline{\Lambda}$ can be produced. See, also,  \cite{Krisch:1985dg}, and Refs.\ there.

Another  idea is to polarise an antiproton by nuclear scattering on nuclei, a method known to work in the case of protons. Antiprotons appear, however,  as weakly polarised when scattered off carbon \cite{Klempt:2002ap,Martin:1988wv}. Other nuclei might be more favourable. It is even better to
use inelastic scattering of antiprotons, rather than elastic. The reaction is 
$\bar{\rm p}+ {\rm A}\to \bar{\rm p}+ {\rm A}^*$, where ${\rm A}$ is the nucleus
ground-state and ${\rm A}^*$ one of its excited states, with well-chosen quantum
numbers. As stressed years ago by Dover 
\cite{Dover:1984yy,Dover:1985py,Dover:1987cb}, and more recently by Yu-shun et
al.\ \cite{Yu-shun:1999dc}, such inelastic transition ``filters'' specific
spin--isospin components of the antiproton--nucleon interaction. If, \eg, the
role of the tensor force is enhanced, a good transfer of polarisation from a longitudinally polarised nuclear target
target to the antiproton might be expected. A spin-orbit term in the effective
antiproton--nucleus potential (which might result from the tensor component of
the antinucleon--nucleon potential) should also produce a good transverse polarisation
of the scattered antiproton \cite{Yu-shun:1999dc}. This would deserve further
studies.

A ``spin-splitter'' has been studied \cite{Onel:1986eu,Penzo:1996as,Conte:1996aq}. It is basically a Stern-Gerlach device, where an inhomogeneous magnetic field pushes up the spin states $\sigma_z=+1$ and down those with $\sigma_z=-1$. A test  was even conducted with protons the IUCF \cite{Rossmanith:1990da}, which cannot be considered as fully conclusive.

The  ``spin-filter'' has also been much discussed: unpolarised antiprotons are maintained circulating in a ring where they regularly hit a polarised proton target. Then, if $\Delta \sigma_{\rm T}=\sigma(\uparrow\downarrow)-\sigma(\uparrow\uparrow)$ is large enough (as compared to $\sigma_{\rm tot}$, the beam of remaining antiprotons become more and more polarised. Tests have been made with protons in a dedicated ring at Heidelberg. However, the tests were not conclusive enough at the time when LEAR was shut down. Also, this method might work better with  longitudinal polarisation, as $|\Delta\sigma_L|>|\Delta\sigma_T|$
%%\footnote{If $\Delta \sigma_{\rm L}=\sigma(\rightarrow\leftarrow)-\sigma(\rightarrow\rightarrow)$
%in terms of the spin projections on the beam momentum, or as $\sigma(++)-\sigma(+-)$ in terms of the initial helicities, it will be shown that $\sigma-\Delta\sigma_L\ge 2 |\Delta\sigma_T|$,
% thus in the limit of a very large effect on $\Delta\sigma_T$, $\Delta\sigma_L$ should also be very large.}%
at least in some model calculations, but it is easier to keep transverse than longitudinal polarisation in a ring. 

A potential improvement is to use the polarisation of the electron of the
hydrogen target, or even better to combine optimally the electron and proton
contributions by comparing the results from polarised ortho- or para-hydrogen.

A recent variant relies solely on the (known) electron--antiproton interaction and uses a spin transfer from an electron-polarised hydrogen gas target to an antiproton beam. The estimate for the antiproton beam foreseen at the future GSI facility is a polarisation of the order of 0.2--0.4 \cite{Rathmann:2004pm}. However, another study does not confirm this optimistic result \cite{Milstein:2005bx}.

Studies have even be done of arranging a circulation of the antiproton beam within a parallel beam of polarised electrons, but so far the results are not fully convincing.

Note however, that if antiprotons have so far resisted attempts to make them polarised, it could be easier to produce polarised antineutrons. By the the charge-exchange reaction
\ppnn, antineutron beams have been produced at CERN and Brookhaven and used to study antineutron induced reactions. As a large transfer of polarisation is expected, an  antineutron beam with high longitudinal polarisation can result if  the charge exchange reaction is performed on a longitudinally polarised hydrogen target \cite{Richard:1980rp,Richard:REC.au,Dover:1981pp}.
\par \vskip 1cm
\subsection*{Acknowledgements}
We benefited from many discussions with several colleagues on various aspects of spin physics.   We thank M.~Asghar for useful comments on the manuscript.
The hospitality extended to us for a collaboration meeting at ECT*, Trento, is gratefully acknowledged. 

\newpage
\bibliographystyle{ourbibstyle3}
\bibliography{SpinBib}
\listoftables
\listoffigures
\end{document}